\documentclass[onecolumn,authoryear]{els-mrw} % Release this line for NameDate reference Style

\usepackage{amsmath,amssymb,amsfonts,amsthm,makeidx,graphicx}
\usepackage{txfonts}
\usepackage{hyperref} 
\usepackage{helvet}
\usepackage{wrapfig}
\usepackage[rightcaption]{sidecap}
\usepackage{notoccite} % to force proper ordering of references?!
\usepackage{xcolor}
\setcitestyle{numbers}

\usepackage{subcaption}
\usepackage{comment}
\setcitestyle{square} 
\usepackage{ifthen} 
\newboolean{uprightparticles}
\setboolean{uprightparticles}{false} %Set true for upright particle symbols
\usepackage{xspace}
\usepackage{upgreek}

\def\MagUp {\mbox{\em Mag\kern -0.05em Up}\xspace}

\ifthenelse{\boolean{uprightparticles}}%
{

 \def\PDelta      {\ensuremath{\Delta}\xspace}
 \def\PXi         {\ensuremath{\Xi}\xspace}
 \def\PLambda     {\ensuremath{\Lambda}\xspace}
 \def\PSigma      {\ensuremath{\Sigma}\xspace}
 \def\POmega      {\ensuremath{\Omega}\xspace}
 \def\PUpsilon    {\ensuremath{\Upsilon}\xspace}
 \let\oldPi\Pi
 \def\PPi         {\ensuremath{\oldPi}\xspace}

 \def\PB      {\ensuremath{\mathrm{B}}\xspace}
 \def\PD      {\ensuremath{\mathrm{D}}\xspace}

 \def\PK      {\ensuremath{\mathrm{K}}\xspace}
 \def\Ps      {\ensuremath{\mathrm{s}}\xspace}

 \def\thebaroffset{0.0em}
}
{

 \mathchardef\PDelta="7101
 \mathchardef\PXi="7104
 \mathchardef\PLambda="7103
 \mathchardef\PSigma="7106
 \mathchardef\POmega="710A
 \mathchardef\PUpsilon="7107
 \mathchardef\PPi="7105
 \def\PB      {\ensuremath{B}\xspace}
 \def\PD      {\ensuremath{D}\xspace}

 \def\PK      {\ensuremath{K}\xspace}
 \def\Ps      {\ensuremath{s}\xspace}

 \def\thebaroffset{0.18em}
}
\newcommand{\offsetoverline}[2][\thebaroffset]{\kern #1\overline{\kern -#1 #2}}%

\makeatletter
\ifcase \@ptsize \relax% 10pt
  \newcommand{\miniscule}{\@setfontsize\miniscule{4}{5}}% \tiny: 5/6
\or% 11pt
  \newcommand{\miniscule}{\@setfontsize\miniscule{5}{6}}% \tiny: 6/7
\or% 12pt
  \newcommand{\miniscule}{\@setfontsize\miniscule{5}{6}}% \tiny: 6/7
\fi
\makeatother

\DeclareRobustCommand{\optbar}[1]{\shortstack{{\miniscule (\rule[.5ex]{1.25em}{.18mm})}
  \\ [-.7ex] $#1$}}

\def\squark    {{\ensuremath{\Ps}}\xspace}

\def\KorKbar {\kern \thebaroffset\optbar{\kern -\thebaroffset \PK}{}\xspace}

\def\D       {{\ensuremath{\PD}}\xspace}

\def\DorDbar {\kern \thebaroffset\optbar{\kern -\thebaroffset \PD}\xspace}

\def\Dp      {{\ensuremath{\D^+}}\xspace}
\def\Dm      {{\ensuremath{\D^-}}\xspace}

\def\DpDm    {\ensuremath{\Dp {\kern -0.16em \Dm}}\xspace}

\def\B       {{\ensuremath{\PB}}\xspace}

\def\BorBbar {\kern \thebaroffset\optbar{\kern -\thebaroffset \PB}\xspace}

\def\Bd      {{\ensuremath{\B^0}}\xspace}

\def\BdorBdbar {\kern \thebaroffset\optbar{\kern -\thebaroffset \Bd}\xspace}

\def\Bs      {{\ensuremath{\B^0_\squark}}\xspace}

\def\BsorBsbar {\kern \thebaroffset\optbar{\kern -\thebaroffset \Bs}\xspace}

\def\Y#1S{\ensuremath{\PUpsilon{(#1S)}}\xspace}

\def\LorLbar     {\kern \thebaroffset\optbar{\kern -\thebaroffset \PLambda}\xspace}

\def\to                 {\ensuremath{\rightarrow}\xspace}

\def\AT#1     {\ensuremath{A_{\mathrm{T}}^{#1}}\xspace}           % 2

\def\C#1      {\ensuremath{\mathcal{C}_{#1}}\xspace}                       % 9
\def\Cp#1     {\ensuremath{\mathcal{C}_{#1}^{'}}\xspace}                    % 7
\def\Ceff#1   {\ensuremath{\mathcal{C}_{#1}^{\mathrm{(eff)}}}\xspace}        % 9  
\def\Cpeff#1  {\ensuremath{\mathcal{C}_{#1}^{'\mathrm{(eff)}}}\xspace}       % 7
\def\Ope#1    {\ensuremath{\mathcal{O}_{#1}}\xspace}                       % 2
\def\Opep#1   {\ensuremath{\mathcal{O}_{#1}^{'}}\xspace}                    % 7

\newcommand{\aunit}[1]{\ensuremath{\text{\,#1}}}
\newcommand{\tev}{\aunit{Te\kern -0.1em V}\xspace}
\newcommand{\gev}{\aunit{Ge\kern -0.1em V}\xspace}
\newcommand{\mev}{\aunit{Me\kern -0.1em V}\xspace}
\newcommand{\kev}{\aunit{ke\kern -0.1em V}\xspace}
\newcommand{\ev}{\aunit{e\kern -0.1em V}\xspace}

\newcommand{\mevc}{\ensuremath{\aunit{Me\kern -0.1em V\!/}c}\xspace}
\newcommand{\gevc}{\ensuremath{\aunit{Ge\kern -0.1em V\!/}c}\xspace}
\newcommand{\mevcc}{\ensuremath{\aunit{Me\kern -0.1em V\!/}c^2}\xspace}
\newcommand{\gevcc}{\ensuremath{\aunit{Ge\kern -0.1em V\!/}c^2}\xspace}
\def\gsim{{~\raise.15em\hbox{$>$}\kern-.85em
          \lower.35em\hbox{$\sim$}~}\xspace}
\def\lsim{{~\raise.15em\hbox{$<$}\kern-.85em
          \lower.35em\hbox{$\sim$}~}\xspace}

\def\tell1  {TELL1\xspace}
\def\ukl1   {UKL1\xspace}

\newcommand{\lhcborcid}[1]{\href{https://orcid.org/#1}{\hspace*{0.1em}\raisebox{-0.45ex}{\includegraphics[width=1em]{figs/orcidIcon.pdf}}}}

\begin{document}

%%%%%%%%%%%%%%%%%%%%%%%%%%%%%%%%%%%%%%%%%%%%%%%%%%%%%%%%%%%%%%%%
%% the following items are mandatory: 
%% - title
%% - author names
%% - affiliation details
%% - abstract
%% - keywords

%% Precise, concise, and informative description of the focus of this work. Avoid abbreviations and formulae in the title
%\chapter{Article title (template for all chapters in Section 6: Experiment)}\label{chap1}
\chapter{The LHCb Experiment}\label{chap1}

%% All author names and affiliations, and email address for corresponding author
\author[1,2]{Marcel Merk}%
\author[2]{Niels Tuning}%

%\author[1,2]{Third Author}%

\address[1]{\orgname{University Maastricht}, \orgdiv{Department of Gravitational Waves and Fundamental Physics}, \orgaddress{Duboisdomein 30, 6229 GT Maastricht, NL}}
\address[2]{\orgname{Nikhef NWO-i institute}, \orgdiv{b-Physics Department}, \orgaddress{Science Park 105, 1098 XG Amsterdam, NL}}

\articletag{Chapter Article tagline: update of previous edition, reprint.}

\maketitle

%%%%%%%%%%%%%%%%%%%%%%%%%%%%%%%%%%%%%%%%%%%%%%%%%%%%%%%%%%%%%%%%
%% the following item is mandatory: 
%% 100-150 word summary of the chapter
\clearpage
\begin{abstract}[Abstract]
	%Text of your abstract e.g.: We give a pedagogical introduction...(100-150 words)
    %Text of your abstract e.g.: We give a pedagogical introduction...(100-150 words)

We present an overview including the historical motivation, design principles, and experimental methodology of the LHCb experiment. Originally conceived as a dedicated experiment for CP violation and rare decays in the $B$-meson sector, LHCb evolved into a general-purpose experiment for physics in the forward direction at the LHC, while maintaining its core optimization on flavour physics.
We review the key detector components for both the original LHCb set-up as well as its upgrade, with emphasis on design features that enable efficient reconstruction of forward-region events. Experimental techniques specific to the forward spectrometer are discussed, highlighting how detailed detector information is translated into event-level observables used in physics analyses.
We present an overview of LHCb's major physics results on CP violation, rare decays, spectroscopy, long-lived particles, $W$- and $Z$-boson physics and heavy ion physics. In all cases we focus on the conceptual methods, while referring to the literature for detailed discussions. We end this review by comparing LHCb's performance to other experiments and shortly present the concept for a future, second upgrade of LHCb at the High Luminosity LHC.

\end{abstract}

%% 5-10 words that embody the key topics in the chapter. What terms would someone put into a search engine if they were looking for a chapter like this?
\begin{keywords}
 	%please enter 5 keywords as follows:
 	LHCb \sep Flavour Physics\sep CP-violation \sep Rare Decays \sep Spectroscopy
\end{keywords}

%%%%%%%%%%%%%%%%%%%%%%%%%%%%%%%%%%%%%%%%%%%%%%%%%%%%%%%%%%%%%%%%
%% the following item is optional: 
%% - Single figure visually illustrating the key topic/method/outcome described in the chapter
%% E.g., if possible, please show a pic of the experiment described in the chapter.  
\begin{figure}[h]
	\centering
%	\includegraphics[width=7cm,height=4cm]{blankfig.pdf}
%	\caption{Optional: If possible, please show a pic of the experiment described in the chapter.  
%		     Please add here some text explaining the pic...}
 	%\includegraphics[width=16cm]{Figures/Detector/LHCb-Detector-CERN-licence.pdf}
 	\includegraphics[width=14cm, trim=0cm 0cm 0cm 4cm, clip]{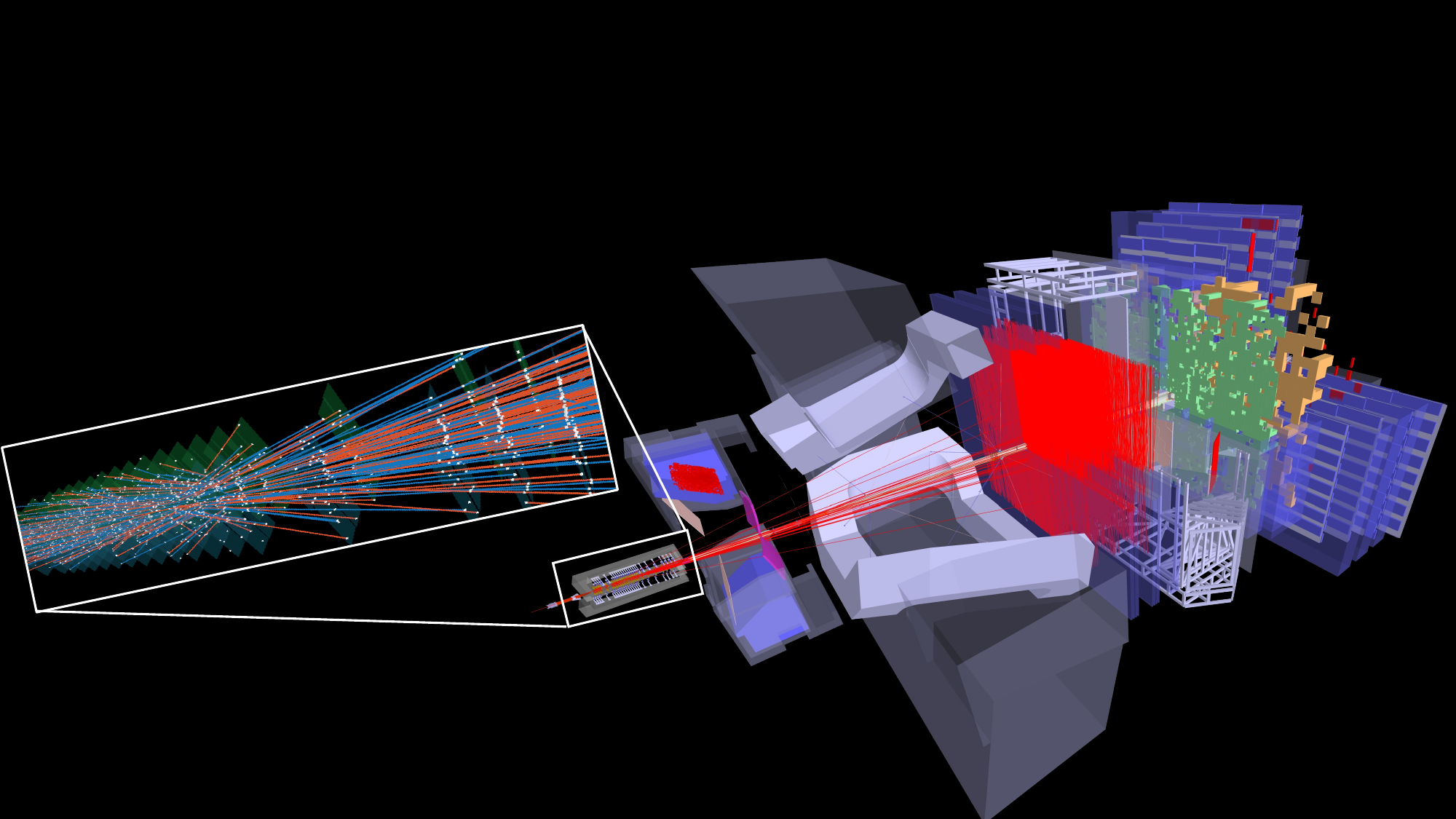} 
	\caption{Display of an LHC event in the LHCb detector. The picture shows the layout of the forward spectrometer with a reconstructed event superimposed. The inset shows a zoom of the interaction region. The two colors of tracks represent particles associated with two $pp$ collisions.}
	\label{fig:titlepage}
\end{figure}

%%%%%%%%%%%%%%%%%%%%%%%%%%%%%%%%%%%%%%%%%%%%%%%%%%%%%%%%%%%%%%%%
%% the following item is optional: 
%% - System of abbreviations/terms/symbols used in the specific field of study/community. List and define
\begin{glossary}[Nomenclature]
	\begin{tabular}{@{}lp{34pc}@{}}
        CP & Charge Parity\\
		LHC & Large Hadron Collider\\
        LHCb & Large Hadron Collider beauty experiment\\
        LHCb-I & Initial LHCb detector used in LHC run-1 and run-2\\
        LHCb-U & Upgraded LHCb detector used in LHC run-3 and forseen for run-4\\ 
        CKM & Cabibbo Kobayashi Maskawa matrix\\
        LoI & Letter of Intent\\
        TP & Technical Proposal\\
        TDR & Technical Design report\\
        VELO & VErtex LOcator\\
        TT & Trigger Tracker or Tracker Turicensis\\
        UT & Upstream Tracker\\
        OT & Outer Tracker\\
        IT & Inner Tracker\\
        RICH & Ring Imaging CHerenkov detector\\
        SciFi & Scintillating Fiber Tracker\\
        ECAL & Electromagnetic CALorimeter\\
        HCAL & Hadronic CALorimeter\\
        MUON & MUON detector\\
        SPD & Scintillating Pad Detector\\
        PS & Pre-Shower detector\\
        HLT & High Level Trigger\\
        RTA & Real-Time Analysis\\
        PID & Particle IDentification\\
        PV & Primary Vertex\\
        SV & Secondary Vertex\\
        BSM & Beyond the Standard Model\\
        FCNC & Flavour Changing Neutral Current\\
        PDG & Particle Data Group\\
%        SME & Standard Model Extension\\
%        GLW & Gronau-London-Wyler\\
%        ADS & Atwood-Dunietz-Soni\\
%        BPGGSZ & Bondar-Poluektov-Giri-Grossman-Soffer-Zupan\\
        LLP & Long-Lived Particle\\
        QGP & Quark Gluon Plasma\\
	\end{tabular}
\end{glossary}

%%%%%%%%%%%%%%%%%%%%%%%%%%%%%%%%%%%%%%%%%%%%%%%%%%%%%%%%%%%%%%%%
%% the following item is mandatory: 
%% List of the key points and topics a reader can expect to learn from this chapter 
\section*{Objectives}
%\begin{itemize}
%	\item Some text on the first topic of the chapter, i.e. what the reader can learn from this article 
%	\item Some text on the second topic of the chapter, i.e. what else the reader can learn from this article
%	\item Some text on the third topic of the chapter, i.e. what else the reader can learn from this article
%	\item ...
%\end{itemize}

%Structure of this Review
This review aims to provide a pedagogical overview of the LHCb experiment at the LHC, including its historical development, experimental setup, data analysis strategies, and highlighting key physics results. The sections of the article are structured around the following objectives:

\begin{itemize}
    \item  The history section reviews the motivation behind the construction of the LHCb experiment and explains the rationale for its unique forward spectrometer design.

    \item The experimental description section explains the functionalities of the subdetector systems, detailing their role in the spectrometer. The set-up for the original LHCb experiment and the upgrade are presented in parallel, discussing how the technological advancements impact the detection performance.

    \item The event reconstruction section explains the main tracking, particle identification and trigger concepts. It illustrates how LHCb's specific detector layout translates into unique strengths for flavour physics analysis in the forward acceptance.

    \item The physics sections provide insight on LHCb's measurement capabilities across the various domains: 
    \begin{itemize}
        \item[-] {\em Flavour Mixing} shows how LHCb measures matter-antimatter oscillations in neutral beauty and charm mesons. 
        \item[-] {\em CP-Violation} presents how all three types of CP violation ({\em direct}, {\em indirect} and {\em interference}) are observed with increasing precision and how these measurements are used to constrain the CKM unitarity triangle of the Standard Model.
        \item[-] {\em Rare Decays} explains how suppressed flavour changing processes are used to search for physics beyond the Standard Model.
        \item[-] {\em Spectroscopy} highlights LHCb's discovery potential for exotic hadrons, including tetraquark and pentaquark states.
        \item[-] {\em Long-Lived Particle Searches} summarizes LHCb's search results on new particles with displaced vertices in the spectrometer.
        \item[-] {\em Electroweak Physics} shows how the forward geometry of LHCb enables precision measurements of the $W$- and $Z$ boson mass and the determination of the weak mixing angle $\sin^2\theta_W$.
        \item[-] {\em Heavy Ion Physics} describes how the forward kinematic reach of LHCb can make complementary contributions to heavy ion physics, complementary to dedicated heavy ion experiments. 
    \end{itemize}
    \item The complementarity with other experiments section compares the strengths and limitations of LHCb in relation to other LHC experiments, as well as to the $e^+e^-$ B-factories. 
    \item The future developments section highlights the potential for flavour physics at the High-Luminosity LHC.
\end{itemize}

%%%%%%%%%%%%%%%%%%%%%%%%%%%%%%%%%%%%%%%%%%%%%%%%%%%%%%%%%%%%%%%%
%% the following items are mandatory: 
%% - Section: Generic experiment description
%% - Section: Physics topics addressed by the experiment
%% - Section: Complementarity with other experiments in the field
%% - Section: conclusion
%%
%% additional sections are optional.
%%

\section{Introduction}
After its discovery in mixing and decay of kaon particles, the puzzle of the underlying origin of CP violation still remained elusive. Theoretical predictions on the kaon parameters $\epsilon$ and $\epsilon '$ are plagued by large hadronic uncertainties hindering numerical comparisons with experiment.
The $B$-meson system offers a more favourable environment: Heavy Quark Effective Theory (HQET) provides a systematic handle on non-perturbative effects, and $B$-meson lifetimes are long enough for high-precision measurements yet short enough to be comparable with the $B^{0}$–$\bar{B}^{0}$ oscillation period. Against this backdrop, the LHCb experiment at the Large Hadron Collider has emerged as the flagship facility for heavy-flavour physics.
This chapter reviews the history, experimental set-up, reconstruction methodology and physics analyses of LHCb. It covers both the original LHCb detector, mainly focused on the physics of $B$-decays, as well as the later upgraded LHCb detector, which developed into a multi-purpose facility.

%\clearpage
\section{History of the experiment}\label{secHistory}
%Please provide a very general and easy to understand introduction to the experiment.

The idea of a dedicated Flavour Physics experiment at the LHC was conceived to provide an answer to the riddle of understanding the matter-antimatter asymmetry. 
The historical development of the LHCb experiment is discussed in detail in Ref.~\cite{LHCb-history}. Here, we present a concise overview, focusing on the physics motivations and the evolution of the spectrometer layout - from the original proposal to the constructed LHCb detector and its subsequent upgrade.

\subsection{First Observations of CP violation}

The discovery of CP violation in 1964 by the experiment of Cronin and Fitch~\cite{Christenson:1964fg} revealed that the symmetry between matter and antimatter in nature is not exact, which came as a complete surprise to the scientific community. Although initially various explanations were proposed to preserve the concept of symmetry, it soon became clear \cite{KLasymmetry:1967} that the physical $K^0_S$ and $K^0_L$ particles are not CP eigenstates, with CP violation occurring in the coherent {\em mixing process} of the neutral kaons $K^0 \Leftrightarrow \bar{K}^0$. The underlying mechanism remained mysterious, prompting new explanations such as the existence of a new superweak force occurring specifically in double strangeness violating ($\Delta S=2$) transitions \cite{wolfenstein:1983}. Later, {\em direct CP violation} was found to occur in $\Delta S=1$ decay processes; initial evidence was reported by the NA31 experiment at CERN \cite{NA31:1988} and later confirmed by the NA48 experiment at CERN \cite{NA48:2002}
and the KTeV experiment at Fermilab \cite{KTeV:2003}. 

In 1973, Kobayashi and Maskawa extended the quark mixing model of Nicola Cabibbo \cite{cabibbo:1963} to include a complex, CP violating phase parameter in the flavour-mixing CKM matrix \cite{KM:ptp:49:652}. However, due to the relatively low mass of kaons, heavy quark effective theory was not equipped to make precise quantitative theoretical predictions to put the origin of CP violation to test. According to the CKM mechanism, CP violating phenomena were expected to occur more strongly in $B$-meson processes, where heavy quark effective theory also allowed to make numerical calculations.

Following the discovery of the $b$-quark in 1977 with the observation of the massive $\Upsilon$ $b\bar{b}$ resonance at Fermilab \cite{bquarkdiscovery}, worldwide experimental efforts shifted focus towards $b$-hadron spectroscopy. The ARGUS experiment at DESY, %\cite{argus-experiment}, 
as well as CLEO and CUSB at Cornell, rapidly expanded the knowledge on the spectroscopy by discovering and measuring the masses of the $\Upsilon(1S)$ to $\Upsilon(4S)$ resonances \cite{kaplan:1996}. In 1983, CLEO reported the first fully reconstructed $B$-meson decays~\cite{Bmeson-discovery}, observed via $B\bar{B}$ pair production from decays of the $\Upsilon(4S)$ resonance, which lies just above the production threshold. After an upgrade of the DORIS-II ring in DESY, the Argus experiment in 1987 observed the phenomenon of $B^0$-$\bar{B}^0$ mixing \cite{Argus-mixing}. The mixing mechanism was similar to kaons but occurring at a much higher rate, with a period roughly equal to the $b$-quark lifetime of about 1.5 ps, as had been measured at the MAC and Mark II experiments at SLAC \cite{b-lifetime-MAC, b-lifetime-MarkII}. 

The more massive $B$-system offers key advantages in comparison to the $K^0$-$\bar{K^0}$ system. First, the $B$ system includes two neutral $B$ mesons: the $B^0$ and $B^0_s$, allowing to test and compare different CKM transitions. Second, $B$-mesons have an intermediate lifetime: long enough to be measured experimentally and short enough to allow for quantum interference phenomena between {\em direct decays} and {\em decays after mixing}. Third and perhaps foremost, the $B$-particles include hundreds of different decay modes, providing an ideal experimental laboratory for testing CP violation predictions. Consequently, CP violation in the $B$-system became a major focus of global research, and was described as the new "Holy Grail" of particle physics.

\subsection{B-meson experiments and the birth of LHCb}

Different experimental approaches were used to produce $B$-mesons: electroweak $e^+e^-$ collisions at $B$-factories (CLEO, BaBar, Belle) via the $\Upsilon(4S)$ resonance; $Z$-boson decays at SLC and LEP; and strong hadronic interactions at Tevatron, HERA, and the LHC. Each method had distinct advantages: $\Upsilon(4S)$ production resulted in exclusive, coherent $B^0$-$\bar{B}^0$ pairs; $Z$-decays produced all types of $b$-hadrons; and hadronic production offered the highest cross-section but also at the cost of a high background rate.

The complementarity of these approaches is demonstrated in the way the $B^0_s$ meson was detected. First evidence for production of $B^0_s$ mesons was obtained by the CUSB experiment in 1990 \cite{Bs-discovery-CUSBII} by studying the photon spectrum from radiative transitions while running on the $\Upsilon(5S)$ resonance. Its existence was confirmed by the ALEPH experiment at LEP in 1992 via observation of the decay $B^0_s\rightarrow D_s^- l^+ \nu X$ decays \cite{Aleph-Bs-Observation}, and finally also via the measurement of the branching ratio of $B^0_s\rightarrow J/\psi \phi$ decays by the CDF experiment at Tevatron in 1993 \cite{CDF-Bs-JpsiPhi}. 

Exclusion limits from LEP at CERN \cite{Combined-LEP-b-physics, Delphi-Bs-Oscillations} suggested a very high mixing frequency of the $B^0_s$-$\bar{B}_s$ meson oscillations. Its measurement would require novel, high-precision silicon vertex detectors to reconstruct decays with sub-millimeter accuracy to determine the flavour-oscillations of the $B^0_s$ decays as a function of their observed decay-times. While the HERA-B experiment at DESY was dedicated to measure the $B^0_s$ oscillations \cite{HERAB-Proposal}, they were first observed instead by the CDF collaboration at Fermilab \cite{CDF-Bs-Oscillations}.

CP violation in the $B$-system was discovered simultaneously by the $B$-factories BaBar \cite{CP-BaBar} at PEP2 in SLAC and Belle \cite{CP-Belle} in KEK, both operating at the $\Upsilon(4S)$ resonance. Unlike CLEO, these $B$-factories collided electrons and positrons in so-called asymmetric mode, producing the $\Upsilon(4S)$, and hence the $B$-meson pair, with a boost inside the detector. This allows for a precise measurement of the difference of the displaced vertices of the two decaying $B$-mesons, and therefore enabling a precise decay-time measurement.

\begin{figure}[htbp]
	\centering
    \begin{subfigure}[b]{0.46\textwidth}
        \includegraphics[width=\textwidth]{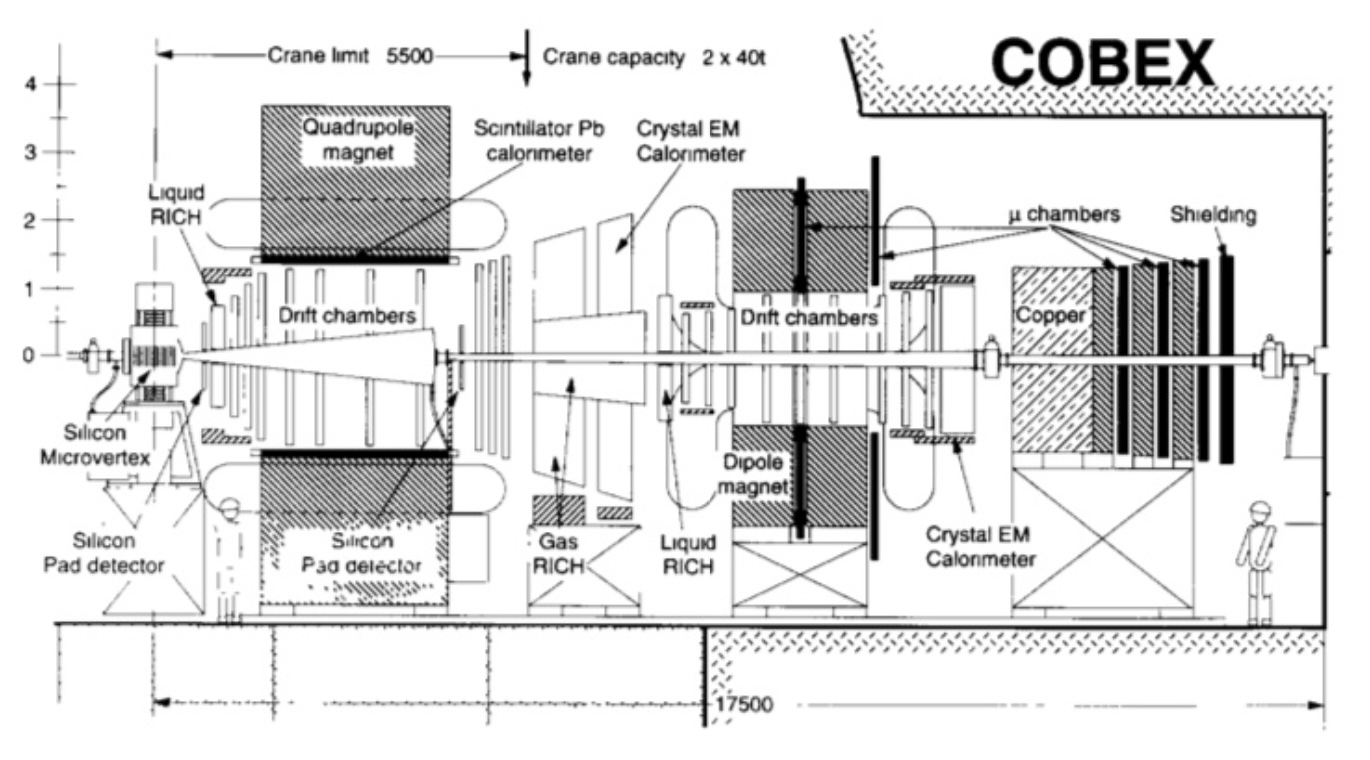}
        \caption*{COBEX: taken from \cite{COBEX-exp}}
    \end{subfigure}
    \hfill
    \begin{subfigure}[b]{0.46\textwidth}
        \raisebox{0.5cm}{\includegraphics[width=\textwidth]{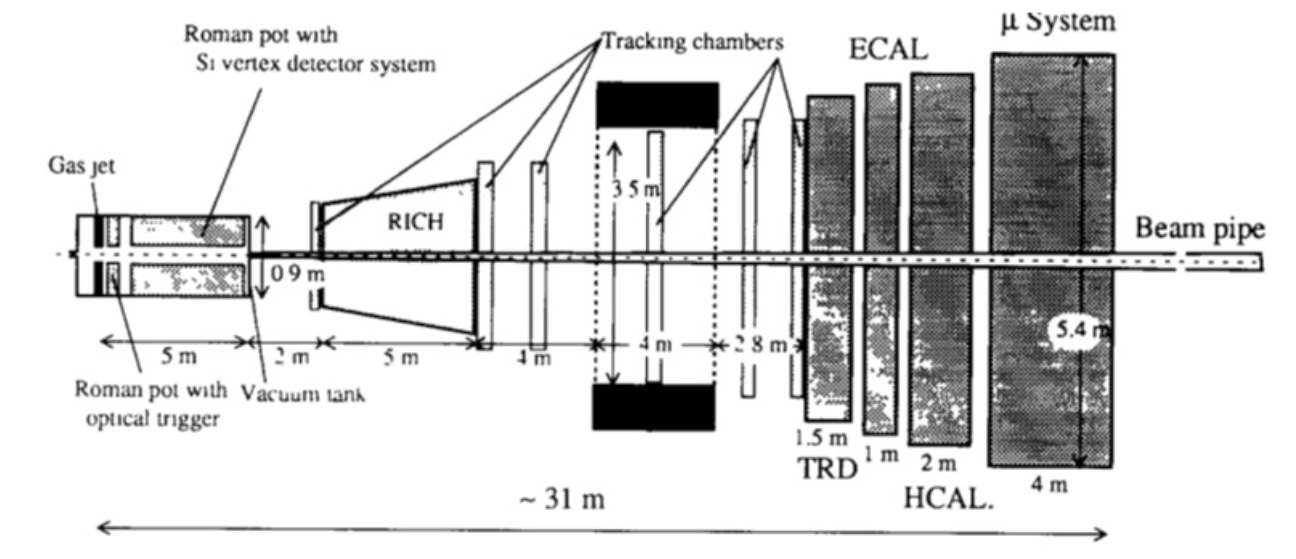}}
        \caption*{GAJET: taken from \cite{GAJET-exp}}
    \end{subfigure}
    \hfill
    
    \vspace*{0.5cm}
    
    \begin{subfigure}[b]{0.65\textwidth}
        \includegraphics[width=\textwidth]{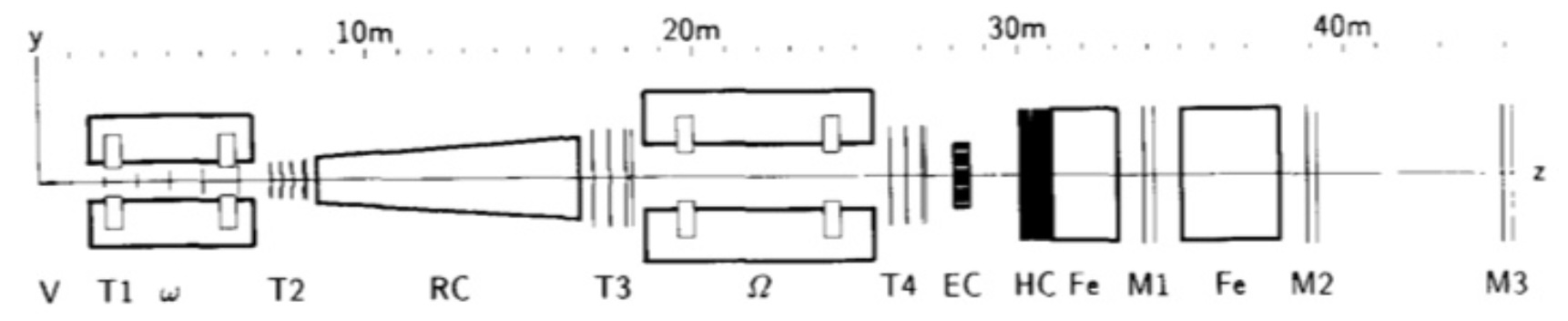}
        \caption*{LHB: taken from \cite{LHB-exp}}
    \end{subfigure} 
    \hfill
    \caption{Schematic representations of the three precursor proposals: COBEX, GAJET and LHB. COBEX was designed as a collider mode experiment, while GAJET used a gas jet into the LHC beam and LHB extracted protons from the beam with a bent crystal.}
    \label{fig:precursors}
\end{figure}

Meanwhile, the high collision energy of proton-proton interactions at the LHC promised abundant beauty particle production. However, achieving high-precision measurements in such a harsh environment posed significant technical challenges, as was experienced by the HERA-B experiment. The $B$ production cross section was low, due to a relatively low collision energy in the fixed target mode.  At the 1992 LHC workshop in Evian, three $b$-physics experiments were proposed (see Fig.~\ref{fig:precursors}): COBEX (a collider mode experiment) \cite{COBEX-exp}, GAJET (an experiment with internal LHC gas jet target) \cite{GAJET-exp}, and LHB (an experiment with extracted LHC beam) \cite{LHB-exp}. In June 1994, the LHC Committee declined to approve any of these proposals individually. Instead, it encouraged the interested groups to form a single collaboration for a collider-mode experiment to exploit the large $b$-production cross section and to include several concepts of the two fixed-target proposals. The new proposal required a convincing trigger strategy, precise charged-particle tracking, and robust particle identification.

The choice of collider-mode operation proved the right one, as the LHC would not have produced enough $b$-hadrons in a fixed target setup in comparison to the $B$-Factories and Tevatron experiments, which both exceeded their initial expectations. In comparison, the fraction of $b\bar{b}$ production relative to inelastic events was around $10^{-6}$ at HERA-B but reached as much as $1/160$ at the LHC, ensuring a relatively low-background beauty-physics program.

\subsection{Initial Design of LHCb}

The LHCb experiment was specifically designed with a focus on precision measurements of CP violation and rare decay phenomena of $b$-particles, as explicitly stated in the first proposal to CERN's LHCC committee \cite{LHCb-LoI}: "A Dedicated LHC Collider Beauty Experiment for Precision Measurements of CP-Violation". 
Due to the high luminosity potential of the LHC and the high collision energy in collider mode, the total $b$-particle production is of the order of $10^{12}$ per running year; unprecedented amounts in comparison to other facilities.  
Despite collider mode operation, the relatively light mass of $b$-particles as compared to the 14 TeV LHC collision energy, leads to predominant production of heavy flavour particle pairs at small angles - boosted forward or backward - with respect to the LHC beams. 
This is due to the large gluon density at low values of Bjorken-$x$, which leads
to asymmetric collisions. 
A proton-proton collision including signal particles and underlying event, results in about 100 particles in the final state.
As a result of the similar experimental conditions as for HERA-B, the newly formed LHCb collaboration designed a forward spectrometer with a geometric design largely based on that of the HERA-B fixed target spectrometer.

Similar to HERA-B, the LHCb detector design includes extensive tracking and particle identification systems, as well as an intricate trigger. The tracking features a silicon microvertex detector that can be inserted very close to the beam line, along with a large downstream tracker extending across a dipole magnetic field for precise particle momentum measurement. The particle identification system includes dedicated Ring Imaging Cerenkov detectors, identifying low-, medium- and high-momentum particles, as well as electromagnetic and hadronic calorimeters for hadron identification, and a muon system for triggering purposes. 

The LHCb detector was designed to operate at an average luminosity of ${\cal L}=2\times 10^{32} \mathrm{cm}^{-2}\mathrm{s}^{-1}$, a factor of 10 lower than the ATLAS and CMS experiments, corresponding to a maximum number of single proton-proton interactions occurring per bunch crossing. The reason was two-fold: first, to limit the high particle density and corresponding irradiation of sensitive detector elements in the forward region, and second, to avoid mismatches of associating secondary $B$ and $D$ decay vertices with the wrong primary production vertex in high pile-up events.
To ensure efficient triggering and reconstruction, the high particle multiplicity of the final state requires fine grained detectors, with inner (high track density) and outer (lower track density) subsystems for tracking, RICHes and calorimetry.

\begin{figure}[htbp]
	\centering
    \begin{subfigure}[b]{0.49\textwidth}
        \hspace*{-0.5cm}\raisebox{1.0cm}{\includegraphics[width=1.05\textwidth]{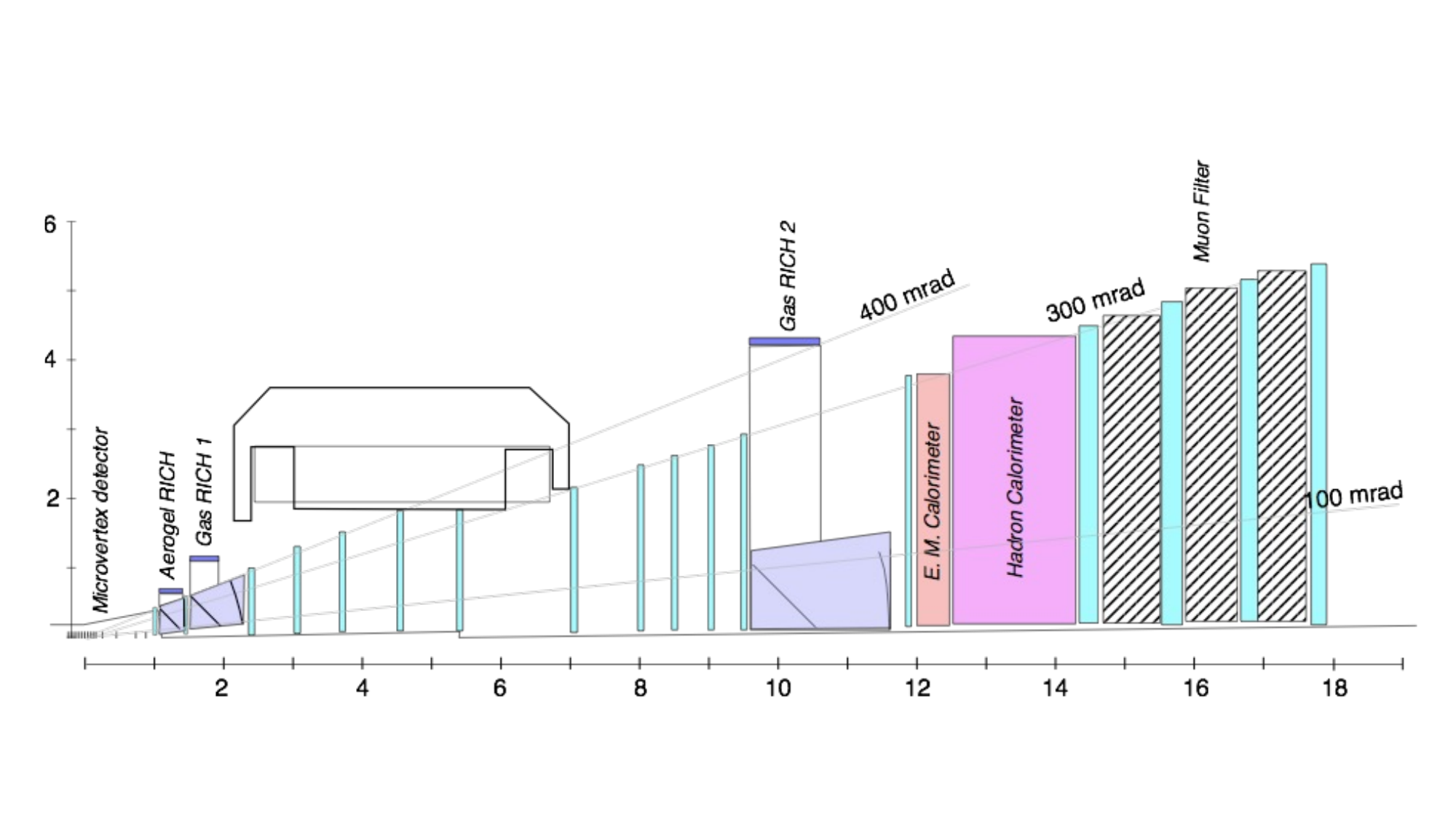}}
        \vspace*{-1.0cm}
        \caption*{LHCb Letter-of-Intent in 1995. Taken from \cite{LHCb-LoI}.}
    \end{subfigure}
    \hfill
    \begin{subfigure}[b]{0.49\textwidth}
        \hspace*{-0.5cm}\raisebox{0.5cm}{\includegraphics[width=1.25\textwidth]{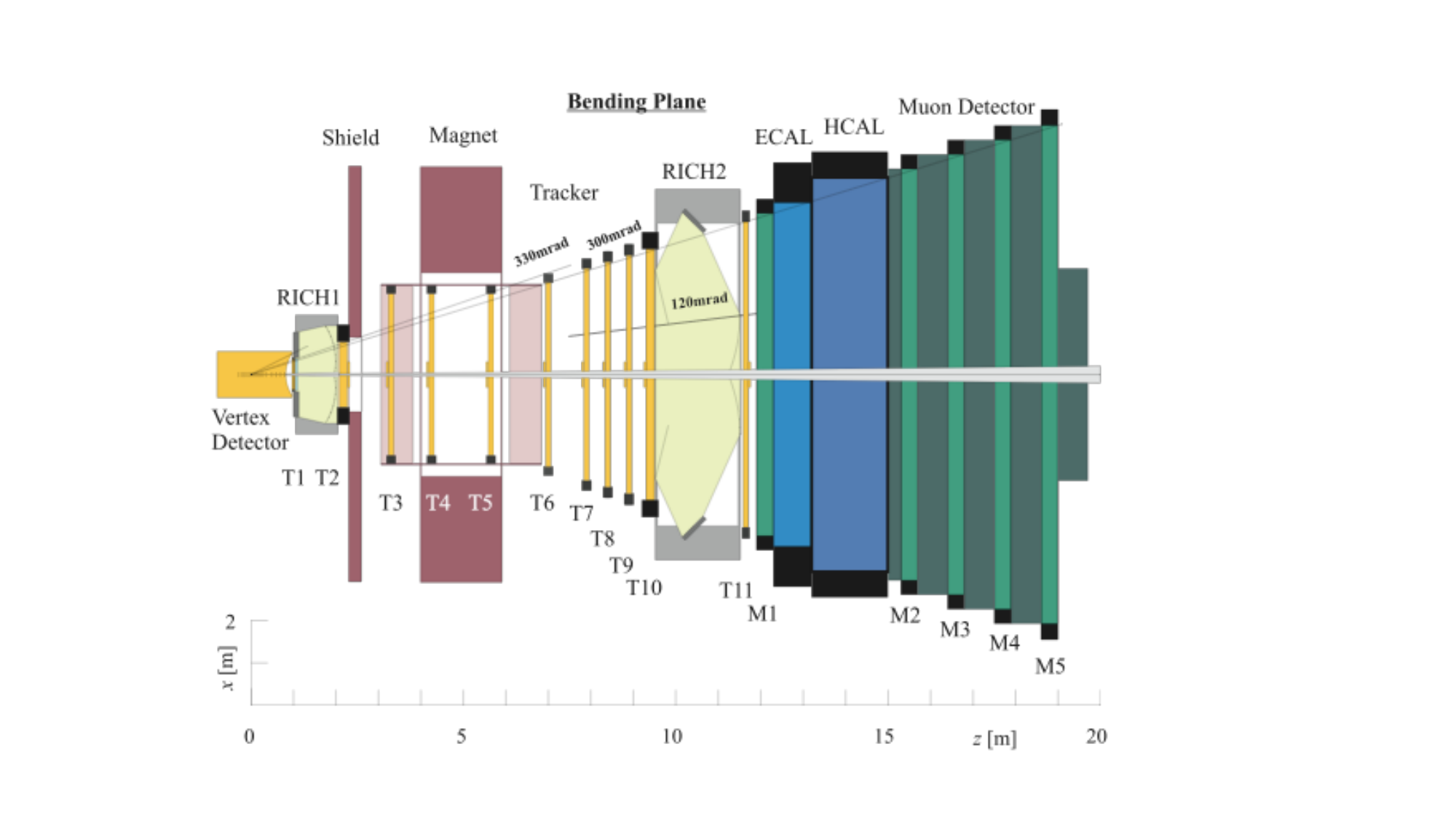}}
        \vspace*{-1.0cm}
        \caption*{LHCb Technical Proposal in 1998. Taken from \cite{LHCb-TP}.}
    \end{subfigure}
    \hfill
    
    %\vspace*{0.5cm}
    
    \begin{subfigure}[b]{0.49\textwidth}
        \hspace*{-1.0cm}\includegraphics[width=1.15\textwidth]{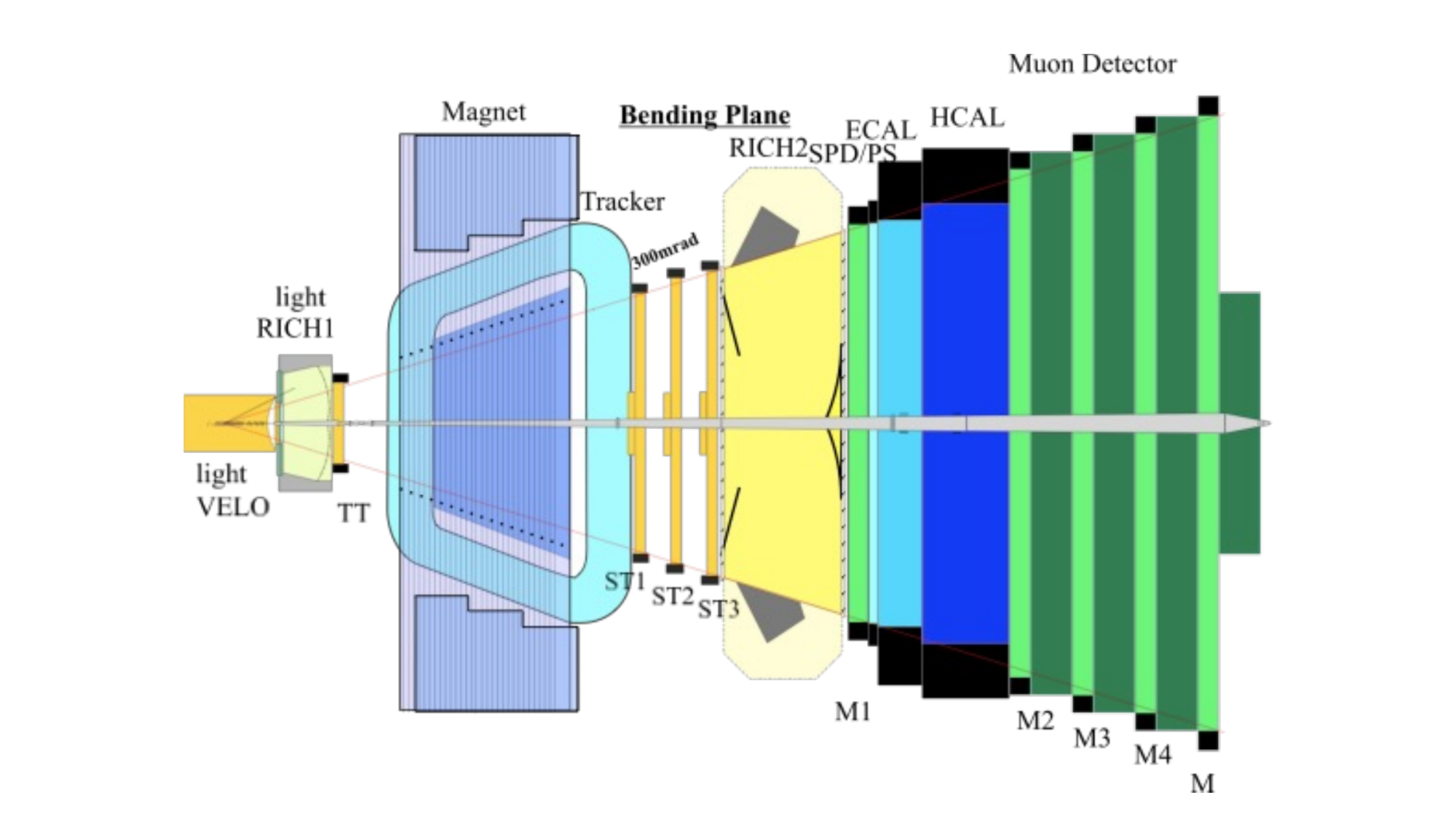}
        \caption*{LHCb Reoptimized Technical Design Report in 2003. Taken from \cite{LHCb-Reopt-TDR}.}
    \end{subfigure} 
    \hfill
    \begin{subfigure}[b]{0.49\textwidth}
        \hspace*{-0.5cm}\includegraphics[width=1.15\textwidth]{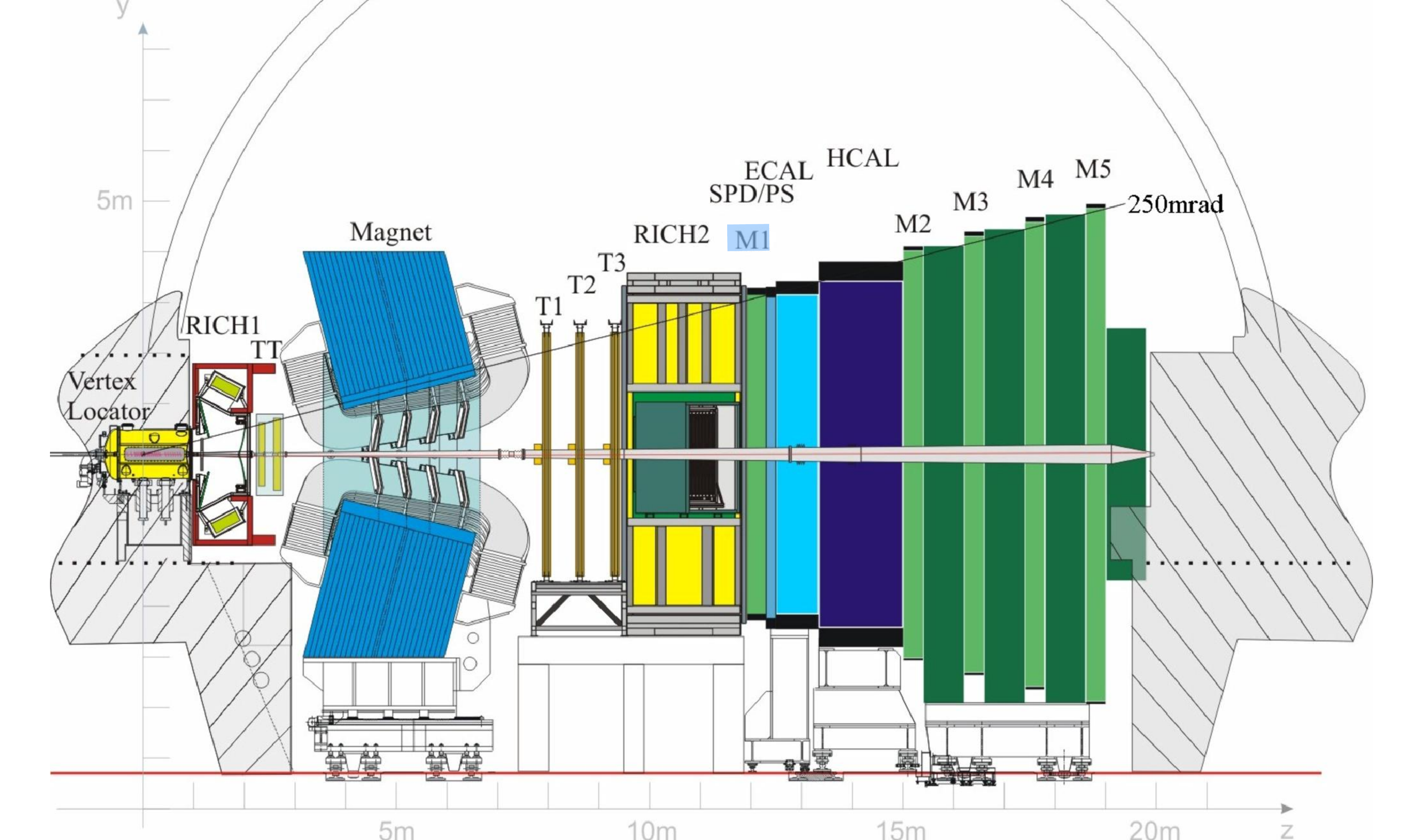}
        \caption*{LHCb Detector at Run-1 \& Run-2. Taken from \cite{LHCb-at-LHC-JINST2008}.}  
    \end{subfigure}
    \caption{The evolution of the LHCb experiment from initial design to construction. The forward spectrometer evolved over time from the Letter-of-Intent, Technical Proposal, Technical Design Report, and final constructed detector.}
    \label{fig:DetectorEvolution}
\end{figure}

Following HERA-B, the central concept of the experiment was originally based on detecting downstream calorimeter and muon trigger seeds, followed by a Kalman filter track-following procedure in the upstream direction towards the vertex detector and finally adding particle identification to the particle candidates. This dictated a spectrometer layout including 11 tracking stations across the tracking volume before, after and inside the large dipole magnetic field, see Fig.~\ref{fig:DetectorEvolution}.

\subsection{Evolution to the LHC Run-1 and Run-2 Spectrometer}

As more precise simulation studies for the Technical Design Reports of the subsystems were carried out, it became clear that, while additional detection redundancy was beneficial, the spectrometer setup contained excessive material thickness. By the time of the Technical Design Report (TDR) of the tracking system \cite{LHCb-OT-TDR}, the material budget across the tracking volume had increased to 60\% of a radiation length and 20\% of an interaction length.

Excess material deteriorates the detection of electrons and photons, increases the multiple scattering of charged particles, and raises occupancies in the tracking stations. A higher fraction of nuclear interaction length leads to more kaon and pion interactions before traversing the entire tracking system. Consequently, the number of reconstructed $B$ decays therefore decreases, even if the efficiency of the tracking algorithm remains high for fully traversing charged particles. These effects collectively contribute to an overall reduction in the number of reconstructed $B$-meson decays \cite{PhDThesis_Hierck}. 

After dedicated studies demonstrated that particle tracks can be extrapolated across the full magnet volume without requiring intermediate stations, a drastic reduction of 11 to 4 tracking stations was adopted in the so-called Reoptimization TDR \cite{LHCb-Reopt-TDR}. At the same time, the number of measurement planes in the vertex detector was reduced, as was the thickness of the individual sensitive silicon layers.  Moreover, the large amount of hits in tracking stations caused by secondary particles could be reduced by replacing an aluminium conical section of the beampipe with a beryllium one. The entire layout of the tracking system was redesigned to adopt a more lightweight configuration nick-named "LHCb-light" at the time. The upstream RICH detector was also adapted by reducing material of the entrance window, employing lightweight mirror materials, and moving support structures outside the active acceptance region. 

\begin{figure}[h]
	\centering
	\includegraphics[width=9.5cm, angle=-90]{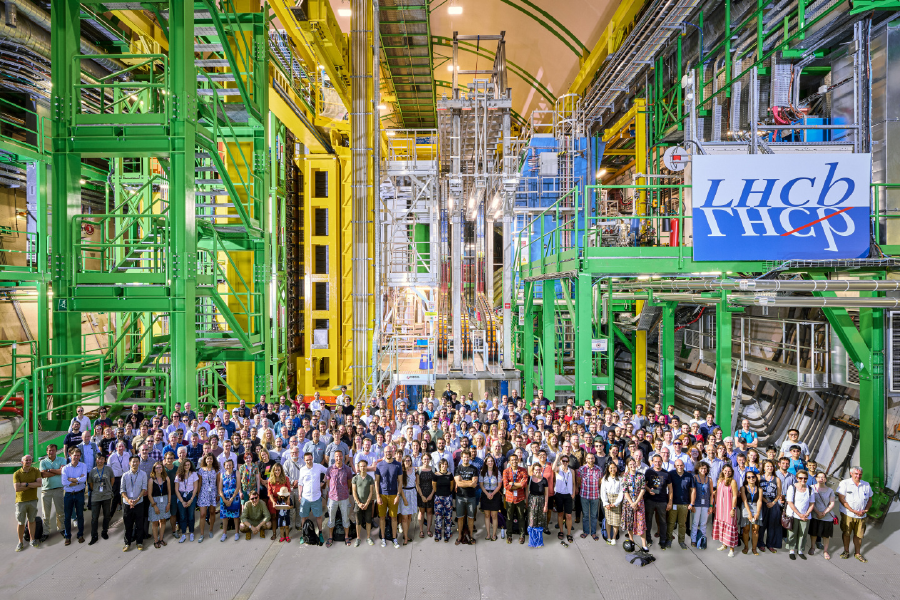}
	\caption{The LHCb collaboration photoshopped into the experimental detector hall. CERN Document Server: CERN-PHOTO-202203-057.} 
	\label{fig:LHCbCollab}
\end{figure}

Fig.~\ref{fig:DetectorEvolution} illustrates the evolution of the spectrometer design: from the LoI in 1994 \cite{LHCb-LoI}, to the TP in 1998 \cite{LHCb-TP}, the Reoptimization TDR in 2003 \cite{LHCb-Reopt-TDR}, to the actual constructed detector. The final LHCb layout includes from left to right a vertex locator ("VELO"),
an upstream Ring Imaging Cherenkov detector ("RICH-1"), an upstream tracker ("TT"), the dipole magnet, a downstream tracker ("T"), a downstream Cherenkov ("RICH-2"), a calorimeter system ("ECAL", "HCAL") and a muon detector ("MUON").
This spectrometer was installed at interaction point 8 of the LHC, in the experimental hall previously used by the DELPHI experiment at LEP. Although the detector has recently evolved into an upgrade, the overall layout has been kept the same. A picture of the final detector, along with the collaboration, is shown in Fig.~\ref{fig:LHCbCollab}. 

Over time, the physics goals of the experiment shifted from confirmation of the CKM paradigm to searching for physics beyond the Standard Model. For CP violation, key objectives included a high-precision CP violation measurement where it was expected to be negligible (in the decay $B^0_s\rightarrow J/\psi \phi$), and comparing the value of the unitarity angle $\gamma$ obtained from tree decays with the one measured in penguin decays. For very rare decays the observation of the forbidden mode $B^0_s\rightarrow\mu^+\mu^-$ became a key motivation, while in rare decays the angular distribution of $B\rightarrow K^{*0}\mu^+\mu^-$ and the photon polarization in $B^0_s\rightarrow\phi\gamma$ became high profile observables. 

The LHCb experiment also expanded its research scope significantly, revealing broader applications of its spectrometer capabilities. The heavy-flavour physics program was extended to include charm physics, a comprehensive QCD spectroscopy program, as well as searches for exotic and long-lived particles, ion-ion collisions, and electroweak and jet physics. Although designed for $b$-physics, LHCb evolved into a versatile forward-direction general-purpose experiment, with its initial specific $b$-physics related design features proving valuable for the broad physics program.

\vspace*{-0.2cm}
\subsection{The LHCb Upgrade}

\begin{wrapfigure}{R}{8.6cm}
	\centering
    \vspace*{-0.3cm}
	\includegraphics[width=8.6cm]{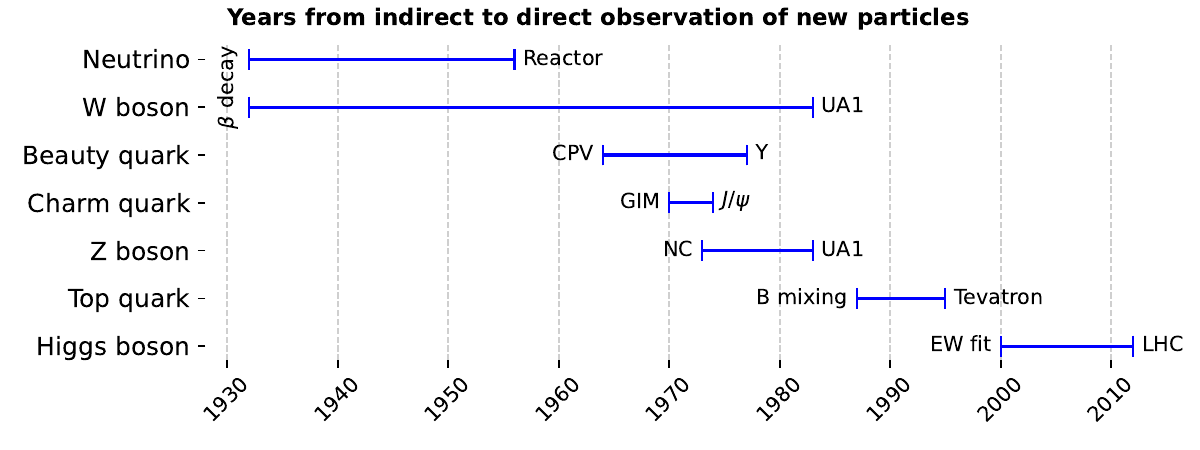}
	\caption{Historically, the signs of the existence of many particles in the Standard Model have been preceded by indirect measurements, illustrating the power of precision measurements in flavour physics. } 
	\label{fig:Indirect}
\end{wrapfigure}
To further expand the capabilities to search for physics beyond the Standard Model, the collaboration proposed an upgraded LHCb detector \cite{LHCb-Upgrade-LoI, LHCb-Upgrade-FrameworkTDR, LHCb-Upgrade-1-JINST2024} for LHC \mbox{Run-3} and \mbox{Run-4}. In this document we refer to this detector as \mbox{"LHCb-U"}, to distinguish from the initial detector, \mbox{"LHCb-I"}. The New Physics searches in Flavour Physics are referred to as {\em indirect searches}, to put it into contrast with the {\em direct searches} for on-shell production of new particles by ATLAS and CMS. Historical examples of indirect first signs of new discoveries are the prediction of the $c$-quark from the suppressed $K^0_L\rightarrow \mu^+\mu^-$ decays (through the GIM mechanism), and the heavy top quark mass, inferred from \mbox{$B^0$-$\bar{B}^0$} oscillations. Similarly, the mass of the Higgs boson was predicted by loop diagrams using electroweak precision measurements at LEP. Fig.~\ref{fig:Indirect} summarizes how historically indirect signs of unseen particles were followed by discoveries later on.

The "LHCb-I" detector operated at $\sqrt{s}=7$ and $8$ TeV in LHC Run-1 (2009 - 2013) and at $\sqrt{s}=13$ TeV in LHC Run-2 (2015 - 2018). 
It was designed to operate at an instantaneous luminosity in the range of $2-5\times 10^{32} \mathrm{cm}^{-2}\mathrm{s}^{-1}$, and indeed collected data at a nominal value of $4\times 10^{32} \mathrm{cm}^{-2}\mathrm{s}^{-1}$. The nominal operation was chosen corresponding to a maximal number of events with a single proton-proton collision, hence minimizing collision pile-up. 

For LHC Run-3 and Run-4, the upgraded detector, LHCb-U, is constructed with higher radiation hardness and finer detection segmentation with improved resolutions, enabling operation at a five-fold increased luminosity of $2\times 10^{33} \mathrm{cm}^{-2}\mathrm{s}^{-1}$. The upgrade is designed such that the overall layout and functionality of the experiment approximately remains the same, whereas the signal statistics, and hence the collision pile-up, increases five-fold. In the detector sections of this document both the main aspects of LHCb-I and LHCb-U upgrade are presented in parallel, whereas the physics section mainly uses results from LHCb-I detector. 

A major upgrade from LHCb-I to LHCb-U is the complete replacement of the hardware trigger system with a full software-based Real-Time Analysis (RTA) system, enabling full detector readout at 40 MHz. This allows the entire event selection process to be implemented in software, leading to a factor of 2–10 improvement in real-time selection efficiencies, depending on the specifics of the decay channel.

The LHCb-U detector still does not exploit the maximum luminosity potential of the LHC, but operates at reduced pile-up conditions. A second upgrade at the High Luminosity LHC era is currently under consideration, targeting a further luminosity increase to $1 - 2 \times 10^{34} \mathrm{cm}^{-2} \mathrm{s}^{-1}$, corresponding to 25-50 simultaneous proton-proton interactions. While the first upgrade focused on finer detector segmentation (e.g., transitioning from strip to pixel detectors) and a versatile trigger with GPUs able
to process events from each bunch crossing, the second upgrade is expected to incorporate high-precision timing measurements into detector elements, effectively transforming 3D detection information into 4D hits.

\clearpage
\section{Generic experiment description}\label{secExperiment}
%Please provide a very general and easy to understand introduction to the experiment.

\subsection{Experimental Challenges and Concept}

\begin{figure}[ht]
	\centering
%	\raisebox{-0.5cm}{\includegraphics[width=7.5cm]{Figures/Detector/hidef_LHCb-upgrade-y.png}}
%  	\hspace*{-8.0cm}\raisebox{3.5cm}{\includegraphics[width=3cm]{Figures/Detector/ForwardAngleBproduction.jpg}}
   %\hspace*{5.5cm}
    \includegraphics[height=5.0cm, trim=10cm 0cm 0cm 0cm, clip]{Figures/Detector/LHCb-Detector-CERN-licence.pdf}
    \hspace*{0.5cm}
    \includegraphics[height=5.0cm]{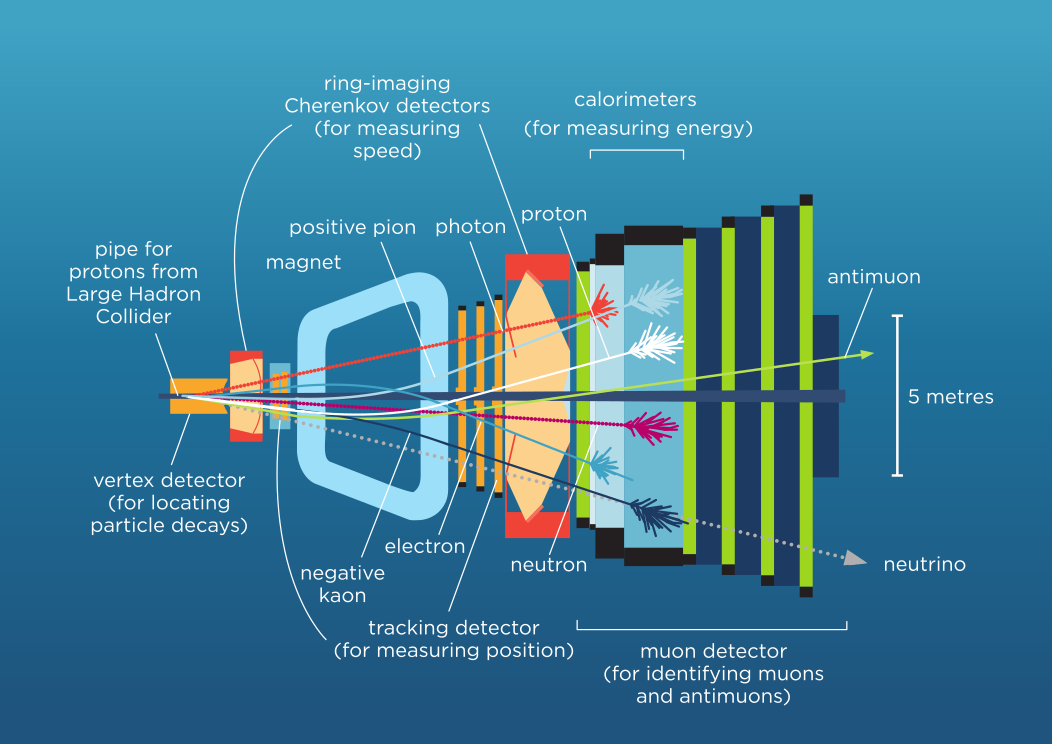}
	\caption{{\em Left:} Layout of the LHCb spectrometer with from left to right the Vertex Locator, RICH-1, Upstream Tracker, Dipole Magnet, Downstream Tracker, RICH-1, Electromagnetic and Hadronic Calorimeters and Muon stations. Credit: CERN.
    %Layout of the LHCb-U detector used in LHC run-3 together \cite{LHCb-Upgrade-1-JINST2024} an inset of the correlated angular production of $b\bar{b}$ mesons. 
    {\em Right:} A schematic depiction of the concepts of LHCb detector systems. Credit: Royal Society Summer Science Exhibition 2016.} 
	\label{fig:LHCbDetector}
\end{figure}

The LHCb-I experiment collected data during LHC Run-1 and Run-2 (2009-2018), while LHCb-U, also known as LHCb-Upgrade1, has been operational since Run-3 (2022-2026). The overall layout of the LHCb-I and LHCb-U detectors remains largely similar, with the primary differences being the upgraded subdetector technologies and the transition from a hardware-based trigger to a fully software-based trigger.
The layout of the generic LHCb experiment is shown in the left panel of Fig.~\ref{fig:LHCbDetector}, 
%Although the experiment underwent a major upgrade before Run-3, the overall physics aims of the upgraded experiment remained the same as before: a multipurpose experiment for precision physics in the forward direction. Broadly speaking, the improved detectors and software trigger allow the upgraded experiment to operate with higher efficiency at a factor of five higher luminosity, while keeping the same overall layout. 
and the conceptual roles of each of the subdetectors in the experiment is illustrated in the right panel of the figure. The spectrometer provides a coverage of $\pm 300$ mrad in the magnet dipole bending (horizontal) plane and $\pm 250$ mrad in the non-bending (vertical) plane, corresponding to an average pseudorapidity acceptance in the range $2 < \eta < 5$. The coordinate system is defined such that the $z$-axis runs along the beam-line towards the downstream direction of LHCb, the $y$-axis points vertically upwards and the $x$-axis is oriented towards the outside of the LHC ring, forming a right-handed coordinate system. 

The experiment is composed of tracking detectors, including the Vertex Locator, Upstream Tracker and  Downstream Tracker, as well as particle identification systems: the Ring Imaging Cerenkov detectors, Calorimeters and Muon stations. Although many individual detector elements and most of the readout system was replaced between LHCb-I and LHCb-U, their primary functionality in the spectrometer remained unchanged. In the following we present the subsystem technologies of LHCb-I and LHCb-U in parallel.
For a comprehensive description of the LHCb-I detector in Run-1 and Run-2, we refer the reader to "The LHCb Experiment at the LHC" \cite{LHCb-at-LHC-JINST2008}, while details of LHCb-U in Run-3 can be found in "The LHCb Upgrade I"~\cite{LHCb-Upgrade-1-JINST2024}.

Due to the high proton beam energies of the LHC, heavy-flavour particles are produced in unprecedented amounts. However, they emerge at small angles relative to the beam line amidst 50 - 100 other particles per single $pp$ collision. 
%The correlated angular distribution of $b\bar{b}$ quark pairs is illustrated in the inset of Fig.~\ref{fig:LHCbUpgrade1}. 
This harsh background environment, combined with the fact that branching ratios of interesting signal $B$-decays range from $\sim 10^{-3}$ (tree-diagram decays) to $\sim 10^{-10}$ (very rare decays), underscores the need for stringent background suppression - both in real time and in offline analysis. Furthermore, analyses of neutral $B$-meson mixing require a precise measurement of the secondary vertex displacement relative to the associated primary vertex (especially in $B^0_s$ mixing), as well as robust flavour tagging capabilities to determine the initial flavour ($b$ or $\bar{b}$) of the signal meson. Additionally, studies of radiative $B$-decays require efficient identification of radiative photons.
Taking it all together, to perform the precision measurements in flavour physics at the LHC the experiment incorporates the following key features:
\begin{itemize}
    \item {\em Tracking}: high resolution charged particle tracking for vertex reconstruction and decay-time measurements, as well as high precision momentum measurement for invariant mass measurements for resonant decay reconstruction,
    \item {\em PID}: efficient particle identification recognizing electrons, pions, kaons and muons, for efficient signal and background separation,
    \item {\em Trigger}: sophisticated and redundant trigger system for Heavy Flavour physics, based on multiple signals, including high transverse electron or hadron energies, transverse momenta for muons, and secondary decay-vertex topologies.  
\end{itemize}
The following summarizes how the detector subsystems of LHCb-I and LHCb-U meet these requirements for an increasing challenging environment in the setups of Run-1, Run-2 and Run-3.

\subsection{Tracking Detectors}

LHCb is particularly well equipped for detection of charged particles. The tracking system consists of the Vertex Detector (or VErtex LOcator, VELO), positioned in close proximity to the proton-proton collision point, a tracking station 
TT (UT) at the entrance of the large dipole magnetic field, and tracking stations 
IT \& OT (SciFi) located directly downstream of the magnetic field,
for LHCb-I (LHCb-U).

The VELO detector is designed to precisely measure the production vertex of unstable particles (primary vertex, PV) as well as their subsequent decay vertices (Secondary Vertices, SV).
Flavour Physics studies at the LHC involve unstable particles with typical proper decay distances of approximately 3~mm for charm, $\sim$7 mm for beauty, and up to $\sim$2 m for strange particles. 
Accurate vertex measurements play a crucial role in suppressing combinatorial background caused by high track multiplicities in underlying events, as well as enabling precise decay time measurements, which are particularly important for $B^0_s$ meson oscillation studies.
The experiment was designed to achieve an average $B$-decay time resolution of approximately 0.035 ps \cite{LHCb-LoI, LHCb-TP}, which was considered sufficient for observing - at that time still unknown - high $B^0_s$-meson mixing frequencies. Given an expected average $B$-meson momentum of around 100 GeV, this corresponds to a decay distance resolution of $\sigma(d) = 200$ $\mu$m.

The upstream and downstream tracking systems measure the deflection of charged particles in the dipole magnetic field, allowing for precise determination of particle curvature across the known magnetic field.
The resolution requirements of the downstream trackers were optimized to match the multiple scattering distortions to obtain optimal momentum resolution performance of the tracker. The hit resolutions of the downstream tracker - approximately 50 $\mu\mathrm{m}$ for the Inner Tracker and 200 $\mu\mathrm{m}$ for the Outer Tracker - enable a momentum resolution of $\delta p / p \sim 0.3\% - 0.4\%$ \cite{LHCb-TP} for a momentum range of 5 - 200 $\mathrm{GeV}/\mathrm{c}$ tracks. This results in typical mass resolutions of about $20 - 30$ MeV/c$^2$ for two-body $B$-meson decays, which is of importance to separate the mass peaks of $B^0$ and $B^0_s$ decays into two muons, as well as distinguish decays of the types $B^0_{(s)}\rightarrow h h'$ where $h$ and $h'$ can be pions and kaons, independent of using particle identification information.

\subsubsection{Vertex Locator}
\begin{wrapfigure}{R}{8.2cm}
	\centering
    \hspace*{-0.3cm}\raisebox{0.8cm}{\includegraphics[height=3.5cm]{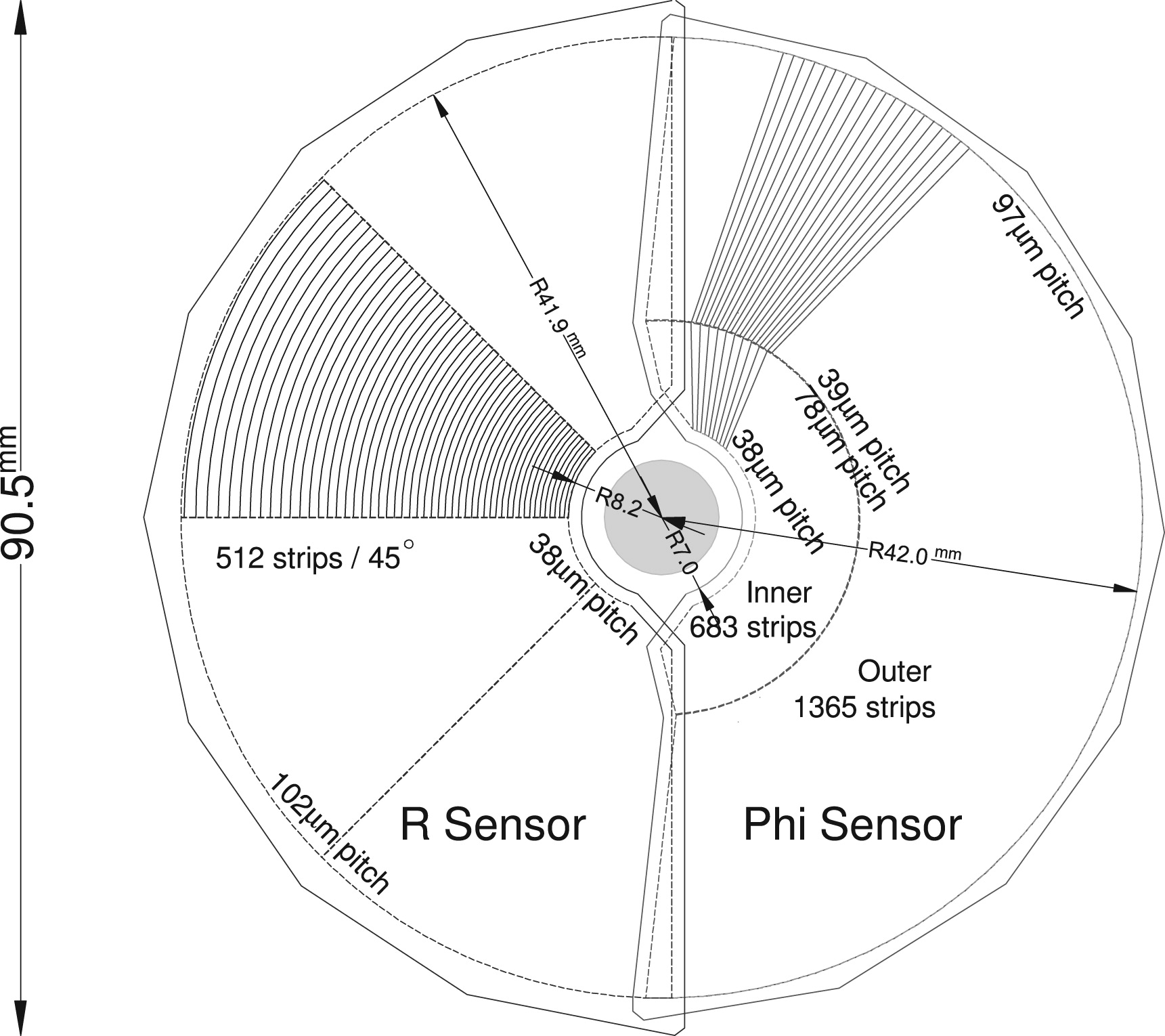}}
    \hspace*{0.2cm}
    \includegraphics[height=4.0cm]{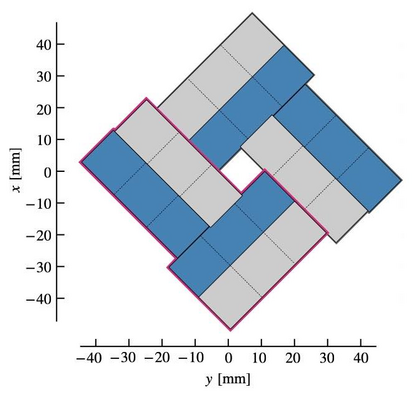}
	\caption{Schematic views of {\em Left:} LHCb-I VELO radial ("R") and azimuthal ("Phi") sensor. Taken from \cite{LHCb-at-LHC-JINST2008}. {\em Right:} VELO LHCb-U pixel sensors on front and back side of modules. Taken from \cite{LHCb-Upgrade-1-JINST2024}.}
	\label{fig:Velo_Layout_XY}
    \vspace{-0.3cm}
\end{wrapfigure}
\begin{wrapfigure}{L}{7cm}
	\centering
    %\vspace*{-1.0cm}
    \includegraphics[height=4.0cm]{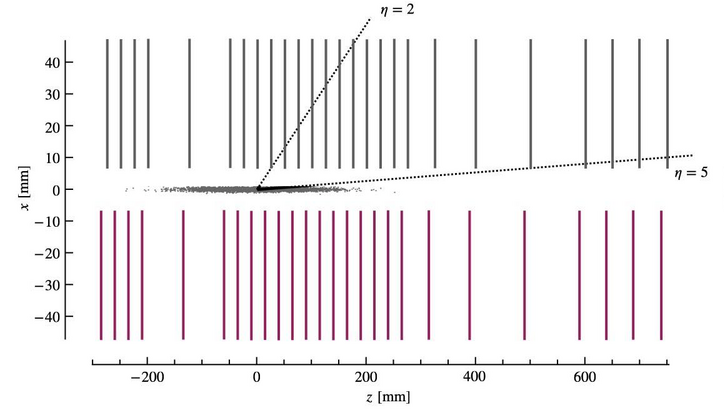}
	\caption{Schematic view of Longitudinal sensor positioning around the interaction region in LHCb-U. Taken from \cite{LHCb-Upgrade-1-JINST2024}.}
	\label{fig:Velo_Layout_Z}
    \vspace{-0.5cm}
\end{wrapfigure}

To accommodate the forward acceptance of LHCb, the Vertex Locator (VELO) is instrumented by a series of stations mounted perpendicular to the LHC beamline, close to the interaction region. The longitudinal size of the interaction region is approximately $\sigma\sim 7.5$ cm along $z$, and the transverse size about $70$ $\mu$m in both $x$ and $y$. In LHCb-I semi-circular silicon strip detectors were used, positioned as close as 8 mm to the beamline, whereas in LHCb-U the strip sensors were replaced by rectangular modules of silicon pixel sensors with enhanced radiation hardness, allowing for a minimal distance of 5 mm to the beam. 
The VELO in LHCb-I was designed with a criterion that each track in the acceptance traverses four stations leading to a design of 23 stations including back-to-back mounted wafers with $R$ and $Phi$ oriented silicon strips, whereas a similar criterion in the LHCb-U VELO led to 26 stations of Silicon pixels, due to a slightly smaller transverse size. The motivation of the $R$ and $Phi$ geometry in LHCb-I was to allow for fast two-dimensional track reconstruction in the $r$-$z$ plane in the trigger using only the hits of $R$-strips. In LHCb-U, 55 $\mu$m sized pixels provide three dimensional hits, enabling improved reconstruction performance in the trigger, and faster pattern recognition. Schematic views of LHCb-I and LHCb-U sensors are shown in Fig.~\ref{fig:Velo_Layout_XY}. The layout of the stations along the beam for LHCb-U is shown schematically in Fig.~\ref{fig:Velo_Layout_Z}.

Due to their proximity to the beam, VELO sensors must withstand high radiation levels and operate at sub-zero temperatures to prevent aging effects. A CO$_2$-based dual-phase cooling system using liquid CO$_2$ evaporation ensures the sensors and electronics remain at sub-zero centigrade temperatures. In the LHCb-I detector, sensors were mounted on graphite-based cooling plates \cite{LHCb-at-LHC-JINST2008}, whereas LHCb-U employs a novel approach where CO$_2$ flows through microchannels embedded in a silicon substrate \cite{LHCb-Upgrade-1-JINST2024}.

To allow safe injection of the LHC beams, the LHCb VELO halves are retracted in between fills to a safe distance of 3 cm. Once stable beam conditions are achieved, the detectors are carefully closed while monitoring backgrounds and beam positions. The closing procedure allows the detector to be positioned such that the beams pass through the center of the detector. The reproducibility of the positioning is better than 10 $\mu$m \cite{Velo-Performance-JINST-2014}. Although the VELO detector operates in vacuum, a thin aluminum encapsulation separates the detector vacuum ($\sim10^{-4}$ mbar) from the beam vacuum ($<10^{-8}$ mbar). This encapsulation also serves as an RF shield to guide the wake field of the passing beams. In LHCb-I it was constructed by plastic deformation of 300 $\mu$m thick aluminum foils, whereas in LHCb-U it was precision-milled from a massive aluminum block and finally thinned via chemical etching to a thickness of 150 $\mu$m. 
Due to a vacuum incident at the startup in 2023 the RF-shield was damaged, such that the VELO in LHCb-U could not be operated in the closed position during 2023 data-taking, affecting the quality of the acquired data. The detector box was replaced and the detector was fully operational at the start of 2024. At the time of writing (2025) it operates at 5 mm from the beam and manages to reach the design specifications enabling high precision tracking in the High Level Trigger. 

\begin{figure}[t]
	\centering
    \includegraphics[height=5cm]{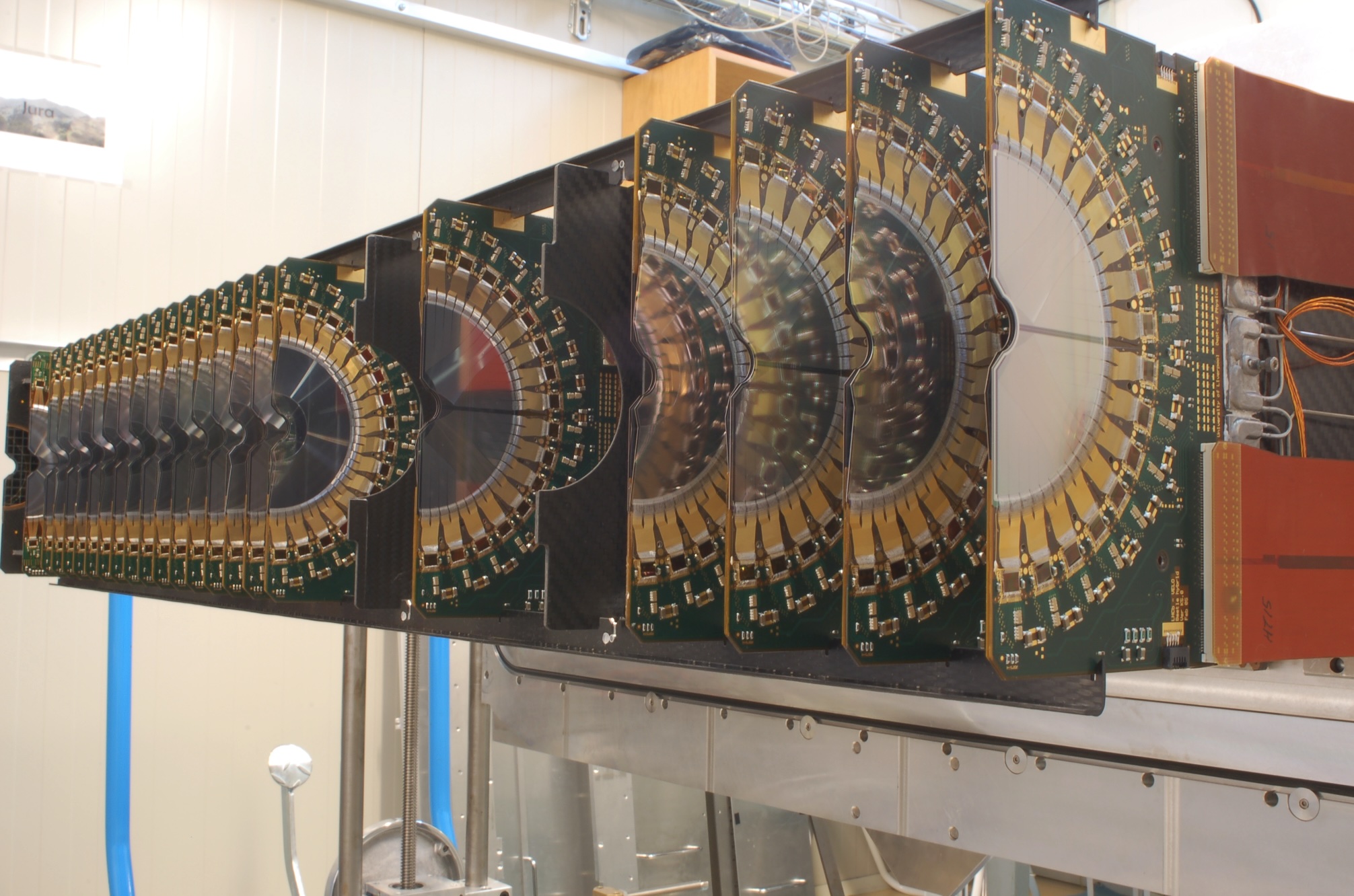}\hspace*{1.0cm}
    \includegraphics[height=5cm]{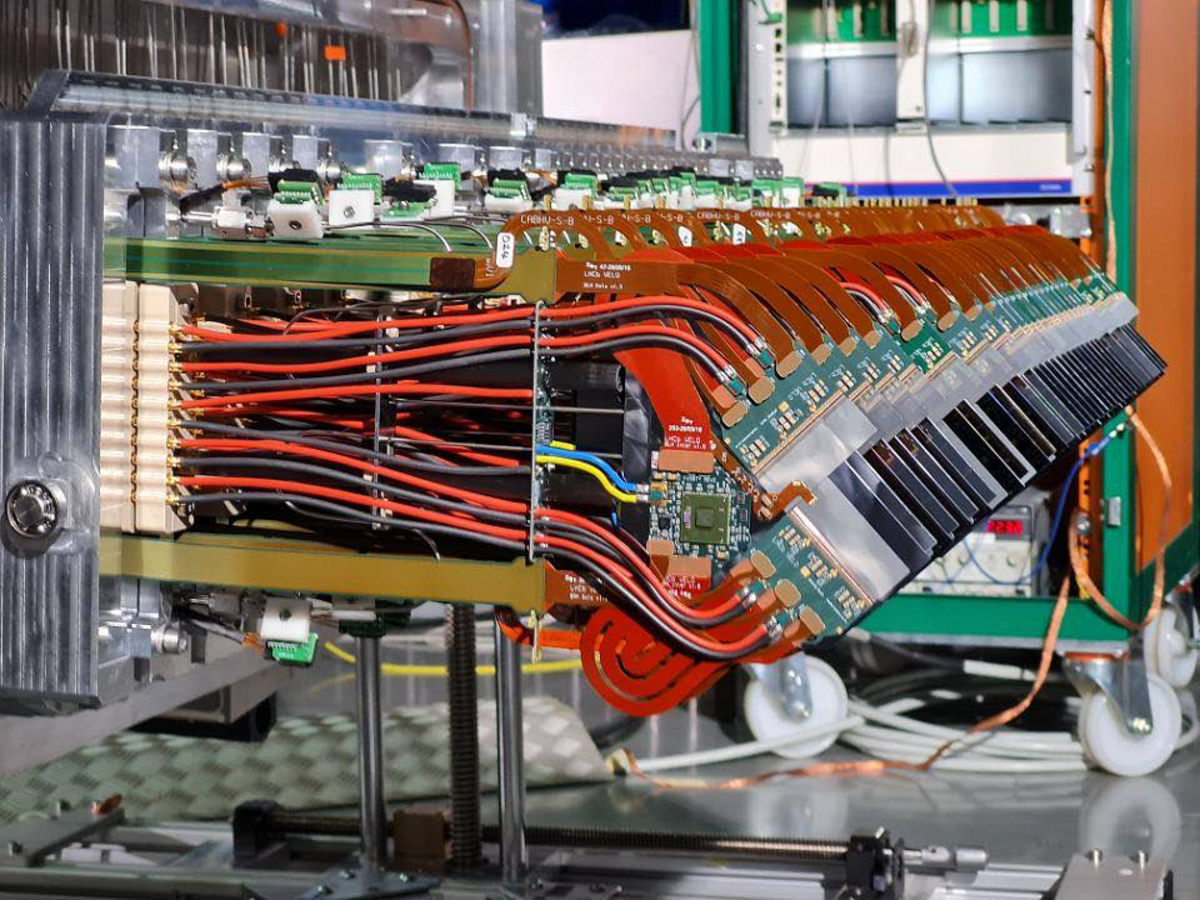}
 	\caption{The VELO half detector prior to installation. {\em Left:} LHCb-I detector with "R"- and "Phi"- strip wafer stations. {\em Right:} LHCb-U detector with pixel stations. Taken from \cite{LHCb-Upgrade-1-JINST2024}.} 
    %\vspace*{-0.3cm}
	\label{fig:Velo_Detector}
\end{figure}
%\vspace*{-0.5cm}

The VELO detector in LHCb-I operated according to specifications during Run-1 and Run-2. The number of inefficient strips was less than 1\% and the number of noisy channels was negligible ($\sim 0.02\%)$. The sensors have a signal to noise ratio of about 20, which was kept stable during the runs by raising the bias voltage to compensate for the gradually rising depletion voltage. The hit resolution of strips closest to the beam varies between 5 - 10 $\mu$m as function of the impinging track angle. Photographs of the LHCb-I and LHCb-U VELO half-detectors prior to their installation are shown in Fig.~\ref{fig:Velo_Detector}.

\subsubsection{Upstream Tracker}

After re-optimizing the LHCb tracking system to include an empty magnet region, a robust tracking station with good resolution, positioned downstream of the VELO and just upstream of the dipole magnet, proved essential. The Silion based upstream tracking station ("TT" in LHCb-I and "UT" in LHCb-U) serves to reconstruct long-living particles ($K^0_S, \Lambda$) that decay outside the VELO volume, as well as low momentum particles that bend out of the fiducial volume and do not reach the downstream tracking system.
Furthermore, making use of the magnetic fringe field between the VELO and this upstream station, an initial particle momentum estimate is obtained that is used to speed up real-time track-finding in the downstream detector by reducing the size of search windows for track extrapolation from VELO towards the downstream detectors.

%\begin{figure}[h]
\begin{wrapfigure}{R}{10cm}
    %\vspace*{-0.5cm}
    \centering
    \hspace*{-0.3cm}\includegraphics[height=6cm]{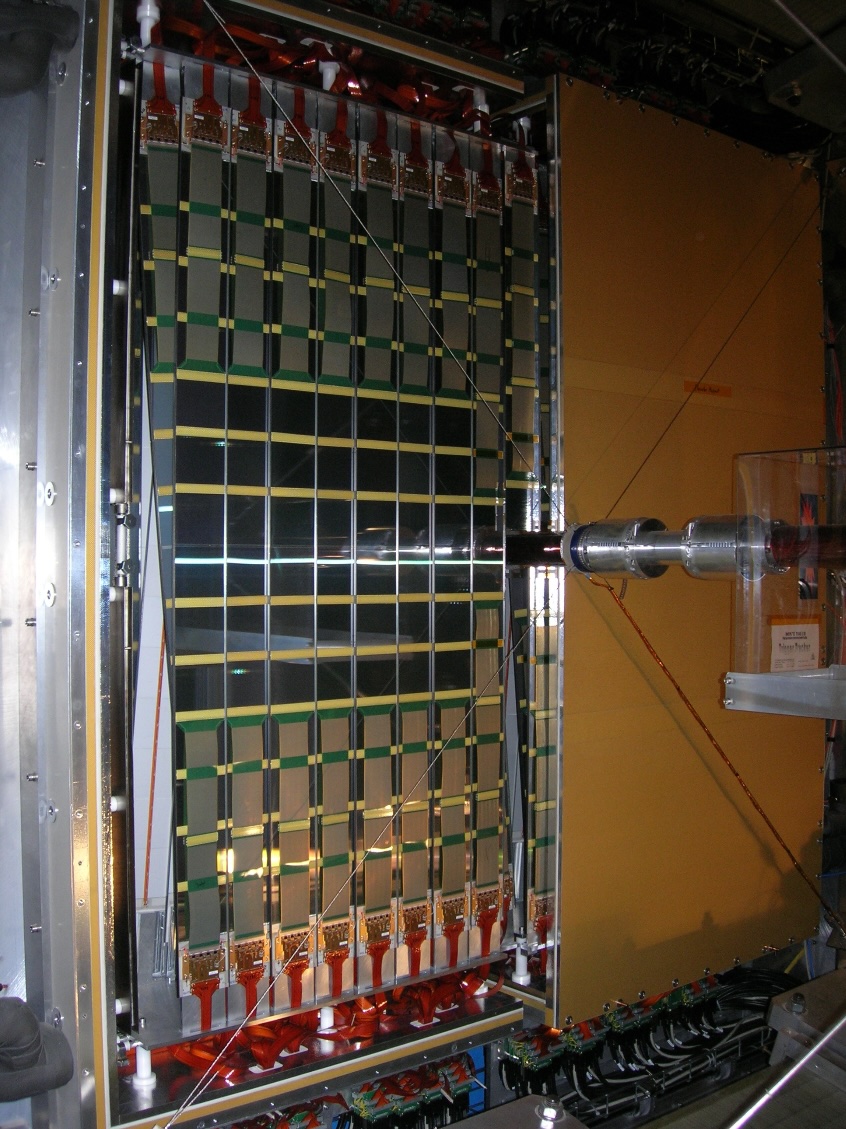}
    \hspace*{0.2cm}
    \includegraphics[height=6cm]{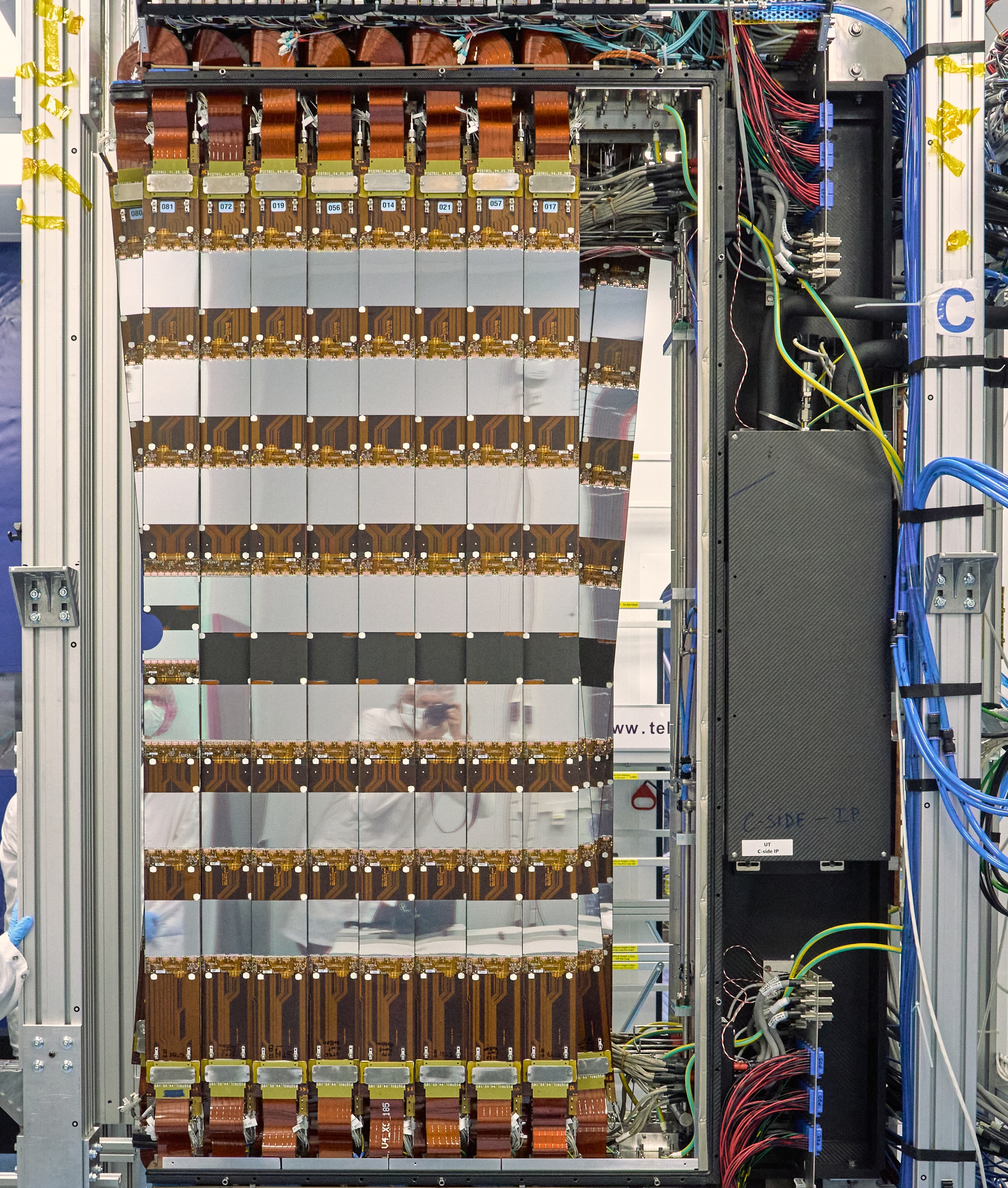}
	\caption{Photographs of {\em left:} the LHCb Trigger Tracker \cite{ST-Installation-NIMA-2010} and {\em right:} the Upstream Tracker (CERN photograph), prior to their installation. } 
	\label{fig:UpstreamTrackerPhotos}
    %\vspace*{-0.5cm}
    \end{wrapfigure}
\begin{figure}[t]
	\centering
    \includegraphics[height=5.5cm]{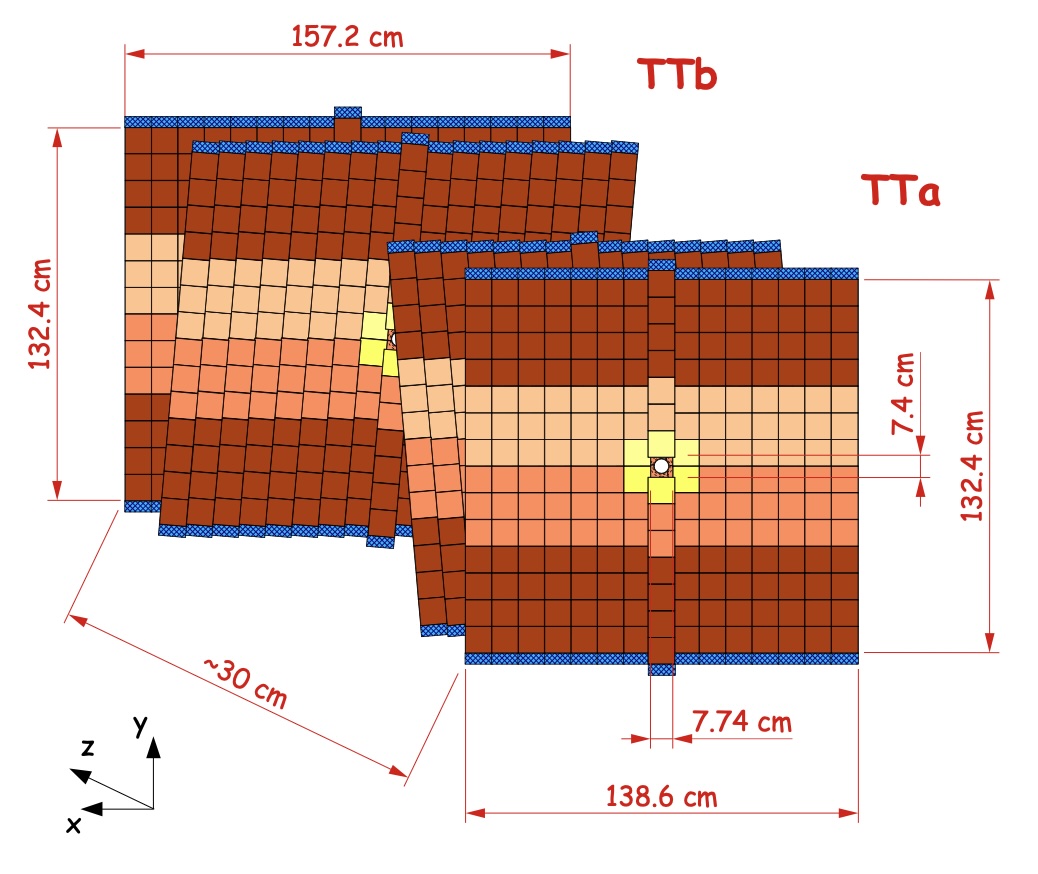}\hspace*{0.5cm}
    \includegraphics[height=5.5cm]{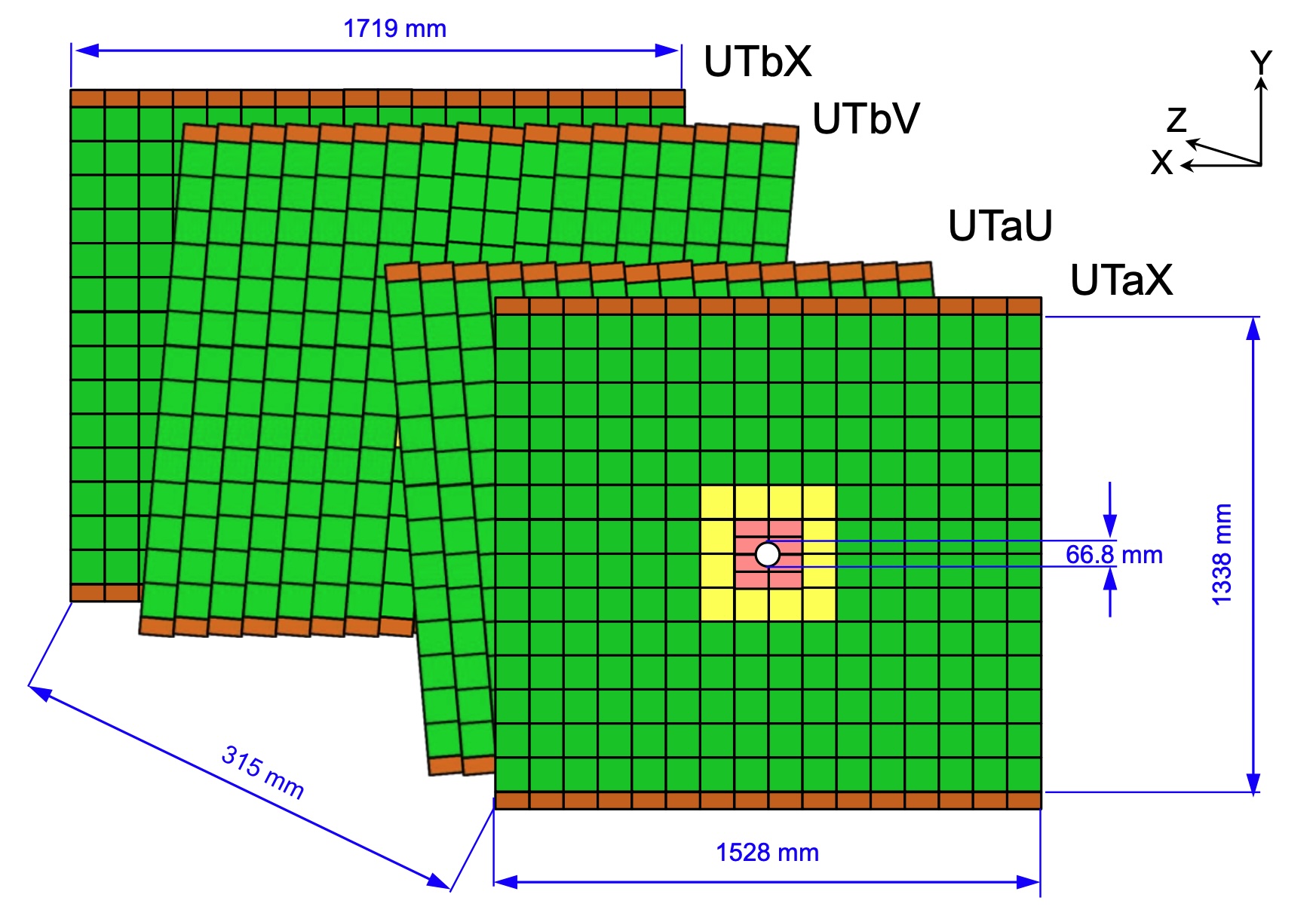}
	\caption{Schematic layout of four detector planes with stereo views $x$-$u$-$v$-$x$ of {\em left:} the LHCb-I Trigger Tracker \cite{ST-Performance-NIMA-2013} and {\em right:} the LHCb-U Upstream Tracker \cite{LHCb-UpgradeTracker-TDR}. } 
	\label{fig:UpstreamTrackerSchematic}
\end{figure}

%\end{figure}
For pattern recognition purposes the tracking station includes four measurement planes, labeled as $x$-$u$-$v$-$x$ with respective stereo angles of $0, -5^o, +5^0, 0$ of the strips with respect to the vertical $y$-axis. The detectors are mechanically constructed in two detector halves $"a"$ and $"b"$, separated by $\sim 30$ cm distance. The setup is illustrated in Fig.~\ref{fig:UpstreamTrackerSchematic}.

In LHCb-I, the upstream tracker ("TT" for "Trigger Tracker" or "Tracker Turicensis") is a silicon strip detector consisting of $500$ $\mu$m thick silicon sensors providing a hit resolution of about \mbox{50 $\mu$m}. The sensors are mounted in ladders of one, two, three or four sensors bonded together depending on their position in the detector and the corresponding particle density (see Fig.~\ref{fig:UpstreamTrackerSchematic}). To accomodate the increased track densities in LHCb-U, the silicon strip sensors were replaced by more fine grained, thinner ($250 \mu\mathrm{m}$), and more radiation hard silicon strip sensors mounted on vertical staves. In addition, the readout electronics were replaced to be able to operate in 40 MHz readout mode. Photographs of both the TT and UT detectors prior to installation are shown in Fig.~\ref{fig:UpstreamTrackerPhotos}.

%\vspace*{-0.5cm}
%\vspace*{-0.5cm}

\subsubsection{Downstream Tracker}

The downstream tracking system reconstructs the particle trajectories traversing the full spectrometer and determines their momenta measuring the particle's deflection in the dipole magnetic field. It consists of three tracking stations using the same $x$-$u$-$v$-$x$ stereo configuration as the upstream tracker, spanning the magnet free region behind the dipole magnet up to the RICH-2 detector.
In LHCb-I, about 98\% of the detection surface is equipped by the Outer Tracker (OT) with 5 mm size drift-tube straw detectors, while the remaining 2\%, surrounding the beam pipe is equipped with silicon strip detectors of the Inner Tracker (IT). A more fine grained technology is used here since this relatively small region experiences high track densities and observes about a quarter of all tracks. 
%In LHCb-U the increased luminosity would lead to too high hit occupancy in the Outer Tracker, which therefor needed to be replaced. To further reduce inactive material of the IT mounting and cooling infrastructure, the OT and IT together detectors were replaced by single Scintillating Fiber Tracker (SciFi) covering the full surface with scintillating fibers of $250 \mu\mathrm{m}$. 

\begin{figure}[h!]
    \begin{picture}(14cm,6cm)(0cm,0cm)
        \put(0cm,0.3cm){\includegraphics[height=5cm]
        {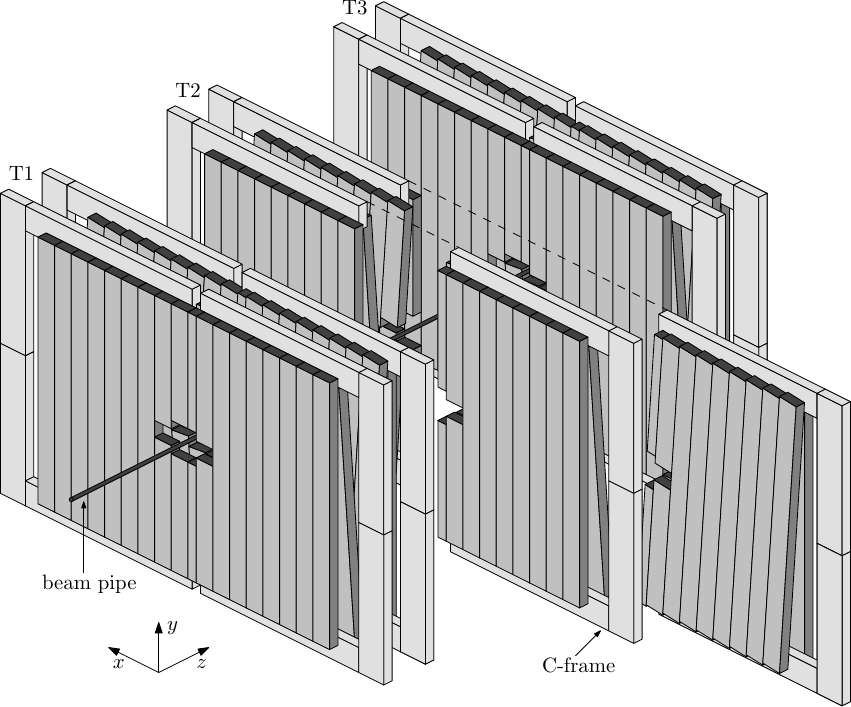}}
        \put(6.2cm,0cm){\includegraphics[height=7.5cm, trim=0cm 0cm 10cm 10cm, clip]{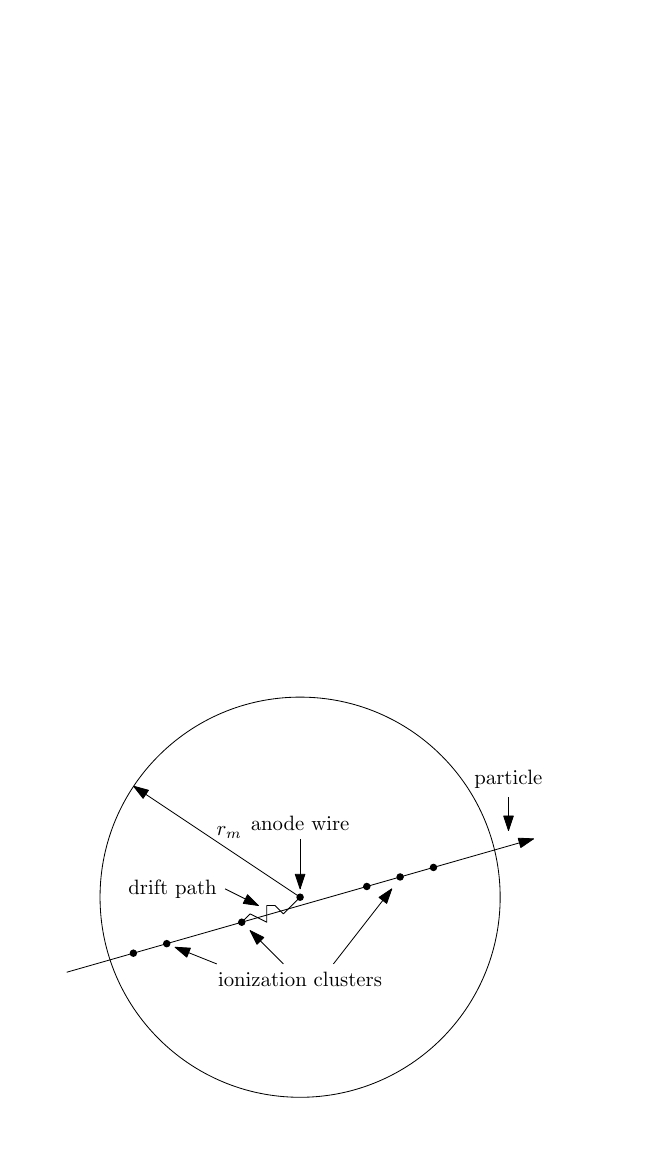}}
        \put(5.4cm,3.3cm){\includegraphics[height=2.2cm]{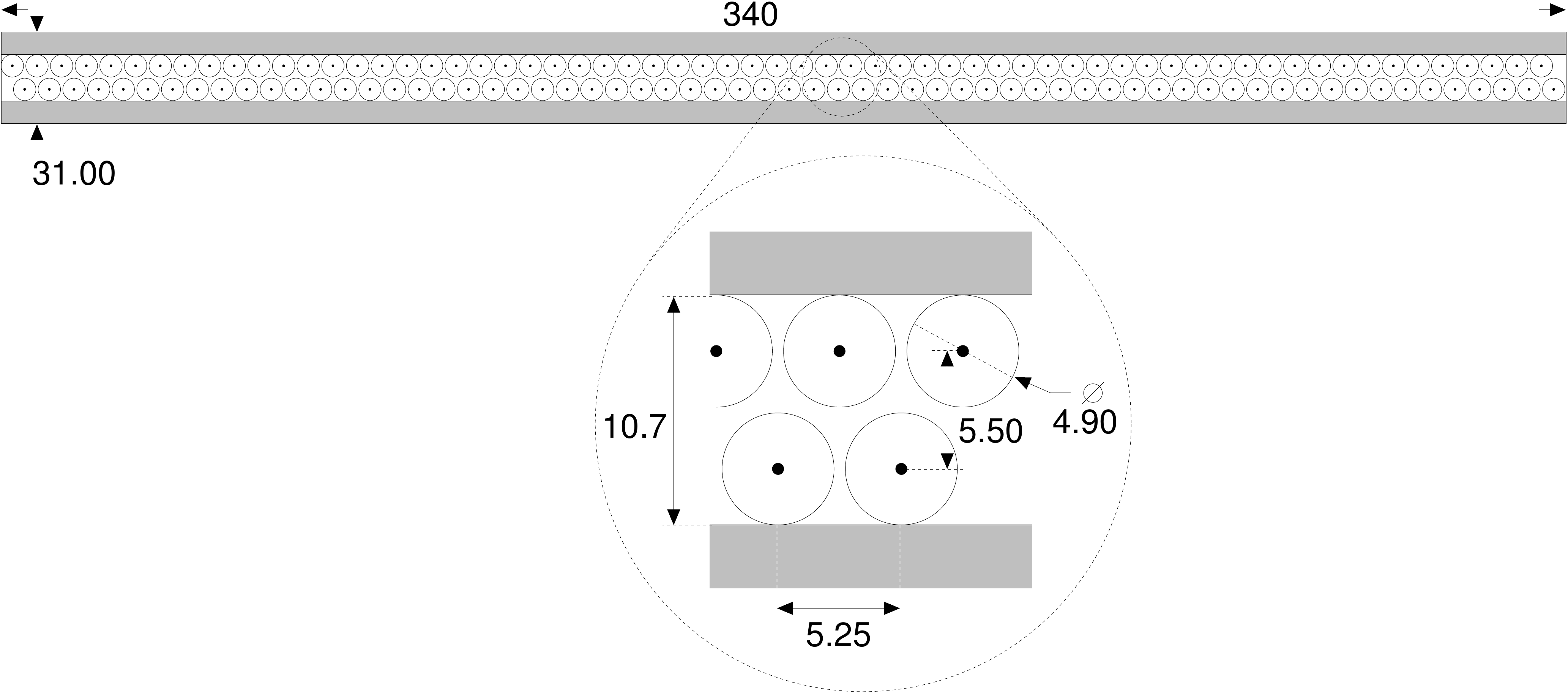}}
        \put(10.7cm,0.5cm){\includegraphics[height=4.0cm]{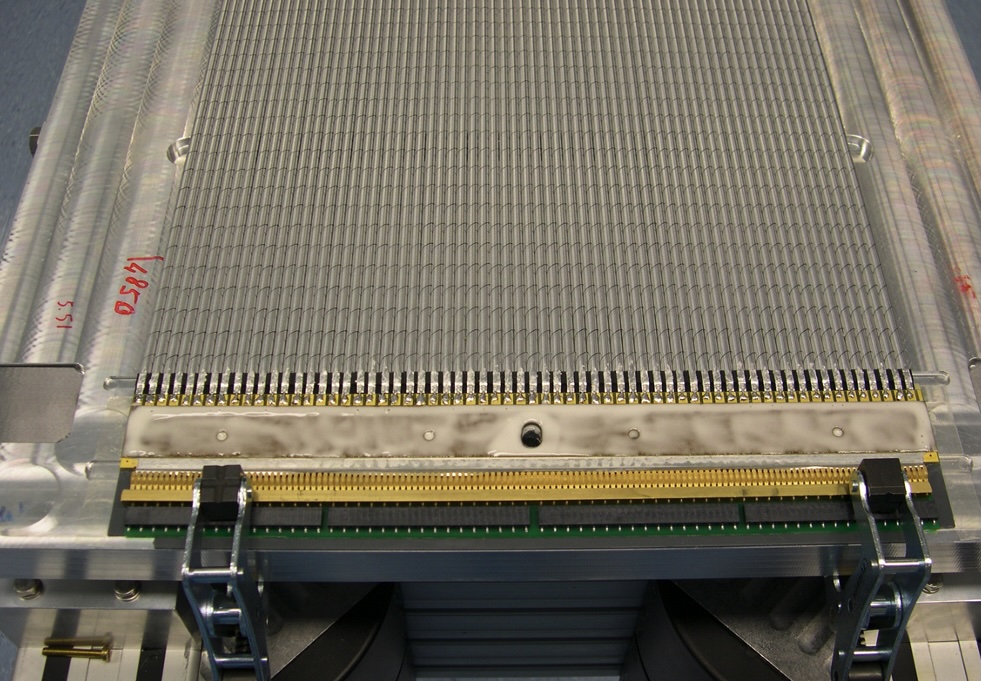}}
    \end{picture}
	\caption{{\em Left:} Isometric view of the three Outer Tracker stations, where one side of the second station is displaced in the opened position. {\em Center:} Detection principles of the straw tube drift detector \cite{OT-Performance-Run2}. {\em Right:} View of the drift tube straws in inside an OT module \cite{LHCb-TrackingSystem-NIMA-2007}.} 
	\label{fig:OuterTracker}
\end{figure}

\begin{figure}[t]
	\centering
    \includegraphics[height=4.5cm]{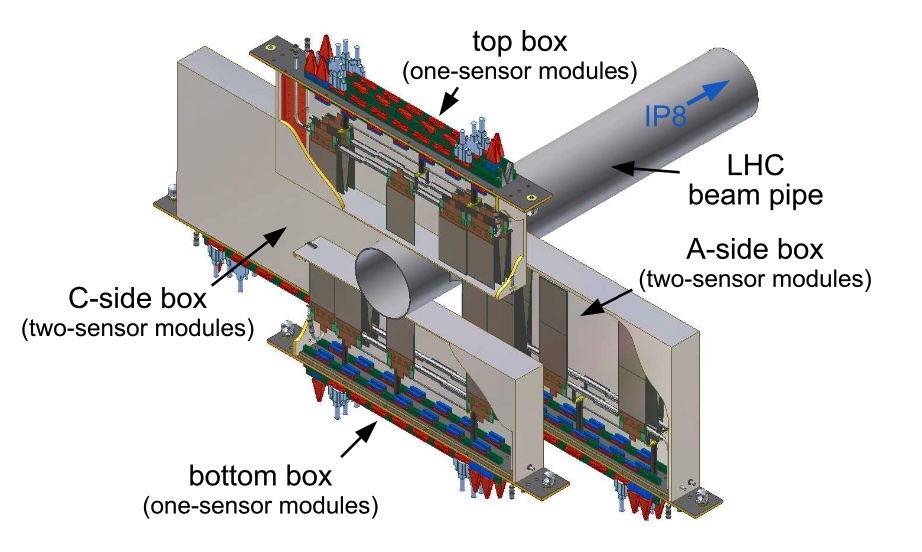}
    \hspace*{1.0cm}
    \includegraphics[height=4.5cm]{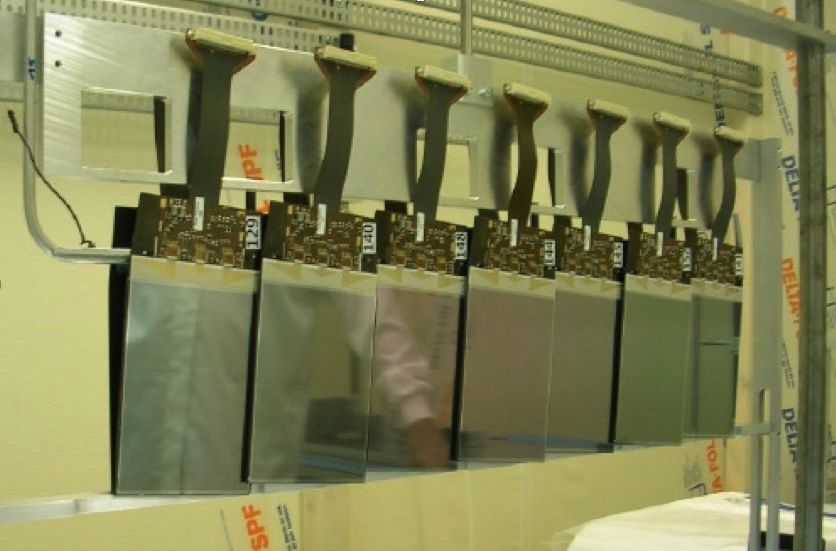}
	%\includegraphics[height=4.2cm]{Figures/Detector/OuterTrackerDeinstallation_cropped.jpg}
%    \hspace*{0.5cm}
% 	\includegraphics[height=4.2cm]{Figures/Detector/LHCb_SciFi_plane.jpg}
	\caption{{\em Left:} Isometric view of a single Inner Tracker detector station consisting of four detector boxes \cite{LHCb-at-LHC-JINST2008} {\em Right:} Photograph of the Si modules during their construction \cite{TrackingPerformance-Tobin-2012}. } 
	\label{fig:InnerTracker}
\end{figure}

The OT straw modules (see Fig.~\ref{fig:OuterTracker}) comprised two staggered monolayers of gas-filled straws measuring the drift time of the ionized electrons reaching the straw cathode. The drift gas mixture, composed of Ar (70\%)/CO$_2$ (28.5\%) and O$_2$ (1.5\%), was chosen to provide a drift distance resolution of $200$ $\mu$m within a maximum drift time of 50 ns. The latter corresponds to two bunch crossings (with 25 ns spacing of the bunches), leading to "spill-over" of registered detector hits from neighboring bunch interactions.
Although the Outer Tracker straws were close to fully efficient, the high channel occupancy in the central region of the detector leads to losses related to multiple tracks traversing single straws. In this region, a finer granularity is provided by the silicon strips of the IT providing a hit resolution of about 50 $\mu$m, significantly better than the surrounding OT. This resolution is needed for the momentum measurement of high momentum particles, which predominantly pass through the inner region. The IT covered a "swiss-cross" shaped region of overall 120 cm wide and 40 cm high, in each of the three tracking stations. An isometric view of one IT station, including four boxes of silicon ladders is shown in Fig.~\ref{fig:InnerTracker} together with a picture of a detector box prior to installation.

\clearpage

\begin{figure}[!ht]
	\centering
%    \includegraphics[height=3.5cm]{Figures/Detector/OTStrawPrinciple.jpg}
%    %\hspace*{0.1cm}
    \includegraphics[height=5.5cm]{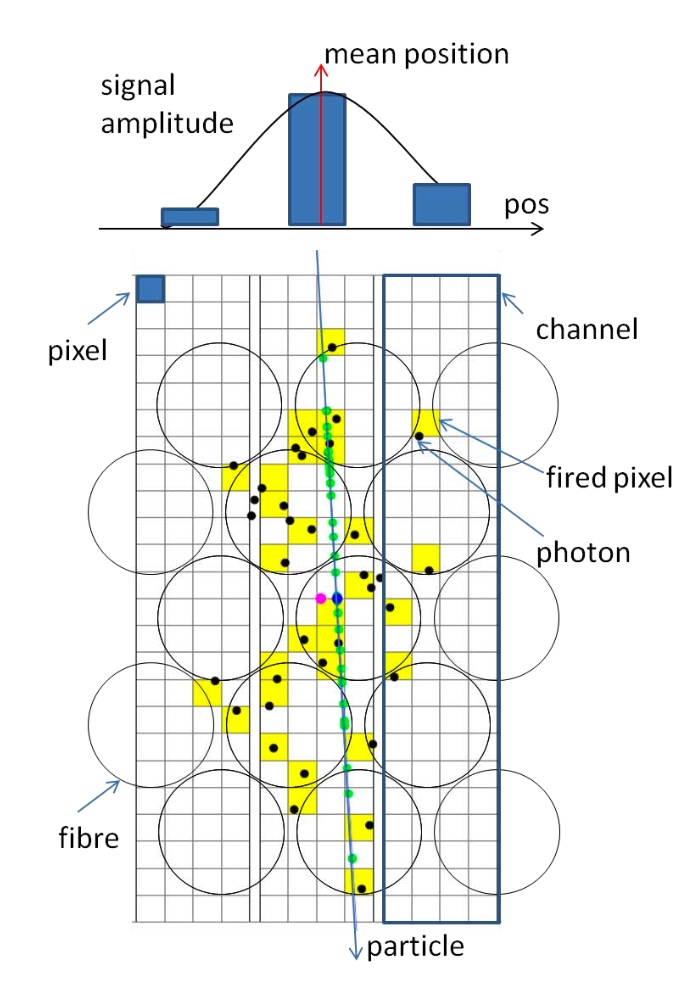}
    \hspace*{0.5cm}
    \includegraphics[height=5.5cm]{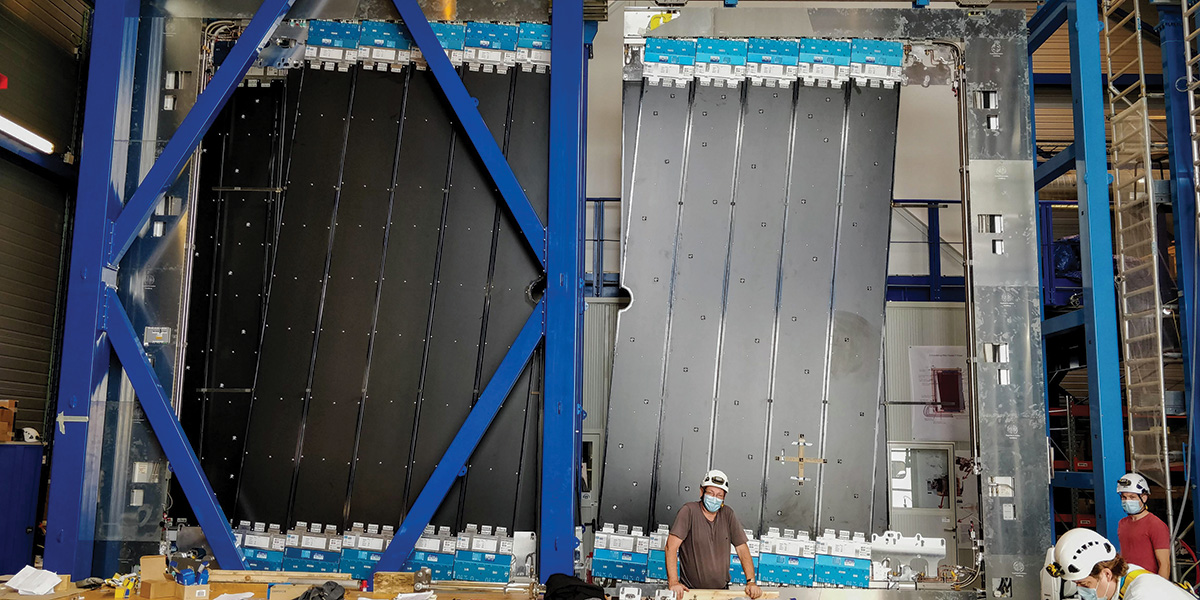}
	\caption{{\em Left:} Detection principle of the fine-grained scintillating fibers read out by SiPMs \cite{LHCb-UpgradeTracker-TDR}. The image schematically illustrates a track traversing the stack of fibers, the generated scintillation photons together with the response of the SiPM. {\em Right:} Photograph of SciFi detector plane prior to installation. The black surface represents the active detector region, the blue ends house the readout electronics and cooling system. Credit B. Leverington.}
    \label{fig:SciFi}
\end{figure}

To accommodate the significantly increased track multiplicity in Run-3, LHCb-U  replaced the OT straw tubes with a scintillating fiber tracker (SciFi), offering 20 times finer granularity. The SciFi detector layers consist of stacked scintillating fibers with a pitch of 250 $\mu$m, read out by silicon photomultipliers (SiPMs) as shown in the left panel of Fig.~\ref{fig:SciFi}, resulting in a photon cluster resolution of about 100 $\mu$m. This single-technology approach eliminates the need for a dedicated IT, reducing material in the acceptance region and improving momentum resolution. Both fiber transparency in high-radiation areas and SiPM radiation hardness were validated for LHCb Run-3 and Run-4, provided the SiPMs are cooled to about $-50^\circ$C. A photograph of the SciFi modules, mounted in their C-shaped frames, is shown in the right panel of Fig.~\ref{fig:SciFi}.

\subsubsection{Dipole Magnet}

\begin{figure}[ht!]
	\centering
    \includegraphics[height=4.5cm]{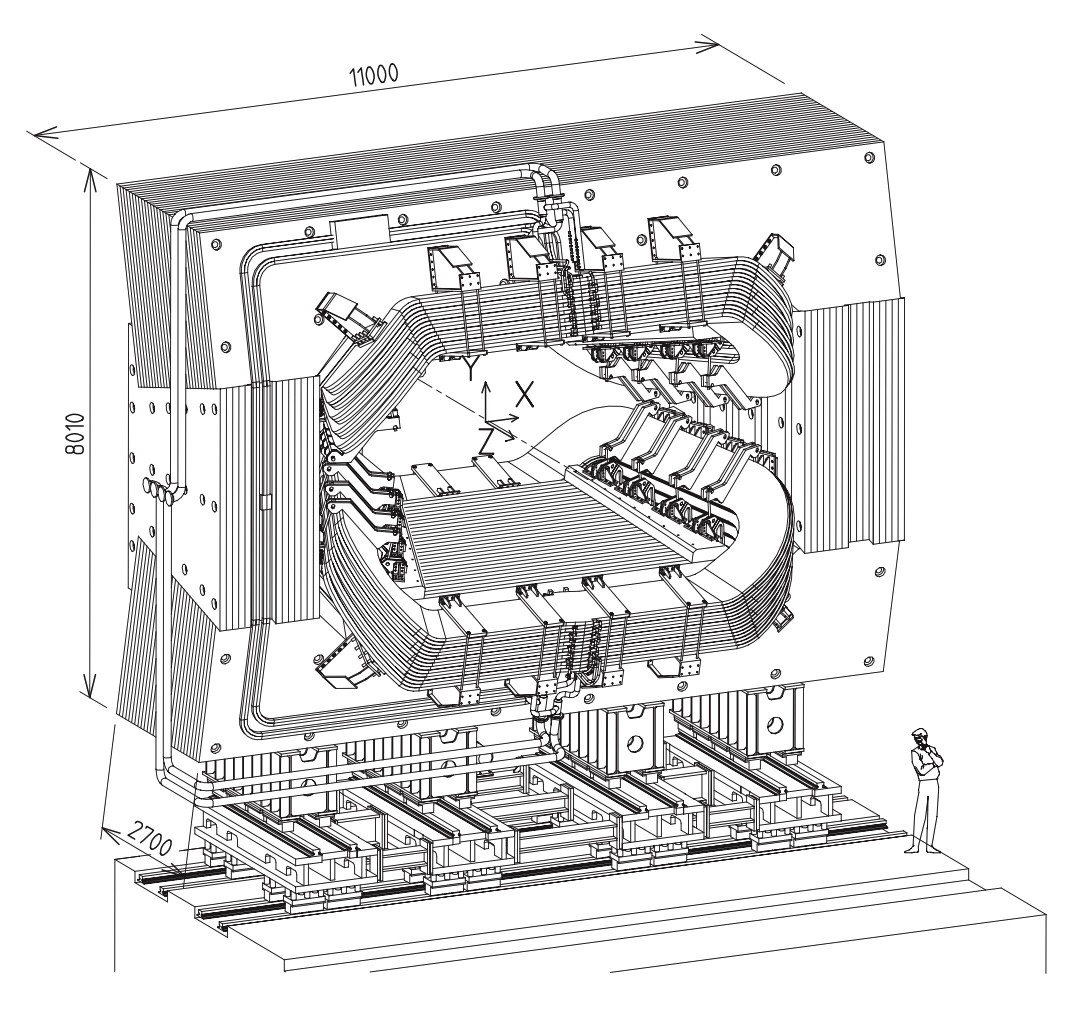}
    \hspace*{0.5cm}
	\includegraphics[height=4cm]{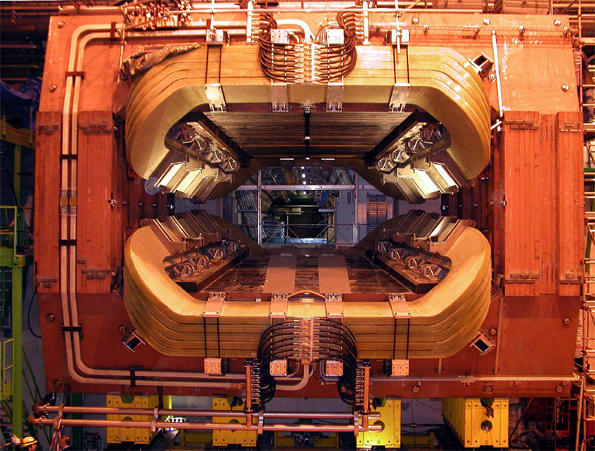}
    \hspace*{0.5cm}
    \includegraphics[height=4.2cm]{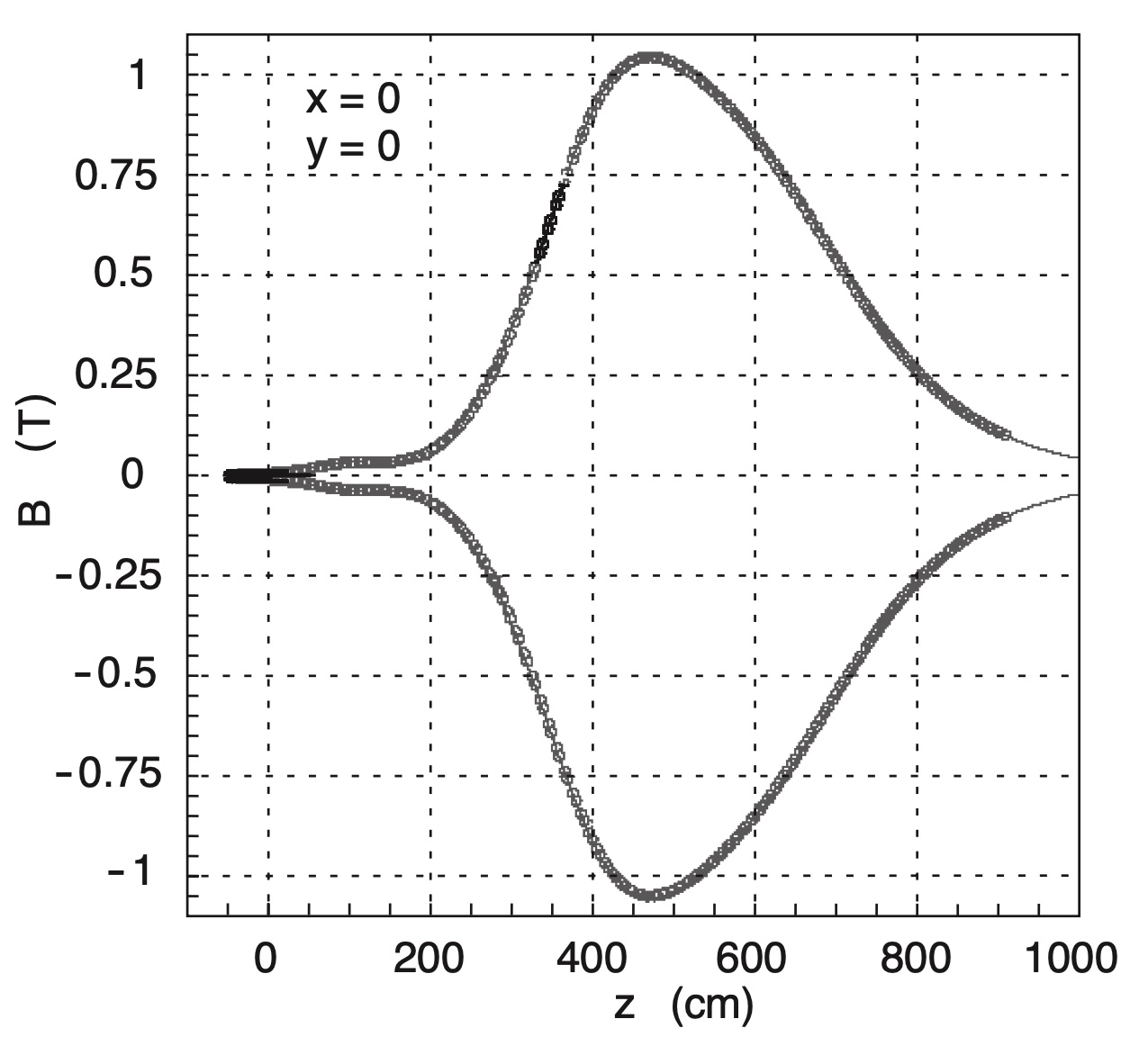}
	\caption{{\em Left:} Schematic view of the LHCb dipole magnet. {\em Center:} Picture of the LHCb magnet after installation. The magnet coil is visible in yellow/lightbrown color and the yoke in red/brown. {\em Right:} Vertical (main) component of the B-field ($B_y$) along the beam axis ($x=y=0$) for both polarities \cite{LHCb-at-LHC-JINST2008}. } 
	\label{fig:Magnet}
\end{figure}

Charged particle tracks are deflected within LHCb's large dipole magnet, which consists of two vertically mirror-symmetric saddle-shaped coils embedded in a massive iron yoke (see Fig.~\ref{fig:Magnet}). The coils are sloped to match the LHCb spectrometer acceptance of 300 mrad $\times$ 250 mrad. The magnet generates a 4 Tm field integral in the vertical direction as observed by particles traversing the full spectrometer.

The field is highly inhomogeneous and contains non-negligible $B_x$ and $B_z$ components in addition to the dominant $B_y$ field. To achieve high precision momentum measurements, the magnetic field has been extensively mapped in three dimensions across the entire tracking volume. The magnetic field is known with a precision of $\mathcal{O}(10^{-4})$, ensuring permille-level momentum resolution.

Particle deflections occur predominantly in the horizontal plane, causing positively  (negatively) charged particles to preferentially be directed towards negative (positive) $x$ of the downstream detector, if the magnetic field points upwards. The direction of the magnet's current can be reversed, generating opposite field polarities. This allows data collection in both configurations, facilitating systematic studies of detector symmetries and charge-dependent effects.

\clearpage

\subsection{Particle Identification Detectors}

Extensive particle identification (PID) is a key feature of the LHCb spectrometer, enabling efficient separation of different particle types.
The Ring Imaging Cherenkov (RICH) detectors distinguish between charged hadrons ($\pi, K, p$), the calorimeter system to identifies electrons and reconstructs neutral particles ($\gamma, \pi^0$), while the muon stations provide muon identification. In LHCb-I the calorimeter and muon systems provided trigger signals for the hardware (level-0) trigger, whereas in LHCb-U this is no longer needed since the full spectrometer is readout and events are (partially) reconstructed at 40 MHz.

\subsubsection{Ring Imaging Cerenkov Detectors}

When a charged particle traverses a medium with a velocity exceeding the speed of light in that medium, it emits Cherenkov radiation at an angle given by: $\cos\theta_C = 1/\beta n$, where $n$ is the refractive index of the medium and $\beta$ the velocity as a fraction of the lightspeed. For velocities above the threshold $v=c/n$ the charged particle starts radiating at small angles and asymptotically approaches $\cos\theta_c=1/n$. 
By utilizing spherical mirrors, photons emitted under a given angle are focused onto a detection plane forming a ring-shaped image around the emitting particle track. The radius of the ring provides a direct measurement of the particle velocity, which together with the particle's momentum, obtained from the tracking system, determines its mass.

\begin{figure}[!ht]
	\centering
  	\includegraphics[width=3.5cm]{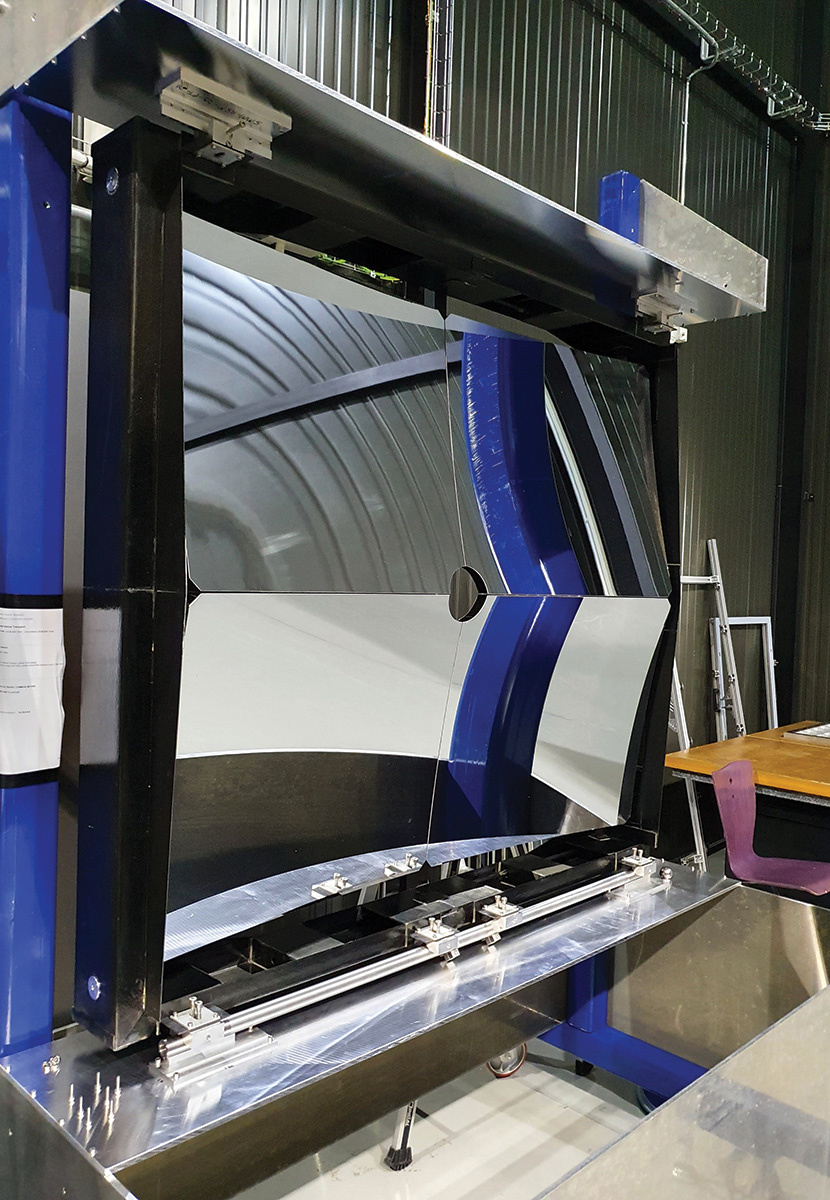}
    \hspace*{1.0cm}
 	\includegraphics[width=9cm]{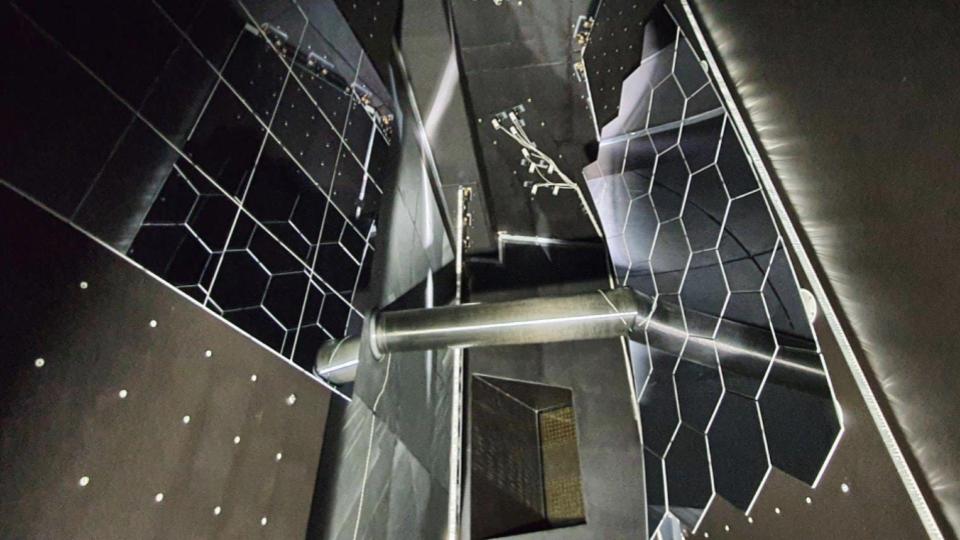}
	\caption{{\em Left:} Photo of the spherical RICH-1 mirrors prior to installation (CERN Document Server).
    %: OPEN-PHO-EXP-2021-010.) 
    {\em Right:} Photo of the spherical RICH-2 mirrors with hexagon shapes on the right and the planar mirrors on the left. At the bottom center the photon detector system can be seen (CERN Document Server).}
    %: LHCb-PHO-MISC-2021-004. }
 	\label{fig:RICH_Photos}
\end{figure}

Identification of pions, kaons and protons over a broad spectrum is achieved with two Ring Imaging Cherenkov (RICH) detectors: RICH-1 for low momentum particles and RICH-2 for high momentum particles. 
Kaon-pion separation is essential for distinguishing final states such as $B^0_{(s)}\rightarrow \pi^+\pi^-, K^+\pi^-,K^+K^-$, as well for background suppression in various analyses, as for example in selecting $B^0_s\rightarrow \phi\phi\rightarrow K^+K^-K^+K^-$ decays. Additionally, in flavour tagging, RICH information is used to identify the flavour of the produced $b$-quark ($b$ or $\bar{b}$), either through the charge of kaons produced close in phase space to the signal $b$-quark ("same-side tagging"), or through the charge of kaons produced in the decay chain of the other $b$-quark ("opposite-side tagging").
Photos of the RICH-1 and RICH-2 detectors prior to installation are shown in Fig.~\ref{fig:RICH_Photos}.

\paragraph{RICH-1}
The RICH-1 detector, positioned upstream in the spectrometer between the VELO and the Upstream Tracker, measures the Cherenkov angle using a C$_4$F$_{10}$ gas radiator for particles up to 60 GeV. 
In the LHCb-I setup, an additional aerogel radiator was included for very low momentum particles, which was removed in the LHCb-U to increase the radiator volume for C$_4$F$_{10}$, enhancing the photon yield, and to reduce material interactions.

The RICH-1 detector is installed around the $25$ mrad conical part of the beam pipe and does not cover the high rapidity region of very high momentum particles. 
The C$_4$F$_{10}$ medium has a refractive index of $n=1.0014$ and an effective radiator length of almost $1$ m, resulting in an average of about 30 Cherenkov photons for relativistic charged particles.

\begin{figure}[!ht]
	\centering
  	\includegraphics[width=3.5cm]{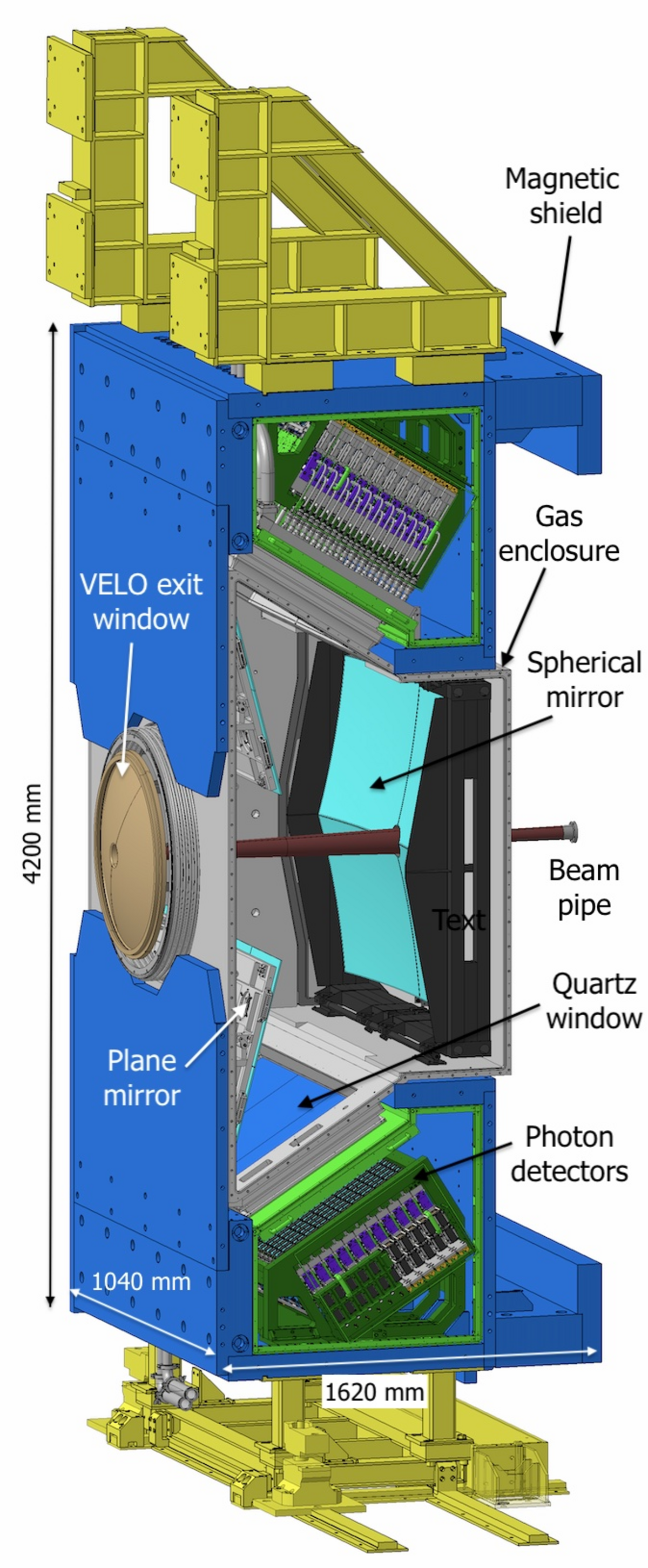}
    \hspace*{1.5cm}
 	\includegraphics[width=7cm]{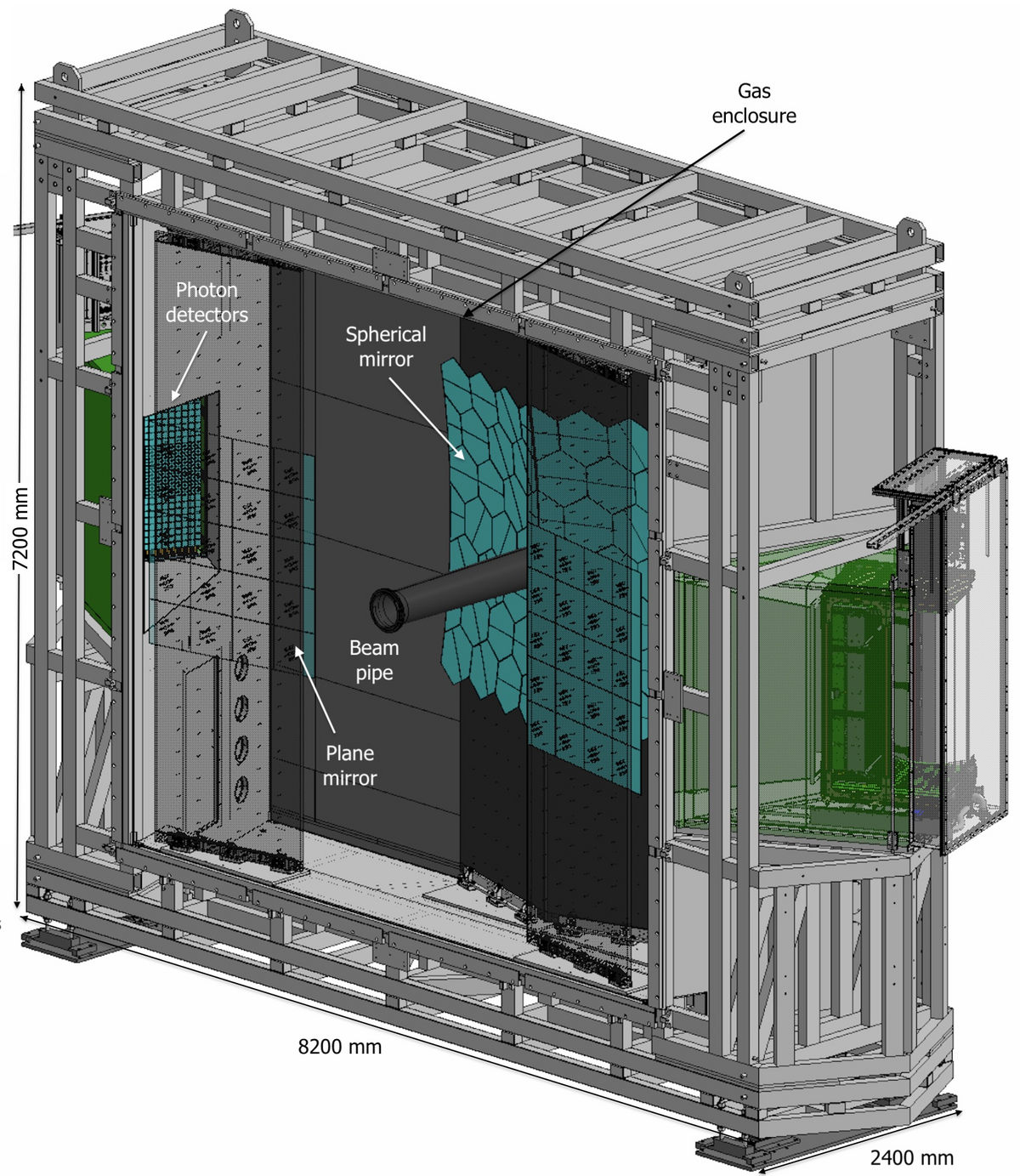}
	\caption{Schematic view of the RICH detectors. {\em Left:} RICH-1, where particles enter at the left side via the VELO exit window. Emitted photons are reflected by the spherical mirror and a planar mirror onto photodetectors at the top and bottom. {\em Right:} RICH-2, where the spherical mirrors bring the photons to the photon detectors at the left and right side. Taken from \cite{RICH-Upgrade-Commisioning-JINST}.} 
    %©2022 IOP Publishing Ltd and Sissa Medialab. All rights reserved. }
 	\label{fig:RICH_Schematic}
\end{figure}

The emitted photons are focused by lightweight spherical mirrors and then directed onto the readout system using planar mirrors, see Fig.~\ref{fig:RICH_Schematic}.  The readout system is positioned outside the fiducial volume and used Hybrid Photon Detectors (HPDs) in LHCb-I, which could operate up to 1.1 MHz readout. For the LHCb-U with 40 MHz readout, the HPDs were replaced by higher-granularity Multi-Anode Photomultiplier Tubes (MaPMTs), which were validated for sufficient radiation tolerance.

\paragraph{RICH-2}

The RICH-2 detector uses CF$_4$ gas as radiator medium, providing particle identification for particles in the momentum range of $15\:\mathrm{GeV} < p < 100\: \mathrm{GeV}$. Its acceptance is optimized for high momentum particles and covers polar angles from 15 mrad up to 120 mrad in the horizontal plane and 100 mrad in the vertical plane. Unlike RICH-1, the RICH-2 mirrors guide the photons to optical detectors located in the horizontal plane, where they are shielded from the magnetic field.
The CF$_4$ medium has a refractive index of $n=1.0005$ and an effective radiator length of almost 2 m, resulting in a total of about 22 Cherenkov photons for $\beta\approx 1$ charged particles.

%\newpage
\subsubsection{The Calorimeter System}

%\begin{wrapfigure}{R}{8.0cm}
%    \centering
%    \includegraphics[height=2.5cm]{Figures/Detector/ECAL-HCAL-LateralSegmentation.jpg}
%	\caption{Lateral segmentation of the ECAL ({\em left}) and HCAL ({\em right}). } 
%	\label{fig:CaloSegmentation}
%    \end{wrapfigure}
The calorimeter system provides identification for electrons, photons and hadrons as well as measures their energies and positions. It consists of an electromagnetic calorimeter (ECAL) followed by a hadron calorimeter (HCAL).
In LHCb-I the calorimeter system was a central part of the level-0 trigger, selecting high $p_T$ electrons and hadrons to reduce the readout rate to 1 MHz while retaining most of the heavy-flavour events. To provide electron-hadron separation, the system included a Scintillating Pad Detector (SPD) and a Pre-Shower (PS) detector, both of which were removed in LHCb-U due to their reduced significance in the all-software trigger environment.

%\vspace*{-0.5cm}
\paragraph{Electromagnetic Calorimeter}

The ECAL allows to trigger on electrons with high transverse energy. To suppress the high background rate from charged pions, longitudinal segmentation of the detector is implemented by installing a Pre-Shower detector (PS) in front of the main section of the ECAL. Additionally, to reject high-$E_T$ $\pi^0$ backgrounds, a Scintillating Pad Detector (SPD) is placed in front of the Pre-Shower detector.

\clearpage

\begin{figure}[t]
	\centering
  	\includegraphics[width=8cm]{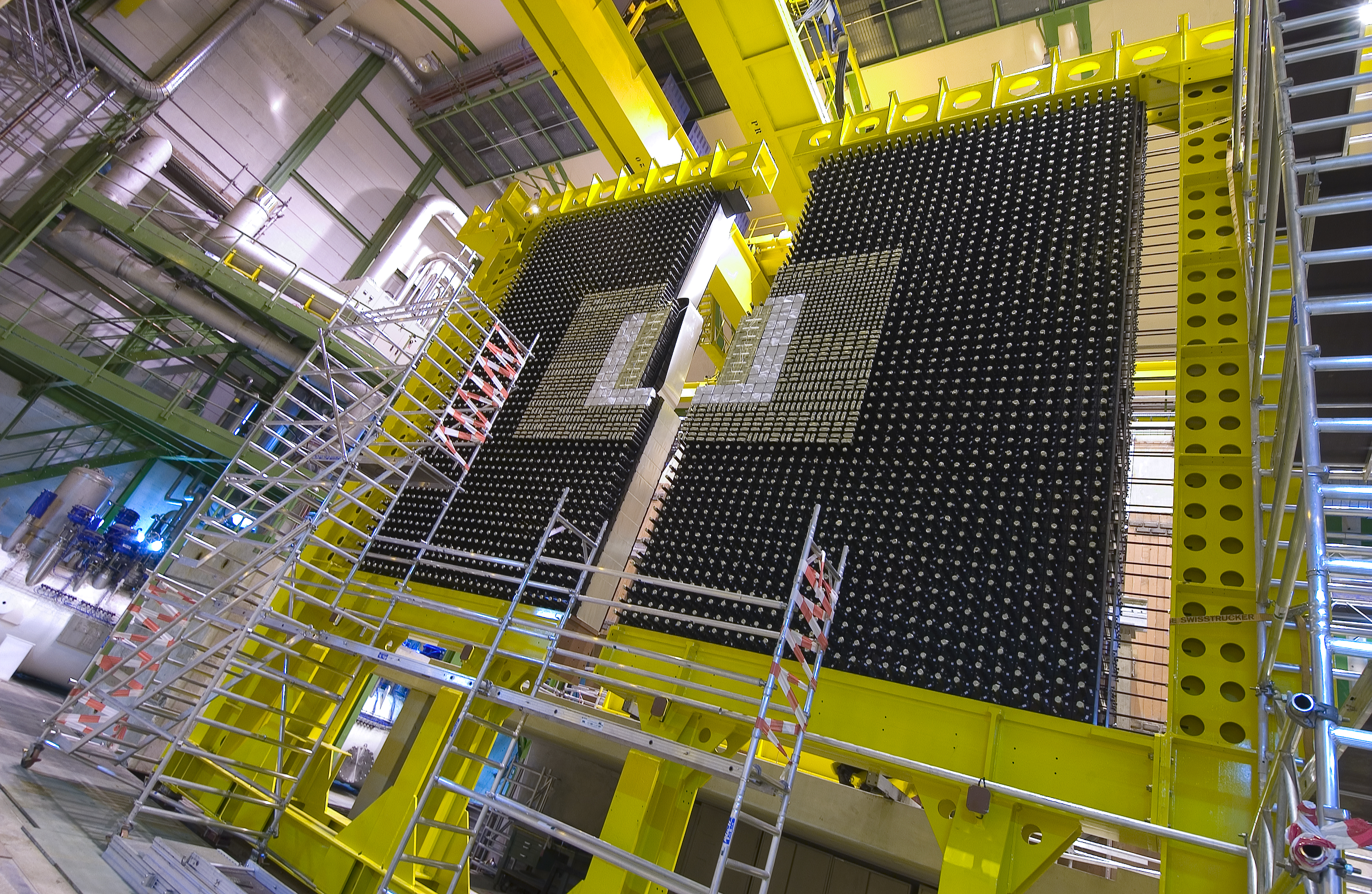}
    \hspace*{1.0cm}
 	\includegraphics[width=6.2cm]{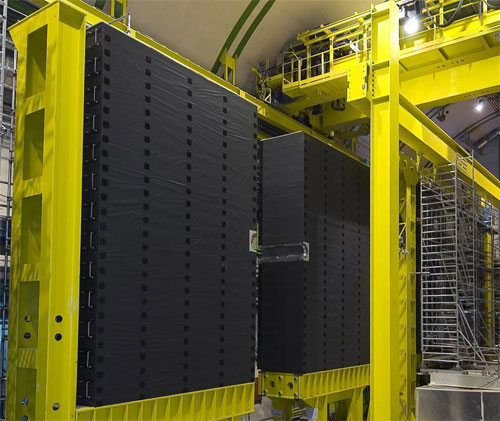}
	\caption{Photos of the calorimeter prior to installation. {\em Left:} The electromagnetic calorimeter, where the higher segmentation towards the center can be seen (CERN Document Server)
    %: CERN-EX-0505010-03). 
    {\em Right:} The hadronic calorimeter with its surface covered by scintillator plates (CERN Document Server).}
    %CERN-EX-0508001-03). }
 	\label{fig:Calo_Photos}
\end{figure}

\begin{wrapfigure}{L}{6.5cm}
    \centering
    %\vspace*{0.5cm}
    \includegraphics[height=4.5cm]{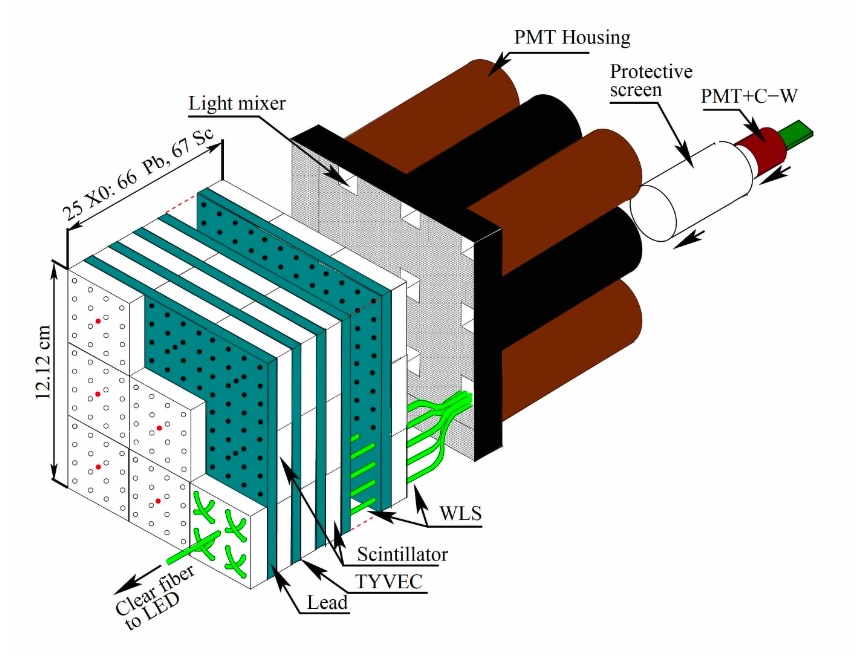}
	\caption{Schematic view of an ECAL cell.  Reproduced from \cite{ECAL-CalibrationPerformance-2020}.} 
    %CC BY 4.0. } 
	\label{fig:ECAL_Cell}
    %\vspace*{0.3cm}
\end{wrapfigure}
The SPD/PS system consists of a lead convertor plate of 2.5 radiation length, sandwiched between rectangular scintillating pads of the SPD and PS, using a projective geometry aligned to the angular acceptance. Test-beam measurements demonstrated an electron-pion separation performance of 99.6\% pion rejection and 92\% electron retention for 20 GeV particles. Since charged particles deposit energy in the SPD while neutral particles do not, the SPD is effective in separating electrons from photons, with a photon misidentification rate of less than 1\% \cite{LHCb-at-LHC-JINST2008}.
In LHCb-U, the SPD/PS system is removed from the setup since the full tracking information is available in the trigger.

The main part of the ECAL is a sampling calorimeter using a lead-scintillator structure referred to as the 'Shashlik' technology, illustrated in Fig.\ref{fig:ECAL_Cell}. Scintillation light is collected via plastic wavelength-shifting fibers and transmitted to photomultiplier tubes (PMTs). The ECAL energy resolution is given by $\sigma_E / E = 10\% / \sqrt{E} \oplus 1\%$, where $E$ expressed in $\mathrm{GeV}$. This translates into a mass resolution of $65 \mathrm{MeV/c^2}$ for $B^0\rightarrow K{^0*}\gamma$ decays and $75 \mathrm{MeV/c^2}$ for $B^0\rightarrow \rho \pi$ decays. 

Due to the strong variation of the hit density as a function of distance from the beamline, the ECAL is divided into inner, middle and outer sections (see Fig.~\ref{fig:Calo_Photos}).
The detector is positioned at 12.5 m from the interaction point and spans the full LHCb acceptance of 300 mrad by 250 mrad. 

\vspace*{0.5cm}
\paragraph{Hadronic Calorimeter}

\begin{wrapfigure}{R}{6.5cm}
    \centering
    \vspace*{-0.7cm}
    \includegraphics[height=4.8cm]{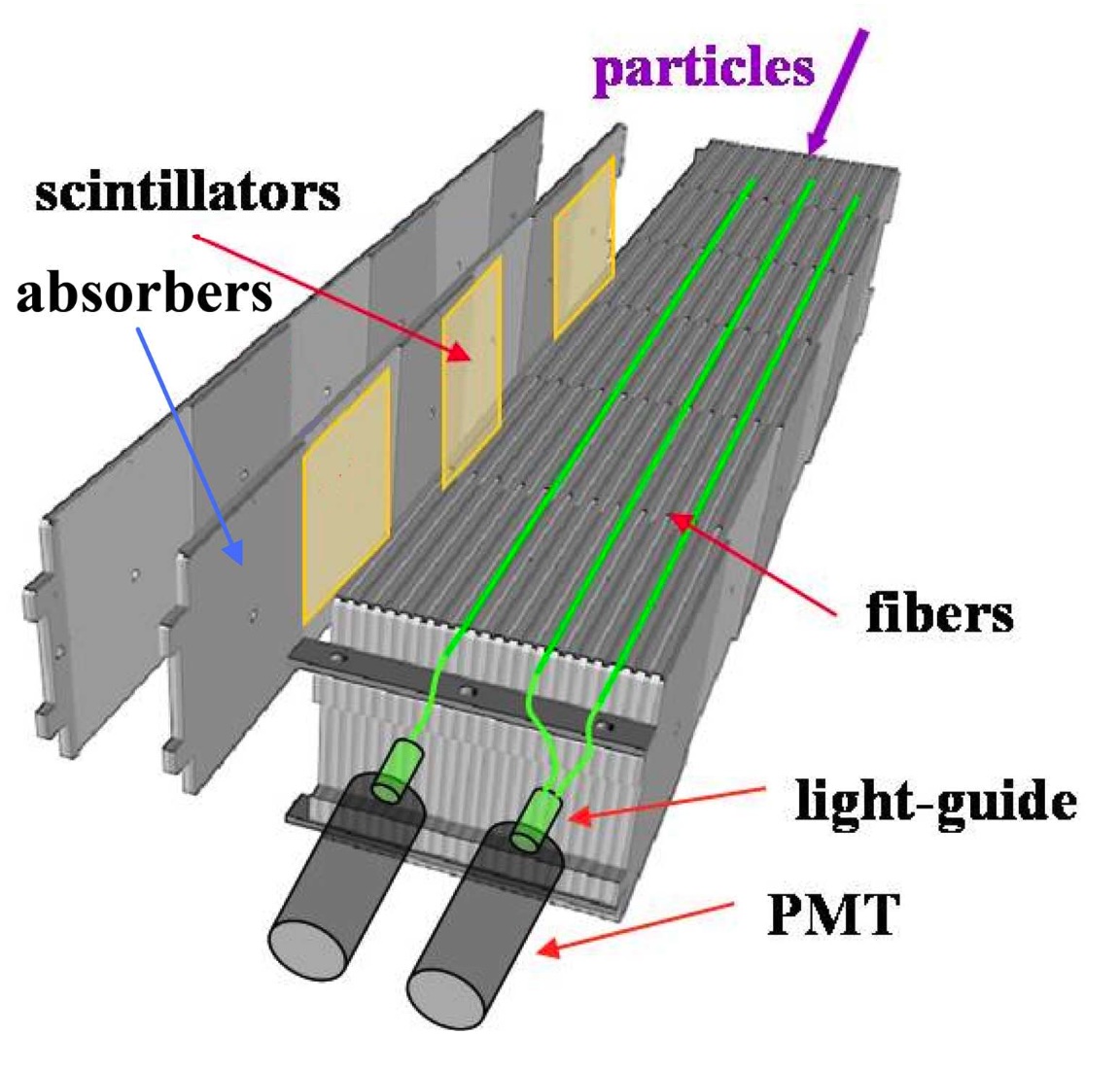}
	\caption{Schematic view of an HCAL cell. Reproduced from \cite{ECAL-CalibrationPerformance-2020}.}
    %CC BY 4.0. } 
	\label{fig:HCAL_Cell}
\end{wrapfigure}
The HCAL is a sampling calorimeter made from iron as absorber and active scintillating tiles, arranged in a structure running parallel to the beam axis (see Fig.~\ref{fig:HCAL_Cell}). Similar to the ECAL, the HCAL is laterally segmented, with finer granularity in the inner region due to the higher particle flux.
Scintillation light from the active layers is collected using wavelength-shifting fibers (WLS) and directed to photomultiplier tubes. The combined ECAL and HCAL system has a total thickness of 6.2 interaction lengths, providing an energy resolution of $\sigma_E / E = \left(69\pm 5\right)\%/\sqrt{E} \oplus \left(9 \pm 2\right)\% $, where $E$ is expressed in GeV.

After collecting an integrated luminosity of 3.4 fb$^{-1}$, a 15\% reduction in attenuation length was observed in the WLS fibers of the inner HCAL cells due to non-uniform radiation damage.
As these inner modules could not be replaced during LHC runs, they were substituted in LHCb-U by tungsten absorber slabs to provide additional shielding for the innermost region of the muon detectors.  This modification had minimal impact on physics performance, as the level-0 trigger was replaced by a full software trigger utilizing complete detector information.

\newpage
\subsubsection{Muon System}

\begin{figure}[!ht]
	\centering
    \includegraphics[width=4.5cm]{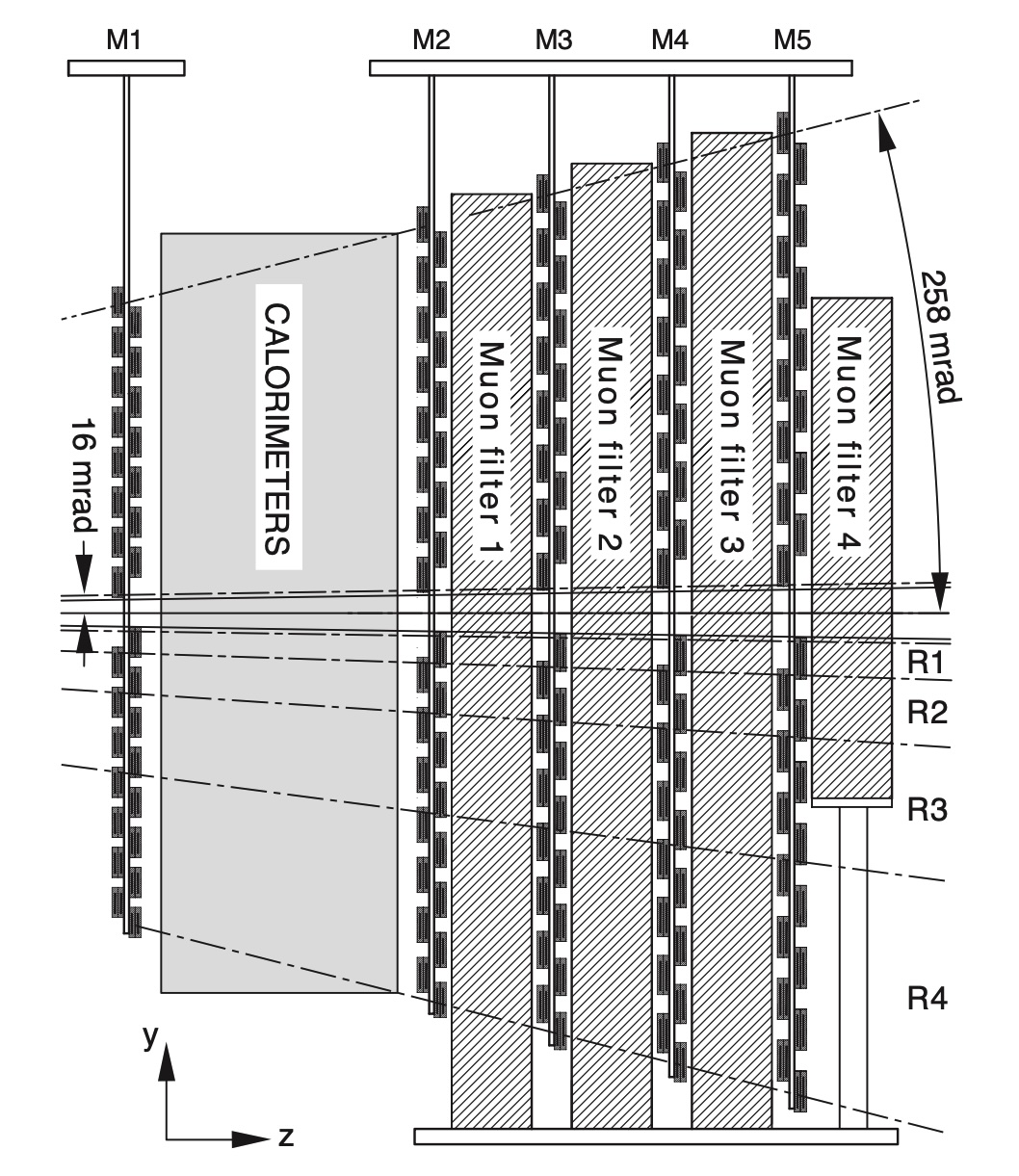}
    \hspace*{1.0cm}
  	\raisebox{0.5cm}{\includegraphics[width=7.0cm]{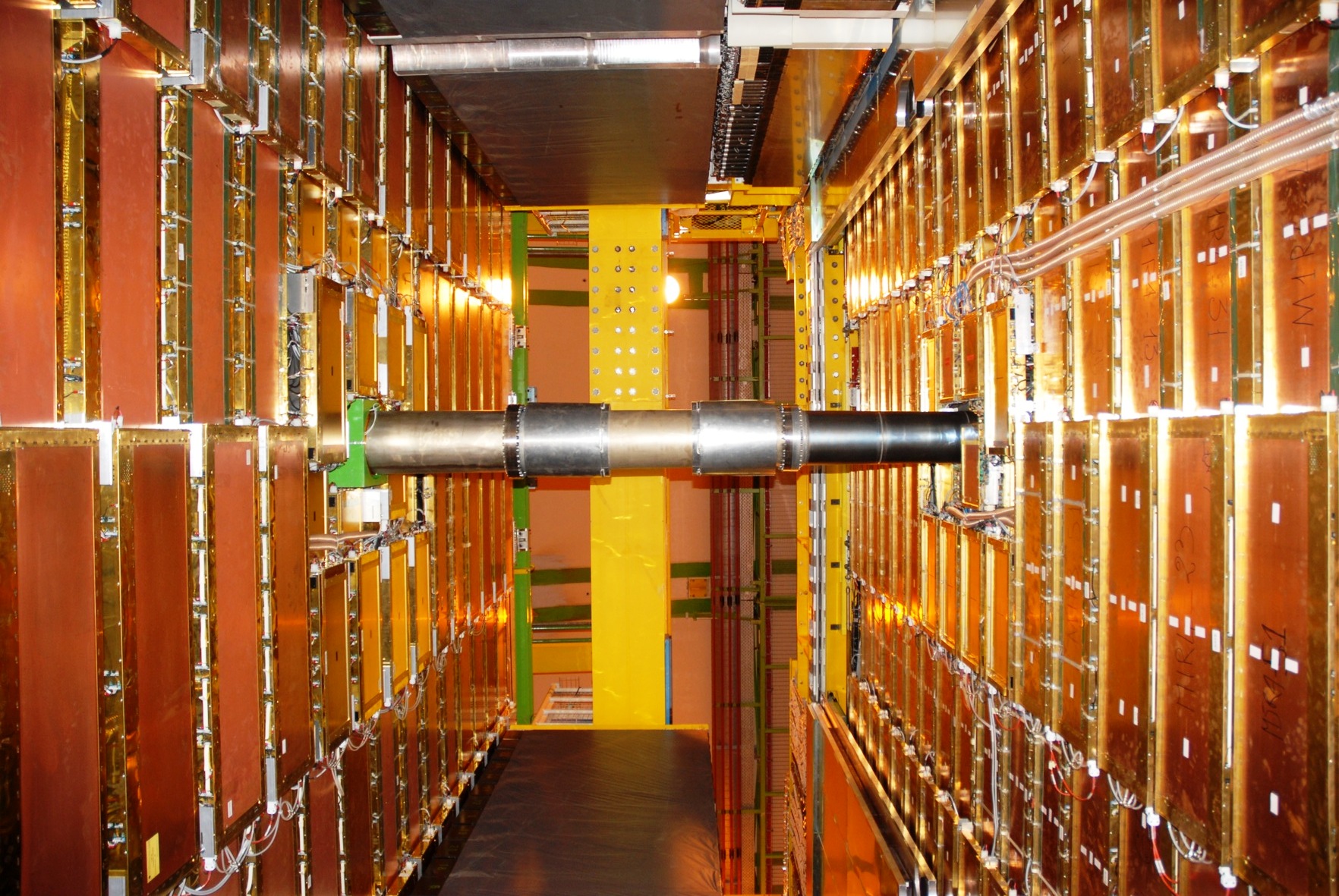}}
	\caption{{\em Left:} Schematic view of the muon system comprising 5 muon stations M1 - M5 \cite{LHCb-at-LHC-JINST2008} {\em Right:} View from the floor of the cavern between the M1 and M2 muon detector planes. In the center of the picture the beampipe is seen (CERN Document Server).}
    %: LHCb-PHO-BEPI-2009-001-1).}  
 	\label{fig:MUON}
\end{figure}

\vspace*{-0.2cm}
Muon identification is essential for both real-time triggering and offline analysis in flavour physics. The LHCb-I muon system consists of five stations (M1-M5), shown in Fig.~\ref{fig:MUON}. Stations M2-M5 are composed of Multi-Wire Proportional Chambers (MWPCs) positioned behind the calorimeter. To improve the transverse momentum estimate of the muon candidate in the level-0 trigger of LHCb-I, an additional station (M1) is installed upstream of the calorimeter, as it measures the muon track before scattering in the calorimeters. M1 in addition uses Gas Electron Multiplier (GEM) technology in the high occupancy central region. It was removed from the setup in LHCb-U as the Upgrade no longer included a hardware trigger.

The muon stations are interleaved with 80 cm thick iron absorbers to filter low energy particles. 
To mitigate efficiency losses caused by front-end electronics dead time under high luminosity conditions, additional tungsten shielding was installed around the beam pipe in front of M2. The MWPCs of the muon system are expected to remain operational throughout the entire LHCb lifetime, 
to at least a total integrated luminosity of approximately 50 fb$^{-1}$.

\subsection{Trigger}

\begin{figure}[!ht]
	\centering
  	\includegraphics[width=11cm]{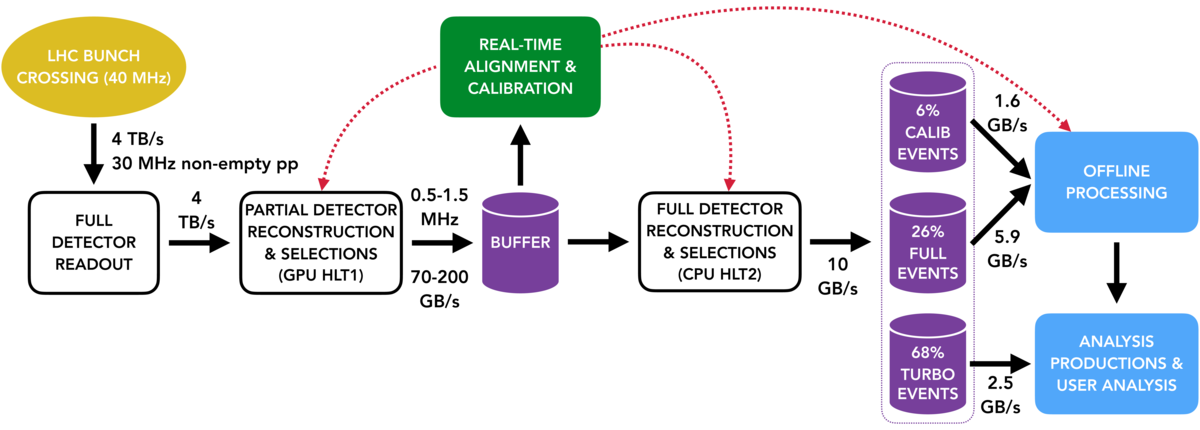}
	\caption{Real-time reconstruction and event selection scheme as used in LHCb-U, taken from LHCb-FIGURE-2020-016.}
    %LHCb collaboration, RTA and DPA dataflow diagrams for Run 1, Run2, and the upgraded LHCb detector, Tech.Rep. LHCb-FIGURE-2020-016, CERN, Geneva(2020).}  
 	\label{fig:RTA}
\end{figure}

\vspace*{-0.2cm}
The LHCb trigger scheme has undergone significant evolution across Run-1, Run-2 and Run-3. In Run-1 and Run-2 (LHCb-I) it included a hardware based level-0 (L0) trigger, processing data from dedicated subdetectors at the 40 MHz LHC bunch crossing rate. The L0 trigger reduced the rate to 1 MHz allocating 450 kHz to a high $p_T$ hadron trigger (HCAL), 400 kHz to single and double muon triggers (MUON) and 150 kHz to electron/photon triggers (ECAL). In Run-1 these selected events were partly reconstructed by the software based High Level Trigger (HLT) and subsequently filtered using exclusive and inclusive selections, resulting in an output 5 kHz.  In Run-2 the high level trigger was enhanced by introducing a two-stage architecture: HLT-1 performed partial reconstruction and selected events based on displaced vertices and dimuon signatures, while HLT-2 executed an offline-like event reconstruction and selection, storing events at a rate of 12.5 kHz. 

For Run-3 (LHCb-U) all front-end electronics were replaced and the hardware trigger was removed. The upgrade enabled real-time event reconstruction and application of inclusive and exclusive software selections at 40 MHz in HLT-1, by using reconstruction in GPU processors. Selected events are subsequently buffered to disk for calibrations and alignment after which HLT-2 performs reconstruction and selection with full offline-level event information. More information is provided in section \ref{subsec:ReconstructionTrigger}.
The scheme for the upgrade trigger, also referred to as Real Time Analysis (RTA) is shown in Fig.~\ref{fig:RTA}.

\clearpage
\section{From Data to Physics}\label{secReconstruction}
%Please provide a very general and easy to understand explanation how you extract the physical relevant information from the experimental data 

The LHCb experiment is primarily designed for precision measurements in flavour physics, with a focus on the reconstruction of decays of $b$- and $c$-hadrons. In particular, the experiment excels in the reconstruction of charged particles while maintaining good capabilities for final states with neutral particles and for performing jet physics analyses. Beyond Flavour Physics, the detector has proven to have versatility to provide state-of-the-art performance for QCD spectroscopy, electroweak physics as well as for heavy ion physics in the forward region. 

Unlike the general purpose experiments ATLAS and CMS, which operate at the LHC's peak luminosity, LHCb operates at a instantaneous luminosity about ten times lower, dictated by the high final-state particle density in the forward acceptance and the proximity of the vertex detector to the interaction point. The upgrade from LHCb-I to LHCb-U had as a main motivation to maintain the excellent resolution capabilities of the original detector at a much higher luminosity of the upgrade, while at the same time improving real time event selection efficiencies. Since the measurement concept remained similar for LHCb-I and LHCb-U, we summarize the methods and performance using unified descriptions and representative values.

Mirroring the detector's hardware setup, the LHCb event reconstruction relies on three core pillars:
\begin{itemize}
    \item {\em Tracking}: High-precision reconstruction of charged particle trajectories for vertex reconstruction and invariant mass determination.
    \item {\em Particle Identification}: Efficient identification of muons, electrons, photons, kaons and pions.
    \item {\em Real-Time Analysis}: Flexible real-time reconstruction of events, enabling fast and efficient selection and background reduction. 
\end{itemize}

\begin{figure}[!ht]
	\centering
    \includegraphics[width=8.5cm]{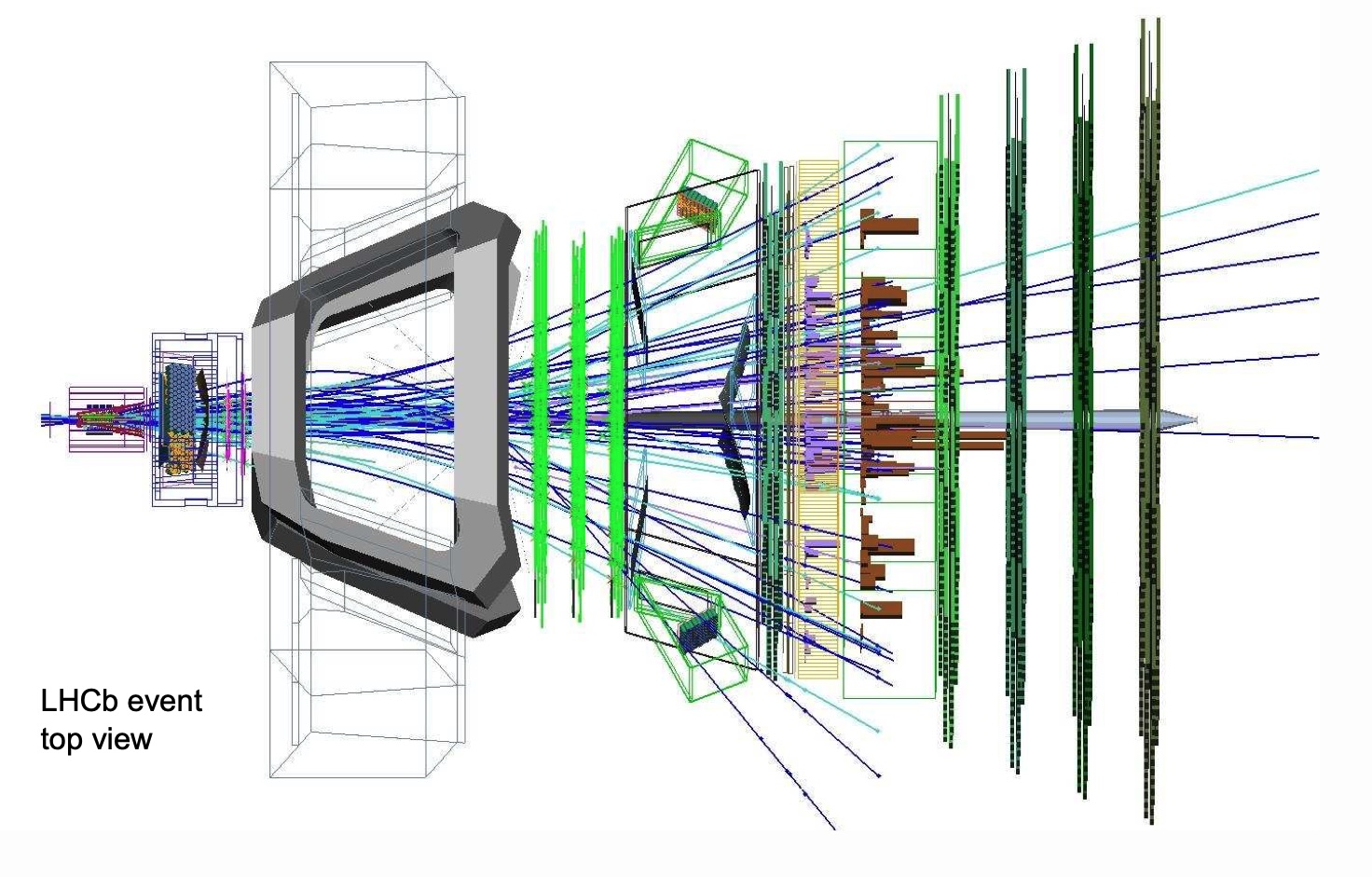} 
    \hspace*{0.5cm}
 	\raisebox{5.5cm}{\includegraphics[width=5cm, angle=-90]{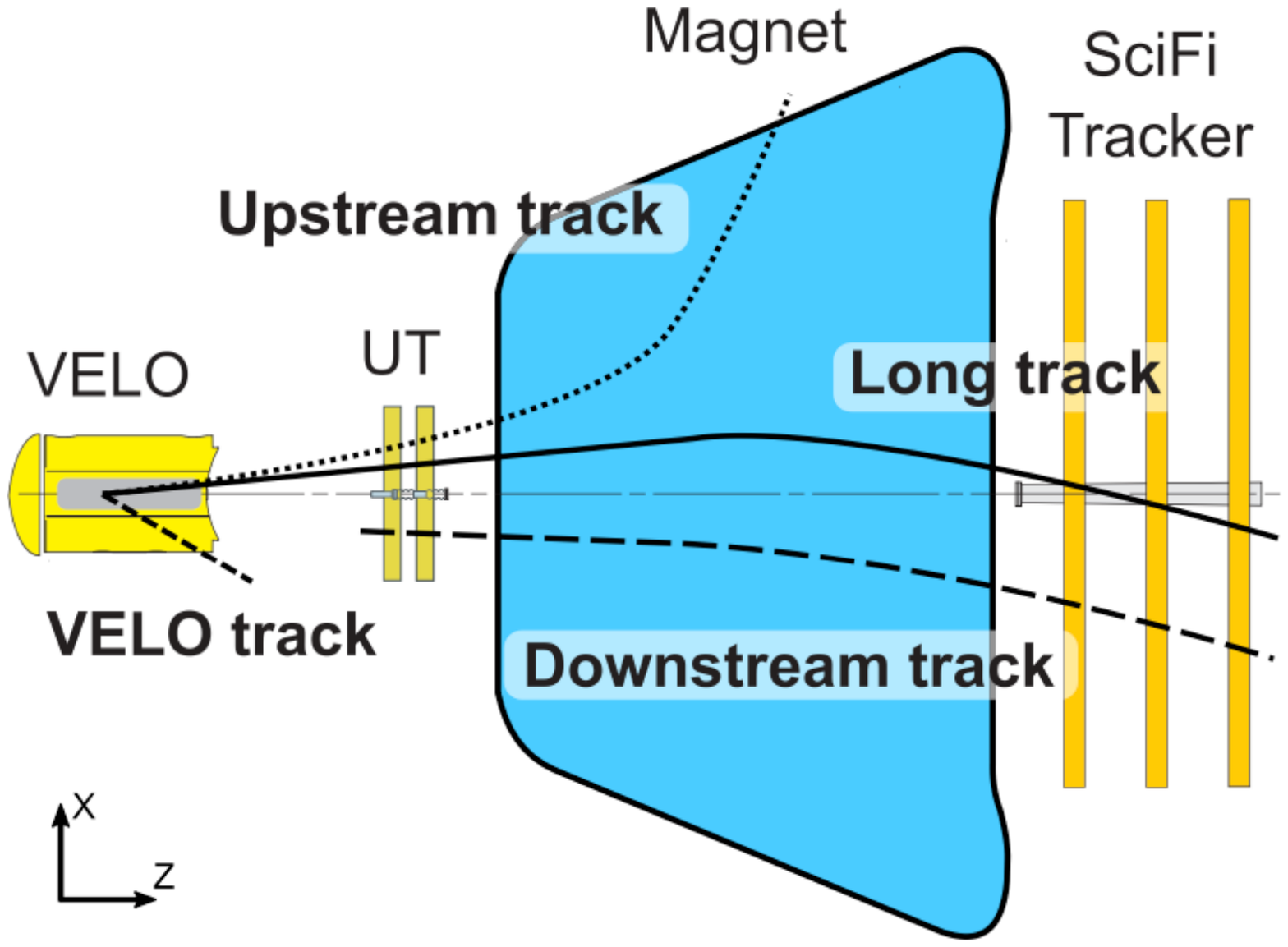}}
	\caption{{\em Left:} Top view display of a reconstructed LHCb event, illustrating bending of the tracks in the dipole magnet field. {\em Right:} The different particle track-types, defined according to their acceptance in the experiment. Picture taken from \cite{FernandezDeclara:2019ycx}.}
	\label{fig:TrackTypes}
\end{figure}

A typical $b\bar{b}$ event, shown on the left in Fig.\ref{fig:TrackTypes}, consists of approximately 70 charged particle tracks when there is a single $pp$ collision (i.e., no pile-up).
These tracks include primary particles directly produced in the $pp$ collision, as well as secondary particles created through interactions of primaries with detector material. Furthermore, a comparable number of neutral particle clusters are observed by the calorimeters. Raw detector hits are processed into reconstructed tracks and associated particle identification (PID) objects, which are then used to form final-state particle candidates. In Run-1 and Run-2 (LHCb-I) the High-Level Trigger (HLT) gradually incorporated more real-time reconstruction capabilities, whereas in Run-3 (LHCb-U) the HLT performs full reconstruction and selection in real-time, significantly improving the physics performance capabilities.

\subsection{Tracking: Track and Vertex Reconstruction}

\subsubsection{Track Finding}

The efficiency of hit detection in the LHCb tracking detectors generally exceeds 98\%~\cite{track-efficiency-JINST2015}, and after extensive calibration and alignment campaigns, the detector resolutions in the LHC environment align with expectations from simulations and test beam studies.

Charge particles follow different trajectories depending on their origin, production angle, and momentum. These are categorized in different {\em track-types}, illustrated in Fig.~\ref{fig:TrackTypes}.  The reconstruction {\em efficiency} of each track type is estimated using Monte Carlo (MC) simulations as well as with data driven methods, and is compared against the {\em ghost rate}, which quantifies the fraction of falsely reconstructed (fake or "ghost") tracks. In the data driven method, the performance is measured directly from LHC data using a "tag-and-probe method" \cite{track-efficiency-JINST2015}, where one track in a two-prong particle decay, the "tag", is used to determine the efficiency for finding the other one, the "probe". The performance values given below correspond to the LHCb-I detector, as they form the baseline for physics analyses presented in this review. The corresponding values for the LHCb-U detector are similar.

\begin{SCfigure}[0.5][!ht]
	\centering
	\includegraphics[width=9cm]{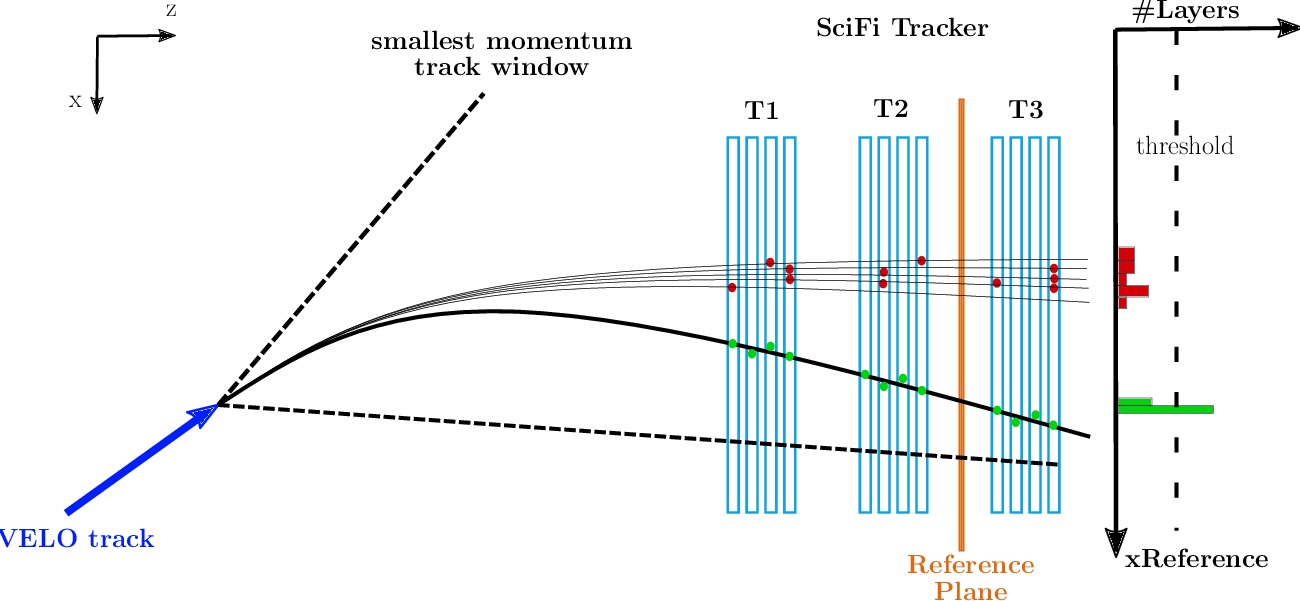} 
	\hspace*{0.5cm}\caption{The principle of Forward Tracking. VELO tracks are extrapolated to the downstream tracker where only hits well aligned with a single momentum estimate provide a peak in the trackfinding histogram.  Picture from \cite{Gunther:2022pdo}.}
    %LHCb-PROC-2022-009 October 26, 2022   Paul Andre Gunther   arXiv:2207.12965v3. }
	\label{fig:ForwardTracking}
\end{SCfigure}

The following track-types are reconstructed:
\begin{itemize}
    \item {\em Long tracks}. These tracks traverse the entire tracking volume and form the basis of most physics analyses. They include segments in both the VELO as well as the downstream trackers. A high reconstruction efficiency is obtained by applying two search methods. The first method first forms VELO track segments and extrapolates them to the downstream tracker, where hits are added with a histogramming method exploiting a consistent momentum estimate of a series of hit-and-track combinations, see the illustration in Fig.~\ref{fig:ForwardTracking}. For use in the HLT it was found that this algorithm is significantly sped up by first confirming VELO segments by hits in the upstream tracker providing an initial momentum estimate, which then allows to reduce the size of the search window in the downstream tracker. The second method matches independently found track-segments of the VELO and downstream trackers for a specific curvature in the magnetic field. Here, the upstream tracker hits play a crucial role in reducing fake track combinations. The efficiency for finding long tracks is $\sim 95\%$, with a corresponding ghost rate fraction of $\sim 9\%$, calculated for tracks with $p>5$ GeV/c \cite{track-efficiency-JINST2015}.
    \item {\em Upstream tracks}. These tracks contain hits in the VELO and the upstream tracker, but not in the downstream tracker. They typically originate from low momentum particles that bend out of the detector acceptance inside the magnet volume. An important type of these particles are slow pions ($\pi_s$) from $D^{*\pm}\rightarrow D^0 \pi_{s}^{\pm}$, where the charge of the slow pion provides a flavour tag for the reconstructed neutral $D$ meson. The efficiency for finding upstream tracks is $\sim 93\%$, with a corresponding ghost rate fraction of $\sim 7\%$, calculated for tracks with $p_T> 0.5$ GeV/c~\cite{UpstreamTracking}.
    \item {\em Downstream tracks}. These tracks originate from decays of neutral particles such as $K^0_S$ and  $\Lambda$, where the produced charged particles do not leave sufficient hits in the VELO to be found as a VELO track-segment. They are reconstructed by adding hits from the upstream tracker to backwards extrapolated track-seeds from the downstream tracker. The efficiency for finding upstream tracks is $\sim 80\%$, with a corresponding ghost rate fraction of $\sim 15\%$ \cite{Stahl-master-thesis-2010}.
    \item {\em VELO tracks}. These are tracks for which no matching hits are found in the upstream or downstream trackers and are often tracks produced with a high polar angle or backward pointing tracks. They are essential for the reconstruction of the primary vertices. The efficiency for finding VELO tracks is $\sim 98\%$, with a corresponding ghost rate fraction of 1-2\% \cite{Velo-Performance-JINST-2014}.
\end{itemize}

%\subsubsection{Track Finding}
%\paragraph{Track Seeding}
%\paragraph{Forward Tracking}
%\paragraph{Track Matching}
%\paragraph{Upstream and Downstream Tracks}

\subsubsection{Track Fitting: Kalman Filter}

The purpose of the track fit is to determine the optimal particle trajectories for precise primary and secondary vertex determination, for calculation of the invariant masses of particle combinations, as well as for prediction of the track angles inside the RICH detectors as input to the photon ring search for particle identification. To do this, both the $x$ and $y$ positions of tracks, as well as their angles are determined along multiple locations along the $z$-axis of the spectrometer, together with the track momentum $p$. 
In LHCb, a reconstructed track is therefore represented by a dynamical track state-vector $\left(x,y,dx/dz,dy/dz,q/p\right)_z$, that evolves as function of its position $z$ along the beam line. The state-vector includes the signed curvature parameter $q/p$, where $q$ is the particle's charge, $\pm 1$, and $p$ its momentum. The choice of the parameter $q/p$ is motivated by its well-defined behavior in the limit of zero curvature, which corresponds to the high-momentum limit $p\rightarrow\infty$. 

\begin{wrapfigure}{R}{8.0cm}
	\centering
\includegraphics[width=8.0cm]{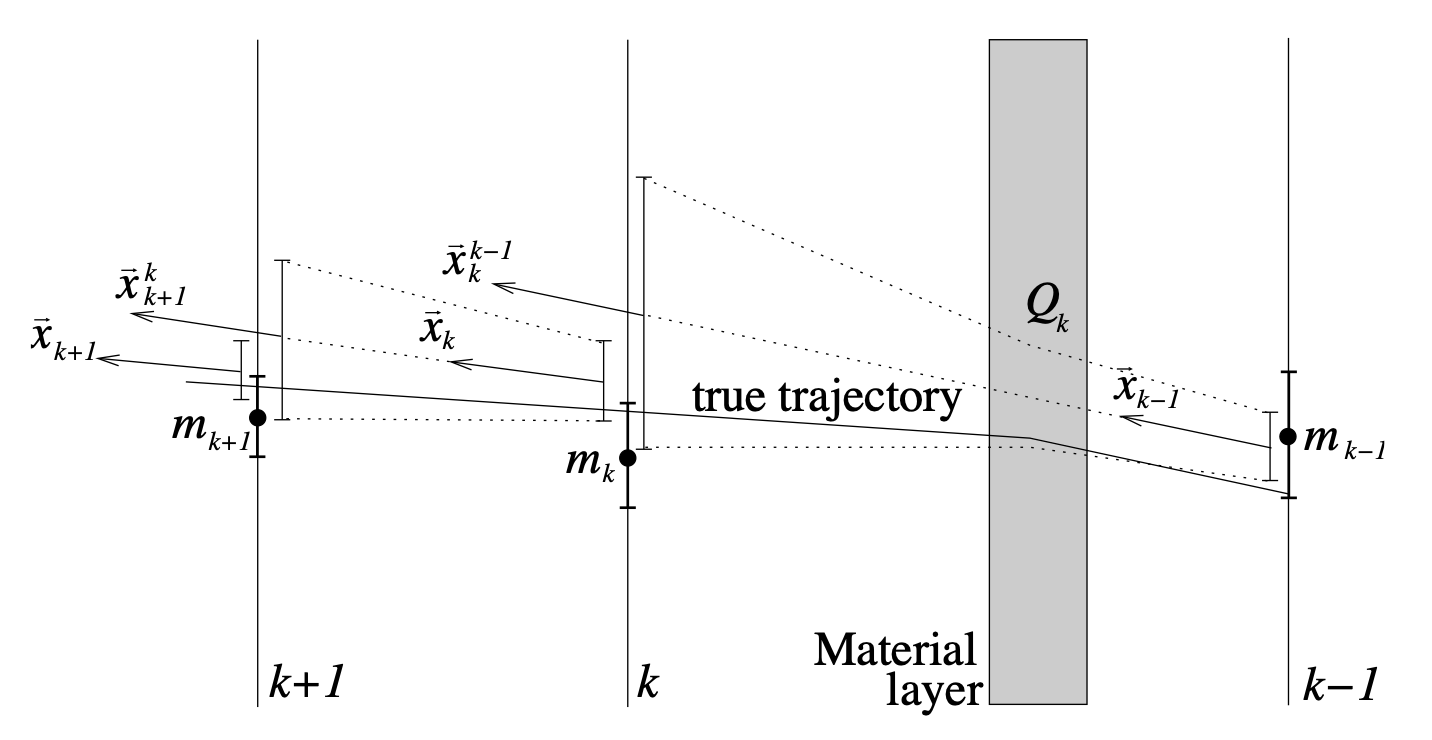} 
	\caption{The principle of the Kalman filter. Taken from \cite{PhDThesis_vanTilburg}.}
	\label{fig:Kalman}
\end{wrapfigure}
Determination of track states in LHCb exploits the Kalman filter method \cite{KalmanFilter}, a computationally efficient method for iteratively incorporating measurements. Unlike traditional global least-squares fitting, which requires the inversion of large matrices, the Kalman filter  incrementally updates the state vector at each detector plane, referred to as a {\em measurement node}.  Correspondingly, insensitive material layers are included as {\em scattering nodes}. 
Mathematically, the Kalman filter is equivalent to a least-squares fit, providing both the optimal track state and the corresponding error covariance matrix at each step. The error covariances are computed using the known precision of the detector hits as well as the detector material with known material thickness expressed as fractions of radiation lengths. Its first application in particle physics was in the DELPHI experiment \cite{KalmanDelphi_1, KalmanDelphi_2} and to-date it is frequently used in many experiments. The mathematical formulation of the filter as applied in LHCb can be found in \cite{PhDThesis_vanderEijk, PhDThesis_Hierck, PhDThesis_vanTilburg}. The principle is illustrated in Fig.~\ref{fig:Kalman}.

The Kalman filter operates in three main steps:
\begin{itemize}
    \item {\em Prediction}: Extrapolating the state vector $\vec{x}$ from a given measurement node $k-1$ to the next one $k$ by the propagation function $f_k$ via $x_k^{k-1}= f_k\left(x_{k-1}\right)$, as well as adding possible propagation noise into the covariance matrix: $Q_{k}^{k-1}=F_k C_{k-1} F_k^T + Q_k$. The transport function $f$ implements track extrapolation across the highly non-uniform magnetic field. The extrapolation exploits a $5^{th}$-order Runge Kutta extrapolation and uses the detailed magnetic field map.
    \item {\em Filtering}: Updating the predicted state with information from the current measurement node $k$. The measurement $m_k$ is compared to the predicted state $x_k^{k-1}$ and the residual is used to make an updated state $x_k$, according to a least squares minimization making use of the Kalman gain matrix. This node now has the optimal state estimate based on the hits used so-far. 
    \item {\em Smoothing}: Refining the state at all nodes by running the filter backward after the forward pass. Reversing the filter after the last added measurement to update all previous nodes provides full track information, resulting in "smooth" track. In practice the smoothing step can be done either by reversing the filter after the final measurement node is reached or, alternatively, with a bi-directional filter connecting states at each hit.
\end{itemize}

For LHCb-U a parametrized Kalman track filter \cite{ParametrizedKalman} is implemented for enhanced speed performance for real time reconstruction in the trigger. It uses a parameterized approach for magnet field extrapolation and for multiple scattering, avoiding time consuming detailed material lookup tables, with only limited cost in precision.

\subsubsection{Momentum resolution and Mass reconstruction}

The track momentum is determined by the reconstructed curvature ($q/p$) of the track state parameter at the particle production vertex. Generally speaking, it is observed as a transverse momentum "kick" ("$p_T^{\mathrm {kick}}$") provided by the integrated magnetic field along the particle trajectory: $p_T^{\mathrm{kick}}=q\int B_y dz $.

\begin{figure}[b]
	\centering
    %\vspace*{-0.4cm}
	%\includegraphics[width=10.cm]{Figures/Reconstruction/ReconstructedTrackResolutions.jpg} 
    \includegraphics[width=7.cm]{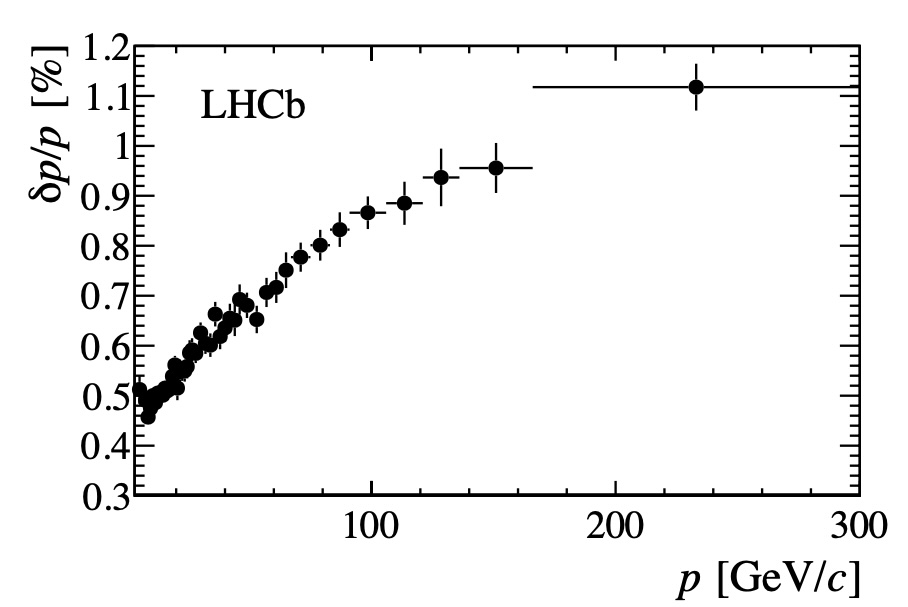} \hspace*{0.5cm}
    \includegraphics[width=7.cm]{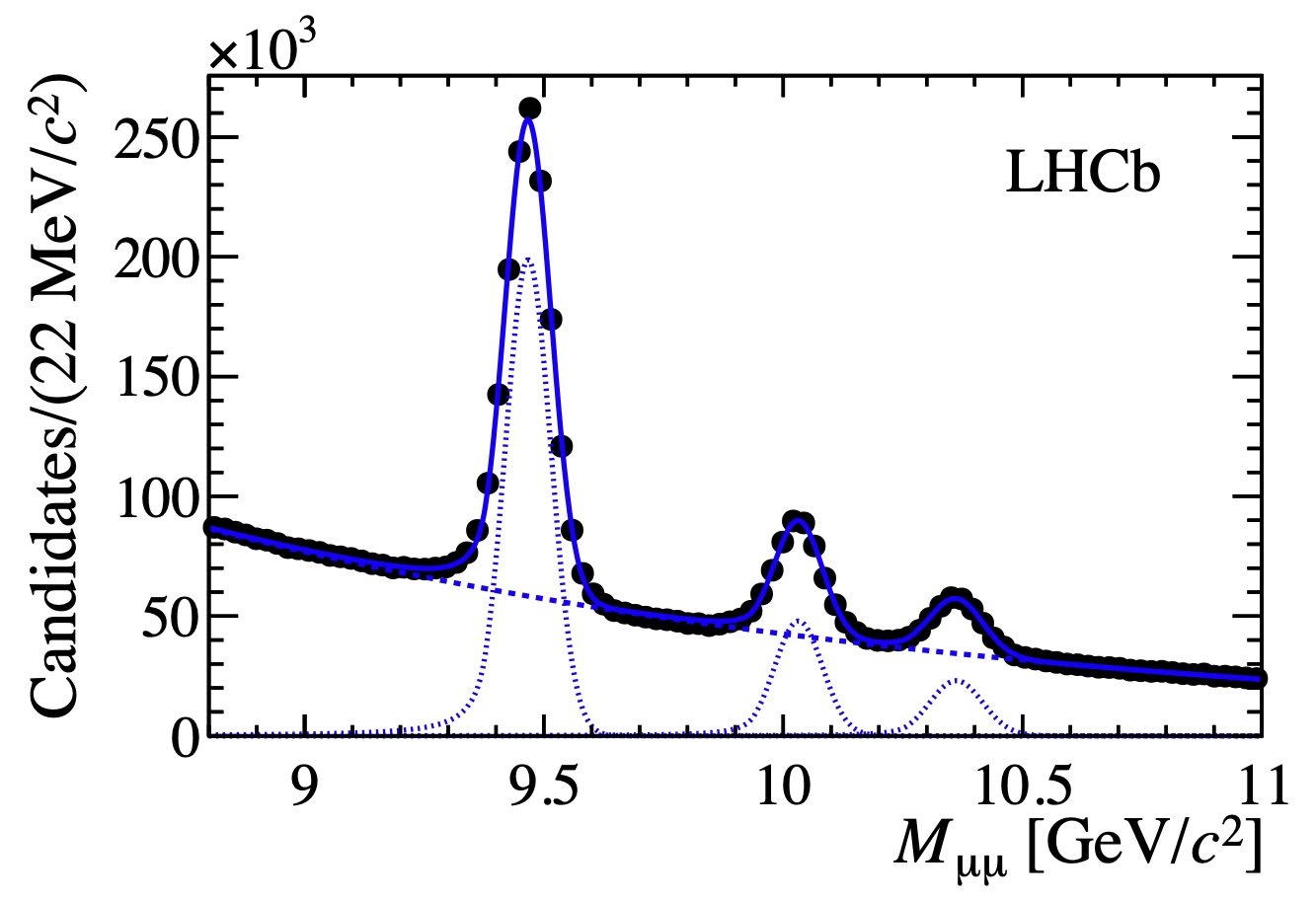} 
	\caption{ {\em Left:} Momentum resolution of the reconstructed muons from $J/\psi$ particle decays expressed as $\delta p/p$ vs momentum $p$, where $\delta p$ is the estimated track error \cite{LHCb-Detector-Performance_2015}. {\em Right:} Reconstructed di-muon mass spectrum where the $\Upsilon(1S)$, $\Upsilon(2S)$ and $\Upsilon(3S)$ resonances are clearly distinguished and their mass resolutions are approximately $50$ {MeV} \cite{LHCb-Detector-Performance_2015}.}
	\label{fig:Momentum_Mass_Resolution}
\end{figure}

The relative momentum resolution, $\delta p/p$, is determined by the precision of track curvature measurements, which benefits from the relatively long measurement lever arm before and after the magnet, as well as from the use of lightweight materials reducing multiple scattering effects.
The downstream tracking stations with $\sim 2\% X_0$ thickness per station and measurement precision of $\sim 100 \mu$m per station lead to a resolution $\delta p/p < 1\%$ over the full momentum range. The left side of Fig.~\ref{fig:Momentum_Mass_Resolution} shows the momentum resolution as function of momentum for reconstructed muons in $J/\psi\rightarrow\mu^+\mu^-$ events.

The precise momentum resolution translates into high mass resolution for reconstructed decay candidates. The di-muon mass spectrum (see the right side panel of Fig.~\ref{fig:Momentum_Mass_Resolution}) illustrates the clear separation of the $\Upsilon$ resonances. For two-body decays, the invariant mass is given by:
%$$
$M_{\mathrm inv}^2= 2p_+ p_- \left(1-\cos\theta\right)$
%$$
where $\theta$ is the opening angle between the two particles. In two-prong $b$-particle decays, the angular uncertainty on $\theta$ is small and the mass resolution is driven by momentum resolution:
$(\sigma_m/m)^2 = \frac{1}{2}(\delta p/p)^2$ .
%$$
%\left(\frac{\sigma_m}{m}\right)^2 = \frac{1}{2}\left(\frac{\delta p}{p}\right)^2\; .
%$$
This relation confirms the consistency between the observed track momentum resolution and the resolution of the $\Upsilon$ mass peaks of about 50 MeV.

\begin{figure}[!ht]
	\centering
    %\vspace*{-0.3cm}
    \hspace{-1cm}
    \includegraphics[width=7.cm]{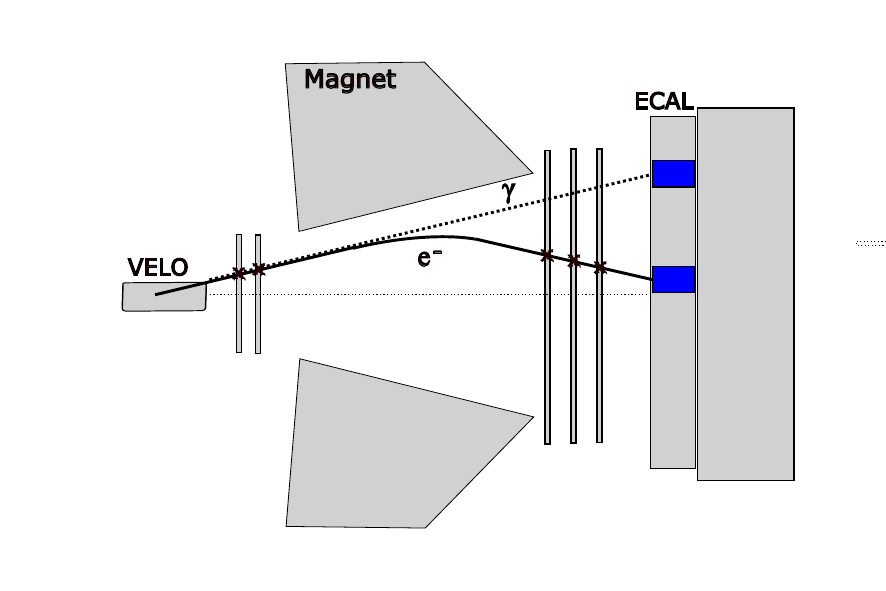} 
    \includegraphics[width=8.cm]{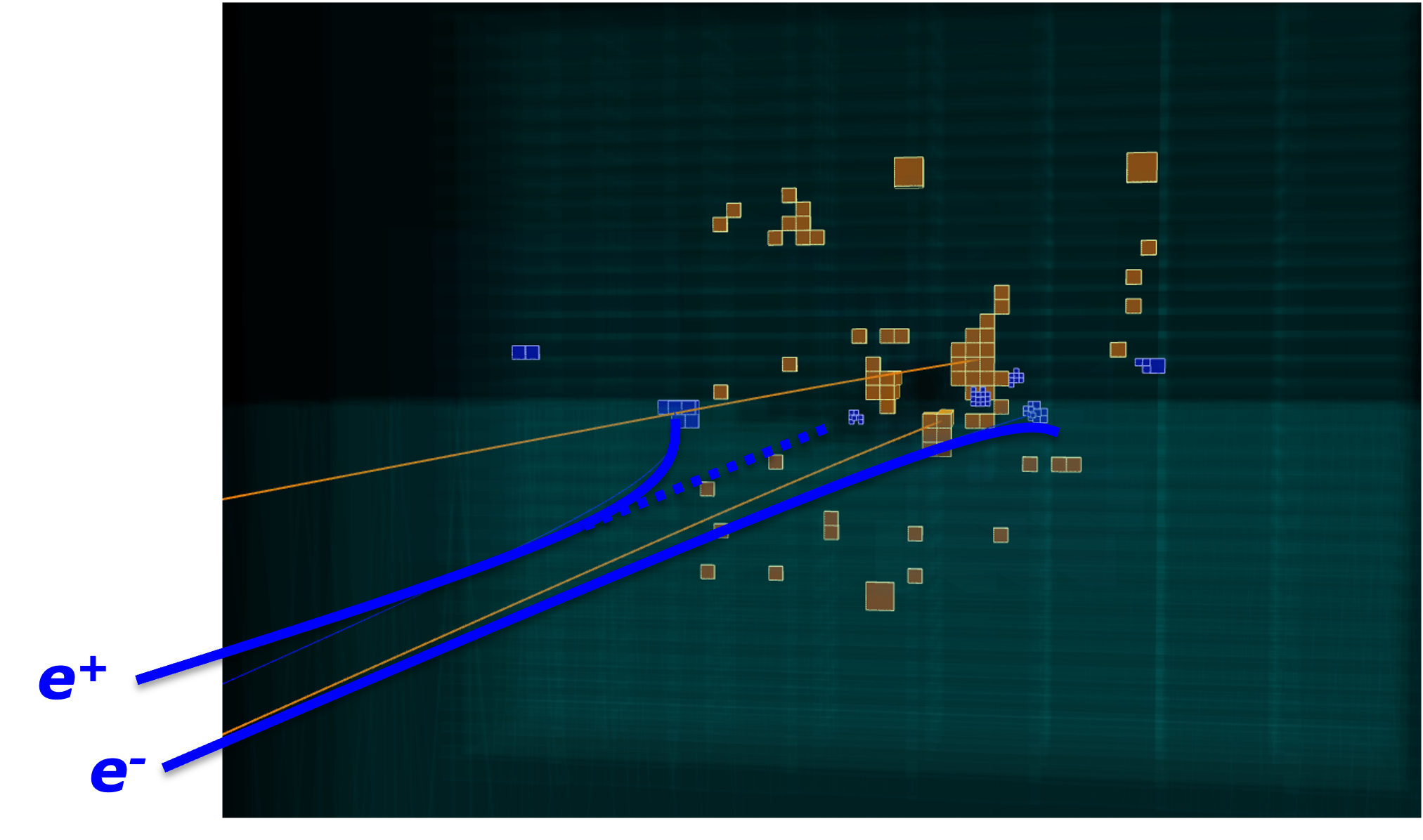} 
	\caption{ {\em Left:} Sketch of the Bremsstrahlung recovery, by adding the energy of the radiated photon to the energy of the measured electron.  {\em Right:} An event display from run 218585, 23 Nov 2018, illustrating the energy deposits in the ECAL and a photon being added to the electron $e^-$.}
	\label{fig:Bremsstrahlung}
\end{figure}

The reconstruction of $V^0$ vertices vertices from long-lived decays, such as from $K^0_s\rightarrow \pi^+\pi^-$ and $\Lambda\rightarrow p\pi^-$ decays, often involves downstream tracks. Since these tracks lack VELO segments, their mass resolution is reduced. While $K_S^0$ candidates reconstructed from long tracks achieve an average mass resolution of $3.5~ {\rm MeV}/c^2$, those including downstream tracks typically have a resolution of $7~ {\rm MeV}/c^2$.

Reconstruction of final states including electrons requires special attention since electrons experience Bremsstrahlung energy losses when traversing detector material, as illustrated in Fig.~\ref{fig:Bremsstrahlung}.
In case a high energy photon is emitted upstream of the magnet, the electron momentum reconstruction is complemented by adding a matching photon cluster along the track direction in the ECAL, when the cluster has $E_T > 75$ MeV. In case the electron radiates downstream of the magnet no correction is needed. In general, decays including electrons have a reduced efficiency compared to minimum ionizing particles, as well as a worse mass resolution: where decays of the type $B^0\rightarrow K^{*0} \mu^+\mu^-$ have an average mass resolution of 40 MeV/$c^2$, the resolution for $B^0\rightarrow K^{*0} e^+ e^-$ decays is typically 140 MeV/$c^2$.

\subsubsection{Vertex reconstruction and decay time measurement}

\begin{figure}[b]
	\centering
    \includegraphics[width=7.cm]{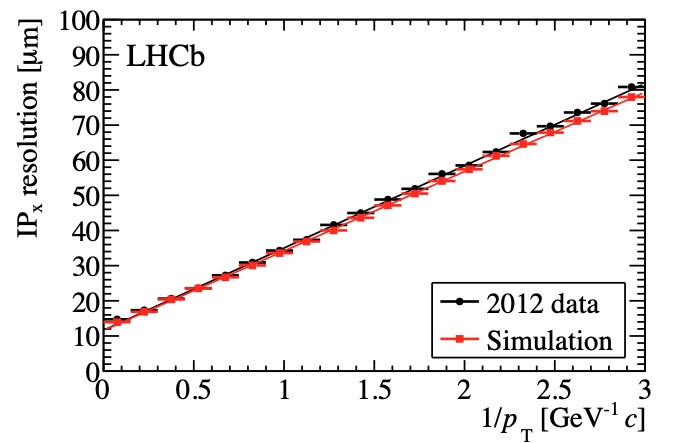} 
    \hspace*{0.5cm}
    \includegraphics[width=7.cm]{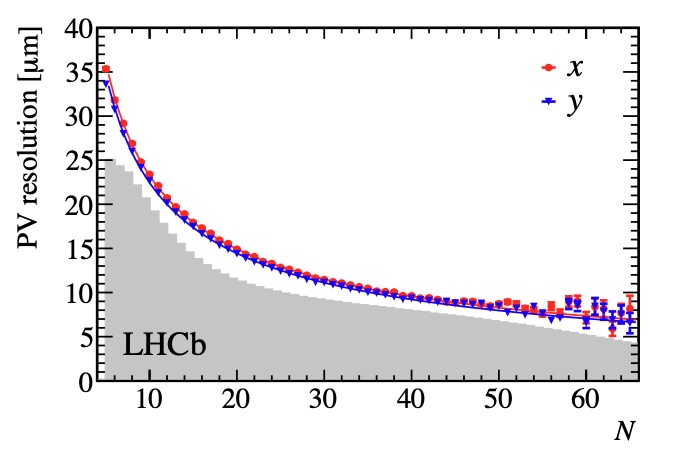}
	\caption{{\em Left:} The resolution of the impact parameter of the particle track to its production vertex as a function of the inverse transverse momentum $1/p_T$. The $\mathrm{IP}_x$ is calculated as the distance in $x$ between the extrapolated track and the vertex at the $z=z_{\mathrm vertex}$ plane. {\em Right:} Precision of the reconstruction of the Primary Vertex as a function of the number of tracks, $N$, for projections in $x$-$z$ plane (red) and $y$ - $z$ plane (blue). The superimposed shaded histogram shows the distribution of the number of tracks, $N$, used. Both plots were made using data from LHCb obtained in 2012 \cite{LHCb-Detector-Performance_2015}.}
	\label{fig:IP_Vertex_Resolution}
\end{figure}

The proton-proton collision point 
%is the production point of many particles in an event and 
is referred to as the primary vertex (PV). The decay point of unstable particles (eg beauty, charm or strange particles) is the secondary vertex (SV). 
%High reconstruction precision is required to assign primary tracks to their production vertex and to separate secondary tracks from them. 
Precise reconstruction of the track states in the interaction region allows for precise vertex reconstruction and powerful separation of secondary tracks from primary vertices. The precision of an extrapolated track is expressed by the impact parameter (IP) resolution: the distance between the extrapolated track state and the particle's actual production point. %As the nearest distance between a point and a line in three dimensions does not follow a Gaussian distribution,
LHCb quantifies this precision as the projected distance in the transverse coordinates $x$ and $y$ in the plane $z=z_{\mathrm vertex}$. The resolution of the measured impact parameter is shown on the left in Fig.~\ref{fig:IP_Vertex_Resolution}.

When there are at least five matching reconstructed track states at the closest position to the LHC beam line, LHCb reconstructs the primary vertices with a precision depending on the number of tracks used in the vertex algorithm. This precision is plotted as function of the number of tracks in the right panel of Fig.~\ref{fig:IP_Vertex_Resolution}. The red and blue points indicate the precision in the $xz$ and $yz$ plane respectively, and the shaded grey area displays the distribution of the number of tracks used in the vertexing algorithm.

By measuring the distance between the primary production vertex (PV) and the secondary vertex (SV), together with the momenta of an exclusively reconstructed final state, the lifetime of a decaying particle is reconstructed using $t =ml/p$,
%$$
%t =\frac{ml}{p}
%$$
where $m$ is the particle's mass, $l$ the measured decay length between PV and SV, and $p$ the reconstructed momentum of the particle, obtained from combining the decay particles.
The precision with which the decay time is measured follows from the uncertainties of the decay distance resolution($\sigma_d$) and momentum resolution $\sigma_p$:
$$
\sigma(t)^2 = \left(\frac{m}{p}\right)^2 \sigma_l^2 + \left(\frac{t}{p}\right)^2\sigma_p^2 
$$
The per-event uncertainty $\sigma(t)$ is reconstructed from the vertex fit and relies on correct tracking covariances obtained with the Kalman filter. The decay time resolution depends on the kinematics of the event candidate but has a typical value around 50 fs, far exceeding the need to resolve the $B_s$-$\bar{B}_s$ flavour oscillations.

For multiparticle decays (e.g. $B\rightarrow D \rightarrow X$) the reconstructed decay-times are determined with a least-squares Kalman fit method developed for the BaBar experiment \cite{DecayChainFit}. It uses a decay chain fit involving multiple decay vertices, allowing for simultaneous extraction of decay time, position and momentum parameters as well as their uncertainties, for all particles in the decay chain~\cite{DecayChainFit}.

%\newpage
\subsection{Particle Identification}

\begin{figure}[htb]
	\centering
    \raisebox{0.3cm}{
    \includegraphics[width=8.8cm]{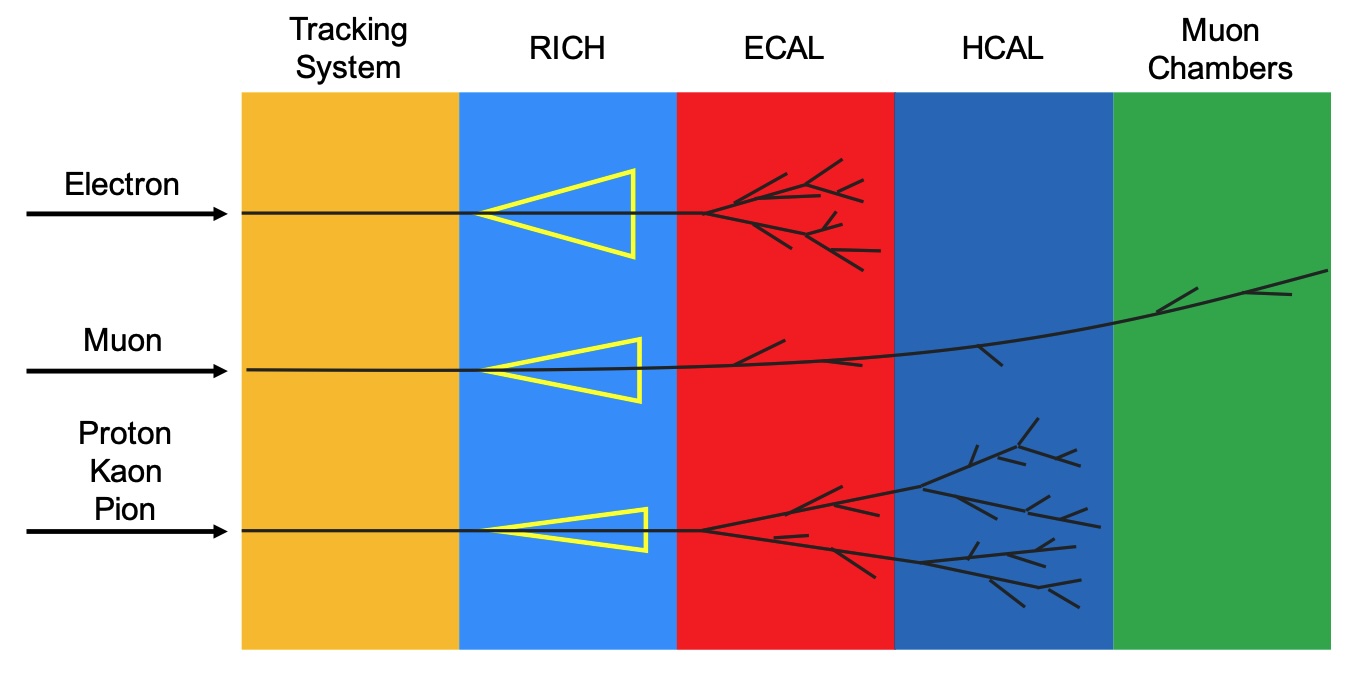}}
    \hspace*{0.8cm}
    \includegraphics[width=6.7cm, trim=0.0cm 0.0cm 0.0cm 12.5cm, clip]{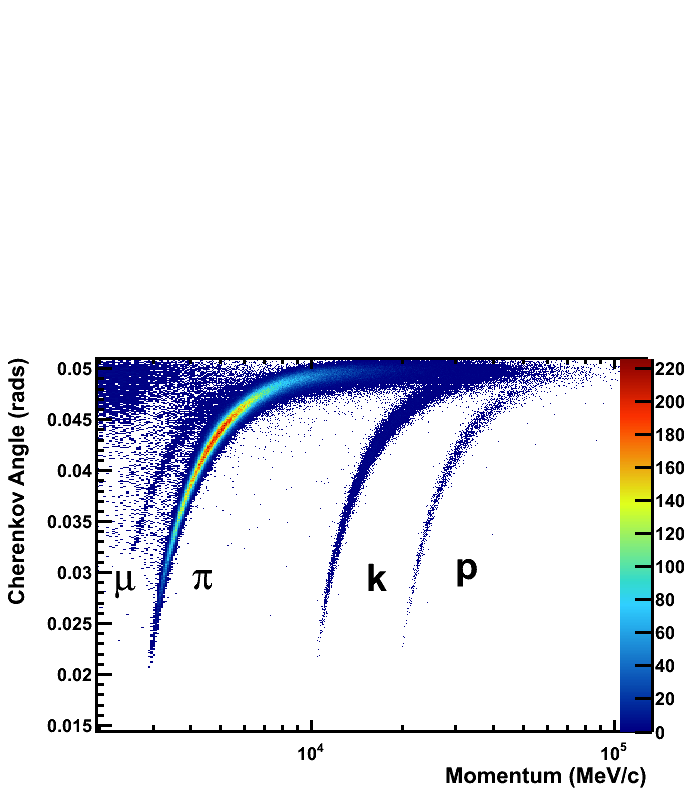}
	\caption{{\em Left:} Schematic illustration of particle identification in LHCb \cite{ProbNN}. {\em Right:} Charged particle identification performance using the RICH detectors~\cite{cavallero2023}.} 
	\label{fig:ParticleID_Schematic}
\end{figure}

%\begin{wrapfigure}{R}{8.0cm}
%	\centering
%    \includegraphics[width=8cm]{Figures/Reconstruction/PID_Schematic.jpg} 
%	\caption{Schematic illustration of particle identification in LHCb \cite{ProbNN}.} 
%	\label{fig:ParticleID_Schematic}
%\end{wrapfigure}
Efficient particle identification is obtained by combining the information of the two RICH detectors, the calorimeter system and the muon system. Fig.~\ref{fig:ParticleID_Schematic} schematically shows how the different sub-detectors provide information to identify electrons, muons, pion, kaons and protons. Charged particles and neutrals are separated since neutrals ($\gamma$, $\pi^0$) leave no hits in the tracking stations. Relativistic charged particles generate Cherenkov rings with a radius depending on their velocity, which, together with the momentum measurement in the tracking detectors, provides a measurement of their mass. The calorimeters use the relative energy deposits in the electromagnetic and hadronic systems to separate electrons from charged pions, and finally only muon tracks reach the muon stations. Each of the sub-detector measurements provide a PID-likelihood normalized on a default particle assumption: $\Delta\mathrm{log}{\cal L}\left(X-\pi\right)$ for the RICHes, $\Delta\mathrm{log}{\cal L}\left(X-h\right)$ for the calorimeter, and $\Delta\mathrm{log}{\cal L}\left(X-\mu\right)$ for the muon system. This allows to combine the PID information into a global particle identification likelihood. Alternatively, by including more individual sub-detector observables, a multilayer neural net was trained to provide a single probability, $\mathrm{ProbNN}_X$, for each particle hypothesis $X$. 
The $\mathrm{ProbNN}_X$ variable is based on six binary one-layer artificial neural networks implemented in the TMVA library~\cite{TMVA:2007ngy} where each network is trained to separate a given particle type from all others~\cite{ProbNN}.

\begin{figure}[!ht]
	\centering
    \includegraphics[width=5.3cm]{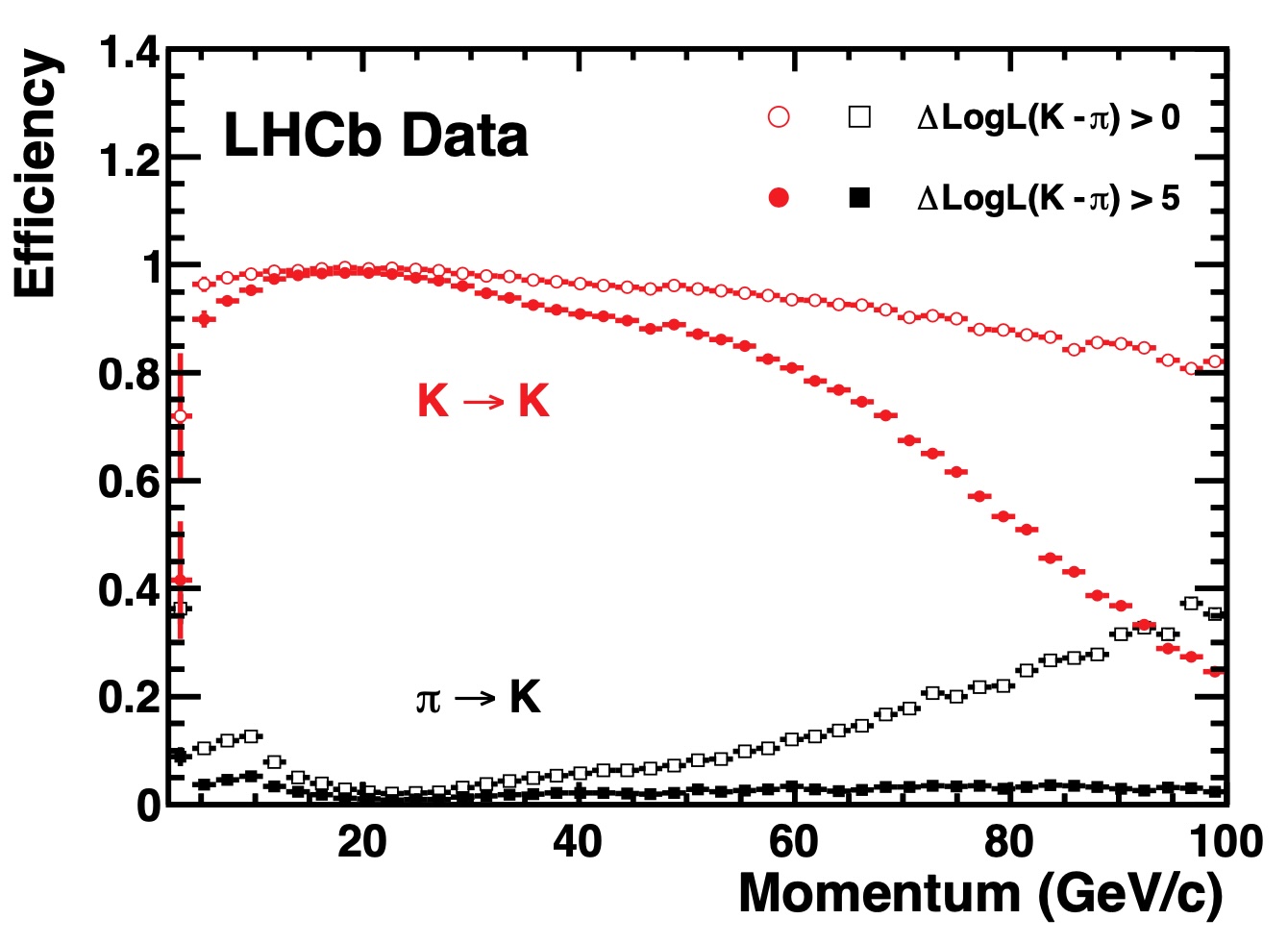} 
    %\hspace*{0.2cm}
    \includegraphics[width=5.5cm]{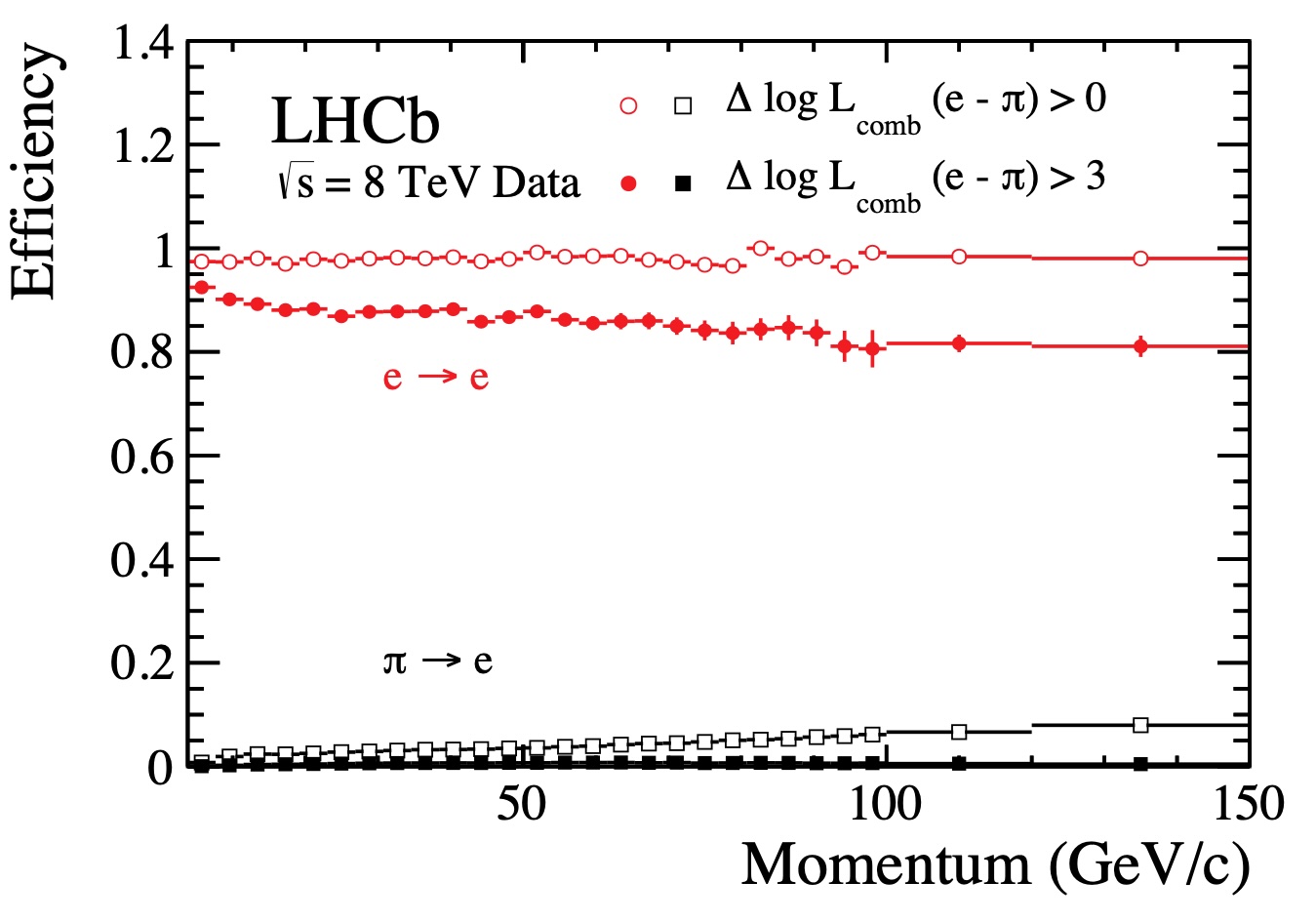} 
    %\hspace*{0.2cm}
    \includegraphics[width=5.5cm]{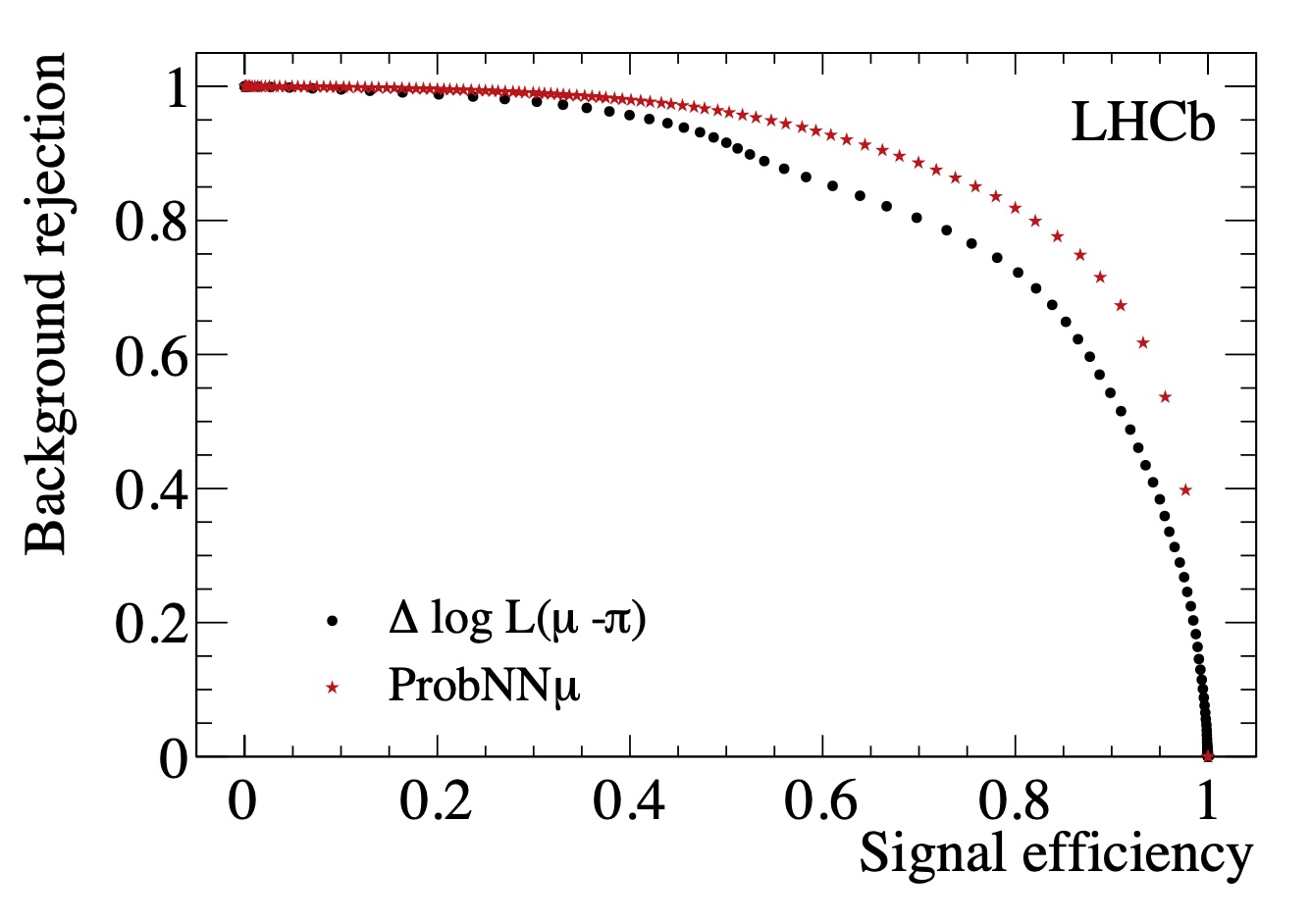} 
	\caption{Particle identification using global PID variables.  {\em Left:} Performance of pion - kaon separation for two values of the $\Delta\mathrm{log}{\cal L}\left(K-\pi\right)$ variable, {\em Center:} Performance of electron - pion separation for two values of the $\Delta\mathrm {log}{\cal L}\left(e-\pi\right)$ variable, {\em Right:} comparing background vs efficiency for muon identification as given by the $\Delta\mathrm{log}{\cal L}\left(\mu-\pi\right)$ and the $\mathrm{ProbNN}_\mu$ variables.}
    %seperation value kaon identification efficiency together with pion misidentification efficiency as function of track momentum. The plot shows the efficiency values for two decision criteria: $\Delta{\mathrm log}{\cal L}\left(K-\pi\right)>0$ and $\Delta{\mathrm log}{\cal L}\left(K-\pi\right)>5$. }
	\label{fig:ParticleID_Variables}
\end{figure}

%\begin{wrapfigure}{L}{7.5cm}
%	\centering
% 	\includegraphics[width=6.8cm, trim=0.0cm 0.0cm 0.0cm 12.5cm, clip]{Figures/Reconstruction/RICH_performance_CKAnglevsMom_NoTheory_jun2011.png} 
%	\caption{Particle Identification performance using the RICH detectors. Plot on Run3 performance extracted from Giovanni Cavallero talk, 11th Hadron Collide Physics conf, 22-26 May 2023, Belgrade.}
%	\label{fig:RICHPerformance}
%\end{wrapfigure}
%\vspace*{-0.5cm}
To separate photons and merged $\pi^0$ particles, a multi-layer perceptron neural network method is used, trained to distinguish energy depositions from photons and $\pi^0$'s in the ECAL, achieving 95\% efficiency for photon identification while rejecting 45\% of the $\pi^0$ candidates \cite{LHCb-Detector-Performance_2015}. 
The PID performance is estimated both with Monte Carlo simulations as well as from real LHC data where, similarly as in tracking, use is made of the "tag-and-probe" method. Here, a given calibration decay with low background is reconstructed ("tag") from kinematic variables without making use of the PID information of one final state particle ("probe"). The performance of the PID identification of that particle is obtained by comparing the correct identification with the information directly obtained from the detector. In Fig.~\ref{fig:ParticleID_Variables} the performance for kaon-pion separation is shown making use of $D^{*+}\rightarrow D^0(\rightarrow K^-\pi^+)\pi^+$ decays, for electron-pion separation using $B^+\rightarrow J/\psi(\rightarrow e^+e^-)K^+$ decays, and for muon-pion separation using $\Sigma^+\rightarrow p \mu^+\mu^-$ decays. The latter plot shows that the $\mathrm{ProbNN}_X$ variable is here somewhat more powerful than the $\Delta\mathrm{log}{\cal L}$ method.  

%\newpage
\subsection{Trigger and Real Time Analysis}

Flavour physics events are characterized by moderate transverse momentum particles produced in $b$- or $c$- hadron decays with a large forward boost, where also the particle density originating from both the underlying event and overlapping minimum-bias collisions (pile-up) is high.
As a result, sophisticated trigger strategies are required to efficiently select signal events while suppressing background.
To achieve this, the LHCb trigger system relies on software-based reconstruction, minimizing the dependence on coarse-grained hardware signals. This ensures both flexibility and the potential for future algorithm upgrades.

\begin{figure}[!th]
	\centering
    \includegraphics[height=6cm]{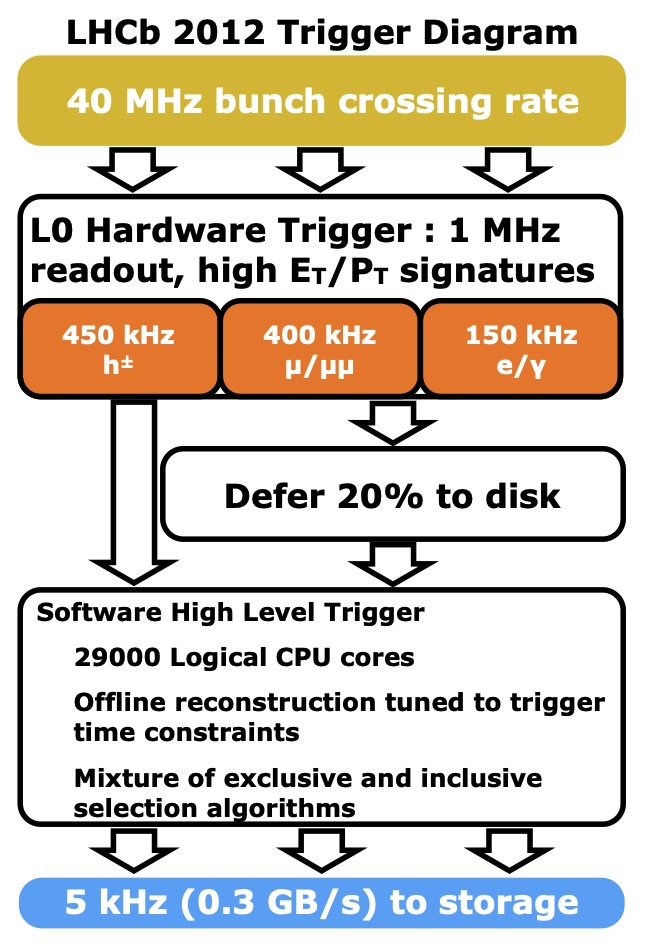}
    \hspace*{1.0cm}
    \includegraphics[height=6cm]{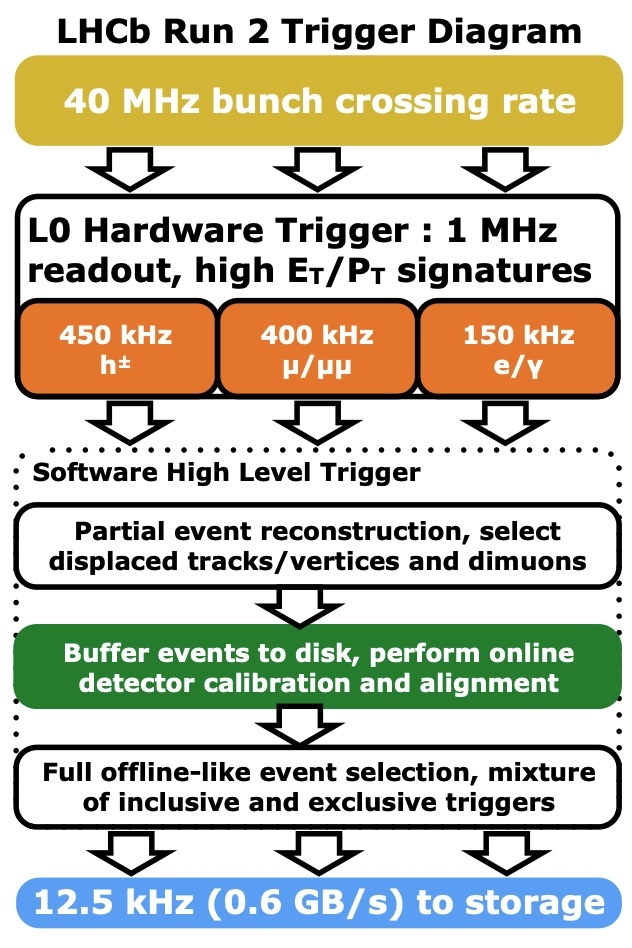}
    \hspace*{1.0cm}
    \includegraphics[height=6cm]{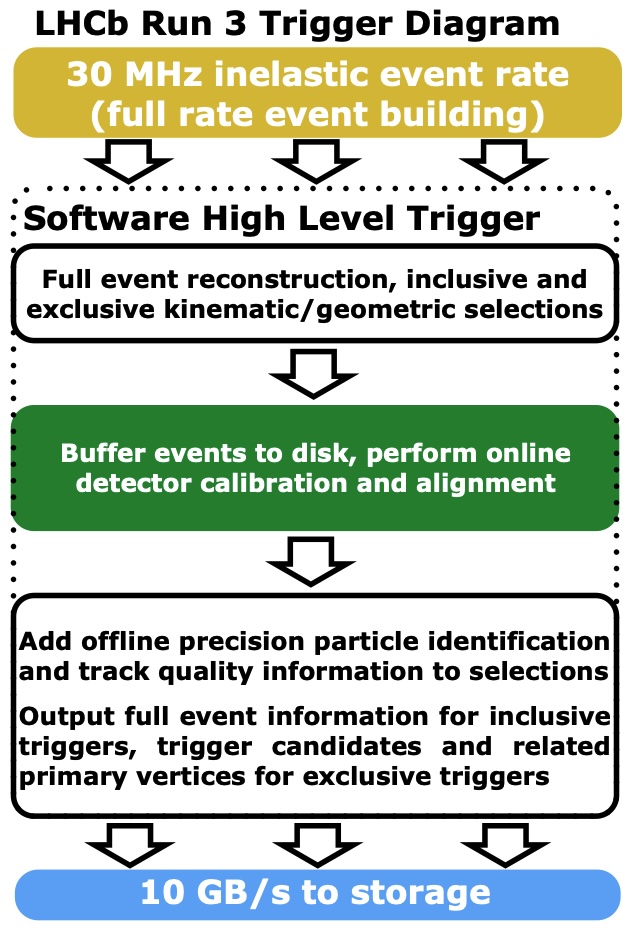}
    \caption{Diagrams for the LHCb Trigger data-flow in Run-1 ({\em left}), Run-2 ({\em middle}) and Run-3 ({\em right}).  Source: LHCb-FIGURE-2020-016.}
    %taken from https://cds.cern.ch/record/2730181/files/LHCb-FIGURE-2020-016.pdf .}
    %(left Source: doi:10.1088/1742-6596/762/1/012046}
	\label{fig:Trigger_DataFlow}
\end{figure}

Fig.~\ref{fig:Trigger_DataFlow} illustrates the evolution of the trigger scheme used in Run-1, Run-2 (LHCb-I) and Run-3 (LHCb-U). In LHCb-I a two-component trigger was used: a hardware level-0 (L0) and a software based High Level Trigger (HLT). In LHCb-U the L0 trigger was eliminated and all bunch crossing events are processed in a full software HLT.

\subsubsection{Trigger in LHCb-I} \label{subsec:ReconstructionTrigger}

\paragraph{Level-0 Trigger}
Taking into account the nominal luminosity at the LHCb interaction point and the total inelastic $pp$ cross section, the 40 MHz bunch crossing rate results in an event rate of about 12 MHz at a luminosity of $2 \times 10^{32}$cm$^{-2}$s$^{-1}$. In LHCb-I the L0 trigger reduces this event rate to a maximum acceptable rate of 1 MHz into the HLT, where the output rate to storage is further reduced to 5 kHz in Run-1 and 12.5 kHz in Run-2. The L0 makes use of the SPD, PS, ECAL, HCAL and MUON detectors and selects events that include high transverse energy signals of the ECAL (150 kHz for $e$ with SPD hits and $\gamma$ without SPD hits), the HCAL (450 kHz for hadrons) and high transverse momentum muon tracks (400 kHz). In addition, two dedicated $R$-strip stations in the VELO detector (the Pile-Up veto detectors) operate at 40 MHz readout and were initially intended for rejection of high pile-up events, but were used for online luminosity calculation.
%, and for event multiplicity information. 
To make maximum use of the compute power of the event filter farm of the HLT, about 20\% of the L0 triggered events are temporarily stored in a disk buffer, to be further processed by the HLT during inter-fill periods. This approach, known as the deferred trigger, also provides redundancy, preventing dead time in intermittent interruptions in the downstream data-flow.

\begin{figure}[t]
	\centering
    \includegraphics[width=14cm]{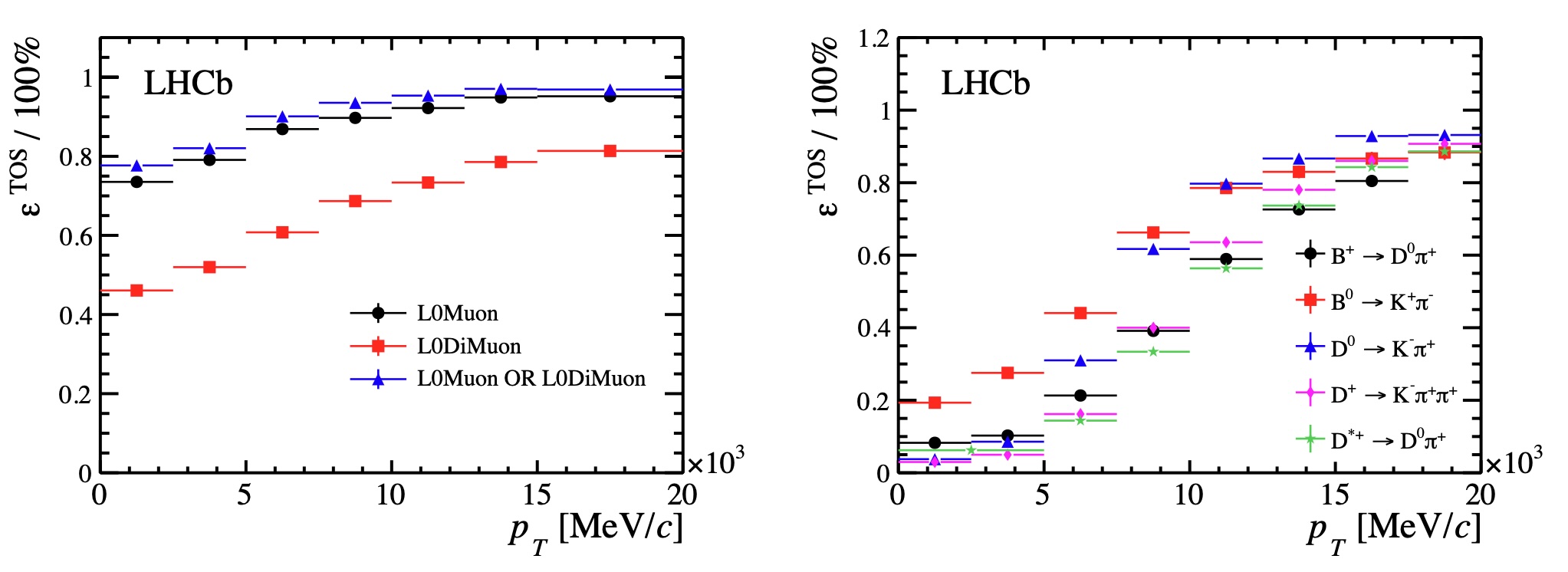} 
	\caption{Level-0 efficiency in LHCb-I Run-1 {\em Left:} L0 muon efficiency for $B^+\rightarrow J/\psi(\rightarrow \mu^+\mu^-)K^+$ events.  {\em Right:} L0 hadron efficiency of several hadronic $B$ or $D$ decay events, plotted as function of the $p_T$ of the decaying $B$ or $D$ meson. Taken from \cite{LHCb-Detector-Performance_2015}.} 
	\label{fig:Level0_Efficiency}
\end{figure}

The L0 trigger records the trigger decision for each signal object, which corresponds to a final state particle. This enables a data-driven determination of trigger efficiencies by analyzing offline selected events where no trigger requirement has been made on the any of the decay products of the signal.
This is referred to as Trigger-Independent-of-Signal (TIS) events, contrary to Trigger-On-Signal (TOS) events, in which the signal particles have been used in the L0 trigger decision. Using events that are both triggered on TIS and TOS the trigger efficiency is calculated as $\varepsilon=N(\mathrm{TOS}\&\mathrm{TIS}) / N(\mathrm{TIS})$. They are evaluated for benchmark muon and hadron triggered channels, as shown in Fig.~\ref{fig:Level0_Efficiency}. 

The muon trigger considers two muon candidates and selects events with either a single muon passing a $p_T$ threshold, or events with the product of the two highest $p_T$ muons passing a threshold. The figure on the left shows the performance of each as well as their combination. The figure for the hadronic trigger on the right indicates that its efficiency strongly depends on the transverse momentum criterium applied in the trigger selections. As expected, the $B$-decays have a higher efficiency than the $D$-decays, and more specifically the two-prong $B$-decays are more efficient than the four-prong decays, eg. $D^+\rightarrow K^-\pi^+\pi^+$ decays.

\begin{figure}[!ht]
	\centering
    \includegraphics[width=6.8cm]{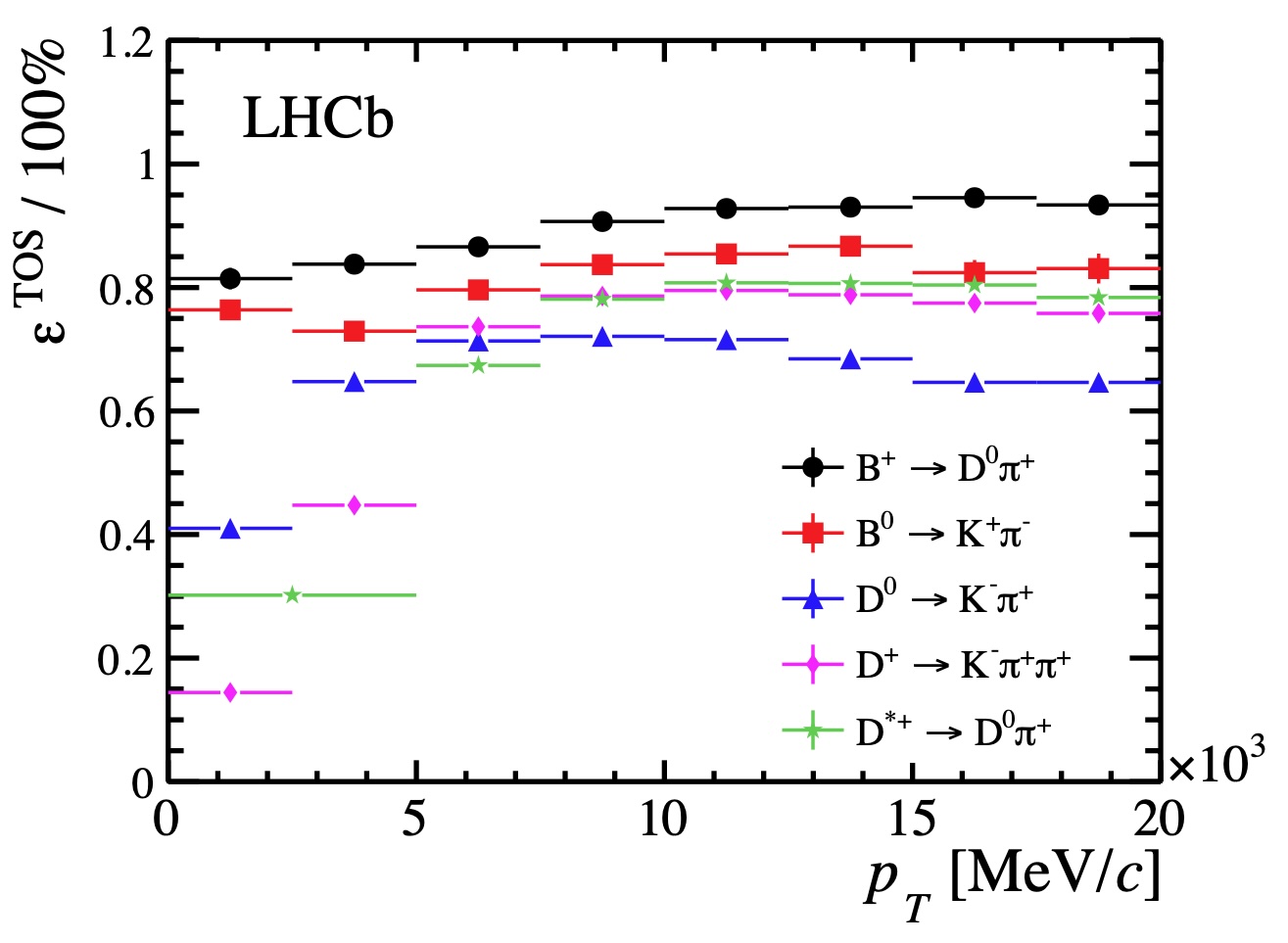} 
    \hspace*{0.3cm}
    \includegraphics[width=6.8cm]{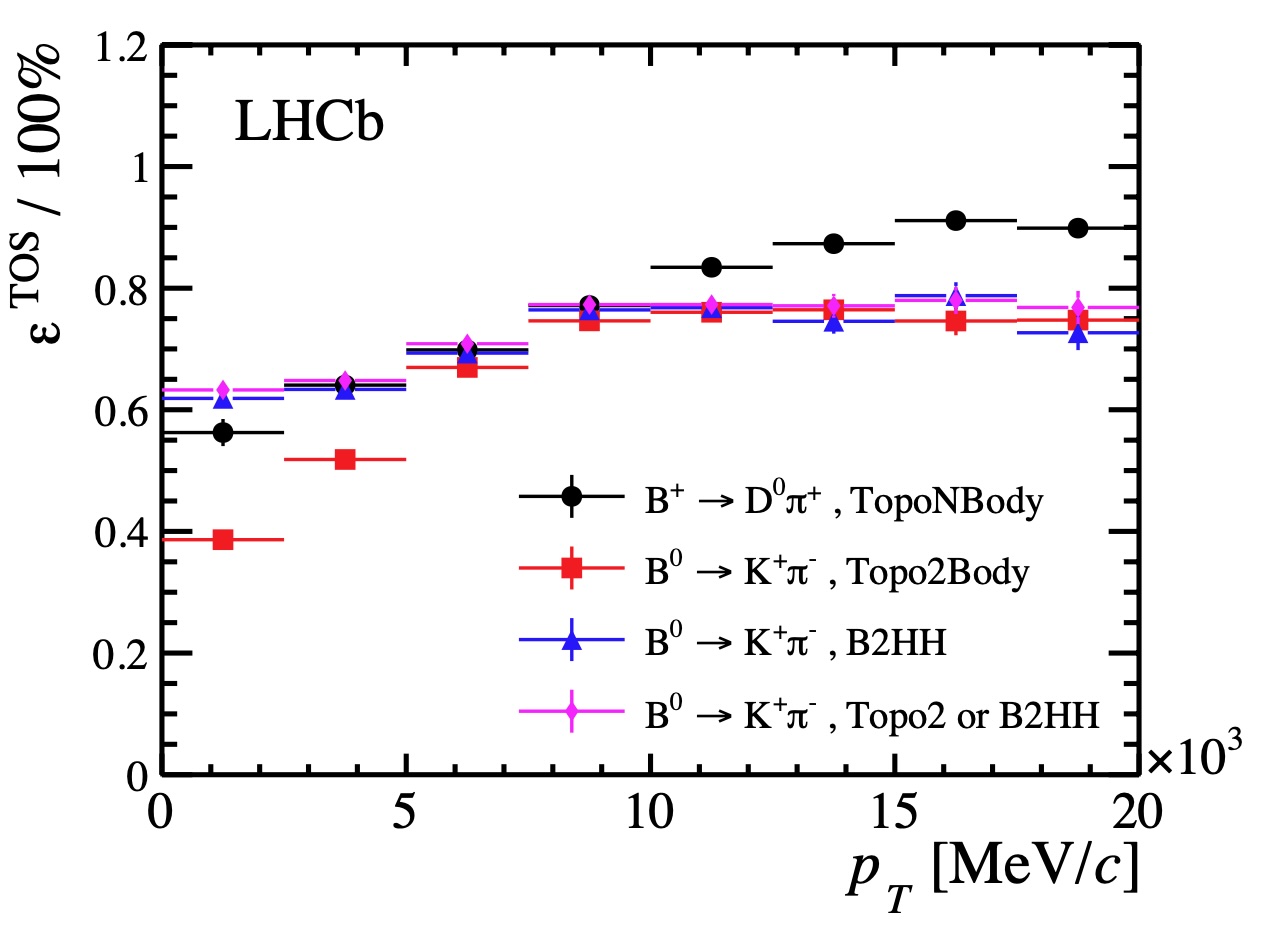} 
	\caption{HLT-1 efficiency in LHCb-I Run-1. {\em Left:} Inclusive track trigger performance shown as TOS-efficiency versus $p_T$ for various beauty and charm modes.  {\em Right:} HLT-2 TOS-efficiency for inclusive $B$ decays and for the exclusive $B^0\rightarrow K^+\pi^-$. Taken from \cite{LHCb-Detector-Performance_2015}.}
	\label{fig:HLT_Efficiency}
\end{figure}

\paragraph{High Level Trigger}

The HLT software trigger reconstructs events in an offline-like manner but tuned for real-time constraints. HLT is split in two stages: HLT-1, which performs a partial (primarily tracking) event reconstruction, and HLT-2, which performs event reconstruction and selects events based on signal decays. 
The HLT evolved significantly from Run-1 to Run-2, incorporating increasingly more sophisticated software based features over time. 

In Run-1, HLT-1 reconstructed VELO tracks and created primary vertices (PVs) if they contained at least five associated tracks. To be considered as candidate for forward $b$- or $c$-hadron decay tracks,  VELO tracks are required to have a significant impact parameter distance to any PV (typically $IP > 100$ $\mu$m). 
HLT-1 searches for continuations of VELO tracks in the downstream trackers, albeit limited to tracks with momentum criteria of typically $p_T> 0.3$ GeV and $p>3$ GeV. For Run-2 the VELO tracking performance was improved to offline quality, and the forward tracking was extended to include all tracks, by first confirming VELO track extrapolations into the TT station to provide an initial moment estimate, and subsequently searching for T-station continuations.
Following the tracking algorithm sequence, particles are constructed by associating them with calorimeter clusters and muon tracks. Neutral particles ($\gamma$, $\pi^0$) are built starting from the energy clusters in the calorimeters. 

HLT-2 uses the reconstructed particles and performs event reconstruction including a mixture of inclusive and exclusive event selections. A key feature of HLT-2 are the "topological" trigger lines, including partial reconstruction of inclusive N-body $B$-decays, making use of a secondary vertex reconstruction separated from the primary vertex. In addition to the inclusive lines an extensive list of exclusive event candidates are selected. The right side of Fig.~\ref{fig:HLT_Efficiency} shows examples of the efficiency of the inclusive topological lines and the exclusive line of $B^0\rightarrow K^+\pi^-$ events.

\begin{figure}[t]
	\centering
    \includegraphics[width=8cm]{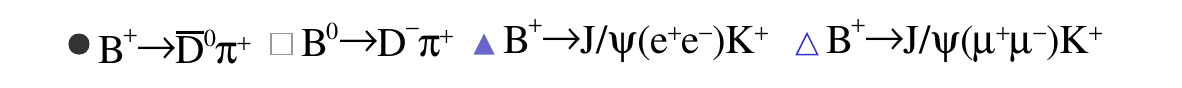}
    \hspace*{-0.5cm}
    \raisebox{0.1cm}{\includegraphics[width=6.5cm, trim=6.5cm 0cm 0cm 1cm, clip]{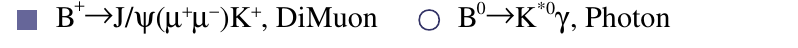}}\\
    \includegraphics[width=5.cm]{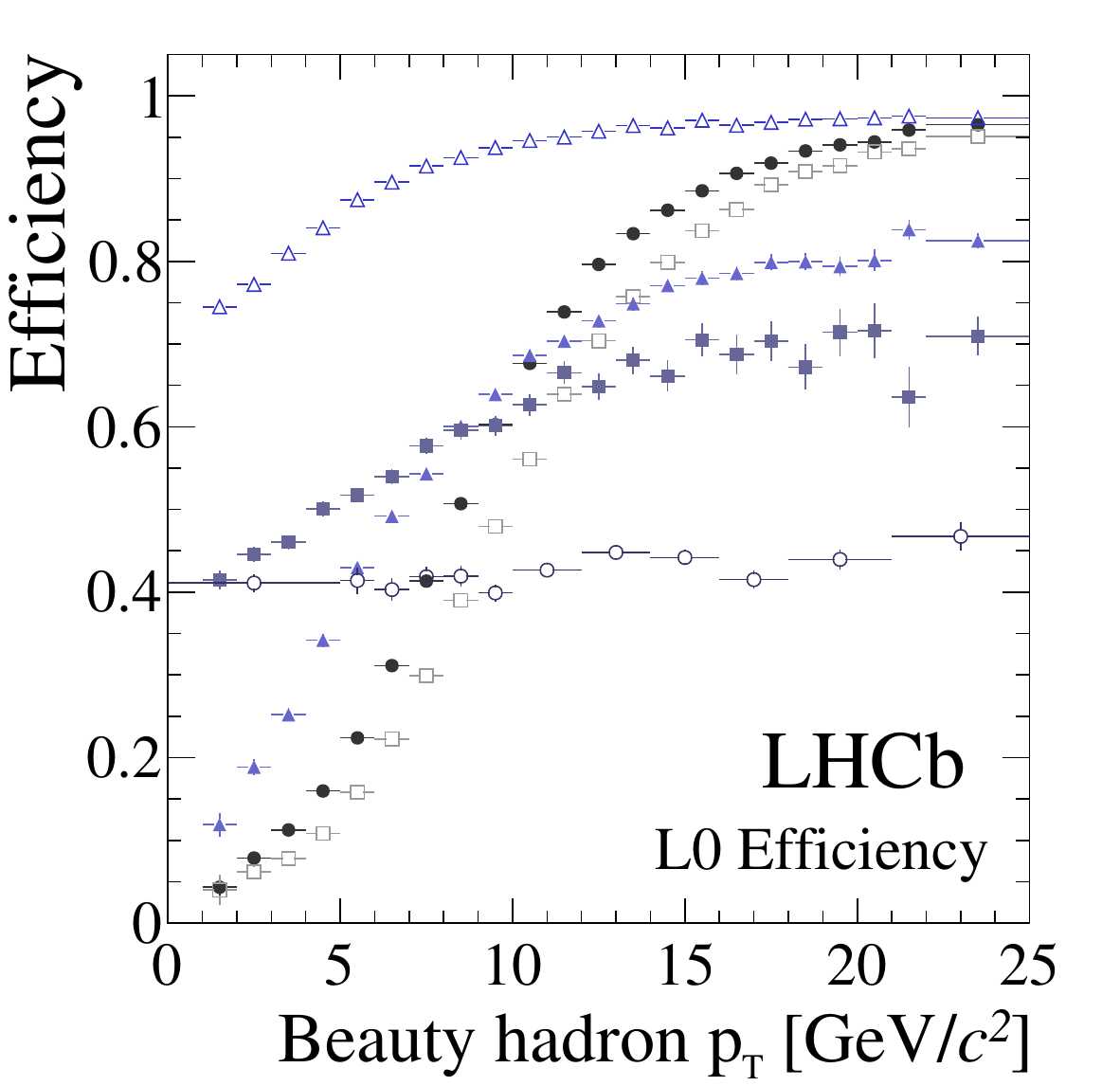}
    \includegraphics[width=5.cm]{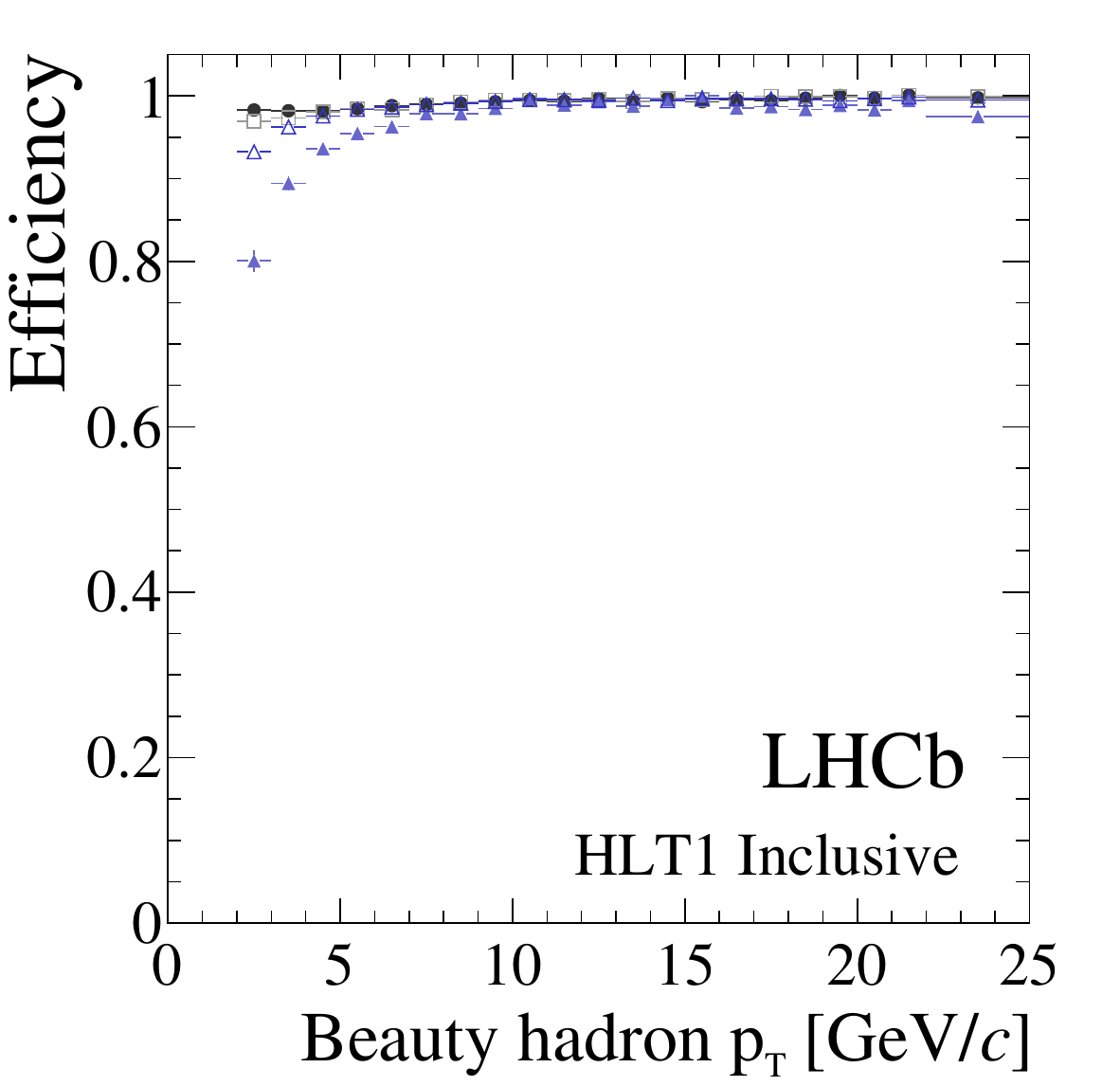} 
    \includegraphics[width=5.cm]{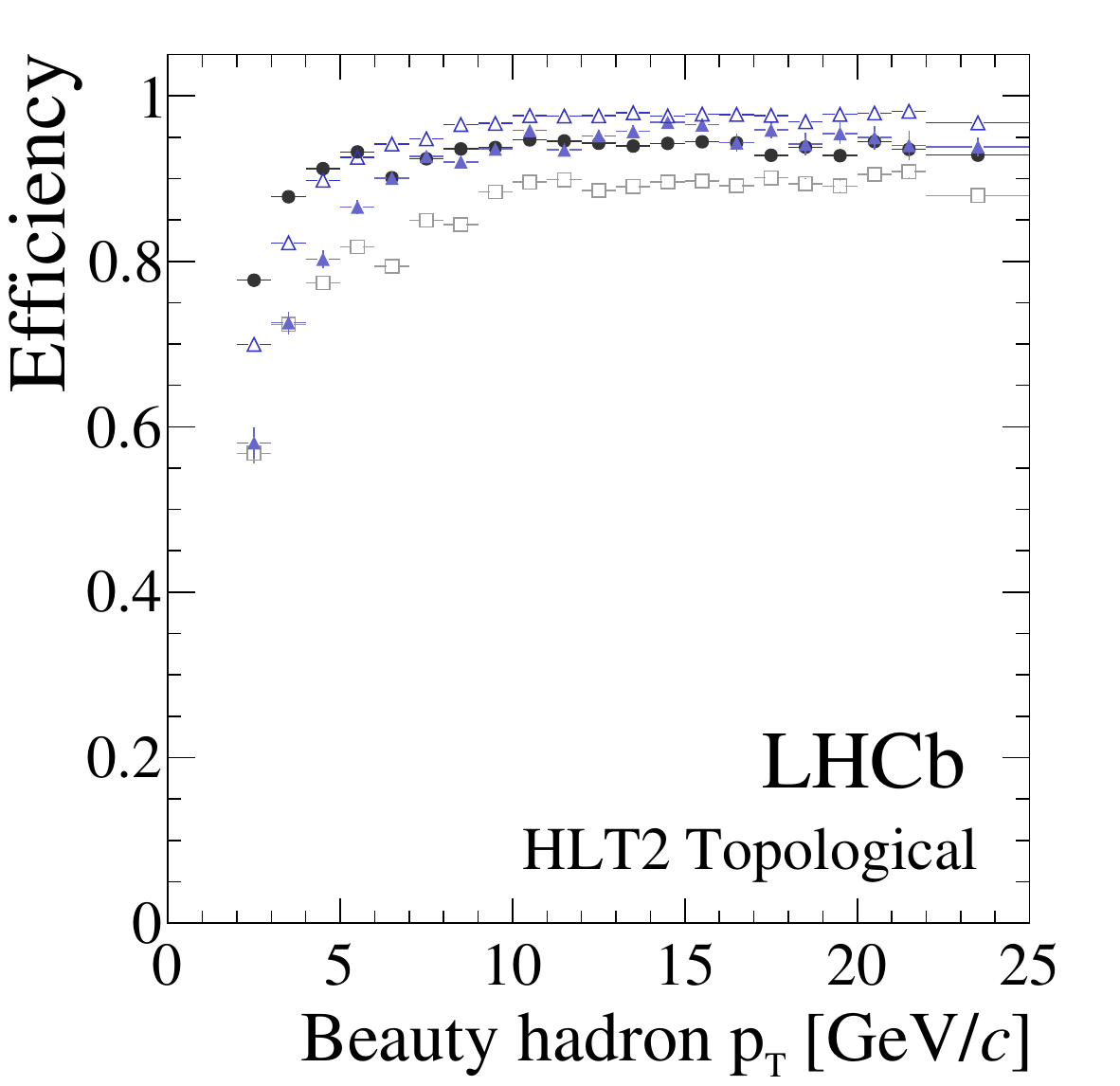} 
	\caption{Trigger efficiencies in LHCb-I Run-2 for various $B$-decay channels. All efficiencies are obtained using the TISTOS method. 
    % $\mathrm{} Eff}=(N_{TIS\&TOS})/(N_{TIS})$. 
    {\em Left:} Performance of L0 as function of $p_T$ for various $B$-decays. {\em Center:} Performance of inclusive HLT-1 line as function of $p_T$. {\em Right:}  Performance of the topological HLT-2 line as function of $p_T$~\cite{LHCb-DP-2019-001}.}
	\label{fig:Trigger_Efficiency_Run2}
\end{figure}

In Run-2 the deferred trigger strategy was applied for all events by explicitly splitting HLT-1 and HLT-2 with a buffer between them. The buffer allows for online detector calibration and alignment, enabling more precise track and event reconstruction before the HLT-2 selection.
Furthermore, various upgrades were introduced for Run-2, laying the foundation for the real-time-analysis approach exploited in LHCb-U. HLT-2 follows a identical strategy as offline, reconstructing all tracks and adding particle identification information. It implements inclusive trigger lines for beauty and charm events, as well as many exclusive trigger lines leading to a total of about 400 trigger selections. A major change was to no longer store raw data, but instead store reconstructed information for selected particles in the events in the so-called "Turbo" stream, leading to a larger bandwidth in event storage to disk.
The performance of the Run-2 trigger is illustrated in  Fig.~\ref{fig:Trigger_Efficiency_Run2}.

%\begin{figure}[t]
%	\centering
%    \includegraphics[width=10cm]{Figures/Reconstruction/HLT1_Beauty_Legend.pdf} \\
%    \includegraphics[width=6.0cm]{Figures/Reconstruction/HLT1TotalEff_Beauty_PT.pdf} 
%    \includegraphics[width=6.0cm]{Figures/Reconstruction/HLT1TotalEff_Beauty_TAU.pdf} 
%	\caption{HLT-1 efficiency in LHCb-I Run-2. {\em Left:} Performance of inclusive HLT-1 as function of $p_T$ for various $B$-decays.  {\em Right:}  Performance of inclusive HLT-1 as function of decay-time for various $B$-decays~\cite{LHCb-Trigger-Run2}.}
%	\label{fig:HLT1_Efficiency_Run2}
%\end{figure}

%\begin{figure}[!ht]
%	\centering
%    \includegraphics[width=10cm]%{Figures/Reconstruction/HLT2TotalEff_Beauty_Legend.pdf} \\
%    \includegraphics[width=6.0cm]{Figures/Reconstruction/HLT2TotalEff_Beauty_PT.pdf} 
%    \includegraphics[width=6.0cm]{Figures/Reconstruction/HLT2TotalEff_Beauty_TAU.pdf} 
%	\caption{HLT-2 topological trigger efficiency in LHCb-I Run-2. {\em Left:} Performance as function of $p_T$ for various $B$-decays. {\em Right:} Performance shown as function of decay-time for various $B$-decays~\cite{LHCb-Trigger-Run2}.}
%	\label{fig:HLT2_Efficiency_Run2}
%\end{figure}

\subsubsection{Trigger in LHCb-U}

%\begin{figure}[h]
%	\centering
%    \includegraphics[width=12cm]{Figures/Reconstruction/LHCbOverallDataFlow.jpg} 
%	\caption{LHCb Upgrade dataflow. Make something similar to this. Source: taken from starterkit talk }
%	\label{fig:LHCbUOverallDataFlow}
%\end{figure}

One of the primary motivations behind the LHCb upgrade detector was to significantly enhance the trigger performance while accommodating a fivefold increase in LHC luminosity. In particular, the goal was to maintain the already high efficiency of muon triggers while improving the efficiency of hadronic triggers. 
Higher luminosity would require stricter L0 thresholds on calorimeter objects, nullifying the luminosity gain.
Achieving higher efficiency required a fundamental change in the trigger strategy: removing the hardware-based L0 trigger and implementing a fully software-driven selection capable of reconstructing and analyzing all detector data in real time,
using impact parameter information. This transition, known as Real-Time Analysis (RTA), enables full event reconstruction at the 40 MHz collision rate, ensuring that every non-empty beam crossing is analyzed with high precision.

The approach from Run-2 with a buffer inbetween HLT-1 and HLT-2 for intermediate calibration and alignment is maintained. This two-stage approach allows to reduce the event rate by a factor of 20 in both and HLT-1 and HLT-2, resulting in an overall factor 400. The reduction in HLT-1 is obtained by an inclusive selection, reconstructing all tracks and primary vertices. 

In HLT-1, for long tracks a relative momentum resolution of better than 1\% (in comparison to the offline resolution of 0.5\%) is obtained as well as a precise impact parameter measurement accompanied by its uncertainty. Displaced tracks are subsequently identified as muons or non-muons and fitted to two-body secondary vertex candidates. The inclusive selections of the HLT-1 include a displaced single track (or muon trigger) requiring large transverse momentum and significant non-zero impact parameter, a two-track (or two-muon) vertex trigger with similar criteria, or a high-mass di-muon trigger without any displacement requirement. Compared to the generic track triggers the identified muon triggers benefit from lower background and therefore require less stringent $p_T$ and IP criteria.
To achieve the necessary processing speed, HLT-1 is fully executed on GPUs within the event builder nodes, substantially enhancing computational speed.

The HLT-2 runs on a CPU farm and implements full event reconstruction with offline-level precision.
HLT-2 implements about 1.000 selections, each tuned for a specific physics topic. For the majority of events the reconstructed information stored in the "Turbo" stream includes objects or particles specifically requested in that selection algorithm. In total 68\% of the data is written in Turbo mode, 26\% includes storage of full events and 6\% of data is dedicated to specific calibration modes.

The data stored is further optimized using a so-called "Sprucing" step built on top of the Turbo, to make sure the amount of data stored is minimal. 
Each physics group tailors the sprucing system retaining the requested information relevant to their analysis. The sprucing system considers all the requested information and discards redundant information further reducing event size. In case multiple selections request to store information for the same events, the super-set of event information requested will be stored.

\subsection{Flavour Tagging}

\begin{figure}[b]
	\centering
	\includegraphics[height=5cm]{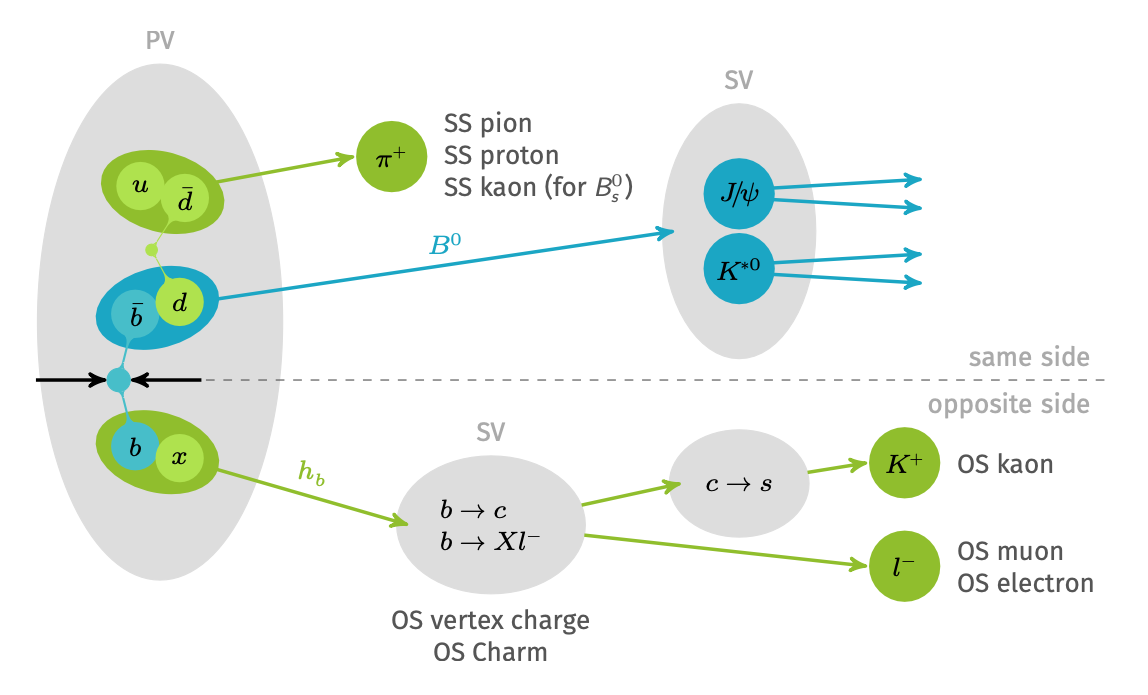} 
    \hspace*{0.5cm}
    \includegraphics[height=5cm]{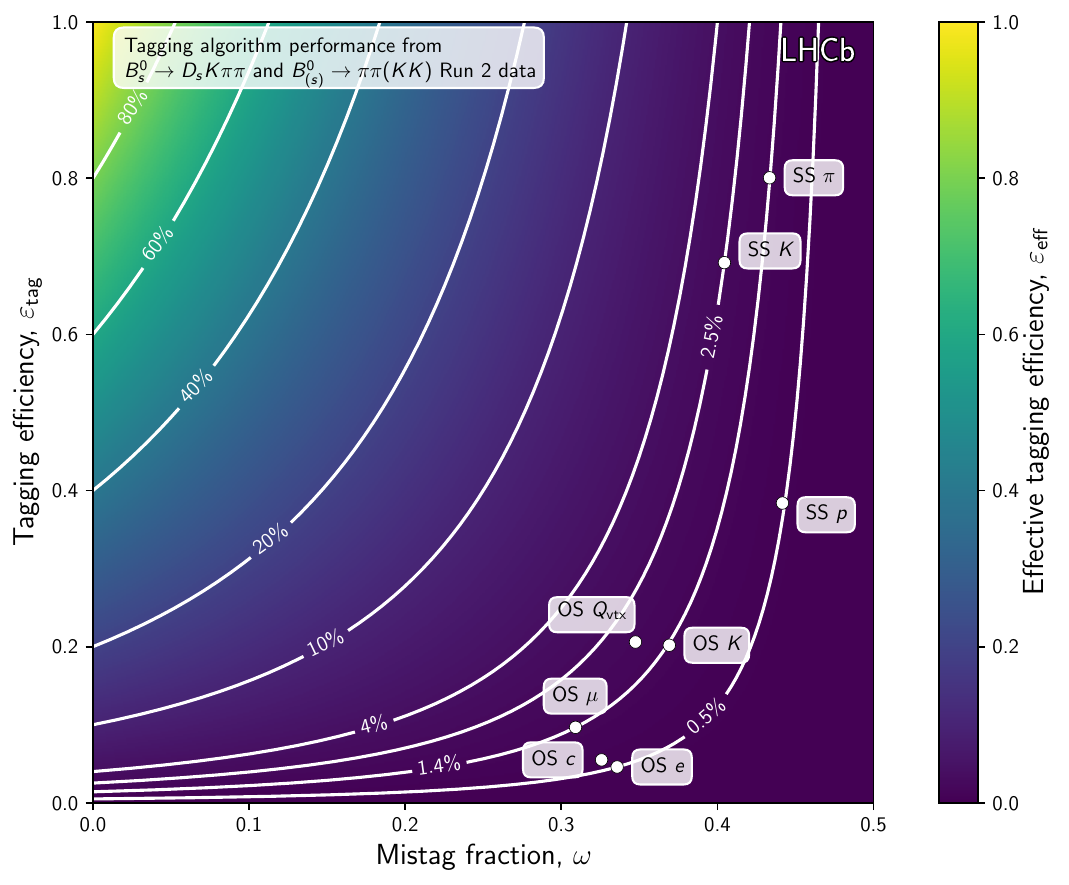}
	\caption{{\em Left:} Overview of LHCb flavour tagging methods~\cite{Fazzini:2018dyq}. The signal $B$-meson decay is indicated in blue, the tagging information is indicated in green. The signal $b$- and opposite $b$-decay particles are displayed in the top and bottom half of the plot. Each of the tagging methods is highlighted. {\em Right:} The eﬀective tagging eﬃciency  $\varepsilon_{\rm tag}$ for diﬀerent algorithms 
    is displayed in the $\omega$-$\varepsilon_{\rm tag}$ plane. The numbers for the OS taggers are extracted from the $B_s^0\rightarrow D_s^\mp K^\pm\pi^+\pi^-$ decays while the numbers for the SS taggers are extracted from $B_{(s)}^0\rightarrow \pi^+\pi^-\left(K^+K^-\right)$ decays. Diﬀerent contours of equal eﬀective tagging eﬃciency values are displayed \cite{FlavourTaggingPerformanceRun2}.}
	\label{fig:FlavourTagging}
\end{figure}

The identification of the production flavour ($b$ or $\bar{b}$) of $B^0$ or $B_s^0$ mesons is essential in analyses involving $B$-meson mixing and time-dependent CP violation. Since the meson’s production flavor must be determined independent from its decay, it must be determined using information present in the rest of the event.
This technique, known as flavor tagging, is fundamentally different at hadron colliders in comparison to the asymmetric $e^+e^-$ $B$-factories BELLE and BaBar. At LHC the produced $b$ and $\bar{b}$ particles hadronize independently on top of an underlying event, whereas at the $B$-factories a $B\bar{B}$ meson pair is produced exclusively in a coherent state such that analyzing the {\em tagging B} decay determines the flavour of the {\em signal B} at the start of its independent oscillation process.

The tagging methods used can be divided in two categories: classical taggers in which the tagging decision depends on a single "tagging" particle, or inclusive taggers in which many observables are used simultaneously, using a machine learning method. A schematic overview of the classical taggers, used for most analyses in Run-1 and Run-2, together with their performance, is given in Fig.~\ref{fig:FlavourTagging}.
The tagging performance is expressed by the combined value of the efficiency, $\varepsilon_{\text{tag}}$ and the probability of making an incorrect tag decision $\omega$, resulting in a per-event effective tagging power $\varepsilon_{\text{eff}}=\varepsilon_{\text{tag}}(1-2\omega)^2$.

\subsubsection{Classical Taggers}
During Run1 and Run2, the physics analyses requiring flavour tag used the classical taggers. The same-side tagging method uses the idea that hadronization in the gluon-string fragmentation process produces a quark nearby in phase space with opposite charge to the light quark in the $B$-meson: i.e. a $\pi^+$ ("SS pion") or proton ("SS proton") for a $B^0$-meson, and a $K^+$ ("SS kaon") for a $B_s^0$-meson. The method has an intrinsic limitation that in about half of the cases the nearest particle is of neutral charge, producing no tag. 

The opposite-side tagging method relies on the fact that a $b\bar{b}$ pair are of opposite flavour. The flavour of the signal $b$-hadron is then inferred from analyzing the decay of the other $b$-hadron. As shown in Fig.~\ref{fig:FlavourTagging} the flavour of the tagging $b$-hadron can be obtained from a reconstructed charm meson ("OS charm"), the total charge of the decay vertex ("OS vertex-charge"), the charge of the kaon in the decay chain ("OS kaon"), or from the charge of lepton ("OS" muon or "OS electron"). The opposite-side tag is diluted by several effects: the tagging $b$-particle may decay (partially) outside the acceptance or, in case it hadronizes to a $B^0$ or $B_s^0$ it may have oscillated and hence provide an incorrect tag for the signal $B$. The result of these five opposite tagging algorithms are combined into an overall tagging decision together and an estimated per-event mis-tag rate $\omega$.

The criteria to select the tagging track uses kinematical, geometrical, and particle identification information and is optimized for each individual tagging algorithm. 
Calibration of the flavour tagging algorithm typically uses a self-tagging signal decay, where the tag can be directly obtained from data, such as the decay $B^+\rightarrow J/\psi K^+$. 
%The performance of the algorithm is expressed as a tagging power $\varepsilon_{\rm eff}=\varepsilon_{\rm tag}\left(1-2\omega\right)$, where $\varepsilon_{\rm eff}$ represents the efficiency that the algorithm provides a tagging decision, and $\omega$ is the probability that it is the wrong tag. 
In case more than one algorithm provides a tagging decision, their combination is used to improve the estimated mis-tag rate $\omega$. Typical effective efficiencies for $B^0\rightarrow D^-\pi^+$ events is $\varepsilon_{\rm eff} = 3.6\% \pm 0.1\%$ for the combination of opposite side tagging algorithms and $\varepsilon_{\rm eff} = 2.4\% \pm 0.1\%$ for the same-side tagging procedure~\cite{FlavourTaggingRun2}.

\subsubsection{Inclusive Flavour Tag}

Instead of deriving the tag decision from a single track, the inclusive flavour tag method uses the entire event information, including kinematic, topological, tracking and PID information from all
tracks to infer the flavour of the signal $b$. The method uses a variable number of inputs since the event multiplicity varies from event to event, and is aimed to provide tagging information in the real-time environment of LHCb-U. An initial implementation \cite{FastInclusiveTagger} uses the DeepSet Neural Network \cite{DeepSetNN}. Due to its inclusive nature the efficiency of the algorithm is 100\% and indicates in simulation studies an effective tagging power about 25\% higher than the classical taggers.

\subsection{Calibration Methods}

In the following chapter the performance of LHCb is presented in terms of physics observables as for example CP asymmetries, decay rates and hadronic resonances. 
Achieving optimal performance of both the trigger and offline analyses relies on a range of calibration procedures. 
These include the calibration of the reconstructed mass scale and resolution, (mis-)identification rates of particle types, and determining the efficiencies of selection procedures.

\paragraph{Momentum, Mass and Decay-time resolution}

A high resolution of invariant mass reconstruction is essential for candidate selection and background rejection, and relies on an unbiased momentum reconstruction.
Since the tracks are fitted across the VELO, upstream and downstream tracking subsystems, this requires a full alignment procedure including the degrees of freedom of many individual detector units. To achieve this, a global track covariance matrix implemented into the Kalman filter allows to determine the large number of alignment degrees of freedom using a large sample reconstructed tracks \cite{Hulsbergen:2008yv, Borghi:2017hfp}.
To verify the correctness of the mass scale known particle masses are used. From low to high mass, in particular two-prong $J/\psi$, $\Upsilon$ and $Z$-boson decays are used. In addition to the central values also the resolution of the mass reconstruction can be determined from data, as was shown in the example if $\Upsilon$ reconstruction in Fig.~\ref{fig:Momentum_Mass_Resolution}.

The reconstruction of $B_s^0$ oscillations relies on correct values of the event-by-event resolutions of the reconstructed decay times. The overall decay time uncertainty can be calibrated by fitting the observed negative-times tail of the reconstructed decay time distribution, which should be due to resolution effects only. This method either uses a dedicated "unbiased" event sample, selected without a decay time requirement. In the latter case the resolution is obtained from random combinations of prompt $\pi^+$ and $D_s^-$ particles.

\begin{figure}[htb]
	\centering
    \raisebox{0.4cm}{
    \includegraphics[height=4.2cm]{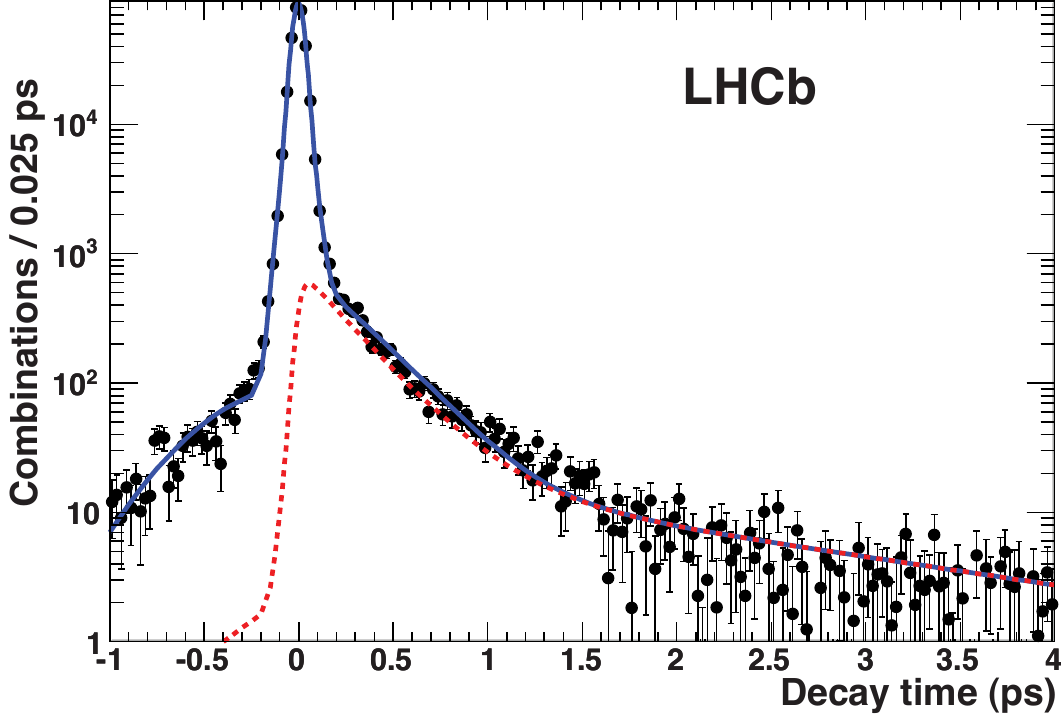}}\hspace*{0.5cm}
    \includegraphics[height=4.7cm]{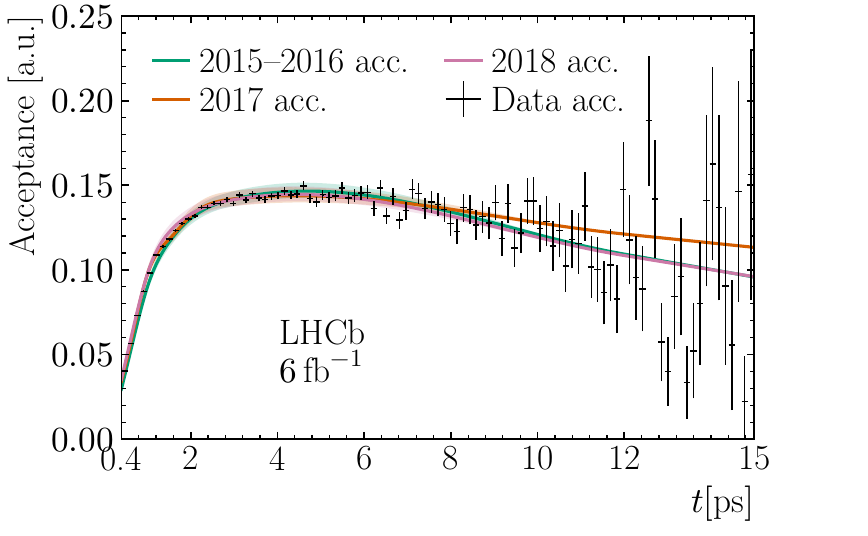} 
	\caption{Decay-time dependent reconstruction effects. {\em Left:} Decay time resolution calibration using the negative tail for $B_s^0\rightarrow J/\psi \phi$ decays~\cite{LHCB-PAPER-2011-031}.
    {\em Right:} Event selection efficiency ("acceptance") as function of decay time for $B_s^0 \rightarrow D_s^- \pi^+$ events~ \cite{LHCb-PAPER-2021-005}.}
	\label{fig:ResolutionAcceptance}
\end{figure}

\paragraph{Efficiencies and Acceptance}

\begin{wrapfigure}{R}{7.0cm}
	\centering
\includegraphics[width=8.0cm]{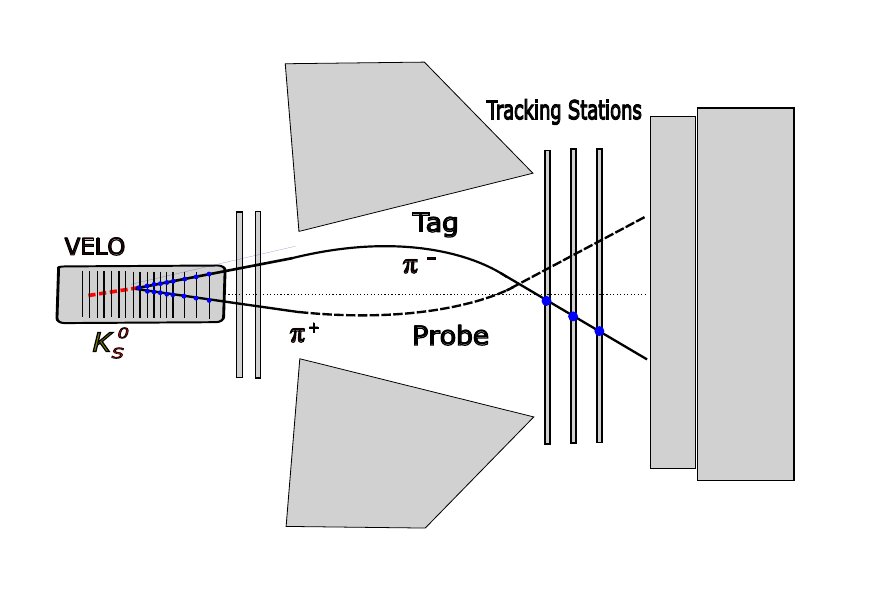} 
	\caption{The principle of the "tag-and-probe" method, to determine the track finding efficiency in the tracking stations.}
	\label{fig:tagandprobe}
\end{wrapfigure}

In measurements of decay rates and branching fractions, knowledge of the selection efficiencies are required. Various data driven methods are used, typically involving a normalization on calibration channels with known branching ratios. Track reconstruction and particle identification efficiencies are found using "tag-and-probe" methods, where not requiring tracking or particle identification information of one particle ("probe") of an otherwise cleanly reconstructed ("tagged") signal decay is used to determine the efficiency for finding the track or determining the identification of the "probed" particle, see Fig.~\ref{fig:tagandprobe}.

The trigger efficiency in LHCb is defined as the efficiency that an event that would be selected in the offline analysis, is actually selected by the trigger.
This efficiency is determined by flagging which specific particle or signature caused an event to be triggered. Using the fact that offline selected events are occasionally also selected on an independent signature referred to as "Trigger Independent of Signal" (TIS) as well as using the signal particles in "Trigger On Signal"(TOS), then the trigger efficiency can be determined the ratio $\epsilon_{\text{trigger}}= N_{TIS \& TOS} / N_{TIS}$.

The main signature separating heavy flavour particles from backgrounds, apart form their mass, is their non-zero decay flight, resulting in a secondary decay vertex associated to the primary collision vertex. The efficiency with which these signatures are selected depends on the actual decay-time for each event. The decay-time dependent efficiency, also referred to as acceptance, must be corrected for each time dependent analysis. For these tagged analyses typically the time dependent acceptance function for the untagged decay rate is compared to the known exponential of the decaying mother particle.  Alternatively, acceptance shapes can be fitted on the data together with the (independent) physics observables. The acceptance curve for $B_s^0\rightarrow D_s^-\pi^+$ events, used in the measurement of the $B_s^0$-$\bar{B}^0_s$ oscillations, is shown in Fig.~\ref{fig:ResolutionAcceptance}. It is often found that the efficiency for events with long decay times reduces.

\paragraph{Magnet Field Reversal}
A powerful cross check on analyses is the inversion of the current in the large dipole magnet, which reverses the direction of the vertical magnetic field. 
A switch of the magnet field is done several times each data-taking year and provides two semi-independent experimental results for a given analysis. As a switch of the magnetic field implies a left-right bending switch of charged particles in the detector, this also provides a cross check on the treatment of local detection efficiencies. Most analyses in LHCb require a consistent result between the "magnet-up" and "magnet-down" data.

%\section{Physics topics addressed by the experiment}\label{secPhysics}
%Please provide a summary of the physics topics addressed by the experiment.

%\subsection{Flavour Physics}\label{subsecFlavour}

\clearpage

\section{Flavour Mixing}\label{secMixing}

The phenomenon of neutral meson oscillations is of fundamental importance for several reasons. 
First, since the mixing process occurs via higher-order diagrams in the Standard Model, 
it is sensitive to potential contributions from new physics that could modify the predicted transition amplitude. 
Second, neutral meson oscillations play a crucial role in the determination of CKM matrix parameters. 
They provide an additional transition amplitude from the initial state of a neutral meson to a given final state, 
which is essential for extracting the relative phase difference between interfering amplitudes, as will be discussed in the next chapter. 
Third, the observation of two neutral kaon particles with vastly different lifetimes and the resulting discovery of CP violation represents a historically significant milestone in particle physics~\cite{cronin:prl:13:138}. 
%It is described in terms of a superposition of $|K\rangle $-states and its quantum-mechanical evolution.
Furthermore, differences in mass (and, consequently, in the available phase space for the final state particles) 
and differences in coupling strengths (via the CKM elements) lead to strikingly different phenomenologies 
across the neutral meson systems, such as $K^0$, $D^0$, $B^0$, and $B^0_s$.

Taking the example of the $B^0$ system, the time evolution can be described as~\cite{Nir:2005js}
\begin{equation}
 i\frac{\partial\psi}{\partial t} = H\psi=(M -\frac{i}{2}\Gamma)\:\psi=
\left(\begin{array}{cc}
M_{11} -\frac{i}{2}\Gamma_{11} & M_{12}-\frac{i}{2}\Gamma_{12} \\
M_{21} -\frac{i}{2}\Gamma_{21} & M_{22}-\frac{i}{2}\Gamma_{22} 
\end{array} \right)
\psi
\label{eq:mixingHamiltonian}
\end{equation}
with $\psi(t)$ expressed in the subspace of $B^0$ and $\bar{B}^0$ as follows
$
\psi(t) = \left(
\begin{array}{c}
B^0\\
\bar{B}^0
\end{array}
\right).
$
We note further that from CPT invariance it follows that particles and antiparticle have equal mass and lifetimes, $M_{11}=M_{22}$ and $\Gamma_{11}=\Gamma_{22}$ and also that $M_{21}=M_{12}^*$ and $\Gamma_{21}=\Gamma_{12}^*$.

\begin{figure}[b!]
	\centering
    \begin{picture}(400,60)(0,0)
        \put(40,0){\includegraphics[scale=0.85]{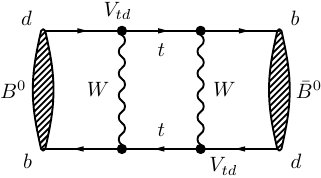}}
       \put(220,0){\includegraphics[scale=0.85]{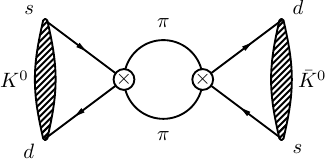}} 
    \end{picture}	
	\caption{{\em Left:} example diagram of off-shell ("dispersive") mixing amplitude $M_{12}$ for $B$ mesons, where the intermediate states consists of off-shell top-quarks. {\em Right:} example diagram of on-shell ("absorbtive") mixing amplitude $\Gamma_{12}$ for $K$ mesons, where the intermediate state consists of on-shell particles.}
%    %Diagrams taken from: Flavour physics in the LHC era - Gershon, Tim - arXiv:1306.4588 - shall we remake them?} 
	\label{fig:mixing-diagrams1}
\end{figure}

The off-diagonal elements consist of 
%two parts shown in fig~\ref{fig:mixing-diagrams}, 
the amplitudes $M_{12}$ and $\Gamma_{12}$,
which describe two possible transitions of the $B^0\rightarrow \bar{B}^0$ process.
The matrix element $M_{12}$ quantifies the short-distance (or dispersive) contribution from the perturbative box diagram, 
which is dominated by the $t$-quark exchange. 
Alternatively, $\Gamma_{12}$ is the long-distance (or absorbtive) contribution, 
which can be seen as the virtual creation of on-shell states. Examples of the $M_{12}$ and $\Gamma_{12}$ amplitudes for $B^0$ and $K^0$ mesons are shown in Fig.~\ref{fig:mixing-diagrams1}

The eigenstates of the Hamiltonian in eq.~\ref{eq:mixingHamiltonian} are the particles with well defined mass and lifetime.
In the $B$ system it is customary to name the mass eigenstates as {\em heavy} $B_H$ and {\em light} $B_L$, whereas in the kaon system they are named according to their observed lifetime as {\em long-lived} ($K_L$) and {\em short-lived} ($K_S$). Using the case of $B$ mesons these mass (and lifetime) eigenstates are expressed as linear combinations of the flavour eigenstates,

\begin{equation}
|B_H\rangle   = p|B^0\rangle - q|\bar{B}^0\rangle \hspace{2.0cm}
|B_L\rangle   =  p|B^0\rangle  + q|\bar{B}^0\rangle  
\label{eq:BHLdef}
\end{equation}
where from diagonalisation of the Hamiltonian (eq. \ref{eq:mixingHamiltonian}) it follows:
\begin{equation}
\frac{q}{p} = \pm \sqrt{\frac{M_{12}^*-\frac{i}{2}\Gamma_{12}^*}{M_{12}-\frac{i}{2}\Gamma_{12}}}\; .
\label{eq:qp}
\end{equation}
The coefficients $p$ and $q$ quantify the relative contribution of the flavour eigenstates 
$|B^0\rangle$ and $|\bar{B}^0\rangle$ to the mass 
%(and lifetime) 
eigenstates.
If $|q/p|=1$ then the two flavour eigenstates $|B^0\rangle$ and $|\bar{B}^0\rangle$
contribute equally, and then the mass 
%(and lifetime)
eigenstate are also CP-eigenstates. 
However, the fact that $|q/p|\ne 1$ for neutral kaons, implies that the 
lifetime eigenstate $K_L$ is not a pure, CP-odd, eigenstate. Instead it contains a small admixture
that allows it to decay to the CP-even $\pi^+\pi^-$ final state, which lead to the discovery of CP violation in 1964.

For the $B$ mesons $\Delta m\equiv M_H - M_L$ is chosen positive by definition. Furthermore, the decay width difference is defined as $\Delta \Gamma \equiv \Gamma_L - \Gamma_H$, such that in the Standard Model the heavy state is expected to have a smaller decay width than that of the light state~\cite{pdg}:
% See Eq.(75.8) in https://pdg.lbl.gov/2023/reviews/rpp2023-rev-b-bar-mixing.pdf
$\Gamma_H < \Gamma_L$.

Using the fact that for $B$ mesons the on-shell contribution in the mixing process is much smaller as the off-shell contribution ($\Gamma_{12}|<<|M_{12}|$), diagonalisation of eq.\ref{eq:mixingHamiltonian} leads to the relations:  
\begin{equation}
\Delta m       \approx  2 | M_{12}|\hspace{2cm}
\Delta \Gamma  \approx  2 |\Gamma_{12}| \cos\phi \; .
\label{eq:DmDGamma}
\end{equation}
%where the approximation explicitly uses the fact that $|\Gamma_{12}|<<|M_{12}|$.
In general there can be a relative complex phase between the on-shell 
and off-shell transition amplitudes $M_{12}$ and $\Gamma_{12}$~\cite{Dunietz:2000cr,Lenz:2006hd}:
\begin{equation}
\phi_{12} = \arg\left( -\frac{M_{12}}{\Gamma_{12}} \right) \equiv \pi + \phi_M - \phi_\Gamma \; .
\label{eq:mixphase}
\end{equation}
As a consequence of this phase difference $\phi_{12}$, the eigenstates of the Hamiltonian become complex linear combinations of the flavour states and are no longer pure CP eigenstates\footnote{
The phase $\phi_{12}$ should not be confused with e.g. the famous phase $\phi_s = \phi_M - 2\phi_{c\bar{c}s}$, 
which is the phase difference between the amplitudes originating from the {\em decay amplitudes} with and without mixing,
where the $B_s^0$ meson decays to $J/\psi\phi$ through the $b\to c\bar{c}s$ transition(see section~\ref{sec:phis}). 
}.
%. Under a CP transformation, these eigenstates do not simply acquire a plus or minus sign, but rather mix into each other with additional phase factors, which means they are no longer pure CP eigenstates.
These mass states propagate in time as:
\begin{equation}
\label{eq:BHLprop}
\left|B_{H,L}(t)\right> = e^{-\left(im_{H,L}+\Gamma_{H,L}/2\right)t}\left|B_{H,L}(0)\right>
\end{equation}

%We get for the time evolution of the state $|\bar{B}^0\rangle $:
%\begin{equation}
%|\bar{B}^0(t)\rangle  = g_-(t) \left(\frac{p}{q}\right)|B^0\rangle  + g_+(t) |\bar{B}^0\rangle 
%\label{eq:P0bart}
%

\begin{figure}[t!]
	\centering
    \includegraphics[scale=0.54]{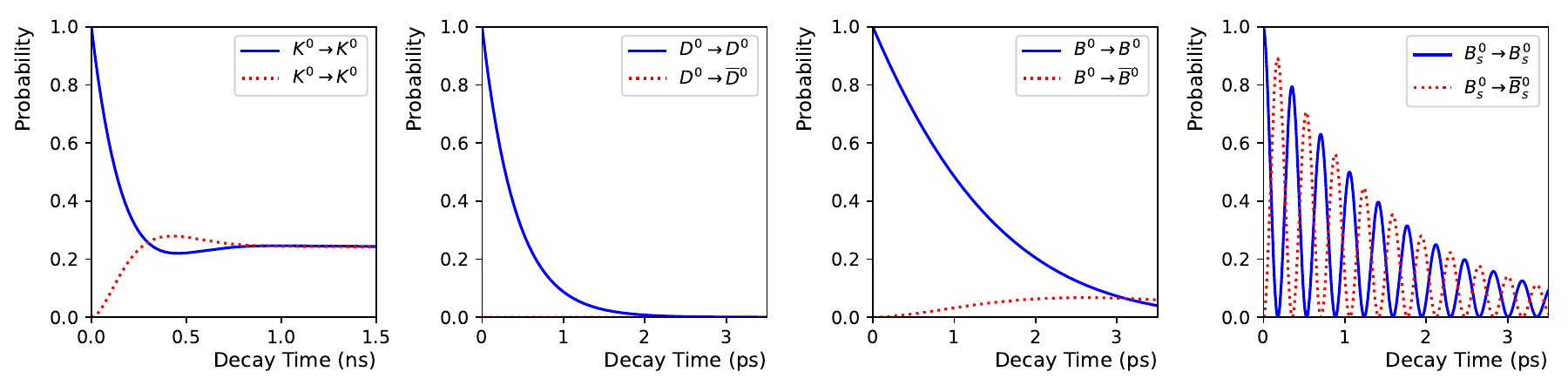}
    \caption{The decay time distributions of the flavour eigenstate of neutral mesons with from left to right: $K^0$, $D^0$, $B^0$ and $B_s^0$. The probability to decay as its original flavour eigenstate (ie. non-oscillated) or as its CP conjugate counterpart after oscillation, are separately indicated by the full blue lines and red dashed lines, respectively.
    The neutral mesons exhibit a different oscillation pattern due to the different values of the lifetime and mass differences. For example, at large lifetime, 50\% of the $K^0$ have been decayed as a $K_S^0$, whereas the remaining 50\% are $K_L^0$ mesons, equally split as $K^0$ and $\bar{K}^0$, and almost all $D^0$ mesons decay before they oscillate. The $B^0$ oscillates and decays at an approximately equal rate, while the $B_s^0$ oscillates several times before decaying.}
	\label{fig:osc}
\end{figure}

The parameters $\Delta m$ and $\Gamma$ are direct observables in the time dependence of flavour tagged decay rates of the neutral $B$ mesons.
Starting from a pure sample of $|B^0\rangle $ particles (e.g. a sample of $B$ mesons produced by the strong interaction with a well-defined quark composition)
the probability of measuring the state $|B^0\rangle$ or $|\bar{B}^0\rangle $ at time $t$ is expressed from  eq.~\ref{eq:BHLdef} and \ref{eq:BHLprop}:
\begin{equation}
\left|\left< B^0|B^0(t)\right> \right|^2 = \frac{1}{2}\left|\frac{p}{q}\right|^2 \left( 
\cosh\frac{1}{2} \Delta \Gamma t + \cos \Delta m t
\right) e^{-\Gamma t} 
\hspace*{0.5cm}\text{and}\hspace{0.5cm}
\left|\left< \bar{B}^0|B^0(t)\right> \right|^2 = \frac{1}{2}\left|\frac{p}{q}\right|^2 \left( 
\cosh\frac{1}{2} \Delta \Gamma t - \cos \Delta m t
\right) e^{-\Gamma t}
\label{Bdecayrate}
\end{equation}
where $\Gamma=(\Gamma_L+\Gamma_H)/2$.
%and  $\Delta \Gamma=\Gamma_L-\Gamma_H$. 
Here $\Gamma$ fulfills the natural role  of decay constant, $\Gamma = 1/\tau$.
%The sign of $\Delta m$ is positive by definition. 
%The heavy state is expected to have a smaller decay width than that of the light state~\cite{pdg},
% See Eq.(75.8) in https://pdg.lbl.gov/2023/reviews/rpp2023-rev-b-bar-mixing.pdf
%$\Gamma_H < \Gamma_L$, and therefore $\Delta \Gamma \equiv \Gamma_L - \Gamma_H$ is expected to be positive in the Standard Model.
%
The dimensionless variables $x$ and $y$ are often used to quantify 
the mixing behaviour, expressing the oscillation rate relative to the lifetime:
\begin{equation}
x =  \frac{\Delta m}{\Gamma}       \hspace{2cm}
y =  \frac{\Delta \Gamma}{2\Gamma} \nonumber 
\end{equation}
The expected decay rates and oscillations are shown in Fig.~\ref{fig:osc} for $K^0$, $D^0$, $B_d^0$ and $B_s^0$ mesons, illustrating the different phenomenology resulting from the interplay different decay rate parameters $\Gamma$, $\Delta m$ and $\Delta\Gamma$. 
The values of the oscillation parameters of these neutral mesons are summarized in Table~\ref{tab:mesons}.
%---------------------------------------------------------------------------------------
\begin{table}[!h]
\begin{center}
\begin{tabular}{lccccccl} 
%\hline
        & m & $\tau=1/\Gamma$ \hspace{0.5cm} $(c\tau)$ & $\Gamma = (\Gamma_L+\Gamma_H)/2$ & $\Delta m$ (ps$^{-1}$) & $\Delta \Gamma = \Gamma_L - \Gamma_H$ & $x$ & $y$\\
\hline
$K^0$-system   &  498 MeV &  {\small $K_S$: 0.09~ns (27mm), $K_L$: 51~ns (15m)} 
                                                    & 5.57 ns$^{-1}$  & 0.005 &                 &      &        \\
$D^0$-system   & 1865 MeV & $0.410$~ps ~ (0.12~mm)  & 2.44  ps$^{-1}$ & 0.010 & 0.031 ps$^{-1}$ & 0.004& 0.0064 \\
$B^0$-system   & 5280 MeV & $1.517$~ps ~ (0.46~mm)  & 0.659 ps$^{-1}$ & 0.507 & 0 ps$^{-1}$     & 0.77 & 0.0005 \\
$B^0_s$-system & 5367 MeV & $1.527$~ps ~ (0.46~mm)  & 0.655 ps$^{-1}$ & 17.77 & 0.087 ps$^{-1}$ & 27.0 & 0.068  \\
\end{tabular}
\caption[vla]{
Approximate values for the oscillation parameters of the various neutral mesons, leading to their characteristic phenomenology and experimental requirements. 
The average width $\Gamma=(\Gamma_S+\Gamma_L)/2$ for the $K^0$ system is practically half the decay width of the $K_S^0$, given
the large width difference.
Using the value $\hbar=6.6 \times 10^{-22}$~MeVs, a mass difference of 1~ps$^{-1}$ corresponds to the tiny
mass difference of 0.66~meV. 
Furthermore, one period of neutral meson oscillations corresponds to $t=2\pi/\Delta m$, so a mass difference of 17.77~ps$^{-1}$ corresponds to
an oscillation period of 0.35~ps, or, equivalently, about five oscillations within one $B_s^0$ meson lifetime.}
\label{tab:mesons}%[htbp]
\end{center}
\end{table}

%---------------------------------------------------------------------------------------

\clearpage

\subsection{\texorpdfstring{$B^0$}{B0} mixing}
\label{sec:BsMixing}
The frequency of the neutral meson oscillations is proportional to 
the amplitude of the box diagram, which is calculated to be (\cite{Buras:1984pq}):
%Consistent with Branco p.211 (8.21):
\begin{equation}
\Delta m_d = \frac{G_F^2 m_W^2 m_B}{6\pi^2} \eta_{QCD} B_B f_B^2 \;  S_0(m_t^2/m_W^2) \; |V_{td}V_{tb}|^2.  
\label{eq:deltam}
\end{equation}
Three parts can be distinguished: a part governed by the strong interaction including the factors $\eta_{QCD}$ (perturbative QCD correction), $B_B$ (non-perturbative bag factor) and $f_B^2$ ($B$ decay-constant), a part governed by the weak CKM factor $|V_{td}V_{tb}|^2$ (see Fig~\ref{fig:mixing-diagrams1}),
quantifying the coupling strength of the weak interaction through the CKM-elements, and finally the kinematic factor $S_0$ denoting the Inami-Lim function~\cite{inami:ptp:65:297,gay:arevns:50:577,Artuso:2015swg}. The latter evaluates the loop integrals in the box diagram for various values of $x=m_q^2/m_W^2$,
$$
S_0(x)=\frac{4x-11x^2+x^3}{4(1-x)^2} - \frac{3x\ln x}{2(1-x)^2}.
$$

In the $B$-system the top and charm quark exchange have similar CKM coupling strengths $|V_{td}V_{tb}|\sim |V_{cd}V_{cb}|$, but because $m_t >> m_c$ the top contribution dominates.
The theoretical uncertainty on $\Delta m_d$ is about 6\%~\cite{King:2019lal} and is dominated by the combined uncertainties of
$\eta_{QCD}$, the bag factor $B_B$, and the decay constant $f_B$.
Note that many of the uncertainties cancel in the ratio $\Delta m_s / \Delta m_d \propto \xi \equiv f_{B_s^0}\sqrt{B_{B_s^0}}/f_{B^0}\sqrt{B_{B^0}}$, which is known to about 0.6\%.
An accurate measurement of the mixing asymmetry oscillations of $B^0$ and $B_s^0$ thus leads to a  determination of the less precisely known CKM-element $|V_{td}|$, which in turn can be scrutinized for its
compatibility with the other CKM elements within the CKM paradigm.

\begin{figure}[!ht]
	\centering
	\includegraphics[width=6cm,height=4cm]{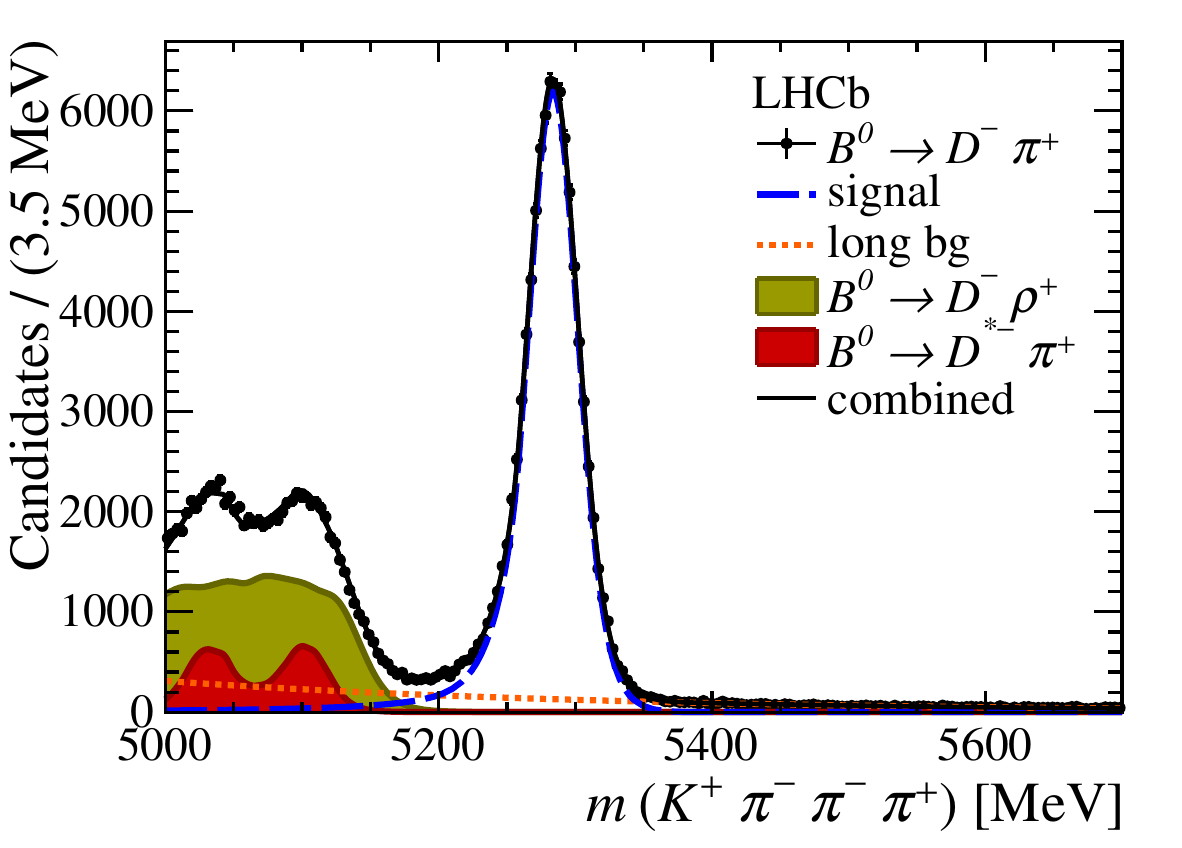}
	\includegraphics[width=6cm,height=4cm]{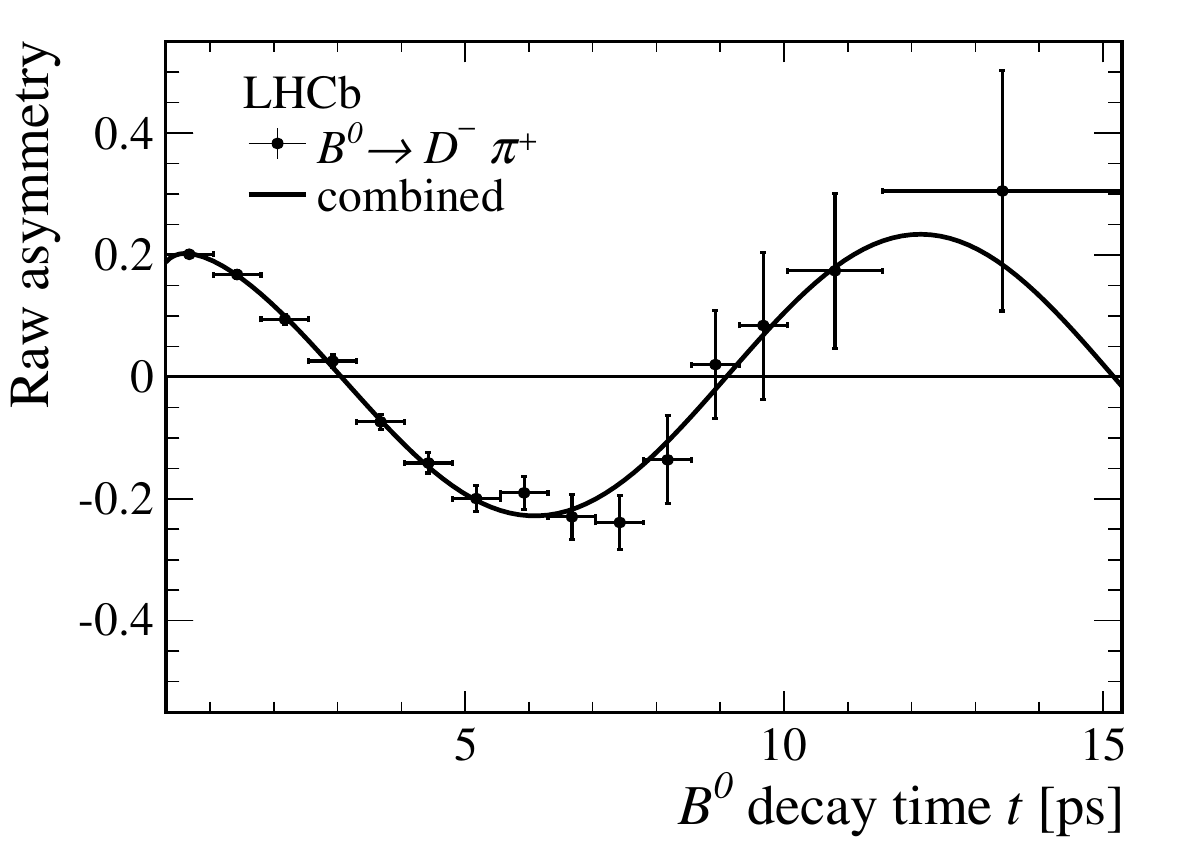} 
	\caption{{\em Left:} The invariant mass distribution of flavour-specific $B^0\to D^-\pi^+$ decays. {\em Right:} The mixing asymmetry of mixed and unmixed events shows the $B^0 \leftrightarrow\bar{B}^0$ oscillation pattern from which the frequency $\Delta m_d$ can be extracted \cite{LHCB-PAPER-2012-032}.
    The amplitude of the oscillations is an indication of the experimental capabilities to determine the $b$-flavour at production.
}
	\label{fig:B0-mixing}
\end{figure}

The mixing asymmetry is measured by comparing the number of mixed and unmixed $B^0$-candidates decaying
to flavour-specific eigenstates such as $D^-\pi^+$, $J/\psi K^+$ or $D^{(*)-}\mu^+\nu$, as a function of the decay time:
\begin{equation}
A_{mix}(t)=\frac{N_{unmixed}(t)-N_{mixed}(t)}{N_{unmixed}(t)+N_{mixed}(t)}= \cos \Delta m_d t \; , 
\label{eq:Amix}
\end{equation}
where unmixed events have a flavour tag that indicates the same flavour at production compared to the flavour at decay. Here, the value for $\Delta\Gamma_d=0$ is assumed in Eq.~\ref{Bdecayrate}.
Fig.~\ref{fig:B0-mixing} shows the measurement of LHCb for hadronic decays $B^0\rightarrow D^-\pi^+$. The experimental mass resolution allows for a clean selection of signal events and the tagging performance allows for a clear oscillation signal. 
The most precise determination by LHCb to date is obtained from 3.0 fb$^{-1}$ of data using semileptonic decays, resulting in 
$\Delta m_d=(505.0 \pm 2.1 \pm 1.0)$ ns$^{-1}$ \cite{LHCb-PAPER-2015-031}, where the first uncertainty is statistical and the second is systematic.

%\clearpage

\subsection{\texorpdfstring{$B_s^0$}{B0s} mixing and \texorpdfstring{$\Delta\Gamma_s$}{DGs} }
\begin{figure}[!ht]
    \begin{picture}(390,130)(0,0)
    \put(-5,10){\includegraphics[trim=0 0 0 4cm,clip,scale=0.26]{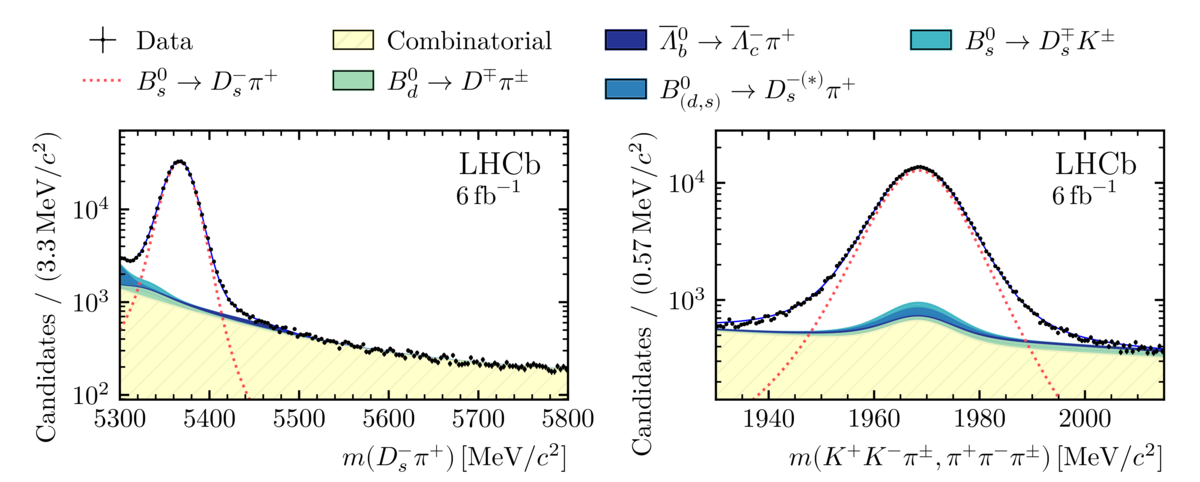}}
    \put(310,0){\includegraphics[scale=0.14]{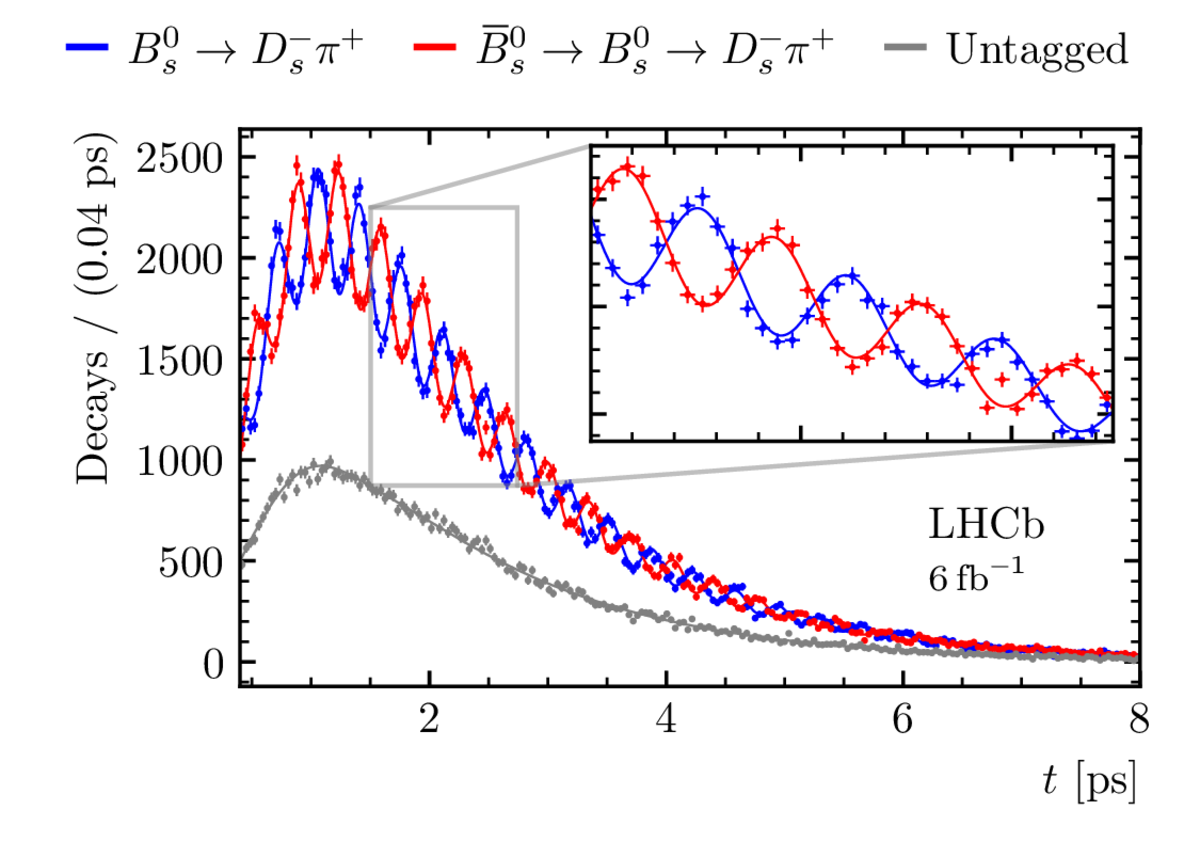}}
    \end{picture}
    \caption{{\em Left:} The invariant mass distribution of flavour-specific $B^0_s\to D_s^-\pi^+$ (and subsequent $D_s^-\to K^+K^-\pi^-$) decays. {\em Right:} The mixing asymmetry of mixed and unmixed events shows the $B_s^0$-$\bar{B}_s^0$ oscillation pattern from which the frequency $\Delta m_s$ is determined \cite{LHCb-PAPER-2021-005}.
    }
	\label{fig:Bs-mixing}
\end{figure}

The $B^0_s$ mixing frequency is determined  analogously to the $B^0$ case, 
by exploiting flavour-specific final states, such as  
$D_s^-\pi^+$ and $D_s^-K^+ \pi^+\pi^-$.
The larger value of $|V_{ts}|$ compared to $|V_{td}|$ leads to 
a higher oscillation frequency, corresponding to a shorter oscillation period of about 0.35~ps.
Fig.~\ref{fig:Bs-mixing} shows that the selected mass peaks for $B_s^0\rightarrow D_s^-\pi^+$ events have low backgrounds and that the decay time resolution is sufficient to resolve the fast $B_s^0$ oscillations, resulting in a measurement $\Delta m_s = (17.7656 \pm 0.0057)$ ps$^{-1}$ \cite{LHCb-PAPER-2021-005}.

%%trim = <left> <bottom> <right> <top>

%\subsubsection{$\Delta\Gamma_s$}
The lifetime (mass) eigenstates of the $B^0_s$ system are to a good approximation pure CP eigenstates, 
due to the fact that $|q/p|\approx 1$.
Decays to CP-even eigenstates such as 
$B_s^0\to J/\psi \eta^{'} (\to \rho^0\gamma)$ with a pseudoscalar $\eta'$ are suitable to measure the 
lifetime of the light eigenstate $B_{s,L}$~\footnote{The long-living lifetime eigenstate $K_L$ and the lighter mass eigenstate $B_L$ are both denoted with the subscript $L$. The fact that they represent the {\em opposite} CP-eigenstates could help as a distorted mnemonic.}.
Decays to the CP-odd eigenstates  such as $B_s^0\to J/\psi f_0(980) (\to \pi^+\pi^-)$ with a scalar $f_0(980)$ are suitable to measure the lifetime of the heavy eigenstate $B_{s,H}$.
The lifetime difference $\Delta\Gamma_s$ is sizeable - as opposed to the very small lifetime difference $\Delta\Gamma_d$ in the $B^0$ system - due to the relatively large branching ratio for the decay to the CP-even final states 
such as $D_s^+D_s^-$, leading to the shorter lifetime of the CP-even $B_{s,L}$ eigenstate.
%Fig.~\ref{fig:DeltaGamma_s} shows the sample of $B_s^0$ decays dominated by the CP-even final state $J/\psi \eta^{'}$, and a sample that is dominated by the CP-odd final state  $J/\psi \pi^+\pi^-$ (with the di-pion invariant mass close to the $f_0(980)$ mass). From these the $B_s^0$ lifetime difference is determined to be $\Delta\Gamma_s = 0.087 \pm 0.012 \pm 0.009$~ps$^{-1}$~\cite{LHCb-PAPER-2023-025}.
A sample of $B_s^0$ decays dominated by the CP-even final state $J/\psi \eta^{'}$, and a sample that is dominated by the CP-odd final state  $J/\psi \pi^+\pi^-$ (with the di-pion invariant mass close to the $f_0(980)$ mass) are used to determine the $B_s^0$ lifetime difference, $\Delta\Gamma_s = 0.087 \pm 0.012 \pm 0.009$~ps$^{-1}$~\cite{LHCb-PAPER-2023-025}.

%\begin{figure}[!ht]
%	\centering
%        \includegraphics[width=6cm,height=4cm]{Figures/Mixing/fig1d-2023-025.pdf}
%        \includegraphics[width=6cm,height=4cm]{Figures/Mixing/fig2d-2023-025.pdf}
%	    \raisebox{0.1cm}{\includegraphics[width=7cm,height=4cm]{Figures/Mixing/DeltaGammas_Value.png}}
%    \caption{
%    {\em Left:} The invariant mass distribution of $B_{s,L}^0$ decays to the CP-even final state $J/\psi \eta^{'}$. 
%    {\em Right:} The invariant mass distribution of $B_{s,H}^0$ decays to the CP-odd final state $J/\psi f_0(980)(\to \pi^+\pi^-)$~\cite{LHCb-PAPER-2023-025}.
%    }
%	\label{fig:DeltaGamma_s}
%\end{figure}

%\clearpage

\subsection{Lorentz violation}
%In quantum field theory the concept of CPT symmetry is fundamentally connected Lorentz invariance, reflecting the fundamental symmetry of spacetime in special relativity. 
%Therefore violation of CPT naturally leads to Lorentz violation \cite{Greenberg:2002uu}. 
%
The Feynman-St\"{u}ckelberg interpretation implies that antiparticles are equivalent to particles that travel backward in time. 
%Applying a large Lorentz boost implies different observer interpretations of processes involving virtual particles and antiparticles, illustrating the connection between Lorentz violation and CPT symmetry.
In quantum field theory the concept of CPT symmetry is fundamentally connected to Lorentz invariance, reflecting the fundamental symmetry of spacetime in special relativity, and violation of CPT naturally leads to Lorentz violation \cite{Greenberg:2002uu}. 
Allowing for CPT violation to occur, the mass 
%(and lifetime) 
eigenstates can be decomposed as follows,
\begin{eqnarray}
\left|B_H\right>  & = & p\sqrt{1-z}\left|B^0\right>  - q\sqrt{1+z}\left|\bar{B}^0\right>  \nonumber \\
\left|B_L\right>  & = & p\sqrt{1-z}\left|B^0\right>  + q\sqrt{1+z}\left|\bar{B}^0\right>  \nonumber
\end{eqnarray}
where
$$
z=\frac{\delta m - i\delta\Gamma/2}{\Delta m + i\Delta\Gamma/2}
$$
is the CPT violating parameter with $\delta m\equiv(M_{11}-M_{22})$ and $\delta\Gamma\equiv(\Gamma_{11}-\Gamma_{22})$, the differences of the diagonal terms in eq~\ref{eq:mixingHamiltonian}.

\begin{wrapfigure}{R}{6.5cm}
	\centering
	\includegraphics[scale=0.16]{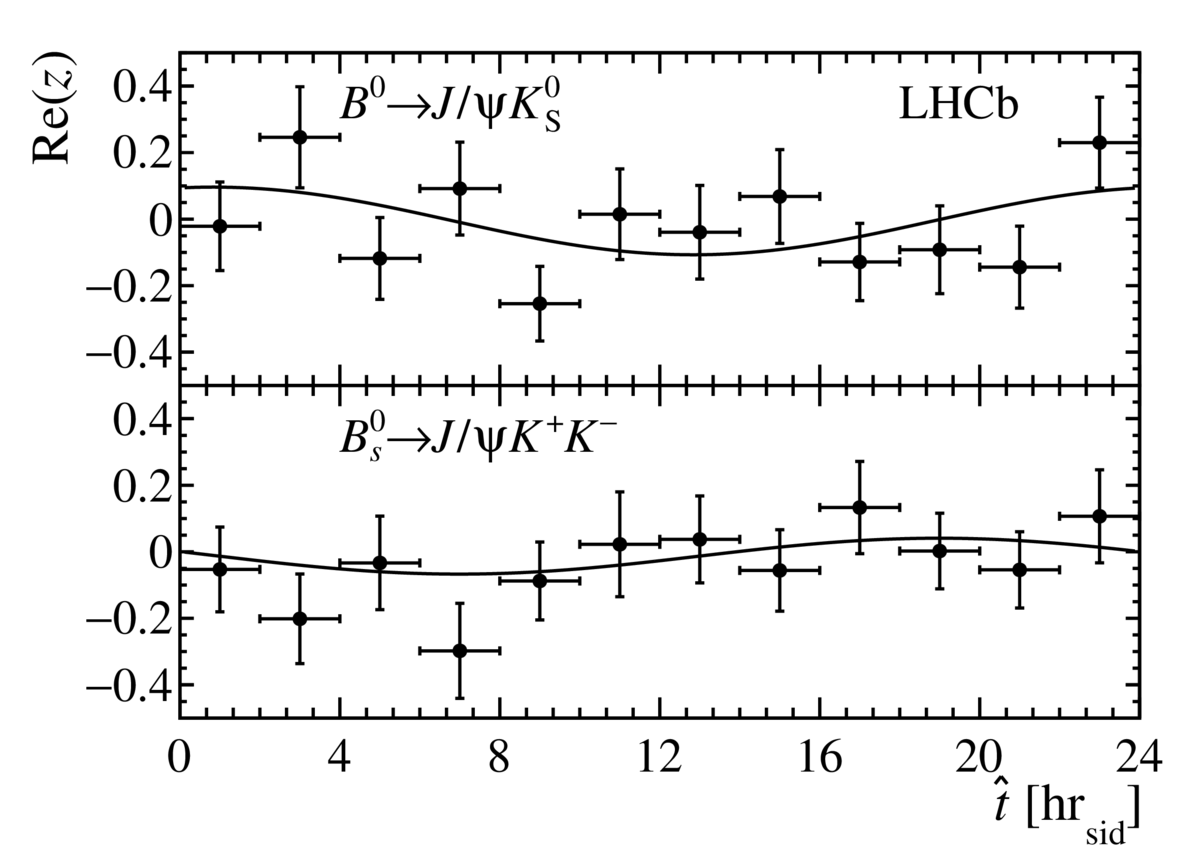}
    \caption{Oscillation rate for both $B^0$ and $B_s^0$ as function of siderial time ~\cite{LHCB-PAPER-2016-005}.}
	\label{fig:LorentzViolation}
\end{wrapfigure}
For the Standard Model Extension (SME) of Colladay and Kostelecky \cite{Kostelecky:1997mh} $z$ introduces a boost and direction dependence of the CPT varying parameter according to 
$$
z=\frac{\beta^\mu \Delta a_\mu}{\Delta m + i\Delta\Gamma/2} \; ,
$$
with $\beta^\mu$ the four-velocity $(\gamma,\gamma\vec{\beta})$ of the $B$-meson, and $\Delta a_\mu$ the space-time SME coefficients that describe CPT and Lorentz violation in the quark sector.

As a consequence, the dimensionless parameter $z$ depends on the four-velocity of the $B^0$-meson in an absolute reference frame, and siderial oscillations may be observed~\cite{Kostelecky:1997mh}.
Since $\Delta a_\mu$ is a real parameter \cite{Kostelecky:1997mh} the relation $\Re(z)\Delta\Gamma=\Im(z)\Delta m$ provides different sensitivities for oscillations with different neutral mesons. For $B$-mesons the fact that $\Delta m >> \Delta\Gamma$ allows to probe the $\Re(z)$ parameter. The dependency of $\Re(z)$ as function of the siderial time for decays of $B^0$ and $B_s^0$ into dominant CP eigenstates is shown in Fig.~\ref{fig:LorentzViolation}.
%\begin{figure}[!b]
%	\centering
%	\includegraphics[scale=0.135]{Figures/Mixing/LorentzViolation.png}
%    \caption{Oscillation rate for both $B^0$ and $B_s^0$ as function of siderial time ~\cite{LHCB-PAPER-2016-005}.}
%	\label{fig:LorentzViolation}
%\end{figure}
Since the $\Re(z)$ parameter scales with $\Delta m$ and the $\Im(z)$ with $\Delta\Gamma$ and $\Delta m >> \Delta\Gamma$, LHCb has measured the expected dominant $\Re(z)$ as function of siderial time. The result is shown in Fig.~\ref{fig:LorentzViolation} and has a result compatible with zero. 

\clearpage

\subsection{\texorpdfstring{$D^0$}{D0} mixing}
The formalism for mixing in the $D^0$ system is identical to the 
$B$-system, but instead of $\Delta m$ and $\Delta\Gamma$ the observables $x$ and $y$ are more commonly used.
%Note that the signs of $x$ and $y$ are difficult to predict and must therefore be determined experimentally.
The determination of the $D^0$ oscillation parameters is experimentally more challenging than for
$B$ mesons, because the oscillation is slow (see Fig.~\ref{fig:osc}) and most $D^0$ mesons will have decayed before they had the chance to oscillate.

\begin{figure}[!ht]  
	\centering
	\includegraphics[width=11cm]{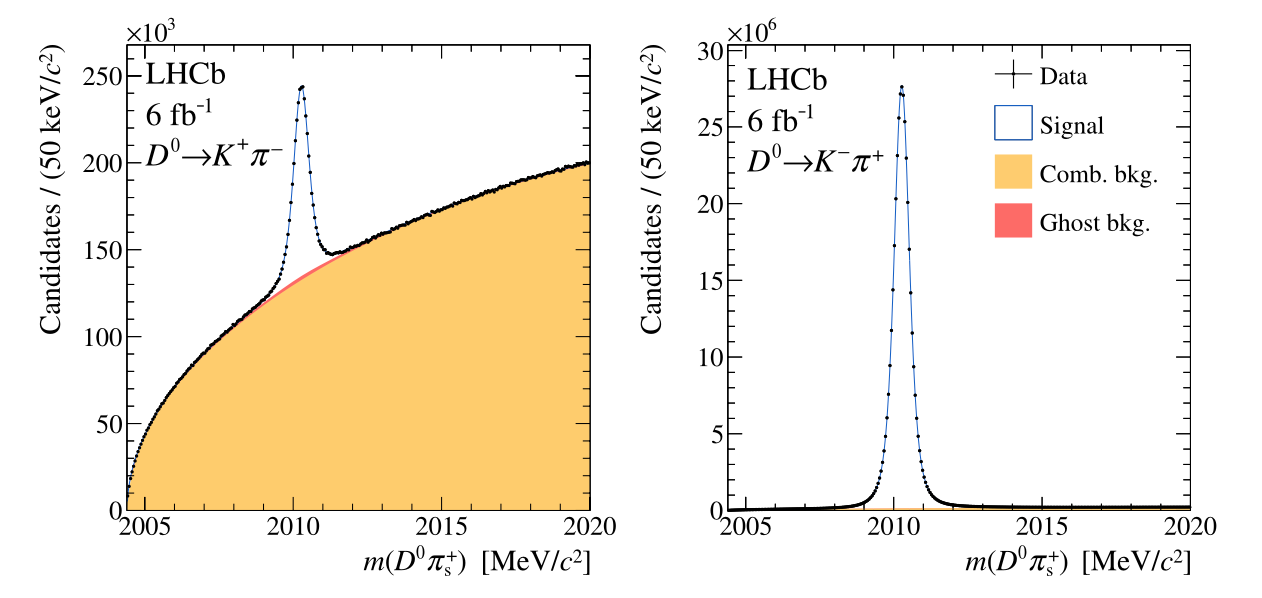}
   	\raisebox{0.2cm}{\includegraphics[width=5.3cm]{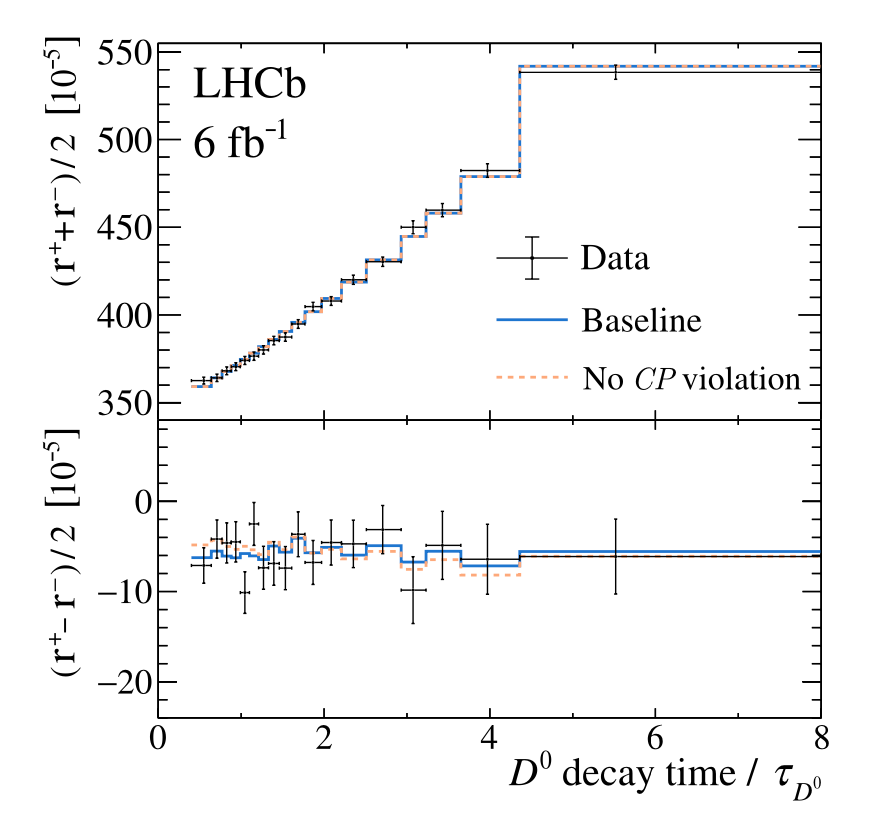}}
    \caption{{\em Left:} The invariant mass distribution of 
    doubly Cabibbo-suppressed decays $D^0\to K^+\pi^-$.
    {\em Middle:} The invariant mass distribution of 
    abundant Cabibbo-favoured decays $D^0\to K^-\pi^+$.
    {\em Right:} The fraction of Cabibbo-suppressed decays increases as a function of decay time due to the mixing process $D^0 \to \bar{D}^0$~\cite{LHCb-PAPER-2024-008}.}
	\label{fig:D-mixing}
\end{figure}

\begin{figure}[!ht]  
	\centering
	\includegraphics[width=7cm]{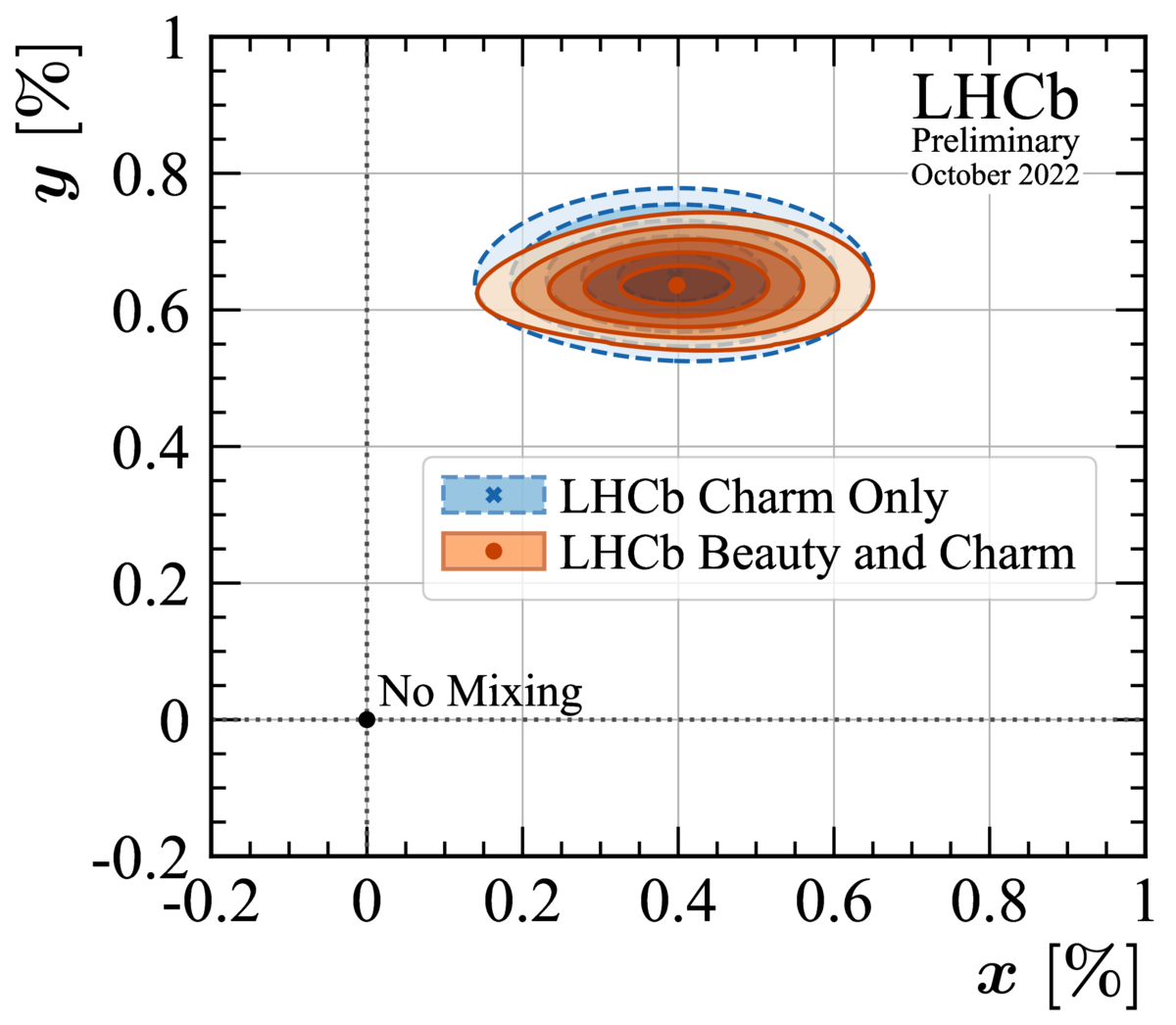}
    \caption{The charm mixing parameters $x$ and $y$ are accurately determined from a fit to multiple results in the charm and beauty sector~\cite{LHCb-PAPER-2021-033}.   }
	\label{fig:D-mixing2}
\end{figure}

The flavour of the neutral $D$ meson at production can be obtained 
either from the charge of the slow pion in the 
decay $D^{*+}\to D^0 \pi^+_s$, or from the 
charge of the muon in the decay $\bar{B}^0\to D^0 \mu^-\nu$.
The use of semileptonic $B$-decays yields smaller sample sizes, but has the advantage that the decay time distribution 
is not distorted by the trigger event selection requiring separation of signal tracks from the primary vertex, thanks to the additional flight distance from the $B$.
Subsequently, the flavour-specific decay mode $D^0 \to K^-\pi^+$
can be used to quantify the number of mixed $D$ mesons as a function of
its decay time.
A small fraction of the neutral $D$ mesons will, however, decay through the doubly Cabibbo-suppressed mode,  ${D}^0 \to K^+\pi^-$. This
fraction will increase as a function of the decay time 
due to the oscillation process $D^0 \to \bar{D}^0$, after which the
$\bar{D}^0$ meson predominantly decays to $K^+\pi^-$. Fig.~\ref{fig:D-mixing} shows the selection of the mixed $D\rightarrow K^+\pi^-$ decays increases as a function of decay time. 
A non-vanishing mass and lifetime difference in the neutral $D$ system 
has clearly been established, see Fig.~\ref{fig:D-mixing2}.

%Mixing parameters:
%\begin{eqnarray*}
%x_{12} & \equiv & 2\left|M_{12}\right|/\Gamma \\
%y_{12} & \equiv & \left|\Gamma_{12}\right|/\Gamma
%\end{eqnarray*}

\clearpage
\section{CP-Violation}\label{secCPViolation}

\subsection{Observing CP violation}

\begin{figure}[!ht]
	\centering
    \includegraphics[scale=0.4]{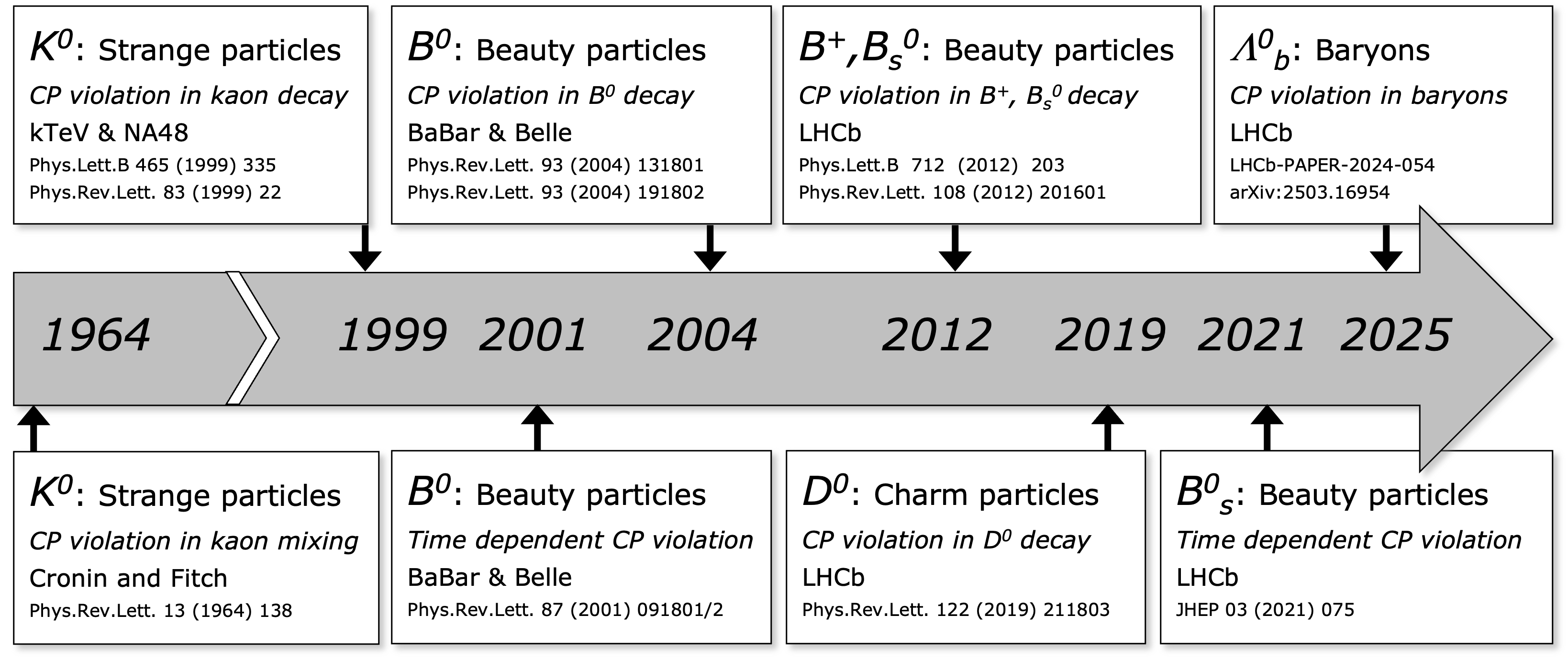}
     \caption{Timeline of the major CP violation discoveries, including recent ones by LHCb.}
	\label{fig:CPVhistory}
\end{figure}

The existence of the phenomenon of CP violation has been well established for various particle decay processes, as is graphically summarized in a timeline in Fig.~\ref{fig:CPVhistory}. 
It has been observed in kaon particle decays both in mixing as well as in decay, in neutral and in charged $B$ meson decays, in $B^0_s$ meson decays, and more recently in $D^0$ decays and in baryons with $\Lambda_b$ decays.
%~\cite{LHCB-PAPER-2011-029}
%~\cite{LHCb:2025ray}. %\cite{LHCb-PAPER-2024-054}.
%

In order to observe a process that violates CP symmetry, the following general requirements need to be fulfilled:
\begin{itemize}
    \item {\em Two interfering amplitudes} (at least) must contribute to the process.
    \item {\em A relative CP-odd phase ("$\phi$")} must be present between the interfering amplitudes due to different complex couplings between the CP-conjugated processes.
    In the SM this phase originates from the 
    weak interaction.
    %for example from the relative phase between the couplings $V_{ub}$ and $V_{cb}$.
    \item {\em A relative CP-even phase} ("$\delta$") must exist between the interfering amplitudes of the process, which does not change sign under CP conjugation. A relative phase shift present due to the strong final state interactions in the two decay amplitudes can play this role, as the strong interaction conserves CP symmetry. It arises from non-perturbative dynamics in the time evolution of a strongly-interacting system \cite{Grossman:2017thq}.
    %\textcolor{purple}{The fact that the relative 'strong' phase is conserved under the CP transformation is non-trivial, and is also known as the strong CP problem.} \textcolor{purple}{$\Rightarrow$ perhaps too much side-remark here?}
    Alternatively, in neutral $B$-mixing a 90$^o$ phase shift between the mass and decay amplitudes in the time development in eq.~\ref{eq:BHLprop} plays the role of the CP conserving phase, leading to the phenomenon of decay-time dependent CP violation. 
    \item {\em The magnitude} of the two interfering amplitudes must be of comparable size to achieve a sizeable effect.
\end{itemize}
CP violation is a pure quantum-mechanical phenomenon that can only be observed in the interference of two coherent amplitudes. We illustrate this with the first observation of CP violation in $B^0_s$ meson decays~\cite{LHCB-PAPER-2013-018}.  
%%The dominant contribution from the Penguin diagram results from $c$ quark exchange. (REALLY??)
%
\begin{figure}[!hb]
	\centering
    \begin{picture}(450,225)(0,0)
        \put(0,50){\includegraphics[scale=0.65]{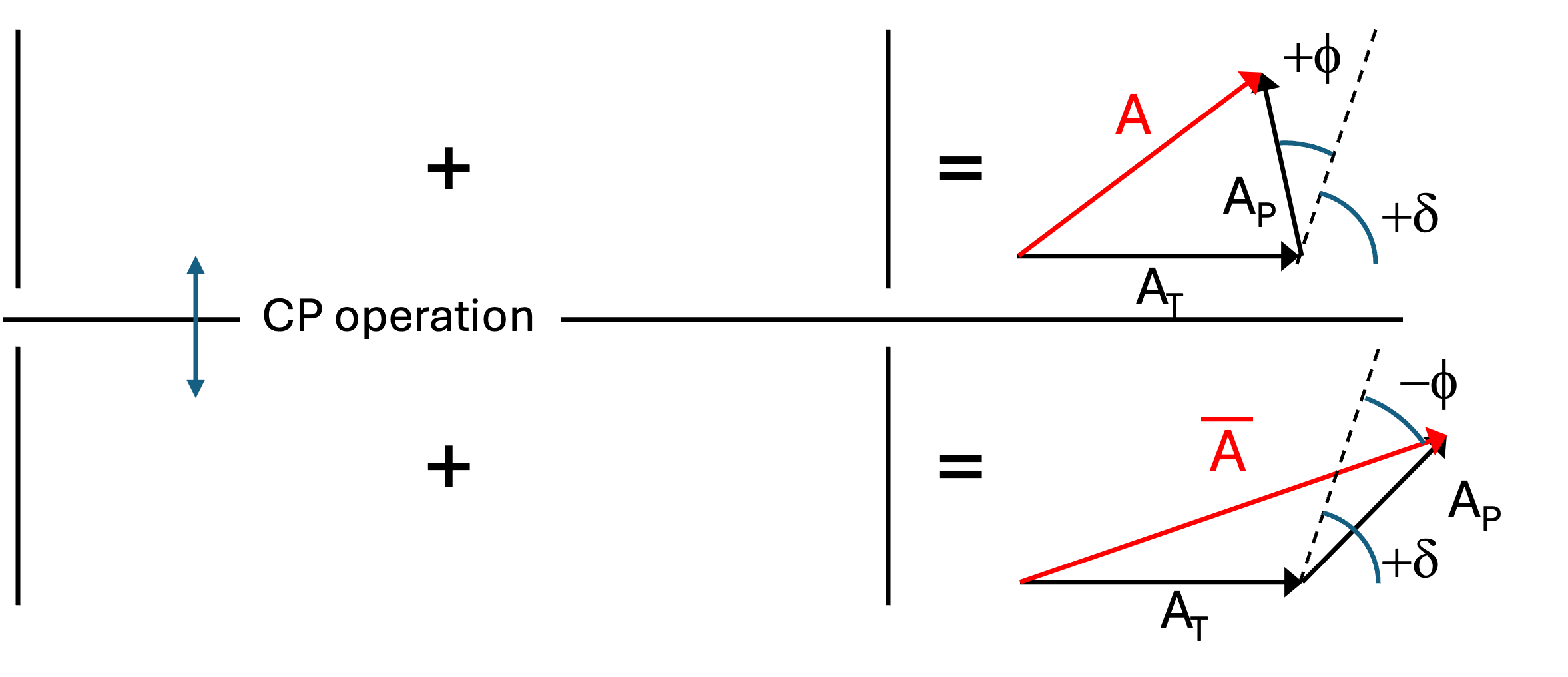}}
        \put(10,135){\includegraphics[scale=0.65]{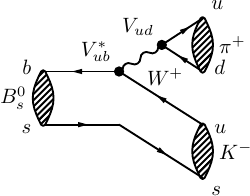}}
        \put(110,135){\includegraphics[scale=0.65]{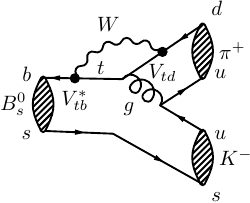}}
        \put(10,56){\includegraphics[scale=0.65]{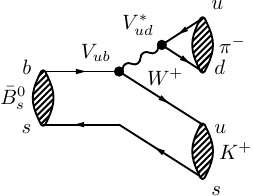}}
        \put(110,56){\includegraphics[scale=0.65]{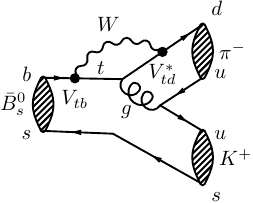}}
       
        \put(345,1){\includegraphics[scale=0.215]{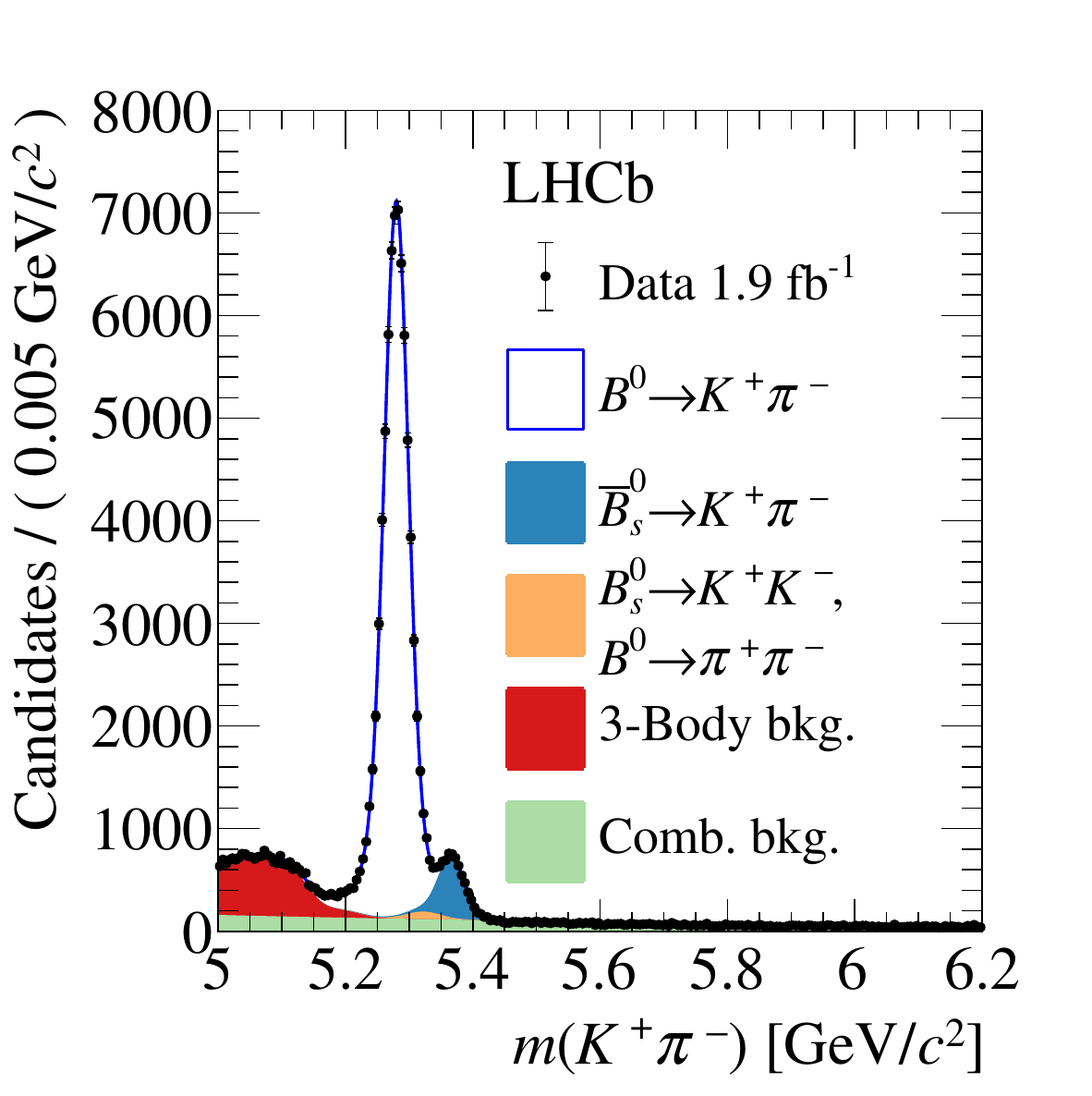}}
       \put(345,114){\includegraphics[scale=0.215]{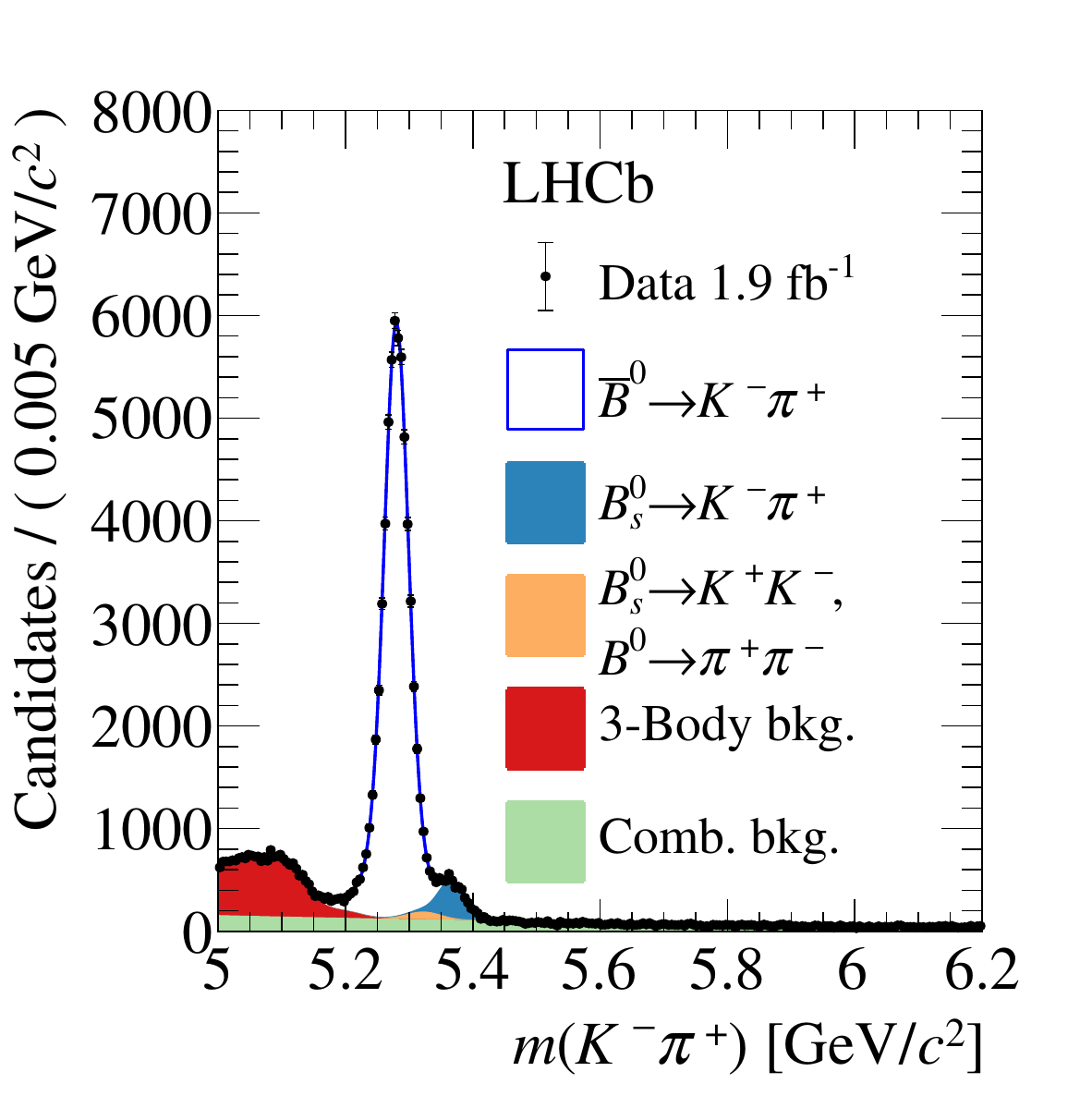}}
     \end{picture}
    \caption{{\em Left:} Illustration how two interfering amplitudes in $B_s^0\rightarrow K^- \pi^+$ decays and their corresponding CP-mirror diagrams result in a different absolute amplitude under CP conjugation, due to the relative CP violating weak phase $\phi$ between the two amplitudes. {\em Right:} The corresponding measurement of LHCb~\cite{LHCb-PAPER-2020-029}.}
    \label{fig:CPinterference}
\end{figure}
%\hspace{0.5cm}
Labelling the decay amplitudes $A$ ($\overline{A}$) of the combined tree and penguin diagrams as $B_s^0\rightarrow K^+\pi^-$ (${\bar{B}^0_s}\rightarrow K^-\pi^+$)  we write:
\begin{eqnarray}
              A   & = & A_T + A_p \: e^{i\phi} \: e^{i\delta}    \\ 
    \overline{A} & = & A_T + A_p \: e^{-i\phi}\: e^{i\delta}   
\end{eqnarray}
such that the difference in decay rate ($R\equiv\left|A\right|^2$) is:
$R - {\overline{R}} = 4 \: A_T \: A_P \sin\phi \: \sin\delta.$
This mechanism is illustrated in Fig.~\ref{fig:CPinterference} together with the individual $K^-\pi^+$ and $K^+\pi^-$ mass distributions observed by LHCb, which have different peak heights, and hence show CP-violation.

Phenomenologically, CP-violating processes are separated in three types.
%\begin{itemize}
%    \item 
    First, CP violation in mixing (sometimes called {\em indirect} CP violation) arises from the interference between dispersive and absorptive mixing amplitudes, and is independent of the final state. 
    This leads to a different transition rate from meson to anti-meson than its reverse process.
    %$\mathrm{Prob}(K^0 \rightarrow \bar{K}^0) \neq \mathrm{Prob}(\bar{K}^0 \rightarrow K^0)$.
    %\item 
    Second, CP violation in decay (also called {\em direct} CP violation) originates from the interference between two decay diagrams and shows up as different decay rates, 
    $A(B\to f)\ne A(\bar{B}\to \bar{f})$. 
    %\item 
    Third, CP violation in the {\em interference} of mixing and decay explicitly includes meson mixing, and is thus decay-time dependent. It involves the interference between mixed and unmixed decay amplitudes, and occurs when the same final state can be reached from $B^0$ as well as $\bar{B}^0$, for example in $B^0$-decays where the final state is a CP-eigenstate,
    $ \Gamma(B^0       {\scriptstyle (\leadsto \bar{B}^0)} \rightarrow f)(t) \neq 
    \Gamma(\bar{B}^0 {\scriptstyle (\leadsto B^0)}       \rightarrow f)(t) $. 
%    The corresponding time-dependendent CP-asymmetry can be expressed as follows
 %   \begin{equation}
%A_{CP}(t) =  \frac{\Gamma_{P^0(t)\rightarrow f} - \Gamma_{\bar{P}^0(t)\rightarrow f}}%
%         {\Gamma_{P^0(t)\rightarrow f} + \Gamma_{\bar{P}^0(t)\rightarrow f}}
% =  \frac{2C_f\cos \Delta mt              - 2S_f\sin \Delta m t}%
 %        {2\cosh\frac{1}{2}\Delta\Gamma t + 2D_f\sinh\frac{1}{2}\Delta\Gamma t}
%\label{eq:ACP}
%\end{equation}
    %where $S_f$ and $D_f$ are the imaginary and real part of the relative amplitude, $\lambda_f=q\bar{A}_f/pA_f$, multiplied by $2/(1-|\lambda_f|^2)$.
%\end{itemize}

%\clearpage
\subsection{CP violation in mixing: "indirect CPV"}
%-----------------------------------------------------

CP violation in mixing was first discovered by Cronin and Fitch \cite{Christenson:1964fg}, which revealed that
the $K_L$ state was not a pure CP-odd eigenstate, but that it contains a small CP-even component. This implies $|q/p|\ne 1$, which in turn implies 
$\mathrm{Prob}(K^0 \rightarrow \bar{K}^0) \neq \mathrm{Prob}(\bar{K}^0 \rightarrow K^0)$, i.e. the transition from $K^0$ to $\bar{K}^0$ occurs at a different rate than the inverse process.
The two interfering amplitudes are the absorptive ('short distance') and dispersive ('long distance') amplitudes ,
which in case of the $B$ systems correspond to the mixing diagrams with internal (virtual) top quarks and
(on-shell) charm quarks, respectively. 
In  $B$-meson systems, CP violation in mixing is studied with the use of flavour-specific final states. 
After all, if the oscillation rate has a preferred direction, then inevitably 
one of the two charge-conjugate final states will appear to be more abundant (provided one starts with equal amounts of $B^0$ and $\bar{B}^0$ mesons),
which is quantified as the flavour-specific asymmetry $a_{fs}$ (or equivalently $a_{sl}$ for semileptonic final states):
\begin{equation}
    a_{fs} = \frac{\Gamma(\bar{B}\to f)-\Gamma(B\to \bar{f})}
        {\Gamma(\bar{B}\to f)+\Gamma(B\to \bar{f})} =
        {\Im\left(\frac{\left|\Gamma_{12}\right|}{\left|M_{12}\right|}\right)} = \frac{\left|\Gamma_{12}\right|}{\left|M_{12}\right|}\sin{\phi_{12}} =
        \frac{\Delta\Gamma}{\Delta m}\tan\phi_{12} ,
\end{equation}
%with $\phi = \phi_M - \phi_\Gamma$ the relative phase between the mixing amplitudes and 
where eq.~\ref{eq:DmDGamma} is used. The phase $\phi_{12}$ is expected to be small for both the $B^0$ and $B^0_s$ case, respectively
$\phi_{12}^d\approx -4.3^o$ and $\phi_{12}^s\approx 0.22^o$~\cite{Lenz:2011ti}.
In practice, neglecting any production and detection asymmetries, the raw untagged asymmetry in semileptonic $B$ decays is determined as
\begin{equation}
A_{raw}=\frac{N(D^-\mu^+)-N(D^+\mu^-)}{N(D^-\mu^+)+N(D^+\mu^-)} \; ,
\end{equation}
which corresponds to $A_{raw}=a_{fs}/2$, where the factor 2 occurs due to the fact that the initial flavour of the $B$ at production is unknown. 

\begin{figure}[!ht]
	\centering
    \begin{picture}(450,150)(0,0)
        \put(30,80){\includegraphics[scale=0.85]{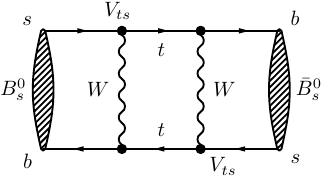}}
       \put(30,15){\includegraphics[scale=0.85]{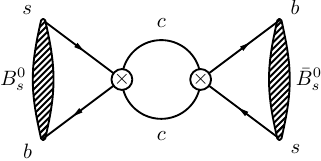}} 
       \put(220,5){\includegraphics[scale=0.17]{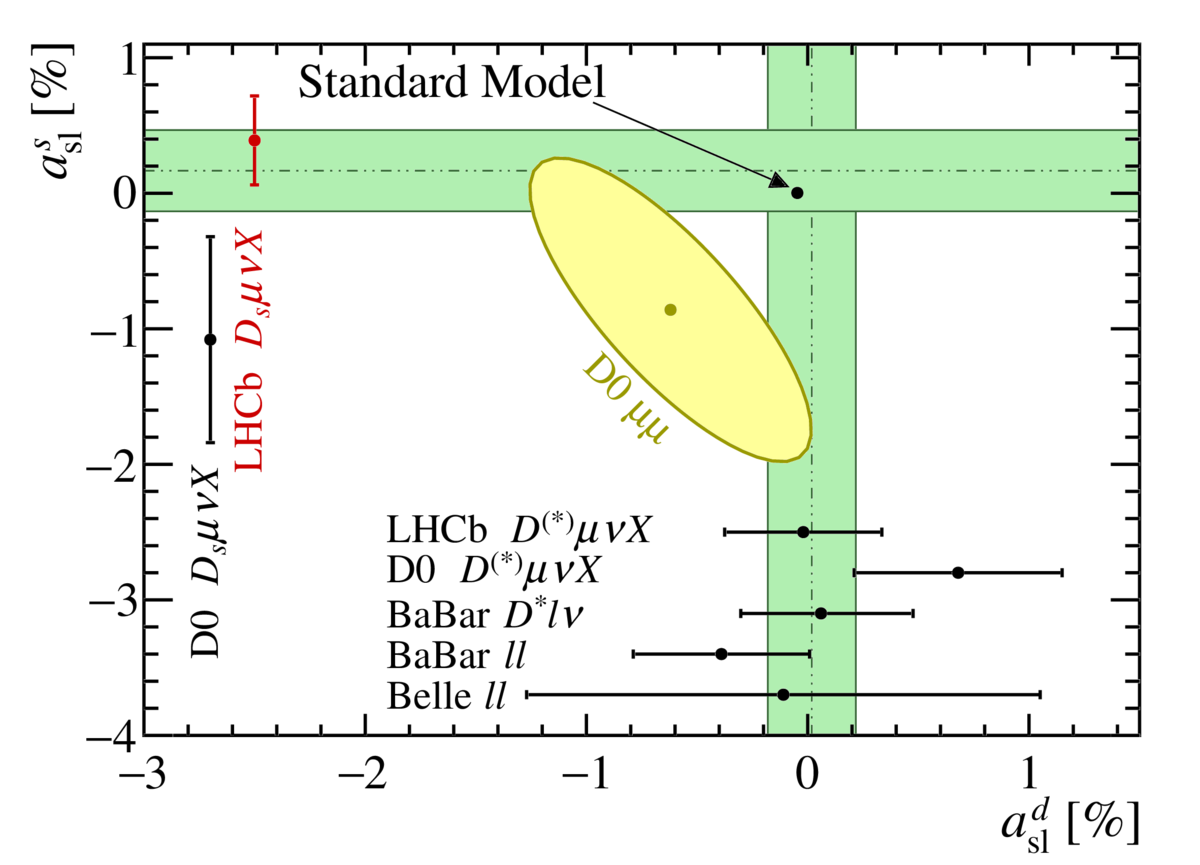}}
    \end{picture}
 \caption{Mixing CP asymmetry for $B_s^0$ mesons. {\em Left:} CP violation in mixing arises from the interference of
 the absorptive and dispersive amplitudes. The absorptive amplitude (bottom) is a long distance effect and is calculated through heavy-quark effective theory, where the effective coupling is indicated with a circle, and is dominated by on-shell charmed intermediate states~\cite{Lenz:2010gu}. {\em Right:} The measured values for
 the flavour-specific asymmetry for the $B^0$ and $B^0_s$ systems \cite{LHCB-PAPER-2016-013}.}
	\label{fig:CP-mixing}
\end{figure}

Fig.~\ref{fig:CP-mixing} shows the contributing diagrams for $B_s^0$ mixing together with the measurement results $a_{fs}\equiv a_{sl}^s$ of the $B_s^0$ plotted versus $a_{fs}\equiv a_{sl}^d$ of the $B^0$. The measurements of LHCb are compared to earlier ones from the D0, BaBar and Belle experiments, and finds no significant CP violation, in agreement with the SM expectation.

%\clearpage
\subsection{CP violation in decay: "direct CPV"}
\label{sec:directCP}
%-----------------------------------------------------

Unlike CP violation in mixing or in the interference of mixing and decay, CP violation in decay is not limited to neutral mesons that oscillate. 
Prominent examples of CP violation in decay involve
charged $B$ mesons and, recently, $\Lambda_b^0$ baryons. 
In addition, CP violation in decay is observed in neutral $D^0$ mesons.

\subsubsection{\texorpdfstring{$B$}{B}-meson decays}

\begin{figure}[!h]
	% 	\includegraphics[width=10cm]{Figures/CPViolation/Gamma_GLW_ADS_Legenda.pdf}\\
%    \centering
    \begin{picture}(450,230)(0,0)
        \put(0,30){\includegraphics[scale=0.8]{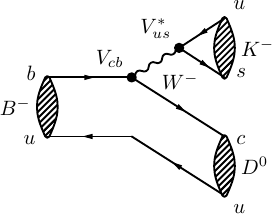}}  
        \put(0,120){\includegraphics[scale=0.8]{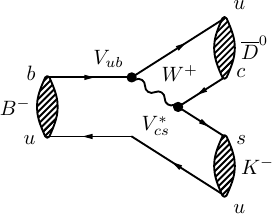}}   
        \put(110,110){\includegraphics[width=6.2cm]{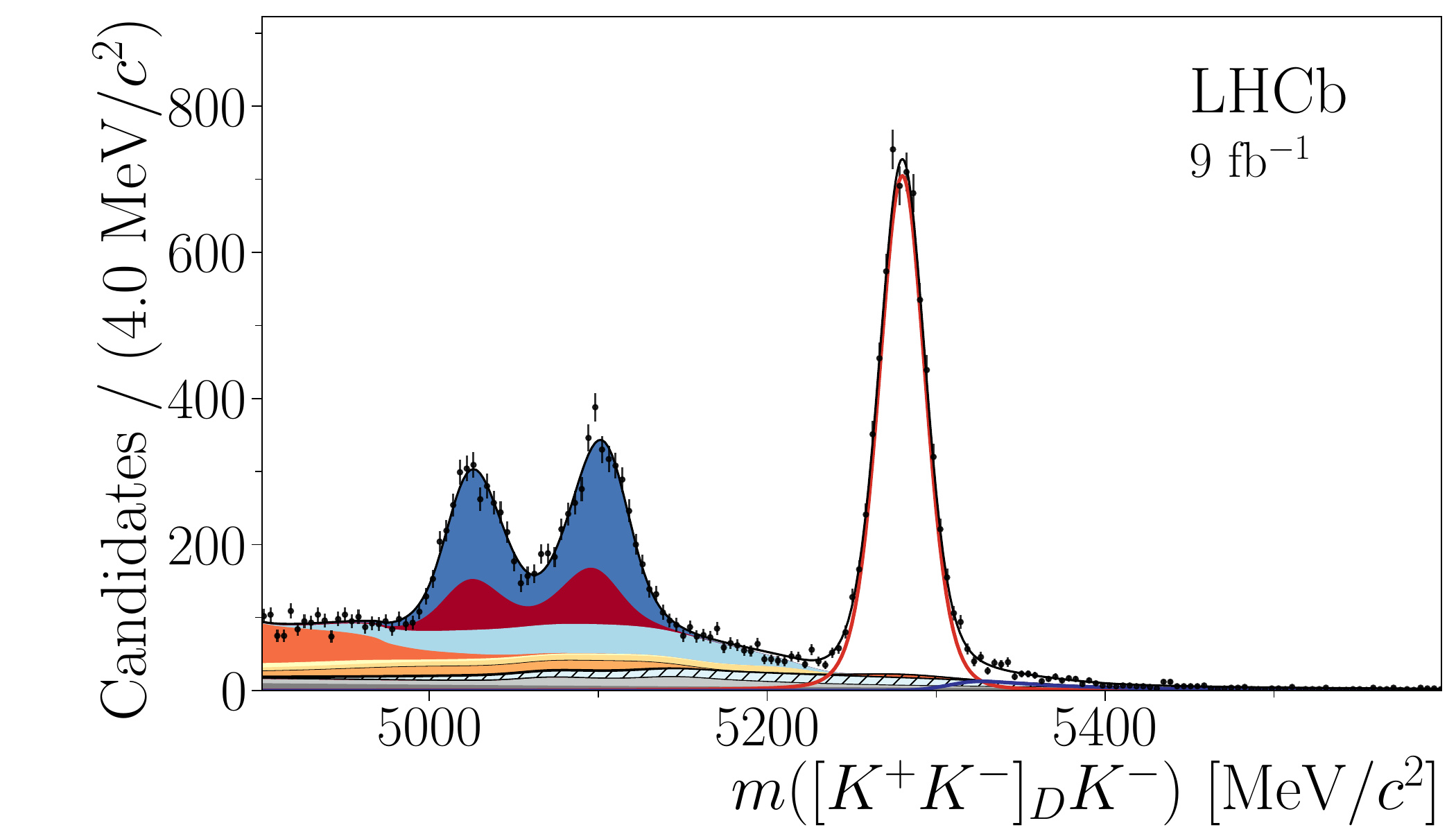}}
 	      \put(285,110){\includegraphics[width=6.2cm]{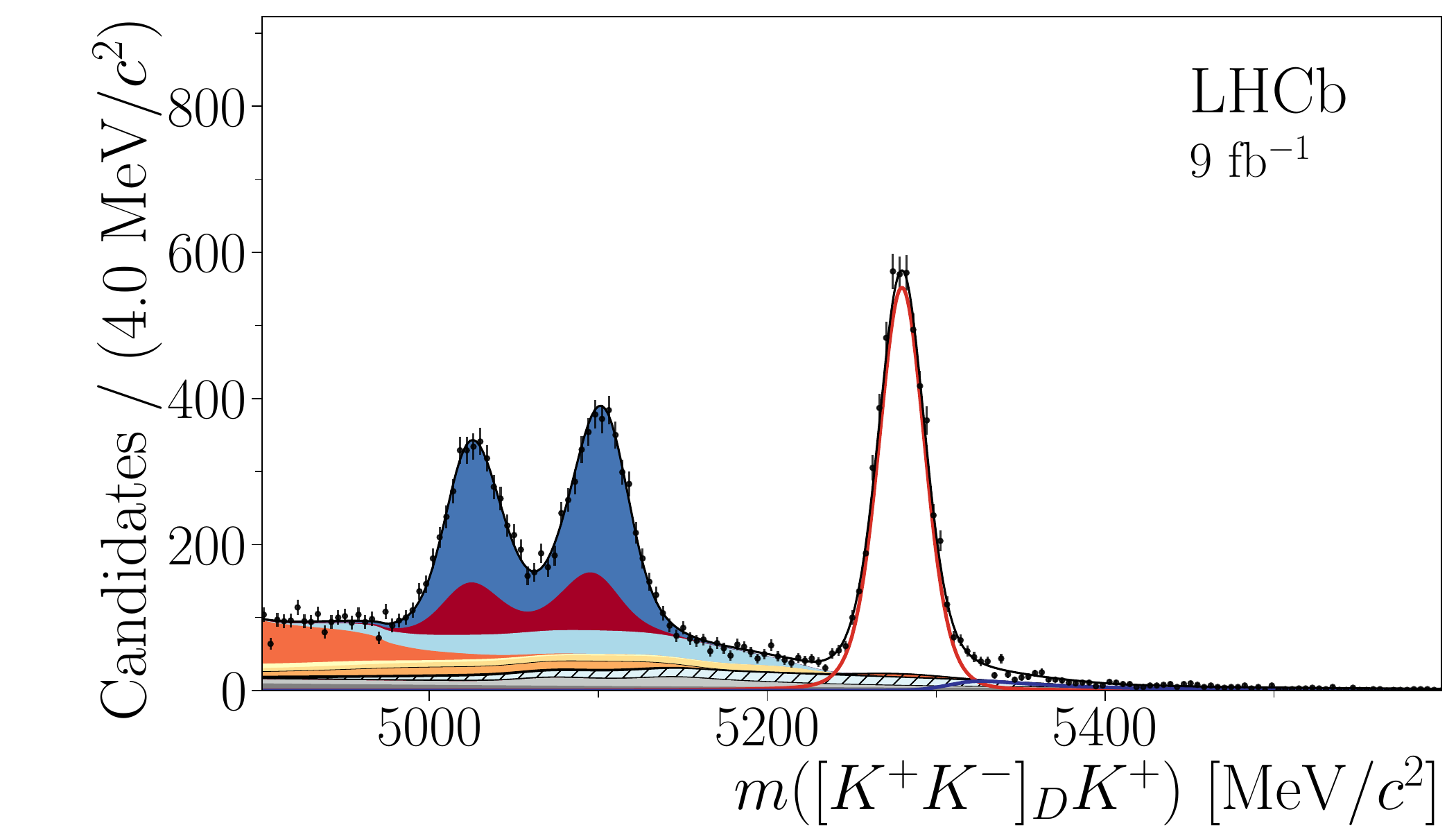}}
  	  \put(110,10){\includegraphics[width=6.2cm]{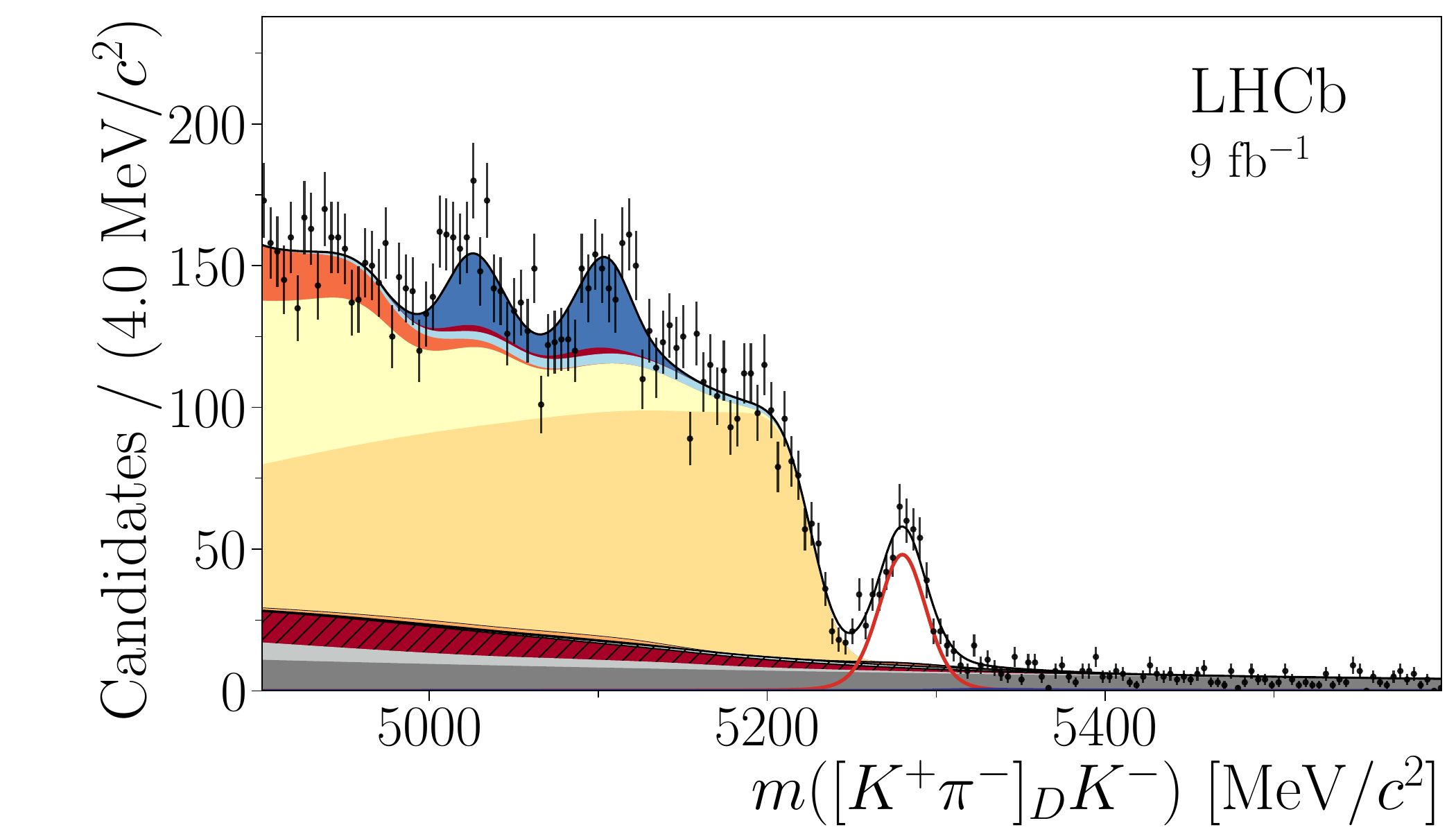}}
   	    \put(285,10){\includegraphics[width=6.2cm]{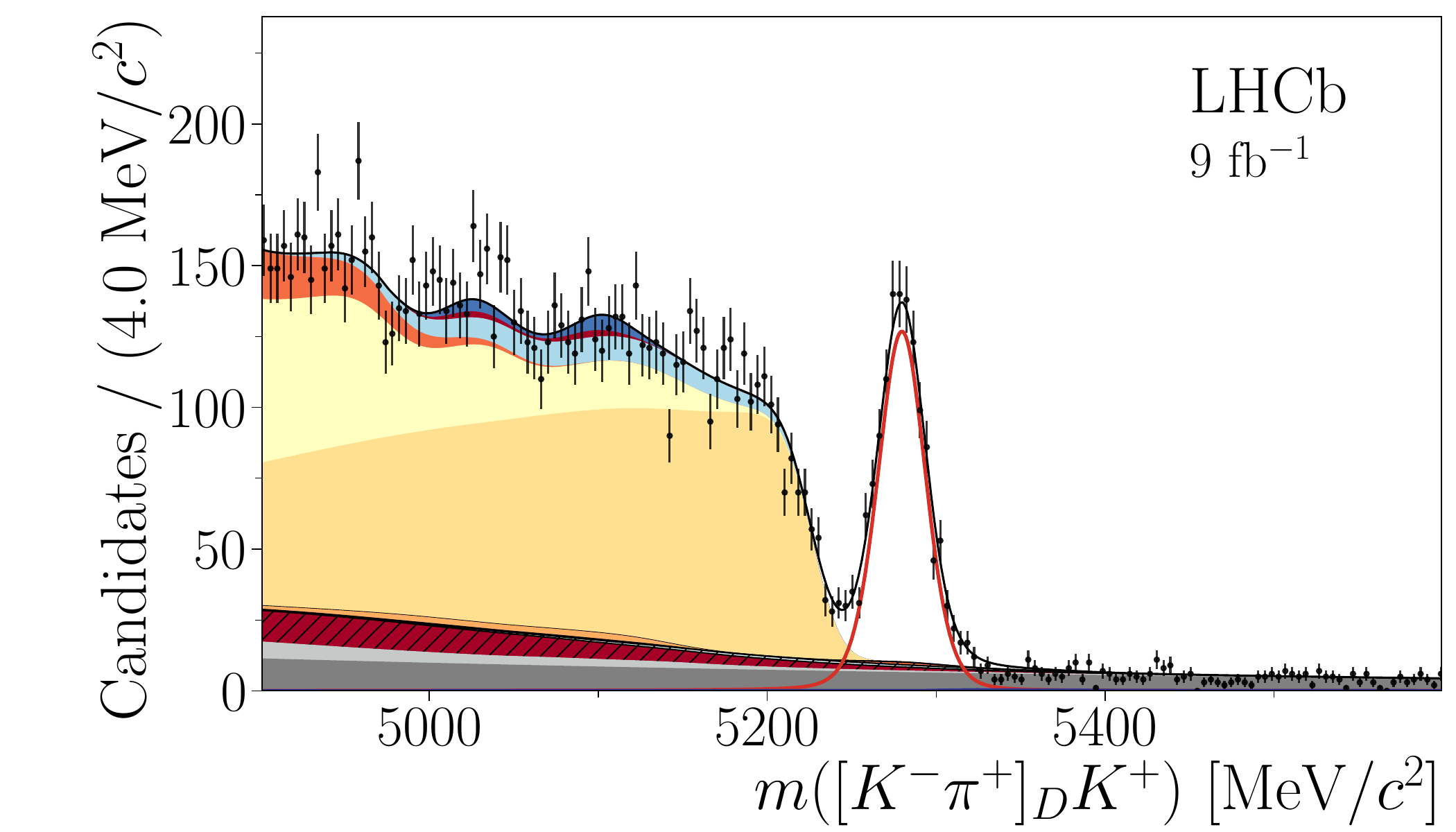}}
	\end{picture}
    \caption{
    {\em Left:} The two interfering diagrams that lead to direct CP violation, due to the relative phase difference between $V_{ub}$ and $V_{cb}$. 
    {\em Right top:} The invariant mass distributions of $B^+\to D(\to K^+K^-)K^+$ and its charge-conjugate show the {\em large event yield} for the GLW modes, whereas 
    ({\em right bottom}) the invariant mass distributions of $B^+\to D(\to K^+\pi^-)K^+$ and its charge-conjugate show the {\em large asymmetry} for the ADS mode due to the similarity of the magnitude of both decay paths \cite{LHCb-PAPER-2020-036}.
    The coloured distributions on the left of the $B$ mass peaks indicate various background
    contributions from $B$ decays with missing particles, so called partially reconstructed backgrounds.}
	\label{fig:CP-gamma-GLW-ADS}
\end{figure}

Many neutral and charged $B$-decays proceed through multiple (interfering) amplitudes,
and thus CP violation is studied in a large variety of decays. 
A prominent example of CP violation in decay is observed in the 
decay rates of $B^+\to DK^+$ compared to $B^-\to DK^-$. 
At first glance in Fig.~\ref{fig:CP-gamma-GLW-ADS} (left) the tree diagram and the colour-suppressed diagram do not interfere,
due to the different $D^0$ or $\bar{D}^0$ flavour in the final state.
However, both the $D^0$ and $\bar{D}^0$ meson can decay to the same final state, and thus resulting in interference between the amplitudes 
%$A(B^+\to \stackrel{\scriptscriptstyle{(-)\hspace{0.3cm}}}{DK^+} \to f_D K^+)$
$A(B^+\to D^0K^+ \to f_D K^+)$
and
$A(B^+\to \bar{D}^0K^+ \to f_D K^+)$.

Using unitarity of the CKM matrix the relative weak phase difference between the diagrams, {$\arg\left[\frac{V_{cb}V_{us}^*}{V_{ub}V_{cs}^*}\right]$}, is found (approximately) equal to angle
$\gamma=\arg\left[-\frac{V_{ud}V_{ub}^*}{V_{cd}V_{cb}^*}\right]$
and a measurement of angle $\gamma$ is obtained by comparing the decay rates involving $B^+$ and $B^-$ as well as intermediate $D$ and $\bar{D}$ states.

The following methods are used to measure the direct CP violation with charged $B$ decays,
depending on the choice of the $D$ final state:
\begin{itemize}
    \item {\em $D$-decays to CP eigenstates} $\pi^+\pi^-$ or $K^+K^-$, known as the GLW method \cite{Gronau:1990ra,Gronau:1991dp}; 
    \item {\em Suppressed $D$-decays}, through the doubly Cabibbo-suppressed decay $D^0\to K^+\pi^-$, known as the ADS method \cite{Atwood:1996ci,Atwood:2000ck};
    \item {\em $D$-decays as multibody self-conjugate modes} like $D^0\rightarrow K_s\pi\pi$, known as the BPGGSZ method \cite{Giri:2003ty,Ceccucci:2020cim};
    \item {\em "Dalitz" method} \cite{Gershon:2008pe} for multibody $B$ decays, such as $B\rightarrow\overline{D}^0 K^+\pi^-$.
    %Example of 2- 3- and 4-body D modes in one paper for $B\rightarrow\overline{D}^0 K^*$: LHCb-PAPER-2024-023
\end{itemize}
Fig.~\ref{fig:CP-gamma-GLW-ADS} shows the LHCb measurements for the GLW method with $B^+\rightarrow [K^+K^-]_DK^+$ decays and the ADS method with $B^+\rightarrow [K^+\pi^-]_D K^+$ decays. The GLW is characterized by a large event yield and relatively small CP asymmetry, whereas the ADS modes involve equally suppressed diagrams with small event yields, but resulting in relatively large interference and CP asymmetries.
In total a few dozen $B$ and $D$ final states have been analyzed with sensitivity 
to the CKM angle $\gamma$, and a combined value and uncertainty of $\gamma= (64.6 \pm 2.8)^o$ has been achieved~\cite{LHCb-PAPER-2021-033,LHCb-CONF-2024-004},
as will be shown in Fig.~\ref{fig:CKM}.

\clearpage
\subsubsection{\texorpdfstring{$D$}{D}-meson decays}

\begin{figure}[!ht]
	\centering
    \includegraphics[width=5.5cm]{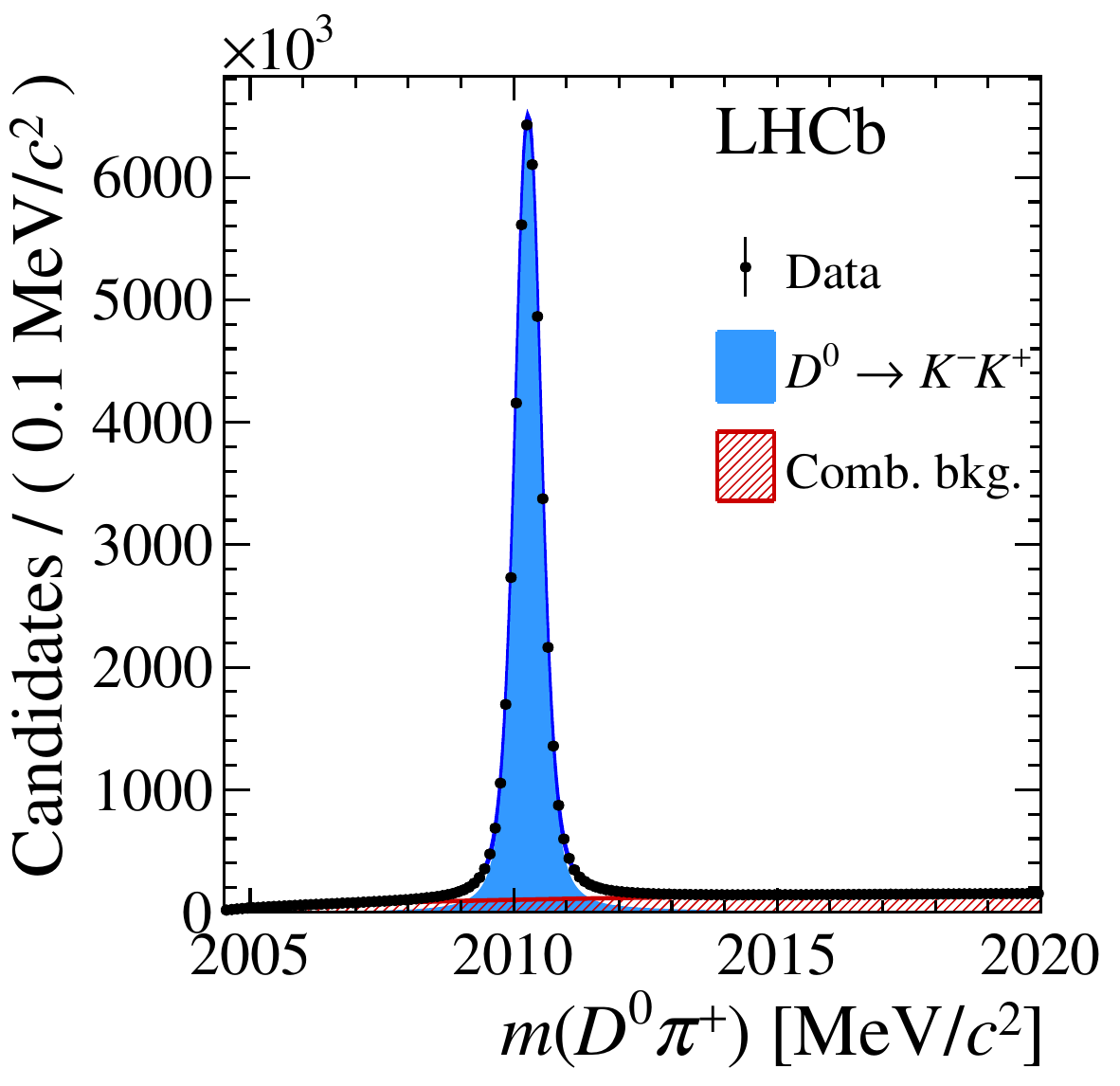} 
    \includegraphics[width=5.5cm]{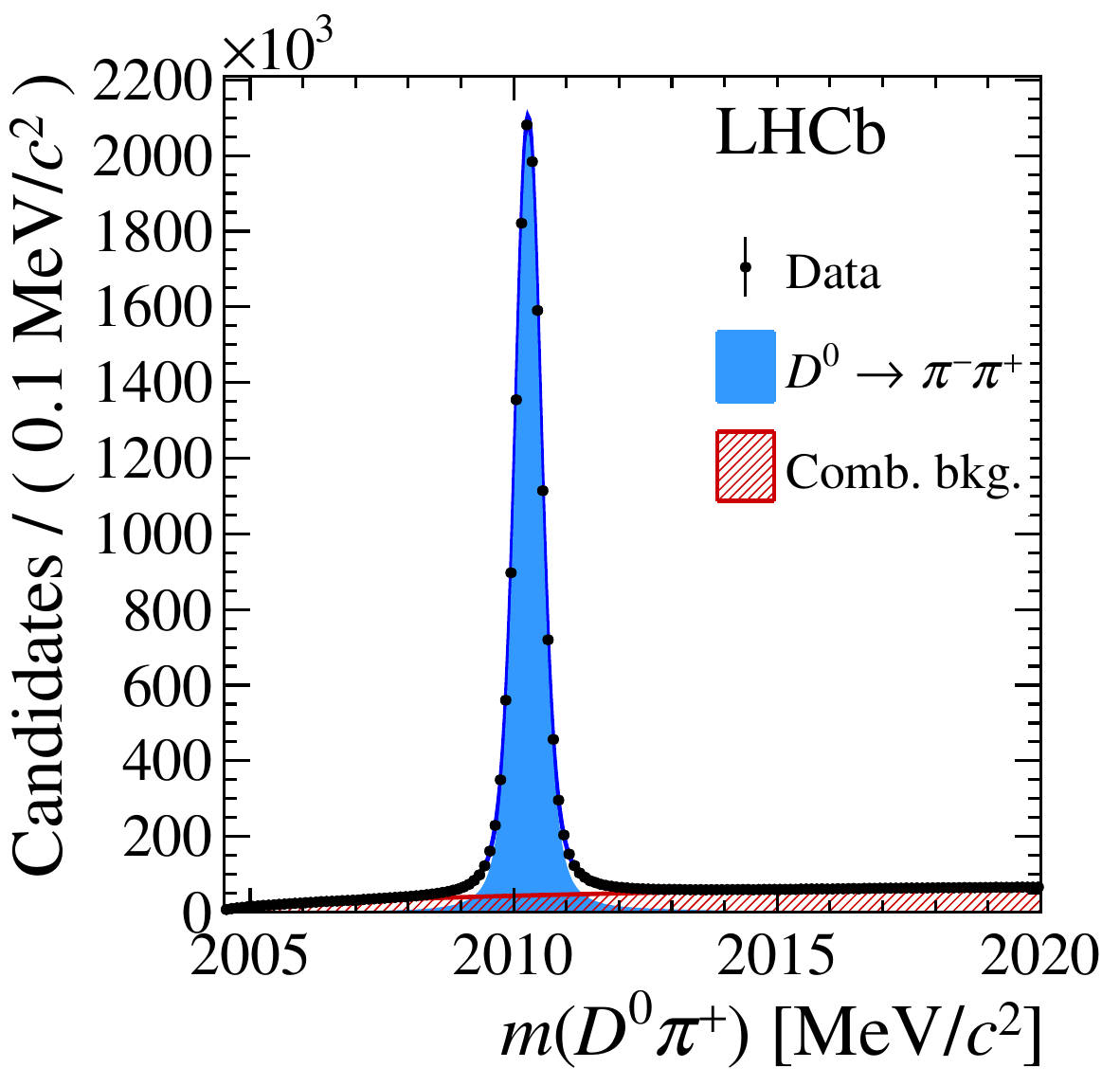}
	\caption{
The invariant distributions of {\em (left:)} $D^0\rightarrow K^+ K^-$ and {\em (right:)} $D^0\rightarrow \pi^+\pi^-$ are shown. These large and pure data samples are used to determine the direct CP violation in both decay modes, and their difference $\Delta A_{CP}$~\cite{LHCb-PAPER-2019-006}.}
	\label{fig:Charm_CP}
\end{figure}

\begin{figure}[!ht]
	\centering
    \begin{picture}(450,170)(0,0)
        \put(10,50){\includegraphics[scale=0.75]{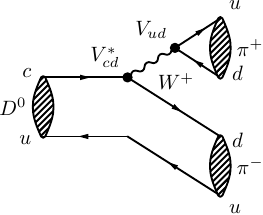}}
       \put(120,90){\includegraphics[scale=0.75]{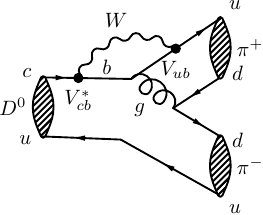}} 
       \put(120,10){\includegraphics[scale=0.75]{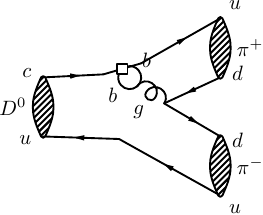}}
         \put(240,5){\includegraphics[width=8cm]{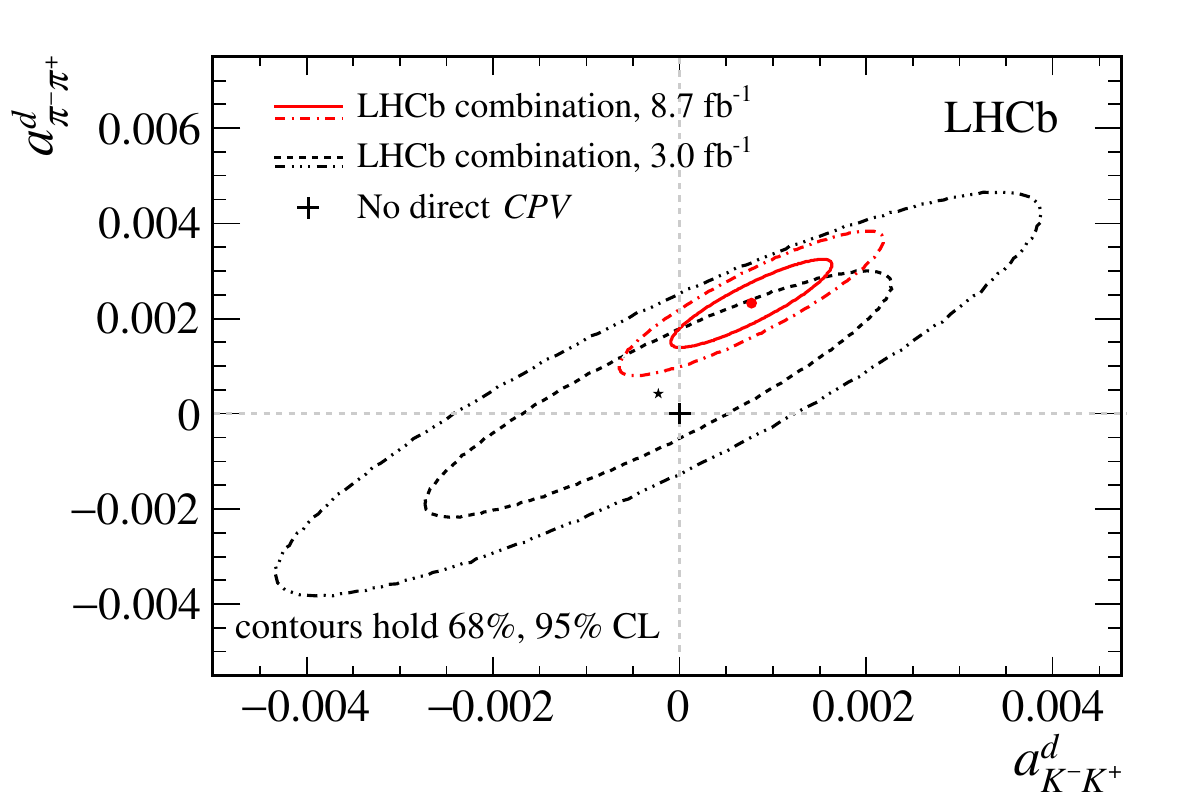}}
    \end{picture}
	\caption{{\em Left:} The tree and penguin diagrams that interfere and can give rise to CP violation in decay. The decay can also proceed through an effective coupling with a $b\bar{b}$ loop in HQET. {\em Right:} The CP asymmetry measured for  $D^0\to K^+K^-$ and $D^0\to \pi^+\pi^-$ decays shows evidence for CP violation in the decay $D^0\to \pi^+\pi^-$~\cite{LHCb-PAPER-2022-024}. }
	\label{fig:Charm_CP2}
\end{figure}

Direct CP violation is also studied in charm decays by comparing the decay rates of $\Gamma(D^0\to f)$ and $\Gamma(\bar{D}^0\to \bar{f})$, where the flavour of the
neutral $D$ meson is either determined from the charge of the slow pion in the 
$D^{+*}\to D^0\pi^+_s$ decay, or from the charge of the muon in the
$B^-\to D^0\mu^-\nu$ decay. 
In $D$ decays, tree and penguin diagrams interfere and can give rise to CP violation in decay. An example of decay diagrams are shown in the left of Fig.~\ref{fig:Charm_CP2}. The penguin diagram is often described through an effective coupling in HQET, as the long-distance effects are important in the low-energy charm system. This is illustrated with the $W$-line contracted to an 4-point vertex, leading to the $b$-quark loop.
The amount of CP violation is expected to be small, and to minimize experimental effects affecting the measured raw asymmetry, a charge-symmetric final state is most advantageous, such as $\pi^+\pi^-$ or $K^+K^-$.
To further reduce detection asymmetries from the reconstruction of the tagging muon or pion, the difference between the CP violation of these final states is measured.
The first observation of CP violation in charm decays is thus achieved comparing the direct CP asymmetries in $D^0\to K^+K^-$ and $D^0\to \pi^+\pi^-$:
$$
\Delta A_{CP} = A_{CP}\left( K^+K^-\right) - A_{CP}\left(\pi^+\pi^-\right)
\hspace*{1.0cm} \text{where} \hspace*{0.5cm} A_{CP}\left( f \right)\equiv\frac{\Gamma(D\rightarrow f) -\Gamma(\bar{D}\rightarrow f)}{\Gamma(D\rightarrow f) + \Gamma(\bar{D}\rightarrow f)}$$
Fig.~\ref{fig:Charm_CP} shows the relatively background-free mass peaks of both decays.
Combining the pion and muon tagged samples, the result is $\Delta A_{CP}=(-15.4\pm 2.9)\times 10^{-4}$, the first observation of CP violation in charm decays.

The individual CP asymmetries for both decay modes separately have been measured 
with the full data sample collected by LHCb between 2011 and 2018, which show evidence for CP violation in the $D^0\to \pi^+\pi^-$ decay \cite{LHCb-PAPER-2019-006}. The results are shown in Fig.~\ref{fig:Charm_CP2}, together with the contributing Feynman diagrams for the $D$ decays into $\pi^+\pi^-$ final state.
Interestingly, the individual CP asymmetries are found to have the same sign,
whereas theoretically one expects them to be of opposite sign~\cite{Piscopo:2024wpd}.

%\clearpage
\subsubsection{Baryon decays}

\begin{figure}[!ht]
\centering
\includegraphics[scale=0.25]{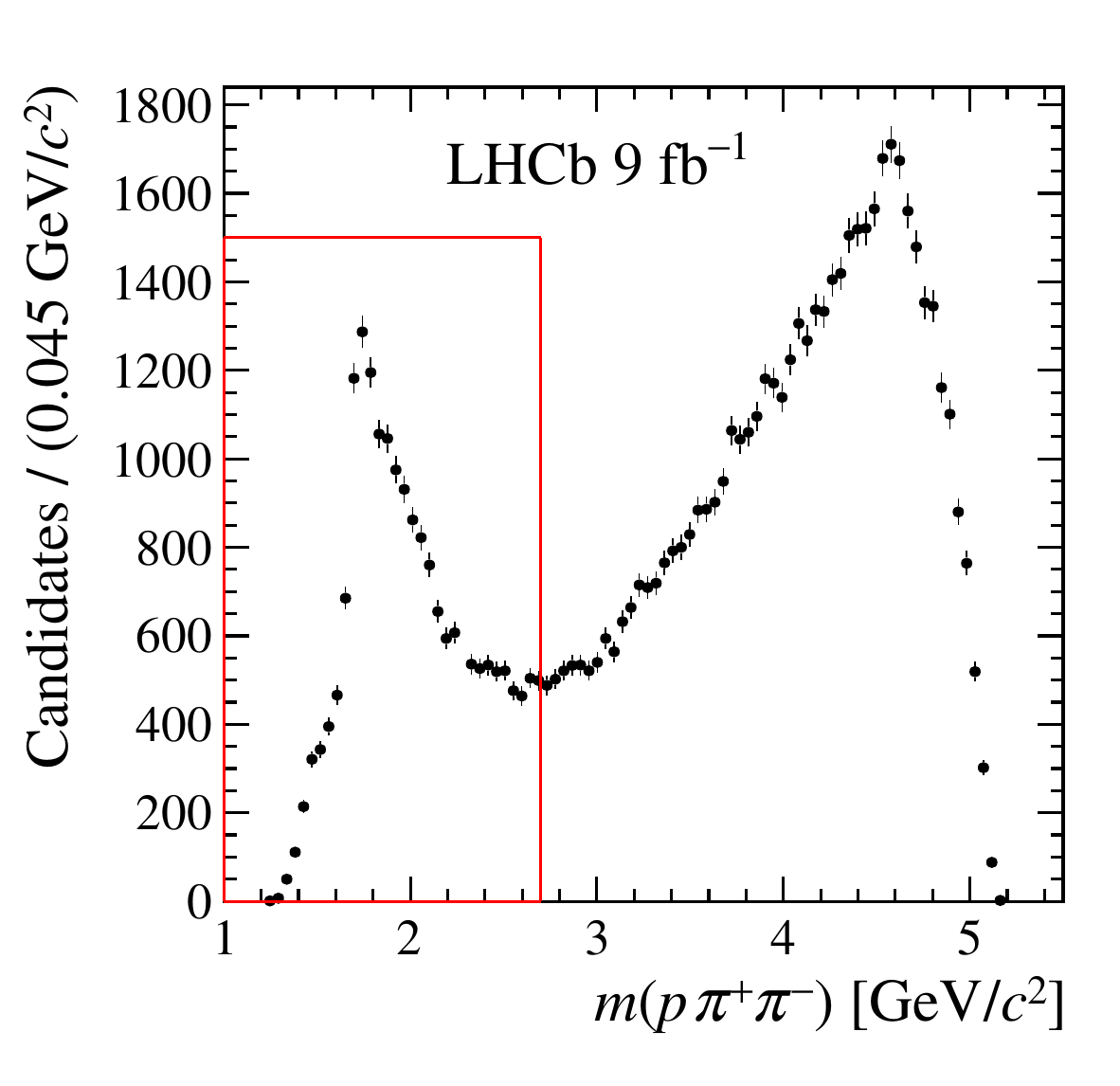}
\includegraphics[scale=0.25]{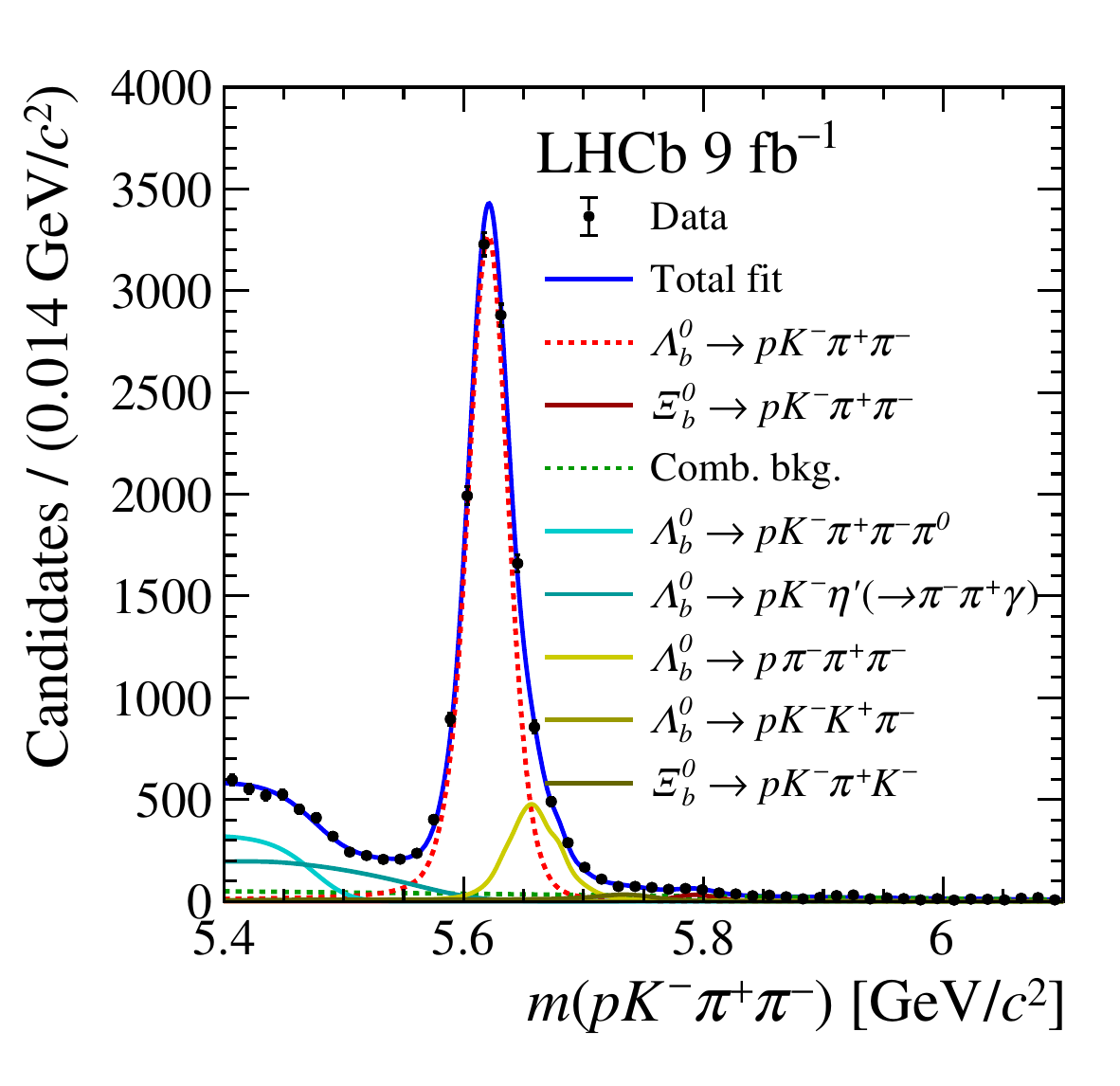}
\includegraphics[scale=0.25]{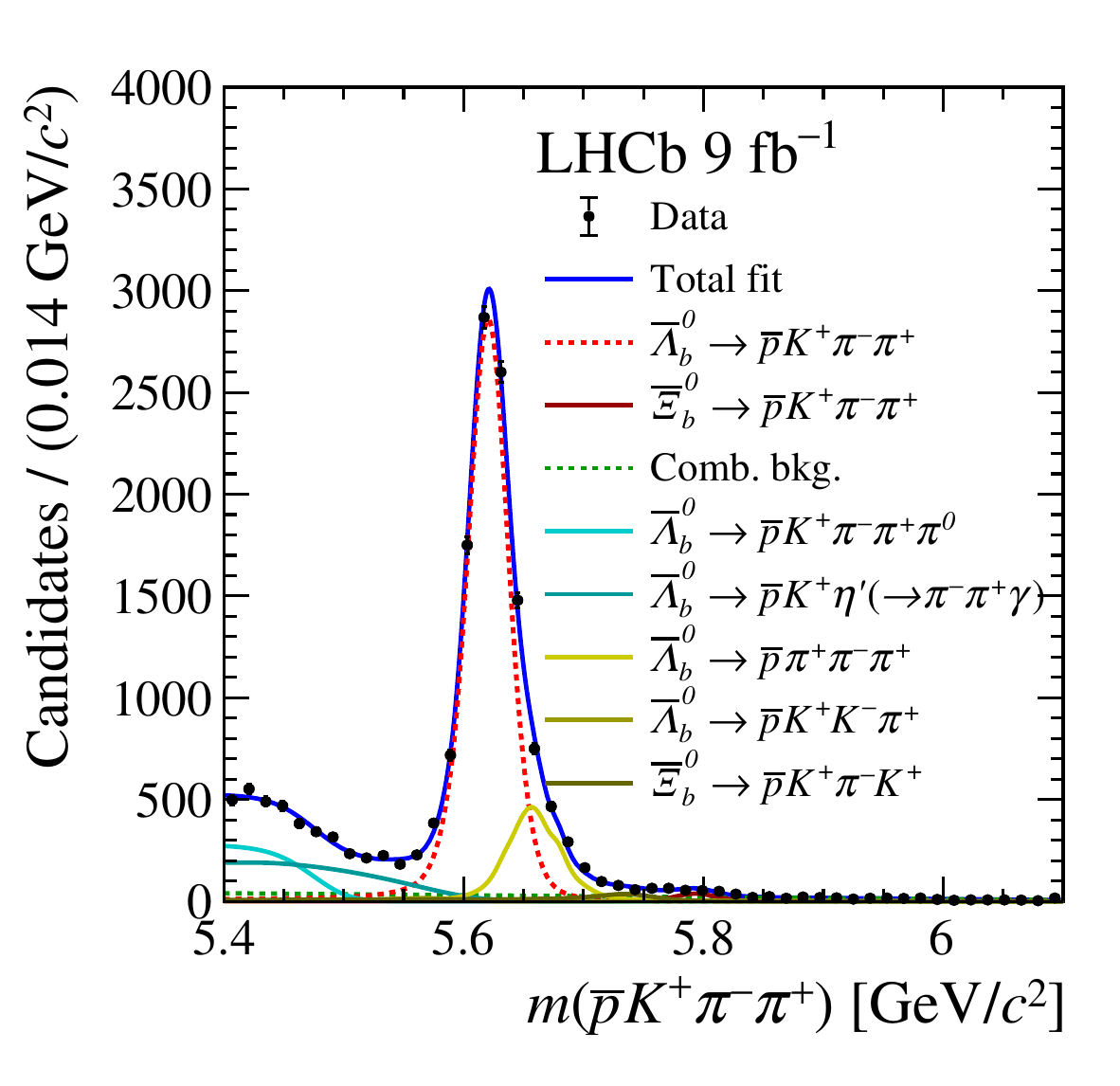}
\caption{CP asymmetry in a selected phase space of $\Lambda_b\rightarrow p K^-\pi^+\pi^-$ and charged conjugate decays. {\em Left:} The selected resonance region of $m(p\pi^+\pi^-)$ with large contributions from excited nucleon resonances. {\em Center:} The observed $m(pK^-\pi^+\pi^-)$ spectrum and {\em (right:)} the observed spectrum of the CP-conjugate decay~\cite{LHCb-PAPER-2024-054}, both for low values of
$m(p\pi^+\pi^-)$ as indicated by the red square in the left figure.}
\label{fig:BaryonCP}
\end{figure}

Until a recent result from LHCb in 2025 \cite{LHCb-PAPER-2024-054}, CP violation in baryonic decays had not been observed. The first observation is seen in the decay $\Lambda_b^0\rightarrow p\pi^+\pi^- K^-$ decay with an asymmetry of $A_{CP}= \left( 2.45\pm 0.46 (stat) \pm 0.1 (sys)\right) \%$, with a non-zero significance of of 5.2 standard deviations. 
The baryonic decay includes several hadronic resonances that play an intricate role in the CP observable involving different relative strengths and strong phases, making the CP mechanism more complicated, but potentially also allowing new constraints to test the CKM paradigm. Fig.~\ref{fig:BaryonCP} shows the observed CP asymmetry of the $\Lambda_b$ decay for a resonance region that shows the largest asymmetry.

%\clearpage
\subsection{CP violation in interference of mixing and decay: "decay-time dependent CPV"}
%-----------------------------------------------------
The property that the $B^0$ meson
can oscillate before decaying is exploited when 
examining final states accessible to both 
$B^0$ and $\bar{B}^0$ mesons, thereby resulting in interfering amplitudes. Since these probability oscillations depend on the decay-time of the $B^0$ meson, the resulting CP violation turns out to be decay-time dependent.
Introducing the complex decay amplitude ratio parameter $\lambda_f=q\bar{A}_f/pA_f$, where $A_f$ and $\bar{A}_f$ represent decay amplitudes for a $B$ and $\bar{B}$ to the final state $f$, leads to the following CP asymmetry:
\begin{equation}
A_{CP}(t) 
 =  \frac{\Gamma_{\bar{B}(t)\rightarrow f} - \Gamma_{B(t)\rightarrow f}}%
         {\Gamma_{\bar{B}(T)\rightarrow f} + \Gamma_{B(t)\rightarrow f}}
 =  \frac{2S_f\sin \Delta m t  - 2C_f\cos \Delta mt }%
         {2\cosh\frac{1}{2}\Delta\Gamma t + 2D_f\sinh\frac{1}{2}\Delta\Gamma t}
\label{eq:ACP}
\end{equation}
where $S_f$ and $D_f$ are the imaginary and real part of the relative amplitude $\lambda_f$ multiplied by $2/(1-|\lambda_f|^2)$.
In general multiple decay diagrams may contribute, leading to a nonzero value of $C_f\ne 0$, implying a non-zero CP asymmetry at decay time $t=0$, and hence a contribution of direct CP violation.
The asymmetry simplifies in case $|q/p|=1$ is assumed - a good approximation in both the $B^0$ and the $B^0_s$ case - and furthermore if the transition is dominated by a single decay amplitude, implying that $|A_f|=|\bar{A}_f|$.
In that case $|\lambda_f|=1$, and with 
$D_f=\Re\lambda_f$, $C_f=0$ and $S_f=\Im\lambda_f$, the asymmetry becomes:
\begin{equation}
A_{CP}(t) = \frac{\Im \lambda_f \sin \Delta m t}%
                 {\cosh\frac{1}{2}\Delta\Gamma t + \Re\lambda_f\sinh\frac{1}{2}\Delta\Gamma t}\; .
\label{eq:ACP-eigenstate}
\end{equation}

Unlike in the $B^0_s$ case where $\Delta\Gamma_s\approx 0.1$, the lifetime difference in the $B^0$ case is small, $\Delta\Gamma\approx 0$, and thus the time-dependent CP asymmetry in  $B^0$ decays further simplifies to
\begin{equation}
A_{CP}(t) = \Im \lambda_f \sin \Delta m t\; .
\label{eq:B0-ACP-eigenstate}
\end{equation}
%\textcolor{purple}{The overall sign of the  time-dependent CP asymmetry depends on the CP eigenvalue of the final state. $\Rightarrow$ Sentence a bit sudden and context lost? Should we include the $\eta_f$ term from the beginning in the $A_{CP}$ definition? Otherwise it appears ad hoc in eq (13). Or we adapt the sentence here or in section 5.4.1 a bit? What do you think, Niels?}
In conclusion, CP violation can occur even when 
CP violation in mixing is conserved, $|q/p|=1$, and when CP symmetry in decay is conserved, $|A_f|=|\bar{A}_f|$, 
namely when the condition $\Im \lambda_f \neq 0$ is satisfied: "CP violation due to interference of mixing and decay".

%\clearpage
\subsubsection{CKM angle \texorpdfstring{$\beta$}{beta}}
The decay $B^0\to J/\psi K_S^0$ is known as the Golden Decay in $B$-physics
due to its clean experimental signature with the decays
$J/\psi \to \mu^+\mu^-$ and $K_S^0 \to \pi^+\pi^-$, and due to its clean theoretical interpretation since it is dominated by a single tree decay diagram.  
The relative amplitude ratio $\lambda_f$, is expressed as
\begin{equation}
\lambda_{J/\psi K_S^0} 
= 
\left( \frac{q}{p}\right)_{B^0} \; \left(\eta_{J/\psi K_S^0} \frac{\bar{A}_{J/\psi K_S^0}}{A_{J/\psi K_S^0}} \right)
= 
-\left( \frac{q}{p}\right)_{B^0} \; \left(\frac{\bar{A}_{J/\psi \bar{K}^0}}{A_{J/\psi K^0}} \right) \;  \left( \frac{p}{q}\right)_{K^0} \; .
\end{equation}
The three parts in this equation correspond to the mixing of the $B^0$-meson, $(q/p)_{B^0}$,
the decay of the $B^0$ or $\bar{B}^0$, $\bar{A}/A$, and the mixing of the $K^0$-meson,  $(q/p)_{K^0}$.
The factor $\eta_{J/\psi K_S^0}$ accounts for the CP-eigenvalue of the final state: 1 or -1.
The $J/\psi$ meson has spin-1 and is CP-even, while the $K^0_S$ meson has spin-0 and is 
(almost) CP-even.
Since the $B^0$ is spin-0, the particles in the final state are produced in a relative angular momentum $L=1$ state. 
As a result, the final state $J/\psi K_S^0$ is CP-odd,
and hence $\eta_{J/\psi K_S^0}=-1$\;\footnote{
An analogous measurement can be performed with the decay $B^0\rightarrow J/\psi K^0_L$,
with $\eta_{J/\psi K_L^0}=+1$.}.

The $B^0\leftrightarrow\bar{B}^0$ mixing is induced by the box diagram, 
with the dominating mass matrix element $M_{12}\propto V_{tb}^*V_{td}V_{tb}^*V_{td}$ (see Fig.~\ref{fig:mixing-diagrams1}).
Neglecting the small term term $\Gamma_{12}$ leads to (see eq.\ref{eq:qp}):
\begin{equation}
\left(\frac{q}{p}\right)_{B^0} = \sqrt{\frac{M_{12}^*}{M_{12}}} = \frac{V_{tb}^*V_{td}}{V_{tb}V_{td}^*}
\label{qoverp}
\end{equation}

\begin{figure}[!ht]
	\centering
    \includegraphics[scale=0.7]{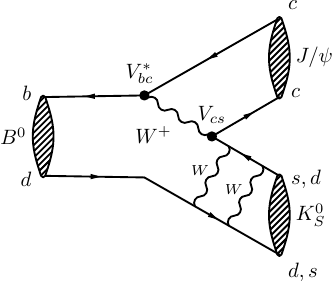}  
    \hspace{1cm}
    \includegraphics[scale=0.7]{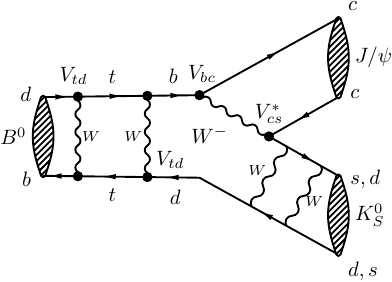}   
	\caption{The two interfering diagrams $B^0\rightarrow J/\psi K^0_S$ and
    $\bar{B}^0 {\scriptstyle (\leadsto B^0)}\to J/\psi K^0_S$ lead to sensitivity of the relative weak phase $\beta$.}
    \label{fig:jpsiks}
\end{figure}
For the ratio of the decay amplitudes one finds (see Fig.~\ref{fig:jpsiks}):
$$
\left(\frac{\bar{A}}{A}\right) = \frac{V_{cb}V^*_{cs}}{V^*_{cb}V_{cs}}.
$$
At this point either a $K^0$ or a $\bar{K}^0$ is produced in the diagram. 
To obtain the CP asymmetry for the physical $K_S^0$ mass eigenstate in the decay $B^0
\rightarrow J/\psi K^0_S$, the
$K^0\leftrightarrow \bar{K}^0$ mixing needs to be taken into account. 
This adds the factor 
$$ 
\left(\frac{p}{q}\right)_K = \sqrt{\frac{M_{12}}{M_{12}^*}} = \frac{V_{cs}V^*_{cd}}{V^*_{cs}V_{cd}}.
$$

\begin{figure}[!ht]
    \begin{picture}(450,125)(0,0)
    \put(0,6){\includegraphics[scale=0.122]{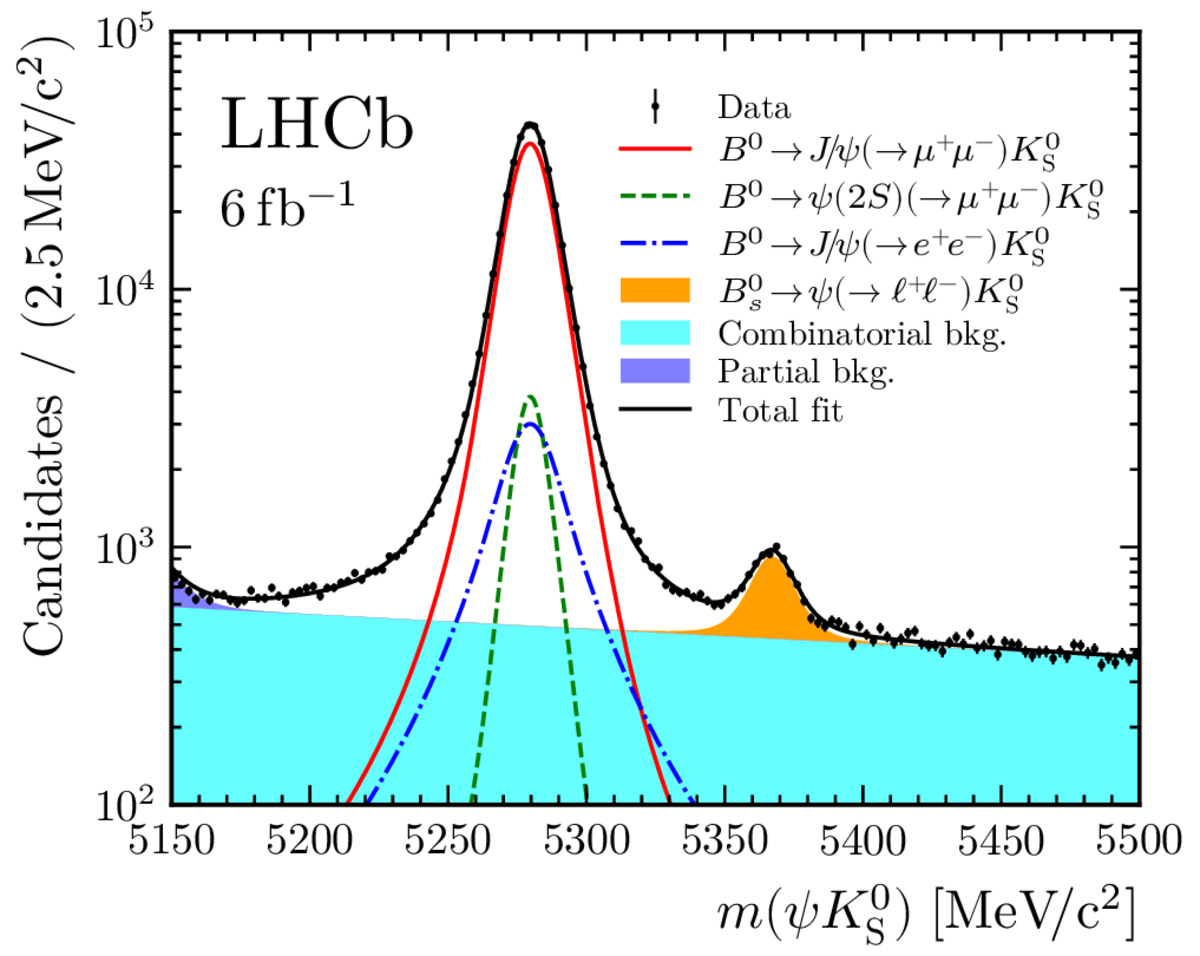}}
    \put(145,5){\includegraphics[scale=0.145]{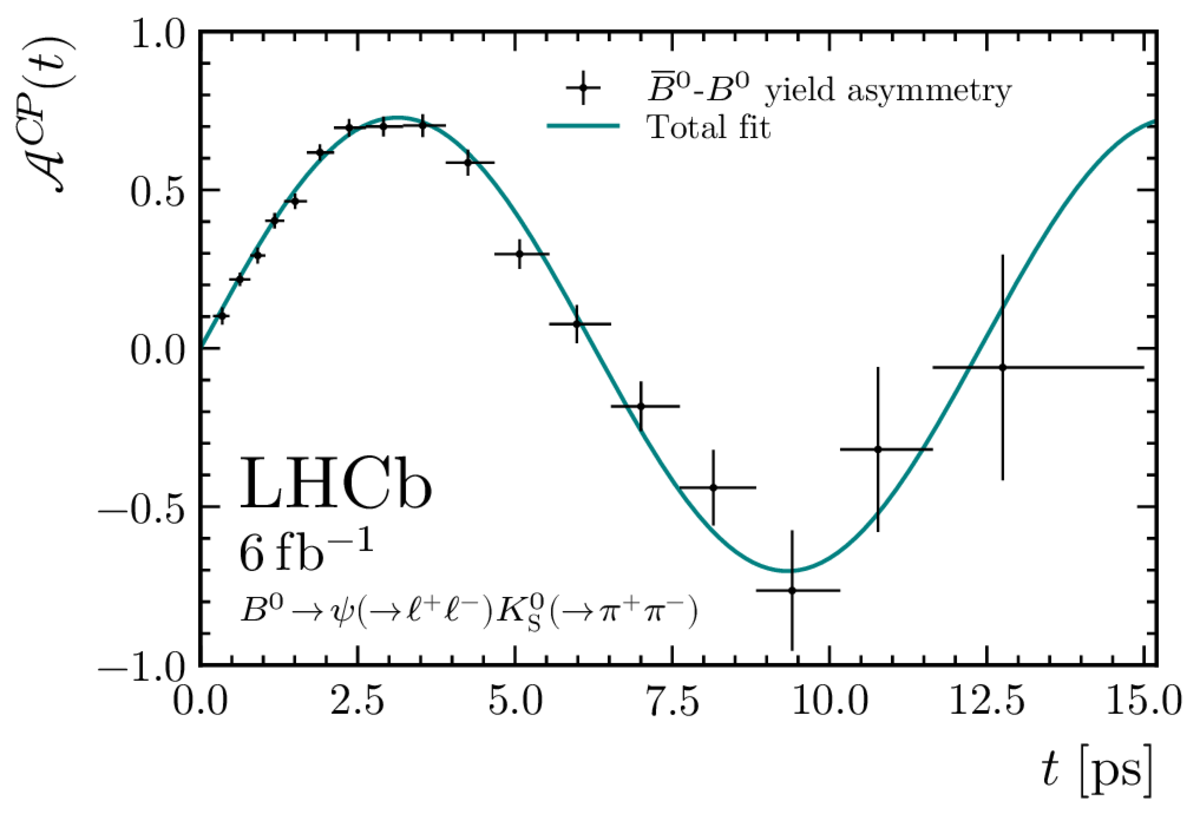}}
    \put(315,0){\includegraphics[scale=0.135]{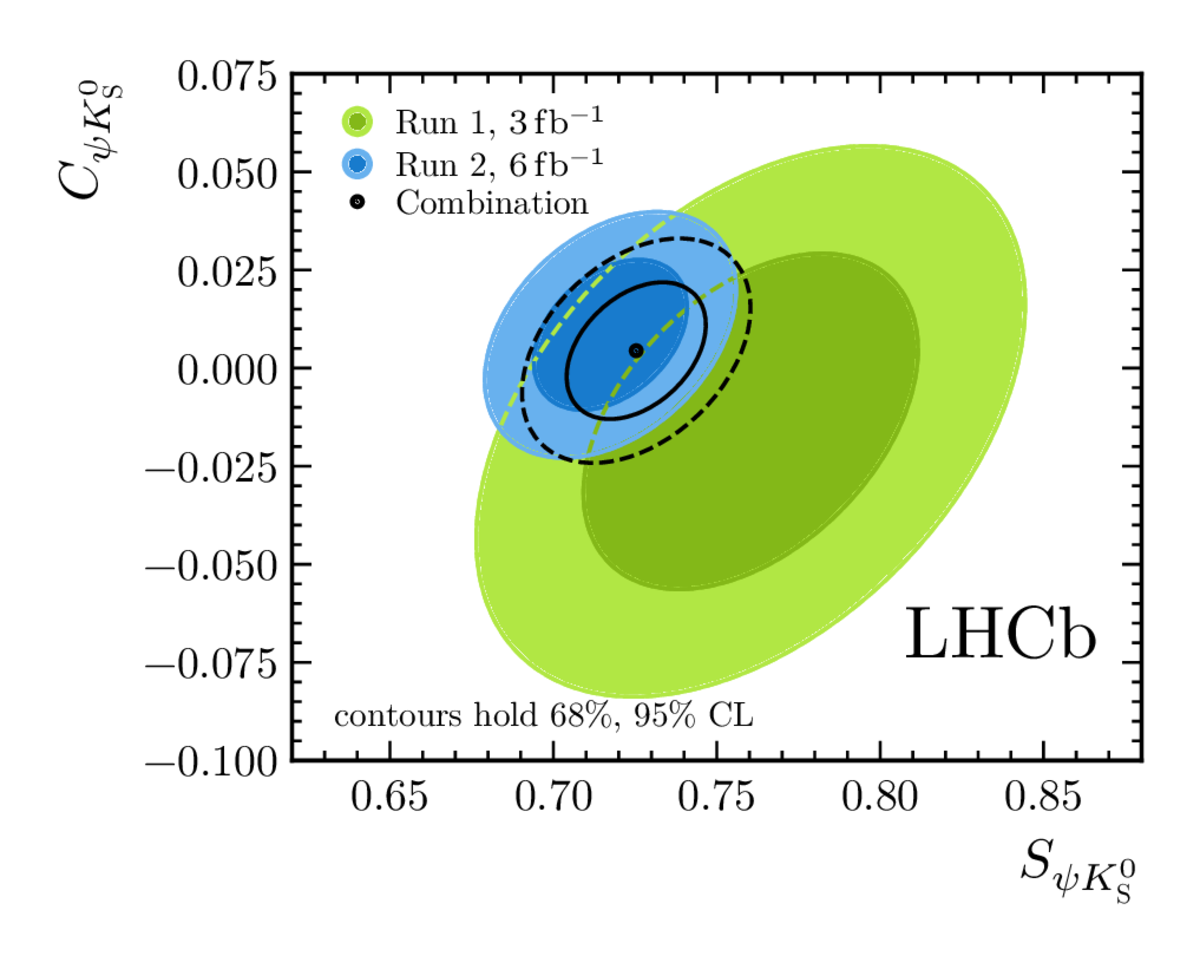}}
    \end{picture}
	\caption{
    {\em Left:} The invariant mass distribution of $B^0\rightarrow J/\psi K^0_S$ decays shows the small contribution of combinatorial background. 
    {\em Middle:} The time-dependent CP asymmetry is shown, where the amplitude of the oscillation quantifies $\sin 2\beta$, see Eq.~\ref{eq:ACP-B0-eigenstate2}. 
    {\em Right:} The contour values of $C_{\psi K_s}$ vs $S_{\psi K_s}$, indicating clear time dependent CP violation and very small direct CP violation in decay~\cite{LHCb-PAPER-2023-013} (with $S_{\psi K_s}=\sin 2\beta$, see Eq.~\ref{eq:B0-ACP-eigenstate}).}
	\label{fig:JpsiKs}
\end{figure}

The imaginary part of the relative amplitudes finally yields the phase difference,
$\Im\lambda_{J/\psi K_S^0} = 
-\sin \left\{ 2 \arg \left(\frac{V_{cb}V^*_{cd}}{V_{tb}V_{td}^*}                           \right) \right\} 
\equiv \sin 2\beta$ and
the CP-asymmetry of the decay $B^0\rightarrow J/\psi K^0_S $ 
is thus 
\begin{equation}
%A_{\mathrm{CP},\;B^0\rightarrow J/\psi K^0_S}(t) = \sin 2\beta \sin(\Delta mt).
A_{\mathrm{CP}}(t) = \sin 2\beta \sin(\Delta mt).
\label{eq:ACP-B0-eigenstate2}
\end{equation}
Fig.~\ref{fig:JpsiKs} summarizes LHCb's measurement of the Golden Decay, 
including a very clean event selection and precisely observed CP-asymmetry. 
The resulting values for the CP-coefficients are $S_{\psi K_s}=\sin2\beta=0.717 \pm 0.013 (stat) \pm 0.008 (syst)$ and $C_{\psi K_s}=0.008 \pm 0.012 (stat) \pm 0.003 (syst)$ \cite{LHCb-PAPER-2023-013}, 
where the latter is consistent with zero, in agreement with the SM.  

\clearpage
\subsubsection{CKM angle \texorpdfstring{$\phi_s$}{phis}}
\label{sec:phis}

\begin{figure}[!ht]
	\centering
   \begin{picture}(450,130)(0,0)
        \put(10,20){\includegraphics[scale=0.65]{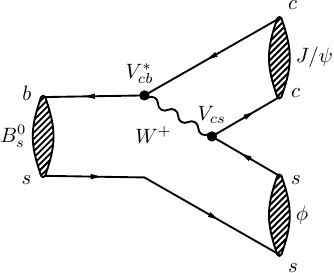}}  
        \put(120,20){\includegraphics[scale=0.65]{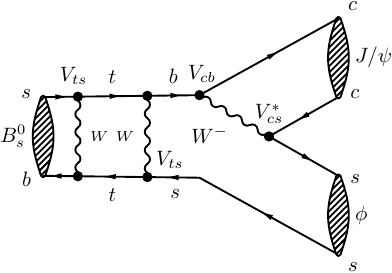} }  
        \put(270,10) {\includegraphics[scale=0.13]{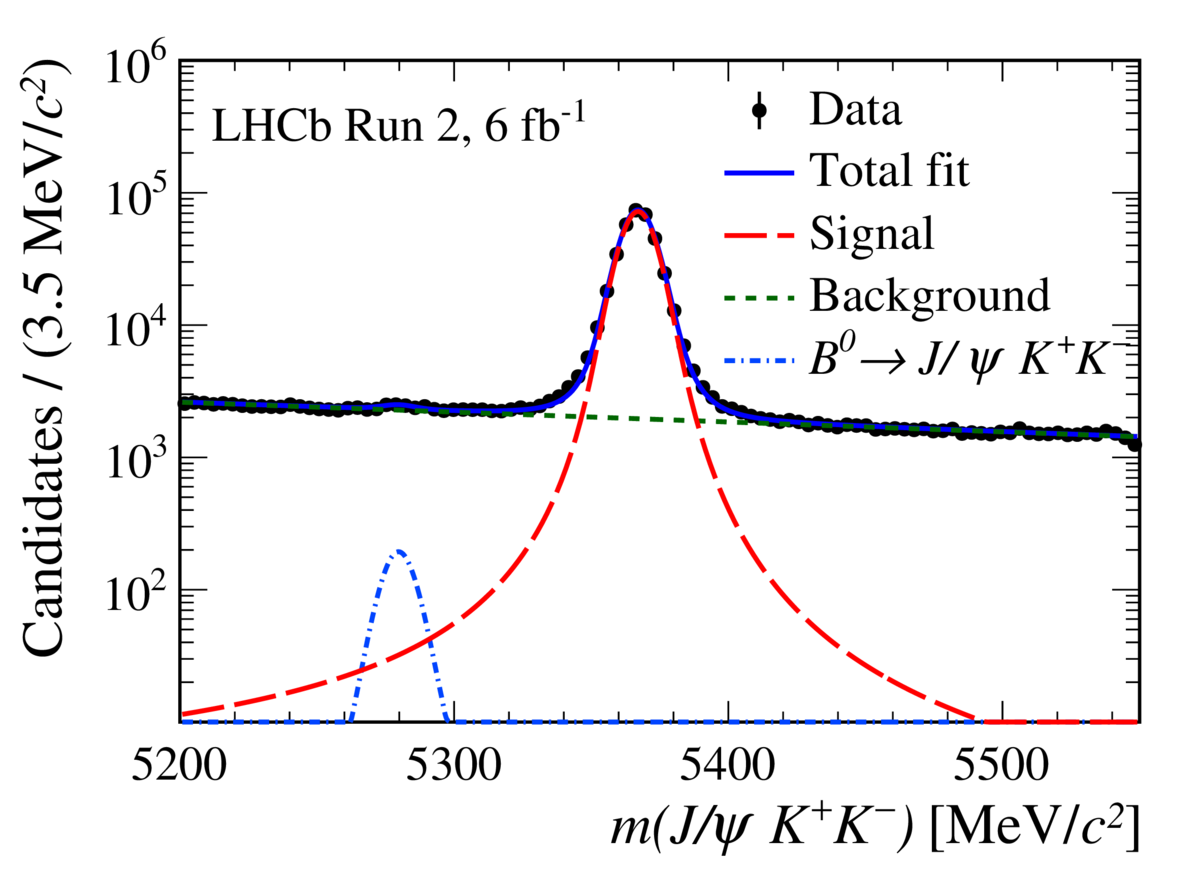}}
     \end{picture}
	\caption{{\em Left:} Diagrams of the decay $B_s^0 \to J/\psi \phi$ with and without mixing. {\em Right:} The invariant mass distribution of $B^0_s\to J/\psi K^+K^-$ candidates, dominated by $B^0_s\to J/\psi \phi$ decays~\cite{LHCb-PAPER-2023-016}. }
    \label{fig:bsjp}
\end{figure}

%\paragraph{The decay $B^0_s\rightarrow J/\psi \phi$}
The decay $B_s^0 \rightarrow J/\psi \phi$ is the $B_s^0$ analogue of the decay
$B^0\rightarrow J/\psi K^0_S$, where the spectator $d$-quark is replaced by an $s$-quark.
However, there are four major differences:
\begin{itemize}
\item  {\em ${V_{ts}}$ vs ${V_{td}}$.} 
  Since the spectator $d$-quark is replaced by an $s$-quark, the CKM-element
  responsible for the CP-asymmetry (in the Wolfenstein parameterization) is now 
  $V_{ts}$, instead of $V_{td}$, see Fig.~\ref{fig:bsjp}. 
  In contrast to $V_{td}$, the imaginary part of $V_{ts}$ is no longer of comparable
  size as the real part and the
  predicted CP asymmetry is therefore predicted to be small, $\arg (V_{ts})\sim \eta\lambda^2$.
\item  {\em No K-oscillations.}
  The CP eigenstate final state contains the mesons $J/\psi$ and $\phi$, and hence we do not need the extra $K$-oscillation step that was required in the $B^0$ system.
\item {\em ${\Delta \Gamma \neq 0}$.} 
  In contrast to the $B^0$ case, the $B_s^0$-system has a non-vanishing value of $\Delta\Gamma$.
  This is caused by the existence of the CP-eigenstate $D_s^{\pm(*)}D_s^{\mp(*)}$, 
  common to $B_s^0$ and $\bar{B}_s^0$,
  with a relatively large branching fraction of around 5\%. 
  Since this is a CP-eigenstate with eigenvalue $+1$ it is only accessible only for the CP-even eigenstate, leading to a lifetime difference for
  $B_{s,H}$ and $B_{s,L}$ with a predicted value of $\Delta\Gamma / \Gamma \sim 0.1$.
  Note that similar situation for the $B^0$ case does not occur, because the branching ratio 
  for $B^0 \rightarrow D^\pm D^\mp$ is Cabibbo suppressed, $A \sim |V_{cd}|$.
\item  {\em Vector-vector final state.}
  The final state now includes two vector-particles with spin-1. 
  The spin of the final state particles $J/\psi$ and $\phi$ can be pointing parallel, 
  perpendicular, or opposite, which need to be compensated by an orbital momentum, of $L=2$, 1 or 0, respectively, to obtain the spin of the initial state, $S_{B_s}=0$.
  As a result, and in contrast to $B^0\rightarrow J/\psi K^0_S$,  
  the final state is not a pure CP-eigenstate but an admixture, where the CP-eigenvalue of an event depends on the orbital momentum given by the factor $(-1)^L$
  in the wave function;
  $$
  \mathrm{CP} \left| J/\psi\phi \right>_L = (-1)^L \left| J/\psi\phi \right>_L
  $$
\end{itemize}
Due to the clean experimental signature, selection of $B_s^0\rightarrow J/\psi K^+K^-$ events is relatively straightforward and provides a very low background data sample, see Fig.~\ref{fig:bsjp}.
The fact that the predicted CP-asymmetry is small in the Standard Model, makes this decay
particularly sensitive to potential new particles.
Any deviation from the Standard Model value would be a sign of New Physics.
The asymmetry for the decay of the $B_s^0$-meson to the common final state $J/\psi\phi$
is given by 
\begin{equation}
A_{CP}(t)
=  \frac{\Gamma_{\bar{B}_s^0(t)\rightarrow J/\psi\phi} - \Gamma_{B_s^0(t)\rightarrow J/\psi\phi}}%
        {\Gamma_{\bar{B}_s^0(t)\rightarrow J/\psi\phi} + \Gamma_{B_s^0(t)\rightarrow J/\psi\phi}}
= \frac{\Im \lambda_{J/\psi\phi} \sin \Delta m t}%
                 {\cosh\frac{1}{2}\Delta\Gamma t + \Re\lambda_{J/\psi\phi}\sinh\frac{1}{2}\Delta\Gamma t}
\label{eq:ACP-bsjp}
\end{equation}
where
\begin{equation}
\lambda_{J/\psi\phi} 
= \left( \frac{q}{p}\Big)_{B_s^0} \; \left(\eta_{J/\psi\phi}\frac{\bar{A}_{J/\psi\phi}}{A_{J/\psi\phi}} \right)
= (-1)^L \; \Big(\frac{V_{tb}^*V_{ts}}{V_{tb}V_{ts}^*}\right) \; \left(\frac{V_{cb}V_{cs}^*}{V_{cb}^*V_{cs}}\right) 
\end{equation}
and
\begin{equation}
\Im \lambda_{J/\psi\phi}  =  (-1)^L \sin \left\{ 2 \arg \left(\frac{V_{cb}V^*_{cs}}{V_{tb}V_{ts}^*}   \right) \right\}  
                          =   (-1)^L \sin 2\beta_s
                          =  + \eta_{J/\psi\phi} \sin 2\beta_s 
\label{eq:sin2bs}
\end{equation}

\begin{figure}[!t]
	\centering
    \begin{picture}(450,120)(0,0)
        \put(-15,0){\includegraphics[scale=0.25]{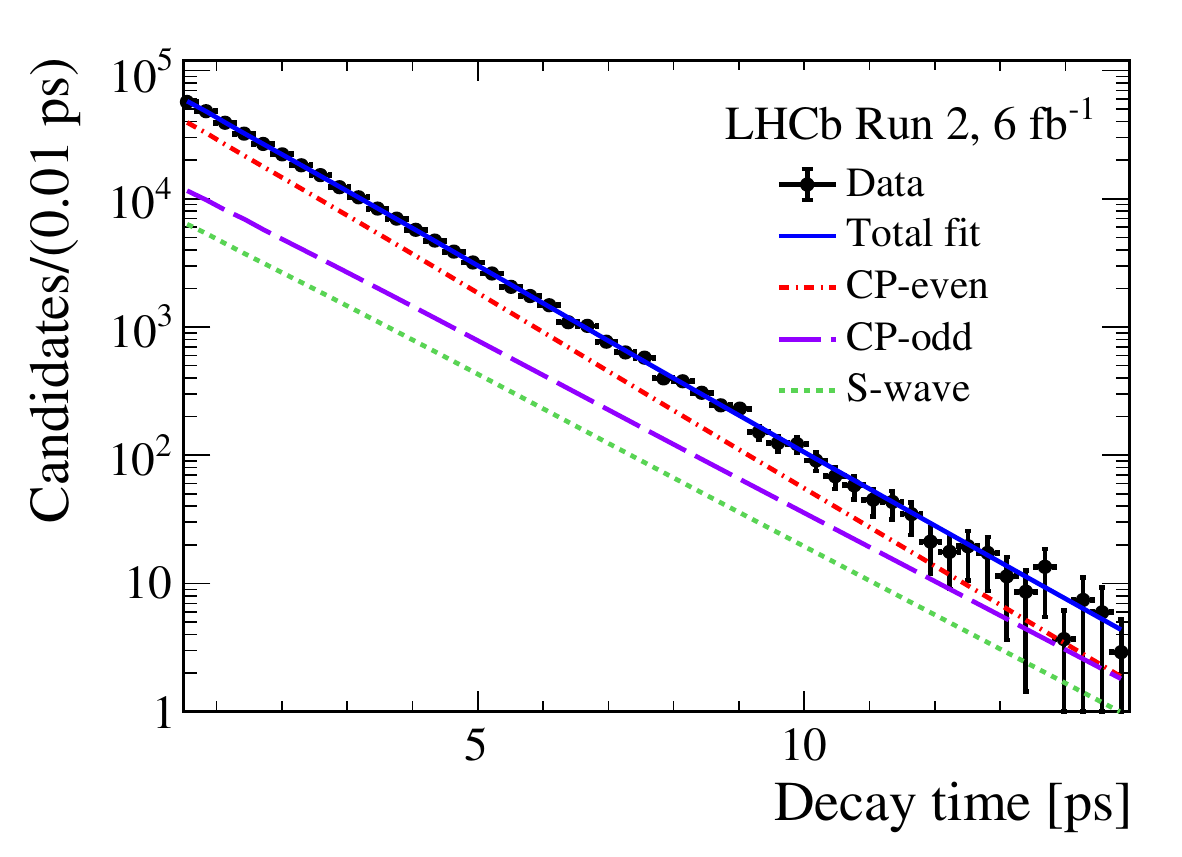}}
 	    \put(125,0){\includegraphics[scale=0.25]{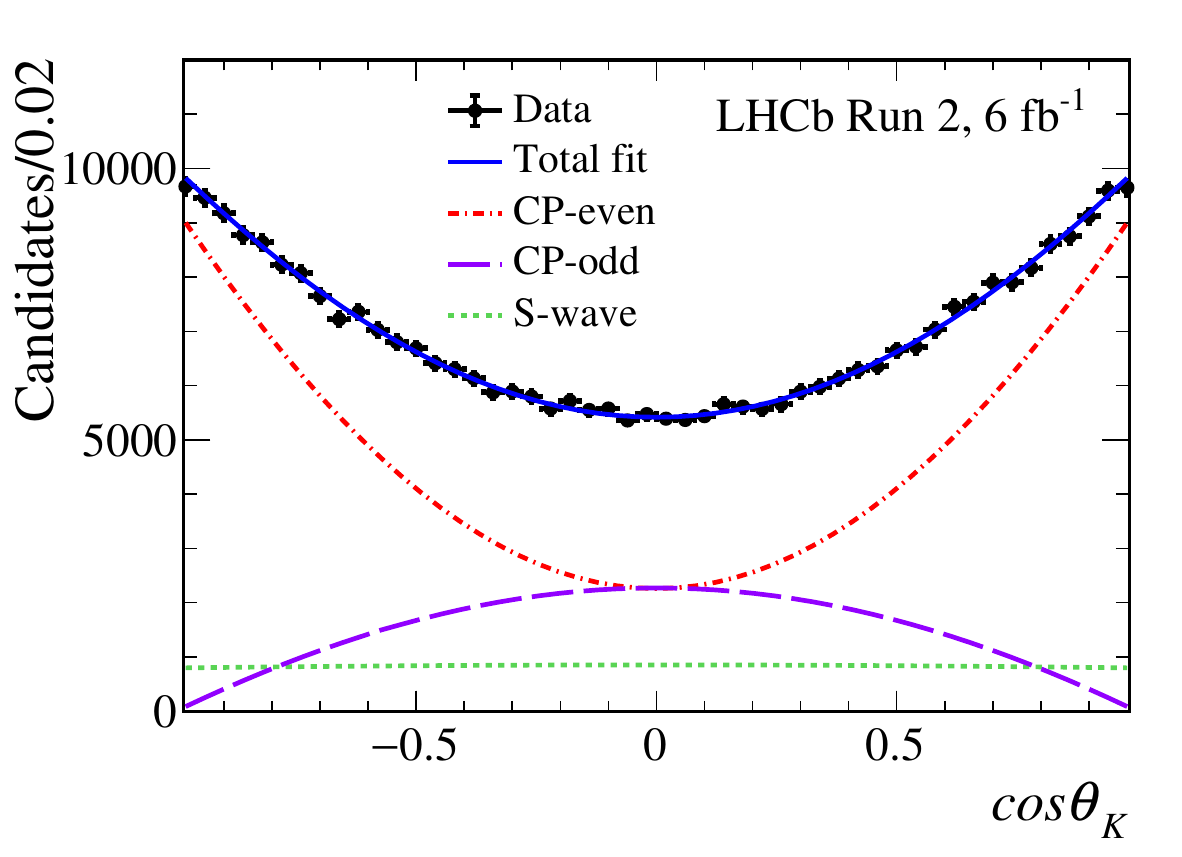}}
  	    %\put(265,130){\includegraphics[scale=0.25,angle=-90]
        \put(265,150){\includegraphics[width=6cm, height=6cm, angle=-90]{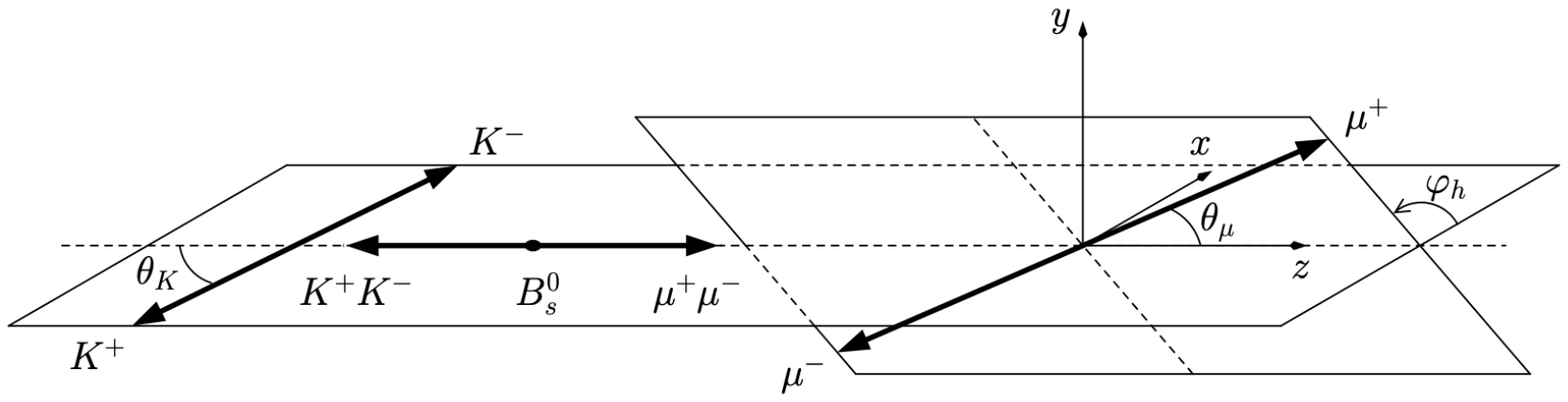}}
    \end{picture}
	\caption{{\em Left:} The decay time distribution of $B^0_s\to J/\psi \phi$ decays shows the two contributions from $B_{s,H}$ and $B_{s,L}$ decays with their characteristic lifetimes. {\em Middle:} The lifetime eigenstates are approximately equal to CP eigenstates (if $|q/p|=1$), which are statistically separated through the simultaneous fit to the decay angles. {\em Right:} Definition of the decay angles $\theta_K$, $\theta_\mu$ and $\phi_h$. \cite{LHCb-PAPER-2023-016}. }
	\label{fig:JpsPhi2}
\end{figure}

\begin{figure}[!ht]
	\centering
    \includegraphics[scale=0.16]{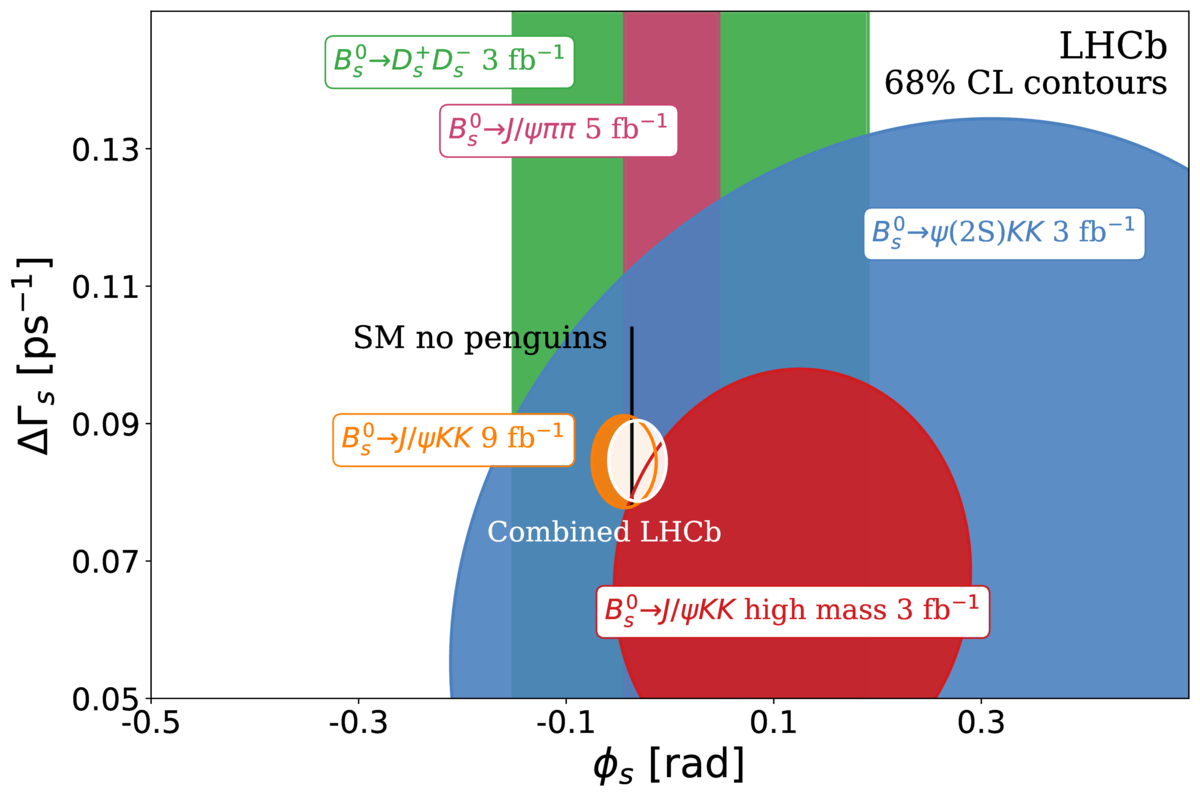}
	\caption{Compilation of LHCb measurements of 
    $\phi_s$ and $\Delta\Gamma_s$ with various final states~\cite{LHCb-PAPER-2023-016}.}
	\label{fig:JpsPhi}
\end{figure}

A relative minus sign
occurs with respect to the $B^0\to J/\psi K^0_S$ case due to the definition of $\beta$ and $\beta_s$: $\beta$ is defined with $V_{td}$
in the denominator, whereas $\beta_s$ has $V_{ts}$ in the numerator.
A complication arises from the above mentioned vector-vector final state.
The contributions from the terms with different orbital momentum, $A_\parallel$, $A_\perp$ and $A_0$,
for values of the orbital momentum of 2, 1 and 0, respectively, need to be
disentangled statistically by examining the angular distributions of the final state particles,
$J/\psi \rightarrow \mu^+\mu^-$ and $\phi\rightarrow K^+ K^-$.
To seperate these CP eigenstates an angular analysis is performed, where three decay angles ($\theta_K$, $\theta_\mu$ and $\phi_h$) are defined as sketched in Fig.~\ref{fig:JpsPhi2}. 
The individual contributions of the CP-even and CP-odd waves are fitted together with an S-wave contribution from non-resonant $K^+K^-$ production, taking into account the angular efficiencies, resulting in a measurement $\phi_s=-0.039 \pm 0.022\: (stat) \pm 0.006\: (syst)\; \mathrm{rad}$, together with a determination of $\Delta\Gamma_s=0.0845 \pm 0.0044\:(stat) \pm 0.0024\:(syst)\; \mathrm{ps^{-1}}$. 
Any $B_s^0$ decay to a CP-eigenstate is sensitive to the interference with and without mixing, and hence can be used to measure $\phi_s$.
Figure~\ref{fig:JpsPhi} shows the measurement contours obtained by LHCb including
the measurements from the decays to the CP-even eigenstate $D_s^+D_s^-$ and
the CP-odd eigenstate $J/\psi\pi^+\pi^-$~\cite{LHCb-PAPER-2023-016}. The $B_s^0$ decays to
the CP-eigenstates are approximately pure lifetime (mass) eigenstates, and hence they
cannot be used to separately measure the lifetime difference $\Delta\Gamma_s$.

%\paragraph{The decay $Bs\rightarrow J/\psi \phi$}
%Most precise measurement. Angular analysis.

%\begin{figure}[!ht]
%	\centering
%	\includegraphics[height=8cm,angle=-90]{Figures/CPViolation/JpsiPhi_HelicityFormalism.pdf}
%	\caption{Helicity Formalism  HFLAV.  LHCB-PAPER-2013-055  }
%	\label{fig:JpsPhi_Helicity}
%\end{figure}

%\vspace*{0.5cm}

\begin{figure}[!ht]
	\centering
%trim = <left> <bottom> <right> <top>
    \begin{picture}(450,130)(0,0)
    \put(10,20){\includegraphics[scale=0.6]{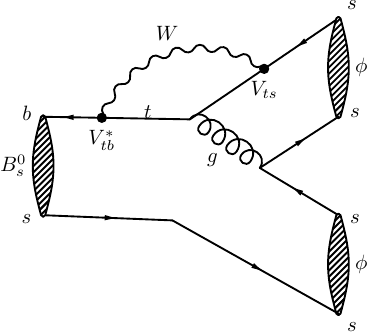}}
    \put(130,20){\includegraphics[scale=0.6]{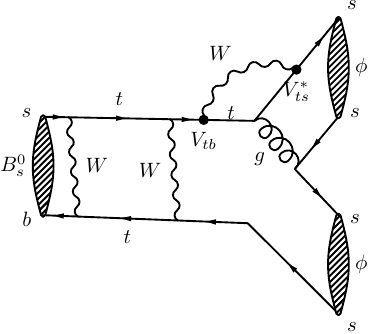}}
    \put(260,10){\includegraphics[trim=9cm 6cm 0.1cm 0.1cm,clip,scale=0.28]{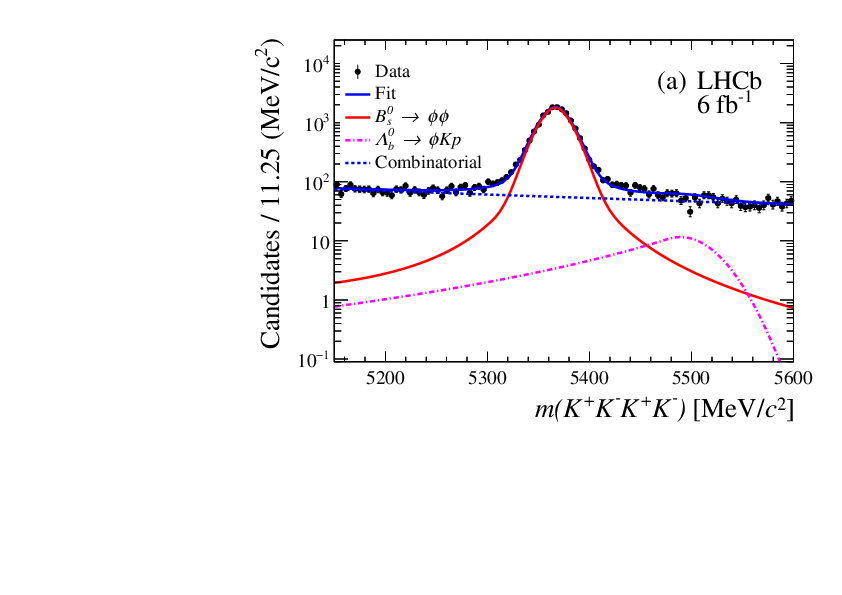}}
%    \put(290,0){\includegraphics[scale=0.32]{Figures/CPViolation/BsPhiPhi_phis.pdf}} 
	\end{picture}
 \caption{{\em Left:} Diagrams of the decay $B^0_s\to\phi\phi$. {\em Right:} The invariant mass distribution of $B^0_s\to\phi\phi$ candidates~\cite{LHCb-PAPER-2023-001}. 
 %{\em Right:} Overview of the measurements of the angle $\phi_s^{ccs}$, which vanishes in the Standard Model 
    }
	\label{fig:phiphi}
\end{figure}

%\paragraph{The decay $B^0_s\rightarrow\phi\phi$}
The decay $B^0_s\to\phi\phi$ is also a CP-eigenstate accessible for the mixed and unmixed $B_s^0$ and probes the weak mixing angle $\phi_s$ in a similar way as  $B^0_s\to J/\psi \phi$.
However, as shown in Fig.~\ref{fig:phiphi}, the decay 
$B^0_s\to\phi\phi$ proceeds through a gluonic penguin diagram with the $b\to s\bar{s}s$ transition, which adds an additional relative weak phase for the CP-conjugate process that compensates the relative weak phase in the box diagram, predicting exactly zero CP asymmetry for this decay in the SM.  
The value $\phi^{sss}_s = -0.042 \pm 0.075 \pm 0.009$~rad is measured~\cite{LHCb-PAPER-2023-001}, 
where any non-vanishing measured CP violation in  $B^0_s\to\phi\phi$ would be a sign of new physics. 
The invariant mass distribution of the selected events in Run-1 and Run-2 is shown in Fig.~\ref{fig:phiphi}.
The LHCb trigger strategy applied from 2024 onwards is particularly advantageous for fully hadronic final states, allowing to access decays with lower transverse momenta, leading to an increased sensitivity for $\phi_s^{sss}$ with the data from Run 3 and Run 4.

\subsubsection{CKM angle \texorpdfstring{$\gamma$}{gamma} with
\texorpdfstring{$B^0_s\rightarrow D_s^\mp K^\pm$}{Bs0->DsK} decays}
\begin{figure}[!ht]
	\centering
    \begin{picture}(450,150)(0,0)
        \put(0,10){\includegraphics[scale=0.57]{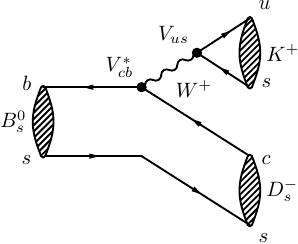}} 
        \put(60,60){\includegraphics[scale=0.57]{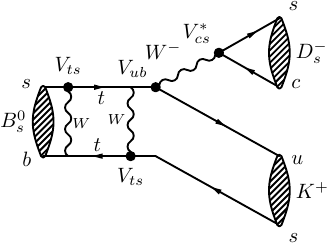}}
        \put(155,5){\includegraphics[scale=0.18]{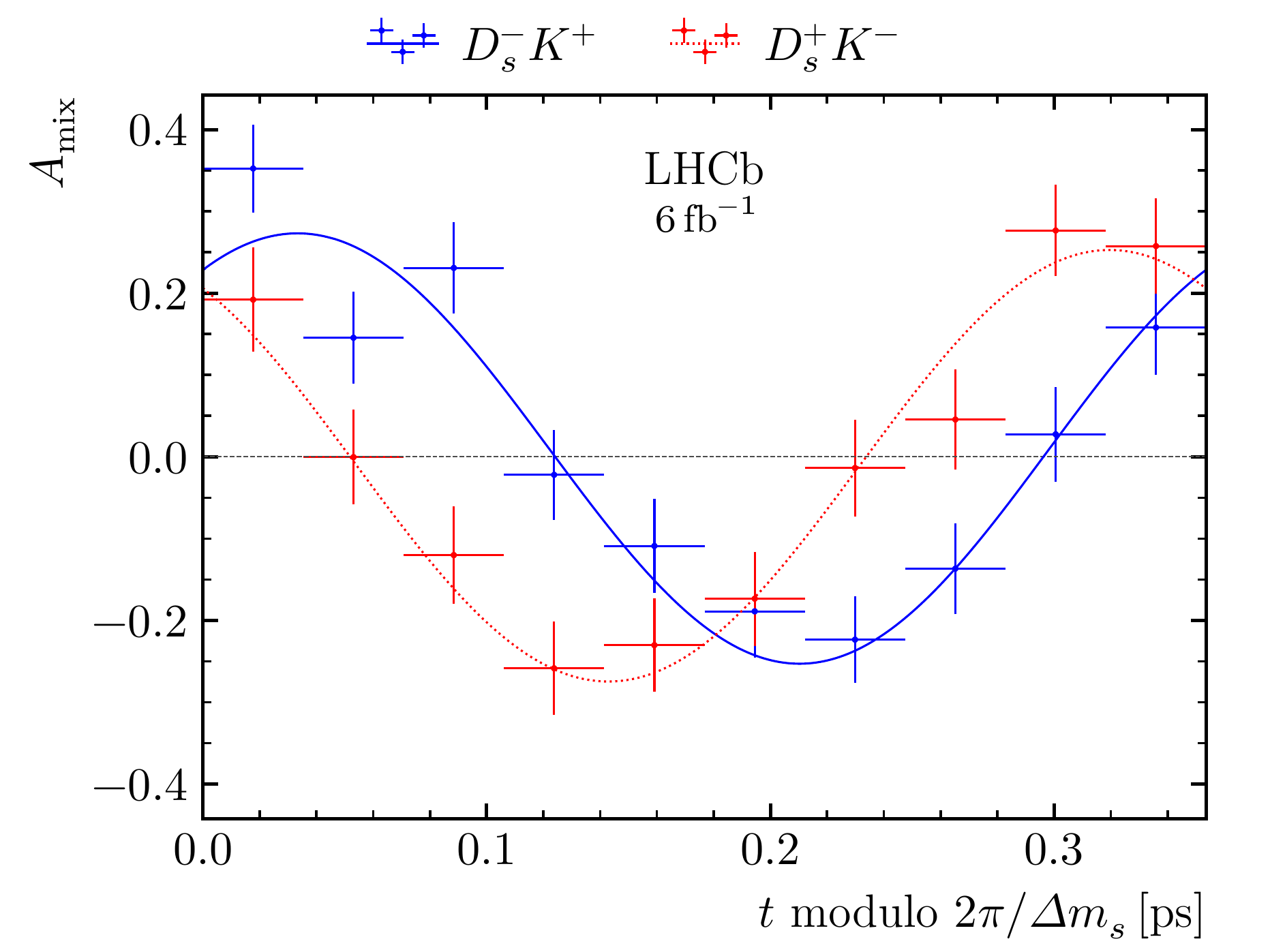}} 
        \put(315,5){\includegraphics[scale=0.35]{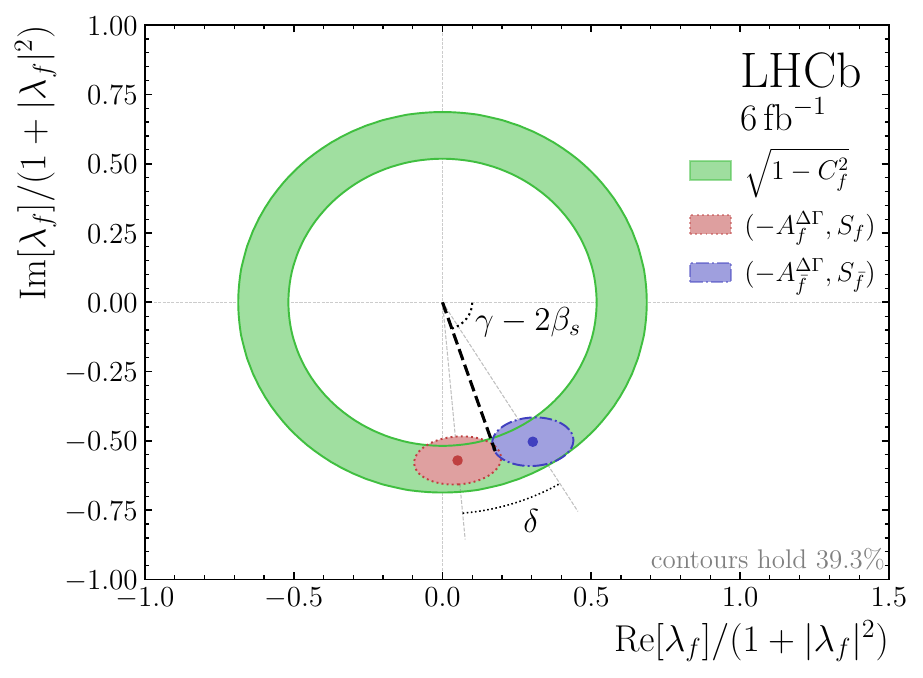}} 
%        \put(150,5){\includegraphics[scale=0.45]{Figures/CPViolation/Fig13-DsK.pdf}} 
%         \put(315,5){\includegraphics[scale=0.17]{Figures/CPViolation/BsDsK_Asym.pdf}} 
         \end{picture}
	\caption{{\em Left:} The decay diagrams for the decay $B_s\rightarrow D_s K$. {\em Middle:} The measured asymmetry of $B_s^0$ and $\bar{B}_s^0$ decays to the $D_s^+ K^-$
    and $D_s^- K^+$ separately~\cite{LHCb-PAPER-2024-020}. 
    {\em Right:} The relative amplitude $\lambda$ can be illustrated in the complex plane. The average phase
    of $\lambda_{D_s^+K^-}$ and  $\lambda_{D_s^-K^+}$ corresponds to the value of $\gamma$,
    whereas the difference quantifies the strong phase difference $\delta$.    
    }
	\label{fig:Bs2DsK}
\end{figure}

The CKM angle $\gamma$ is the relative phase between $b\rightarrow u$ and $b\rightarrow c$ couplings, and as presented in section \ref{sec:directCP}, is measured via direct CP violation. However, it is also measured in interference of decays with and without mixing, using the fact that both a $B_s^0$ and $\bar{B}_s^0$ can decay into the same final states, even though $D_s^\mp K^\pm$ final state is not a CP eigenstate.
The interfering amplitudes $B_s^0\rightarrow D_s^- K^+$ and 
$B_s^0\rightarrow \bar{B}_s^0\rightarrow D_s^- K^+$ have different amplitudes~\cite{DeBruyn:2012jp}, but nonetheless give rise to a time-dependent asymmetry in a similar way as in section \ref{sec:BsMixing}. 
%In order to disentangle the relative CP-violating weak phase 
%($\gamma-2\beta_s$) 
%$\phi_w$ from the CP-conserving strong phase also the pair $B_s^0\rightarrow D_s^+ K^-$ and $B_s^0\rightarrow \bar{B}_s^0\rightarrow D_s^+ K^-$ is studied.
%\textcolor{purple}{As shown in Fig.~\ref{fig:Bs2DsK} the two diagrams differ by a weak phase resulting from that of the box diagram ($\phi_s=2\arg(V_{ts})\equiv 2\beta_s$) and the difference in the decay diagrams ($\arg(V_{ub}$).} 
%
As shown in Fig.~\ref{fig:Bs2DsK}, the amplitude of the decay
$B_s^0 \rightarrow D_s^- K^+$ proceeds proportional to $A_f \sim V_{cb}^*V_{us}$
and the decay $\bar{B}_s^0 \rightarrow D_s^- K^+$ proceeds proportional to
$\bar{A}_f \sim V_{ub}V_{cs}^*$. 
Overall the two paths therefore differ by a weak phase $\phi_w=\gamma-2\beta_s$ including contributions of the decay diagrams ($\gamma = \arg(V_{ub})$) and that of the box diagram ($\phi_s=-2\arg(V_{ts})=-2\beta_s$ in the SM).

Since both the $B_s^0$-decay and the $\bar{B}_s^0$-decay are equally Cabibbo suppressed, they give rise to a sizeable interference and hence large CP asymmetry.
%\textcolor{orange}{Both amplitudes will not only differ by their magnitude, but also by a  relative phase~$\gamma$, originating from the relative phase between $V_{ub}$ and $V_{cb}$. $\Rightarrow$ included above.}
%\begin{equation}
%\frac{A_{D_s^- K^+}}{\bar{A}_{D_s^- K^+}}=\frac{|A_{D_s^- K^+}|}{|\bar{A}_{D_s^- K^+}|}e^{-i\gamma}.
%\end{equation}
% 
Furthermore, the mesons in the transitions 
$B_s^0 \rightarrow D_s^- K^+$
and $\bar{B}_s^0 \rightarrow D_s^- K^+$ are produced differently, and the strong phase $\delta_s$ is introduced, accounting for strong interactions in the final state.
%\begin{equation}
%\frac{A_{D_s^- K^+}}{\bar{A}_{D_s^- K^+}} = 
%\frac{|A_{D_s^- K^+}|}{|\bar{A}_{D_s^- K^+}|}e^{i(\delta_s-\gamma)}.
%\end{equation}
%
Combining the above leads to the following expression:
\begin{equation}
\lambda_{D_s^- K^+} 
\equiv \left(\frac{q}{p}\right)_{B_s^0} \left( \frac{\bar{A}_{D_s^- K^+}}{A_{D_s^- K^+}} \right)
= \left|\frac{V_{tb}^*V_{ts}}{V_{tb}V_{ts}^*}\right| \; \left|\frac{V_{ub}V_{cs}^*}{V_{cb}^*V_{us}}\right|
\left| \frac{A_2}{A_1}\right| e^{i(\delta_s-(\gamma-2\beta_s))} 
\end{equation}
%\textcolor{purple}{Needs few checks on signs!}\\
In order to disentangle the relative CP-violating weak phase 
($\gamma-2\beta_s$) from the CP-conserving strong phase $\delta_s$, also the asymmetry of $B_s^0\rightarrow D_s^+ K^-$ and $\bar{B}_s^0\rightarrow D_s^+ K^-$
are studied, leading to $\lambda_{D_s^+ K^-}$ with an opposite sign for the weak phase.
The measurement of both mixing asymmetries is shown in Fig.~\ref{fig:Bs2DsK}. The fact that they differ is a measure for the CP violation parameter $\lambda_{f}$, the phase of which is shown in the right panel of the figure. 
The observed CP asymmetry using 6 fb$^{-1}$ of run-2 \cite{LHCb-PAPER-2024-020} is combined with that from 3 fb$^{-1}$~\cite{LHCB-PAPER-2017-047} of run-1, and combined with the value of the mixing phase $\phi_s$, to yield $\gamma=\left(81^{+12}_{-11}\right)$~\cite{LHCb-PAPER-2024-020}.

%\paragraph{$B_s\rightarrow K^+K^-$} Skip? NT
%\begin{figure}[h!]
%    \centering
%     \includegraphics[width=5cm,height=5cm]{Figures/CPViolation/BsKK_Mass.pdf} 
%     \includegraphics[width=5cm,height=5cm]{Figures/CPViolation/BsKK_asym.pdf} 
%     \caption{$B_s\rightarrow K K$ LHCb-PAPER-2020-029    arXiv:2012.05319    10.1007/JHEP03(2021)075 }
%     \label{fig:BsKK}
%\end{figure}

%Skip DsK ?
%\end{comment}

%\clearpage
\subsection{The Unitarity Triangle}
\begin{figure}[!ht]
    \centering
     \includegraphics[width=9cm]{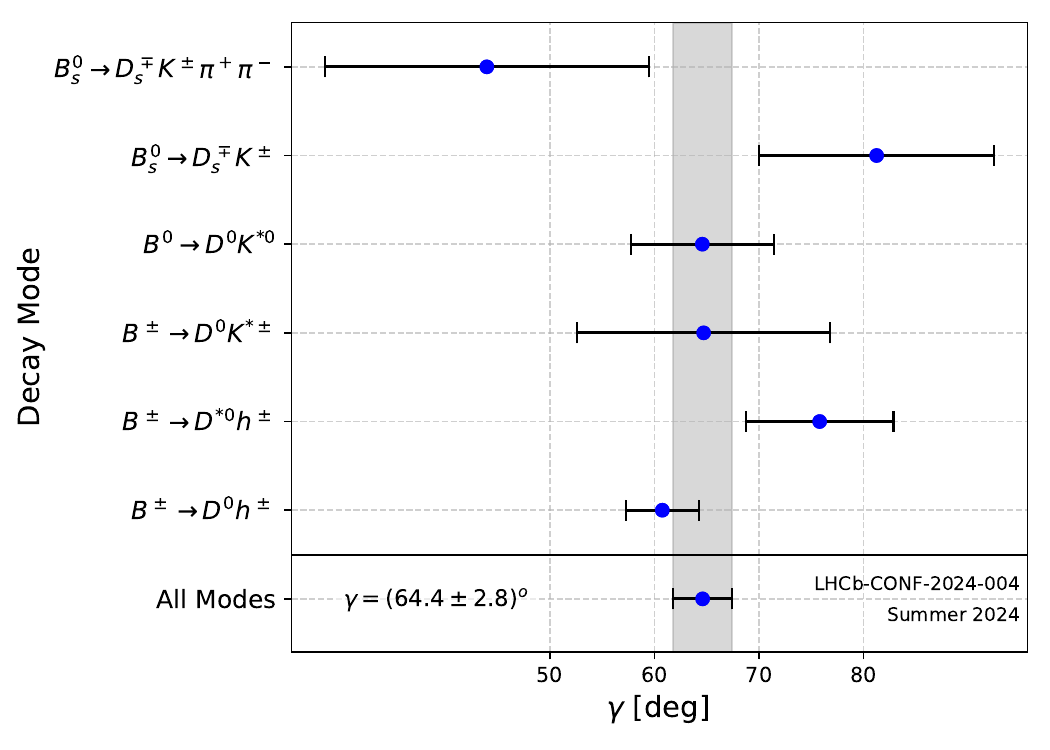} 
     \includegraphics[width=7cm]{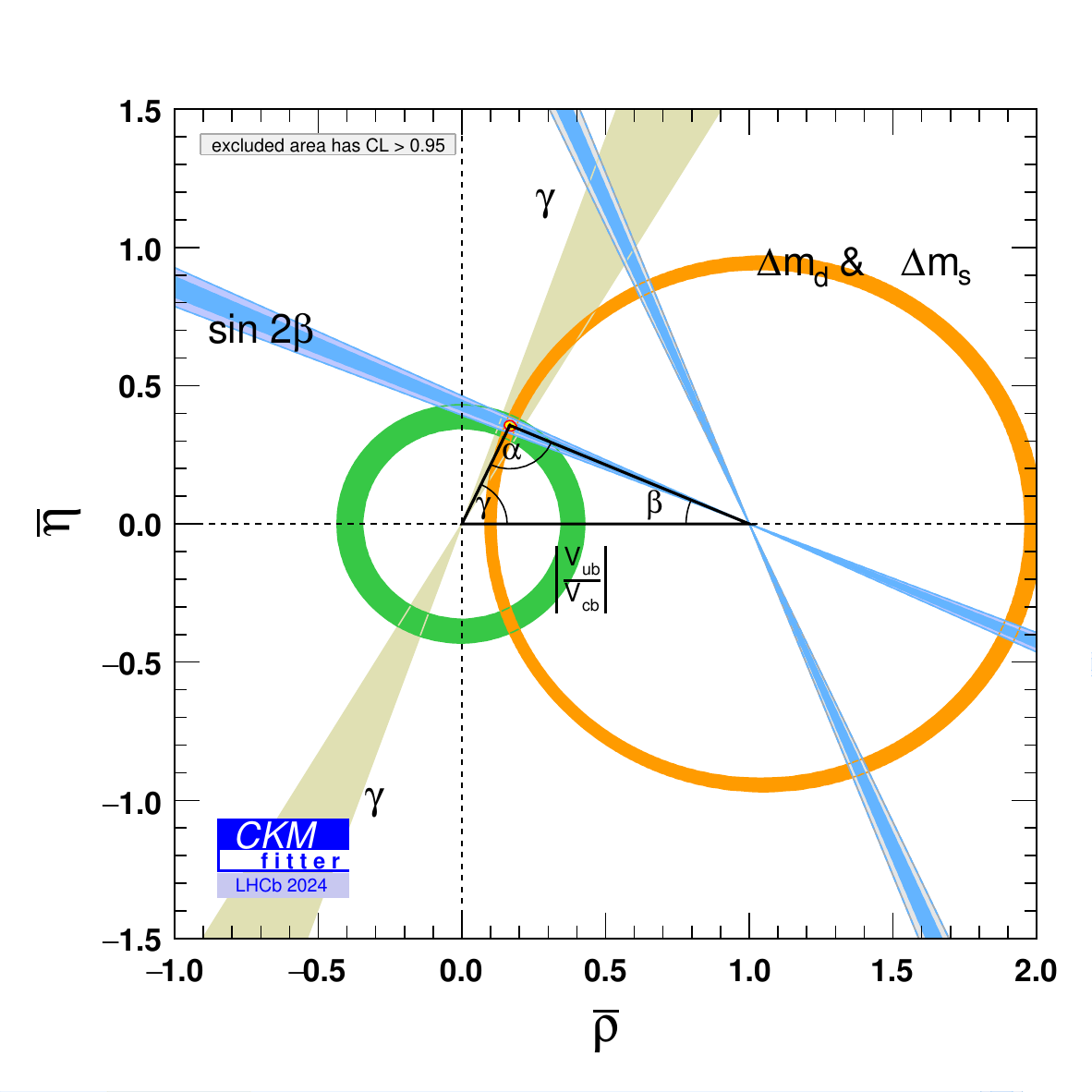} 
     \caption{{\em Left:} The various determinations of the CKM angle $\gamma$ all give consistent result, with an average of $\gamma=(64.6\pm 2.8)^o$ (adapted from~\cite{LHCb-CONF-2024-004}). {\em Right:} Within the current experimental precision, the measurements of $\gamma$, $\sin 2\beta$, $|V_{ub}|$ and $\Delta m_s$ by LHCb give a consistent picture of the CKM paradigm
     (figure courtesy from CKMfitter group~\cite{Charles:2004jd}).}
     \label{fig:CKM}
\end{figure}

%---------------------------------------
In the previous sections LHCb's measurements of various 
CKM angles have been discussed, corresponding to CP violating weak phases. The internal consistency of the CKM paradigm is complemented with the measurement of the $B_{(s)}^0$ oscillation frequency $\Delta m_d$ and $\Delta m_s$, which gives similar experimental constraints on the apex of the unitarity triangle as the CKM angle $\gamma$. LHCb's measurement of $\gamma$ from Run-1 and Run-2 are summarized in the left panel of Fig.~\ref{fig:CKM} and result in a value of $\gamma=64.4^o \pm 2.8^o$~\cite{LHCb-CONF-2024-004}.
Furthermore, the constraints from the measurement of $|V_{ub}|$ can be compared to the information from the CKM angle $\beta$.
The magnitude of $|V_{ub}|$ can be probed by measuring the decay rate of semileptonic decays such as
$B^0\to \pi^-\mu^+\nu$, $B_s^0\to K^-\mu^+\nu$~\cite{LHCb-PAPER-2020-038} or $\Lambda_b^0\to p\mu^-\nu$~\cite{LHCb-PAPER-2015-013}, where the latter two decays are primarily accessible at LHCb.
The right side panel of Fig.~\ref{fig:CKM} shows the bounds on the unitarity triangle determined with LHCb measurements.

The results from the LHCb experiment alone give an impressively
consistent picture of the CKM-paradigm, 
even without considering the measurements from other experiments.
Nevertheless, a continuous experimental and theoretical effort
is required to understand the numerous tensions that still exist today,
such as:
\begin{itemize}
\item the inclusive vs exclusive determination of $|V_{ub}|$ and $|V_{cb}|$~\cite{Martinelli:2023fwm,Gambino:2020jvv};
\item the $K - \pi$ puzzle on the relative size and phase of $B\to K\pi$ decays~\cite{Berthiaume:2023kmp,Fleischer:2018bld};
\item the Cabibbo angle anomaly on the unitarity of $|V_{ud}|^2+|V_{us}|^2$~\cite{Crivellin:2022rhw,Grossman:2019bzp};
\item the $B\to Dh$ branching ratios that are low compared to expectations~\cite{Fleischer:2010ca,Bordone:2020gao,Piscopo:2023opf}.
\end{itemize}
In the next chapter the status of rare decays is discussed, and its
sensitivity to effects beyond the Standard Model.

%\begin{figure}[h!]
%    \centering
%     \includegraphics[width=7cm]{Figures/CPViolation/HFLAV_gamma_Moriond2024.pdf} 
%     \includegraphics[width=7cm]{Figures/CPViolation/HFLAV_CKM_moriond2024.pdf} 
%     \caption{HFLAV Moriond 2024   gamma and CKM  }
%     \label{fig:HFLAV}
%\end{figure}

\clearpage

\section{Rare Decays}\label{secRareDecays}
The class of rare $B$ decays are characterized by low branching ratios, and include the suppressed flavour-changing neutral-current (FCNC) electroweak penguin (EWP) decays of the fundamental process $b\to s\ell^+\ell^{\prime -}$.
These decays can be divided into the semi-hadronic $B \to K\ell^+\ell^-$ decays,
the very rare fully leptonic $B \to \ell^+\ell^-$ decays  and the forbidden 
lepton-flavour violating decays $B \to X \ell^+\ell^{\prime -}$.
In the Heavy-Quark Effective Theory (HQET) framework for $B$-decays the amplitude of a $B$-decay is written as ~\cite{Buras:1997fb}:
\begin{equation}
 {\cal A}\left(B\rightarrow f \right) = \left<f\left|{\cal H}_{\text{eff}}\right|B\right>=\frac{G_F}{\sqrt{2}} \sum_i {\cal C}_i(\mu)\: \left<f\left|{\cal O}_i(\mu)\right|B\right>
 \label{eq:Bdecay}
\end{equation}
where the sum includes an expansion over non-pertubative hadronic matrix elements with operators ${\cal O}_i$ and their perturbative Wilson coefficients ${\cal C}_i$, calculated at renormalisation scale $\mu$, which is typically taken to be the mass of the decaying hadron.
With rare decays the goal is to test whether the Wilson expansion as calculated from the SM, provides a full prediction of these decays or whether there are perhaps deviations with a signature of NP.

\subsection{Semi-hadronic \texorpdfstring{$B \to K\ell^+\ell^-$}{B->Kll} decays}

\begin{figure}[!ht]
    \centering    
    \begin{picture}(450,100)(0,0)
        \put(0,35){\includegraphics[scale=0.6]{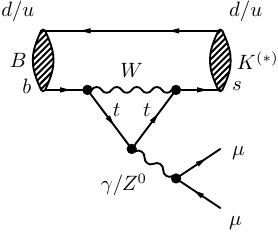}}
        \put(80,40){\includegraphics[scale=0.61]{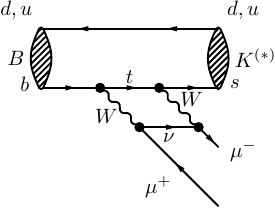}}
        \put(160,0){\includegraphics[scale=0.2]{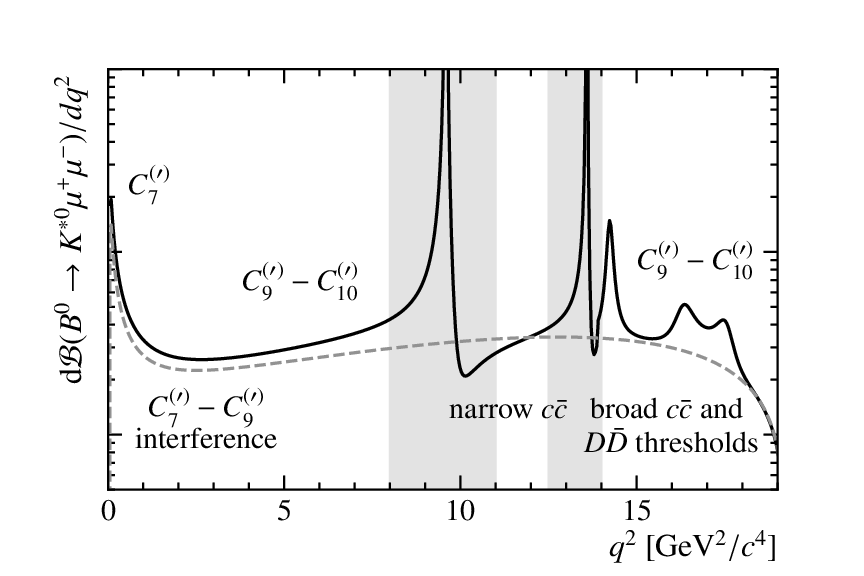}}
        \put(320,0){\includegraphics[scale=0.16]{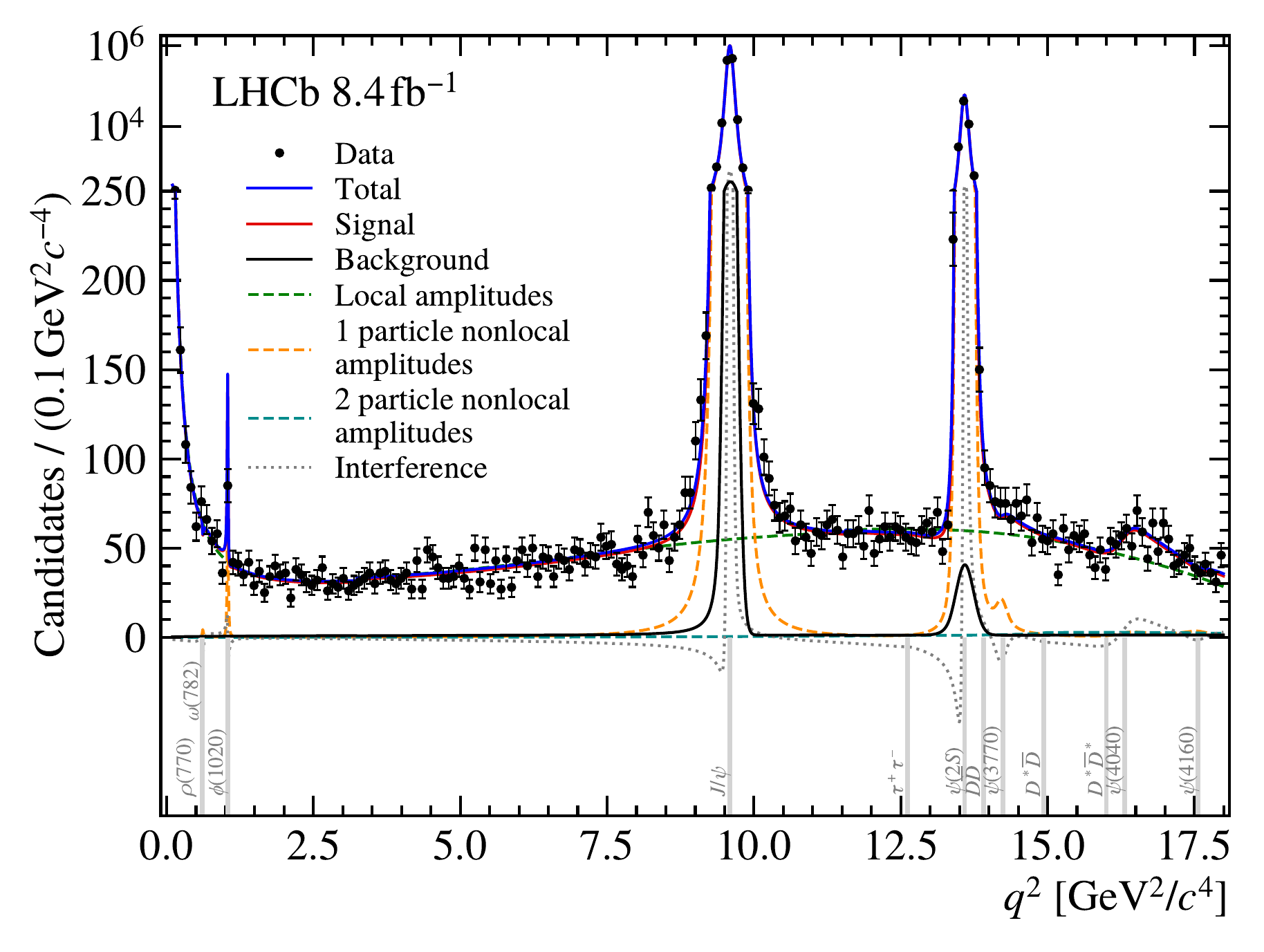}}
    \end{picture}
     \caption{{\em Left:} The diagrams show the decay topologies of the semi-hadronic $b\to s\ell\ell$ processes for $B^0$ or $B^-$ mesons decaying to $K$ or $K^*$ mesons. {\em Middle:} A sketch of the decay rate of $B^0 \to K^{*0}\mu^+\mu^-$ as a function of the di-lepton invariant mass $q^2$~\cite{LHCb-PAPER-2023-033}. 
     {\em Right:} The measured decay rate $q^2$. The photon pole at $q^2\approx 0$ and the charm resonance contributions around $q^2\approx 9 (13.5)$~GeV$^2$ are clearly visible~\cite{LHCb-PAPER-2024-011}. }
     \label{fig:BKStarMuMu}
\end{figure}

The FCNC EWP process $b\to s \ell\ell$, referred to here as semi-{\em hadronic} to distinguish it from the unsuppressed semi-leptonic $b\rightarrow c l \nu$ decays, is studied across a variety of channels. In these the initial $b$-hadron can be either a $B^0$, $B^+$, $B_s^0$ or $\Lambda_b^0$ hadron, depending
on the accompanying spectator quarks, and on the spin of the strange hadron ($K$ or $K^*$).
Examples of the Feynman diagrams are given on the left side of Fig.~\ref{fig:BKStarMuMu}.

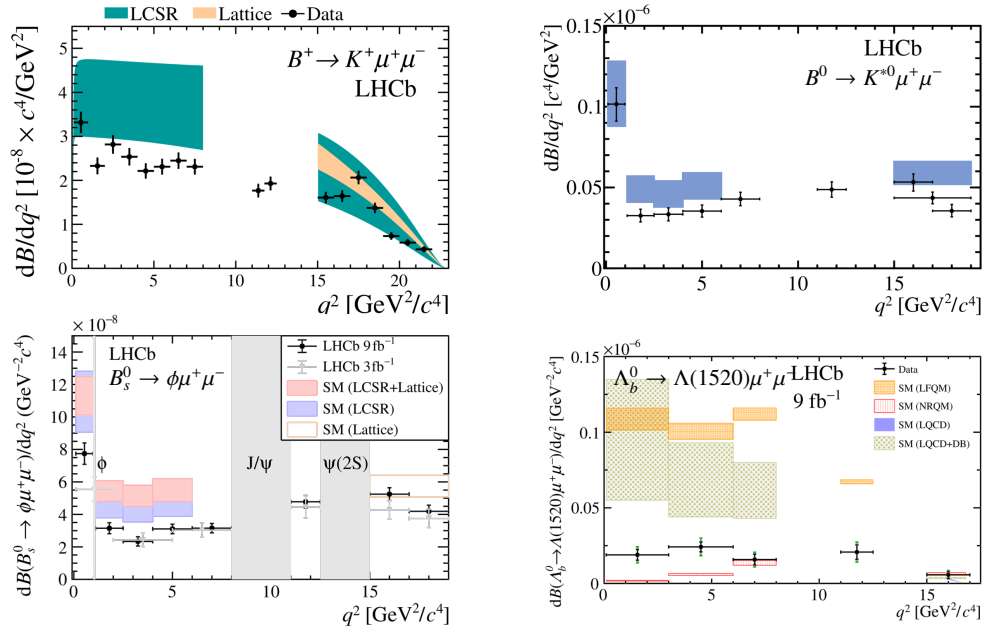
\begin{figure}[!hb]
    \centering   
    \begin{picture}(400,210)(0,0)
        \put(0,117){\includegraphics[scale=0.145]{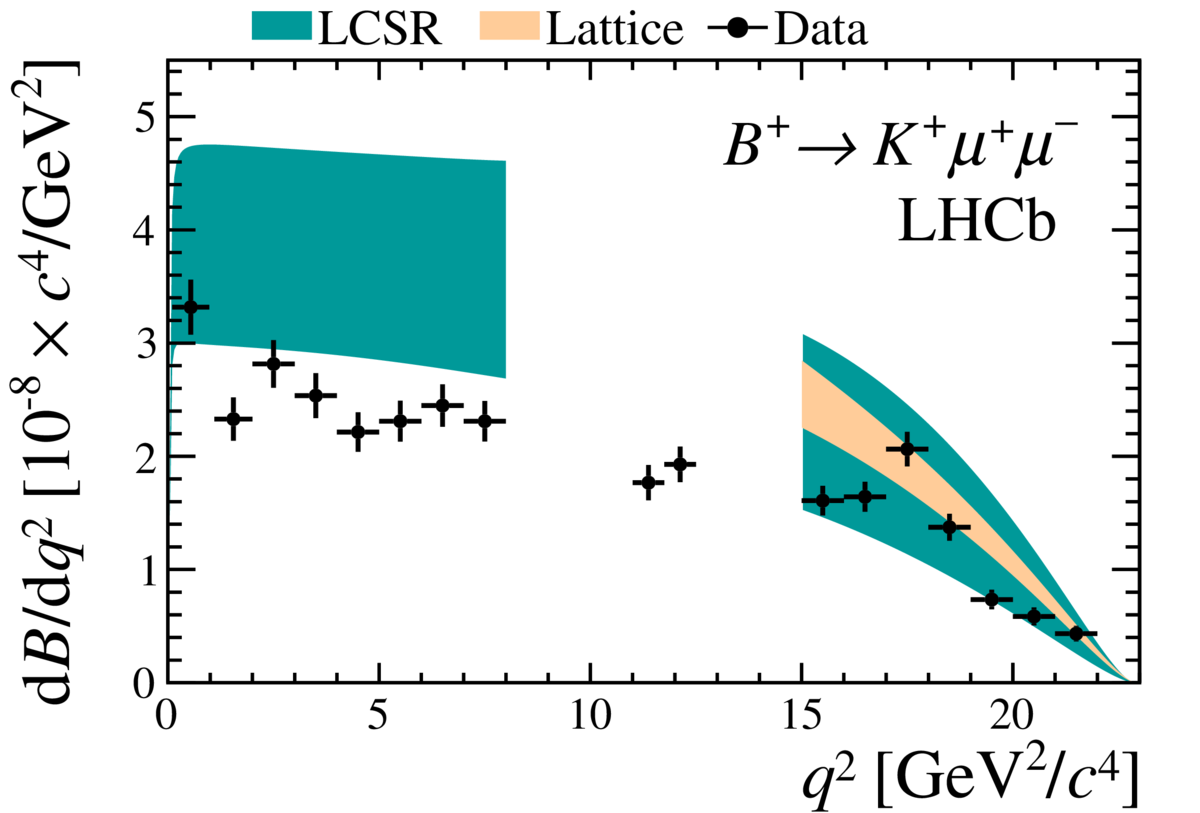}}
        \put(200,117){\includegraphics[scale=0.145]{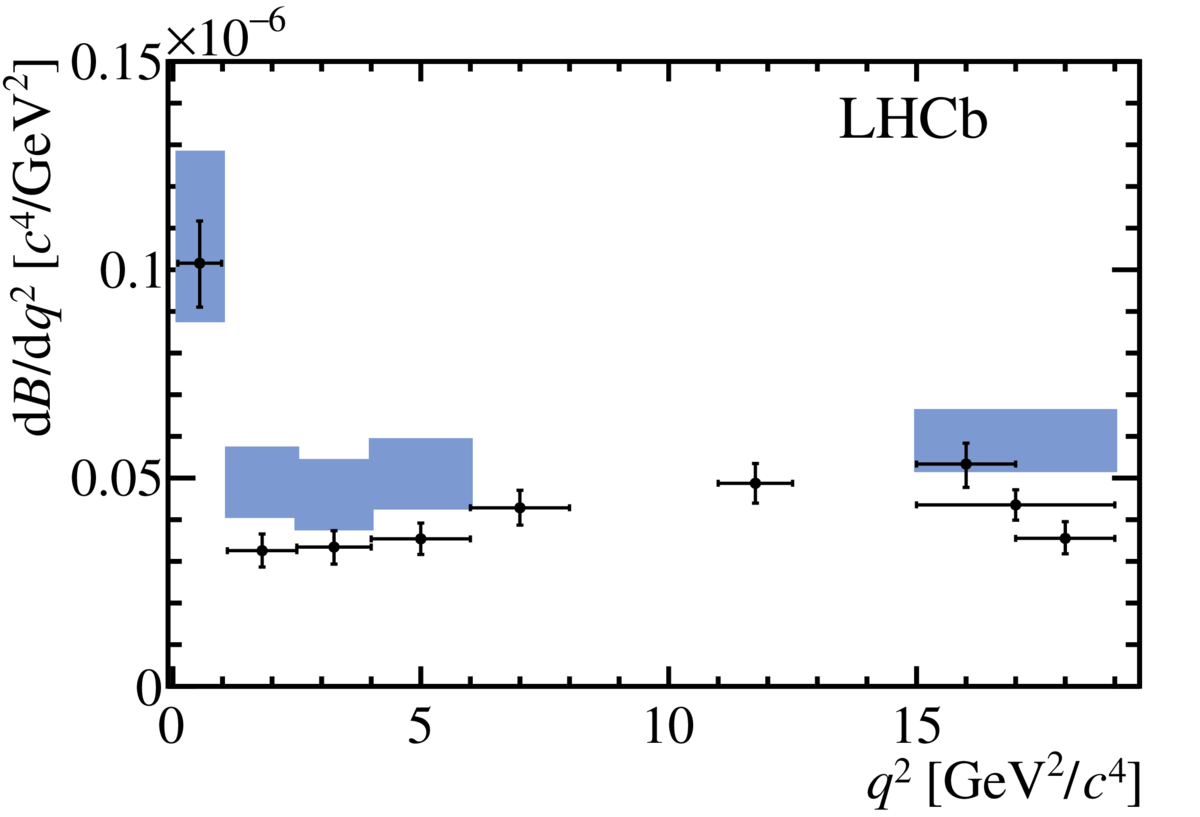}} 
        \put(0,0){\includegraphics[scale=0.145]{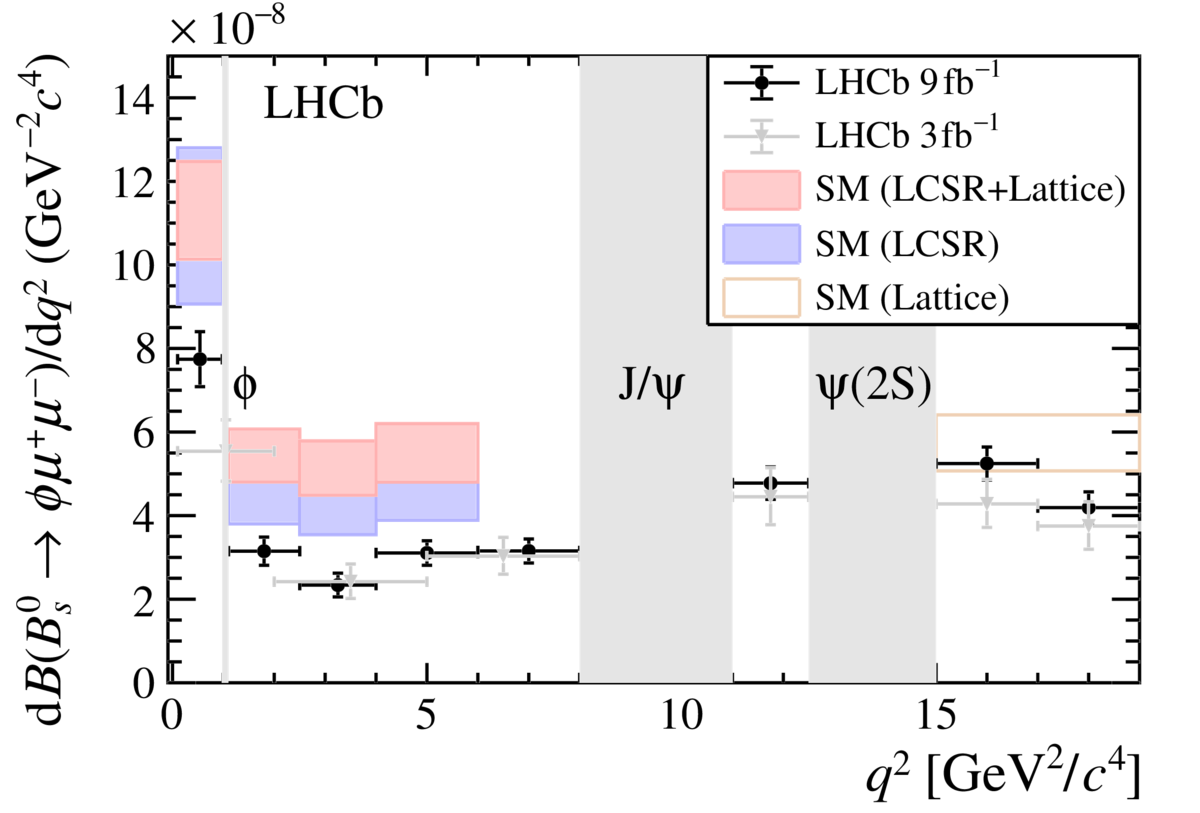}}
        \put(200,0){\includegraphics[scale=0.145]{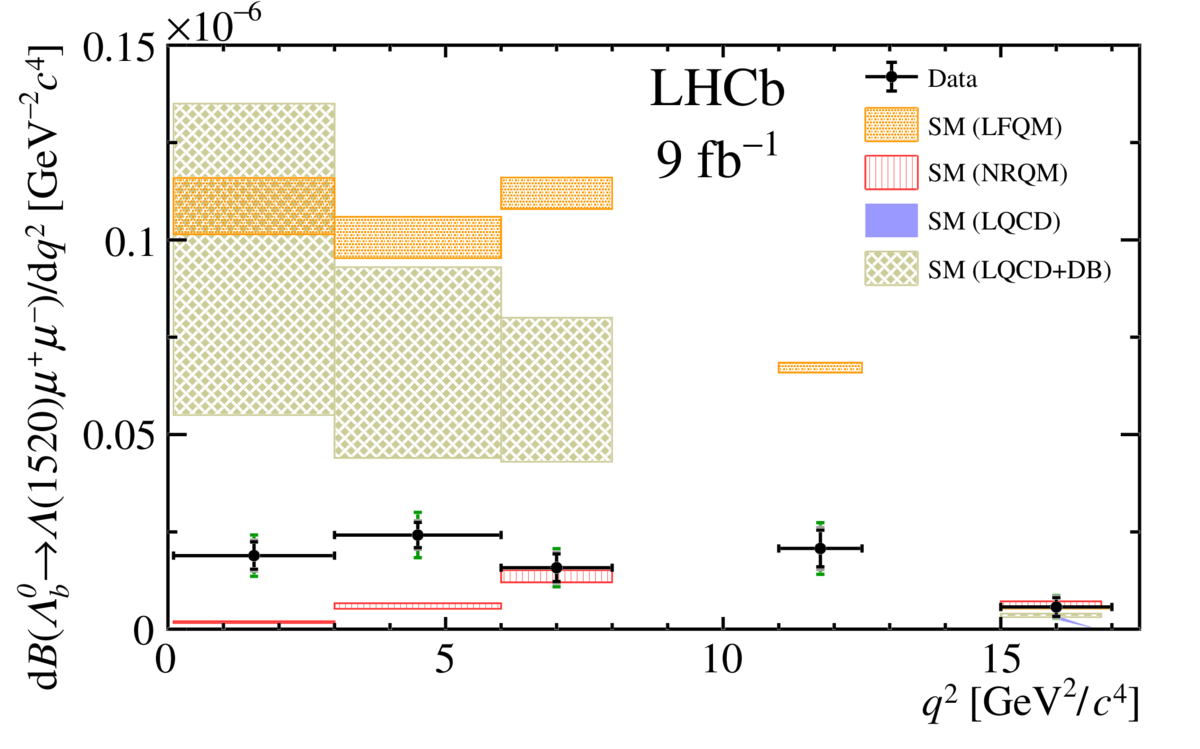}}   
        \put(300,205){$B^0 \to K^{*0}\mu^+\mu^-$}
        \put(38,92){$B^0_s\to \phi\mu^+\mu^-$}
        \put(228,92){$\Lambda_b^0\to \Lambda(1520)\mu^+\mu^-$}
    \end{picture}   
     \caption{Decay rates for the $b\to s \mu^-\mu^+$ process, 
     as a function of the di-lepton invariant mass $q^2$, separately for the hadrons 
     $B^+$~\cite{LHCB-PAPER-2014-006}, 
     $B^0$~\cite{LHCB-PAPER-2016-012}, 
     $B_s^0$~\cite{LHCb-PAPER-2021-014} and 
     $\Lambda_b^0$~\cite{LHCb-PAPER-2022-050}, compared to theory predictions.}
     \label{fig:BRbsll}
\end{figure}

%\newpage
%\begin{wrapfigure}{!t}{7cm}
%	\centering
%    \includegraphics[scale=0.22]{Figures/RareDecays/S5-counting.png}
%	\caption{The $P_5^{'} =S_5/\sqrt{(1-F_L)F_L}$ observable is proportional to the asymmetry of events in the red area compared to the blue area. Similar to the case of Fig.~\ref{fig:JpsPhi2}, the angle $\theta_K$ describes the angle of emission between the $K^+$ and $K^{*0}$ in the di-hadron rest frame, whereas $\phi$ corresponds to the angle between the di-lepton and di-meson planes.}
%	\label{fig:S5}
%\end{wrapfigure}
A sketch of the differential decay rate of this process is shown in the middle panel of the figure and
notably depends on the squared invariant mass
of the lepton pair, $q^2$, as well as on the decay angles of the final state particles. At the lowest values of $q^2$ the lepton pair is collinear and a large contribution
to the decay rate originates from the photon pole, provided the final state kaon is spin-1 to
conserve angular momentum. Above $q^2>7$~GeV$^2$ the contribution from charm quarks becomes sizeable,
which is most notable around the mass of the $J/\psi$ and $\psi(2S)$ resonances. 
The region $1<q^2<6$~GeV$^2$ is thus most interesting to detect contributions beyond
the Standard Model by comparing measurements to theoretical predictions. The measured decay rate is shown in the right side panel of the figure.
Multiple experimental observables are available, such as the decay rate as a function
of $q^2$, the angular asymmetries, and a comparison of the processes with different lepton types in the final state.
As shown in Fig.~\ref{fig:BRbsll} the experimental measurements of the decay rates for the various hadron types all exhibit a deficit compared to the theoretical predictions. However, since the theoretical predictions depend on
long-distance QCD calculations such as lattice QCD or light-cone sum-rules, they are model dependent and hence suffer from relatively large uncertainties.

%$B^+ \to K^{+}\mu^+\mu^-$ branching ratio \cite{LHCb-PAPER-2014-006}.
%$B^0 \to K^{*0}\mu^+\mu^-$ branching ratio \cite{LHCb-PAPER-2016-012}
%$B^0_s\to \phi\mu^+\mu^-$ branching ratio \cite{LHCb-PAPER-2021-014}
%$\Lambda_b^0\to \Lambda(1520)\mu^+\mu^-$ \cite{LHCb-PAPER-2022-050}

The angular asymmetries are theoretically better understood than the absolute decay rates and experimentally inefficiencies cancel in the ratio of event yields. A particularly interesting observable is the forward-backward asymmetry and the $q^2$ value at which this asymmetry vanishes, known as the zero-crossing point $s_0$~\cite{Ali:1999mm,Beneke:2001at}.
Symmetries in the angular distributions of $B^0 \to K^{*0}\mu^+\mu^-$ have
been subject to considerable effort
to construct observables that have small theoretical uncertainties.
The decay can be described in a set of six
clean observables $P_{1,2,3,4,5,6}$ together with the longitudinal polarization fraction of the $K^*$, $F_L$, and the differential
decay rate that depends on the form factors (neglecting S-wave contributions)
~\cite{Descotes-Genon:2012isb}.

\begin{figure}[!hb]
    \centering   
    \includegraphics[scale=0.275]{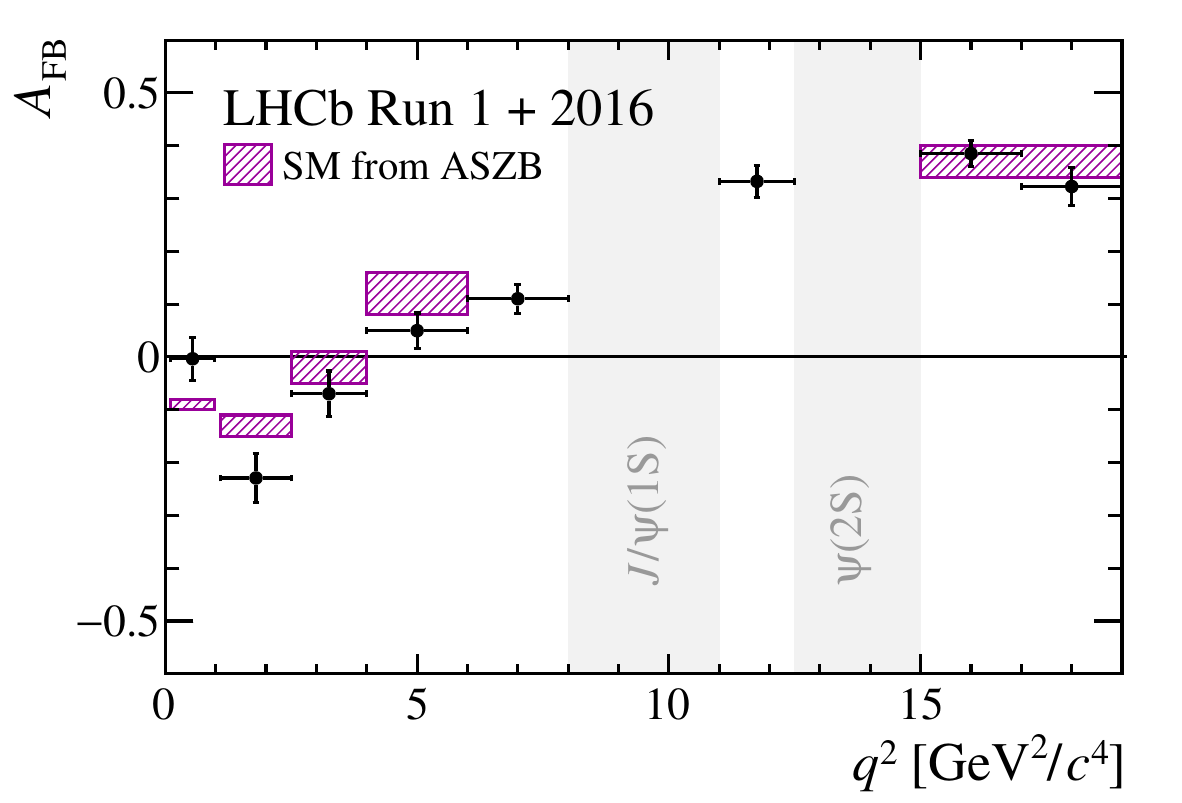}    
    \includegraphics[scale=0.275]{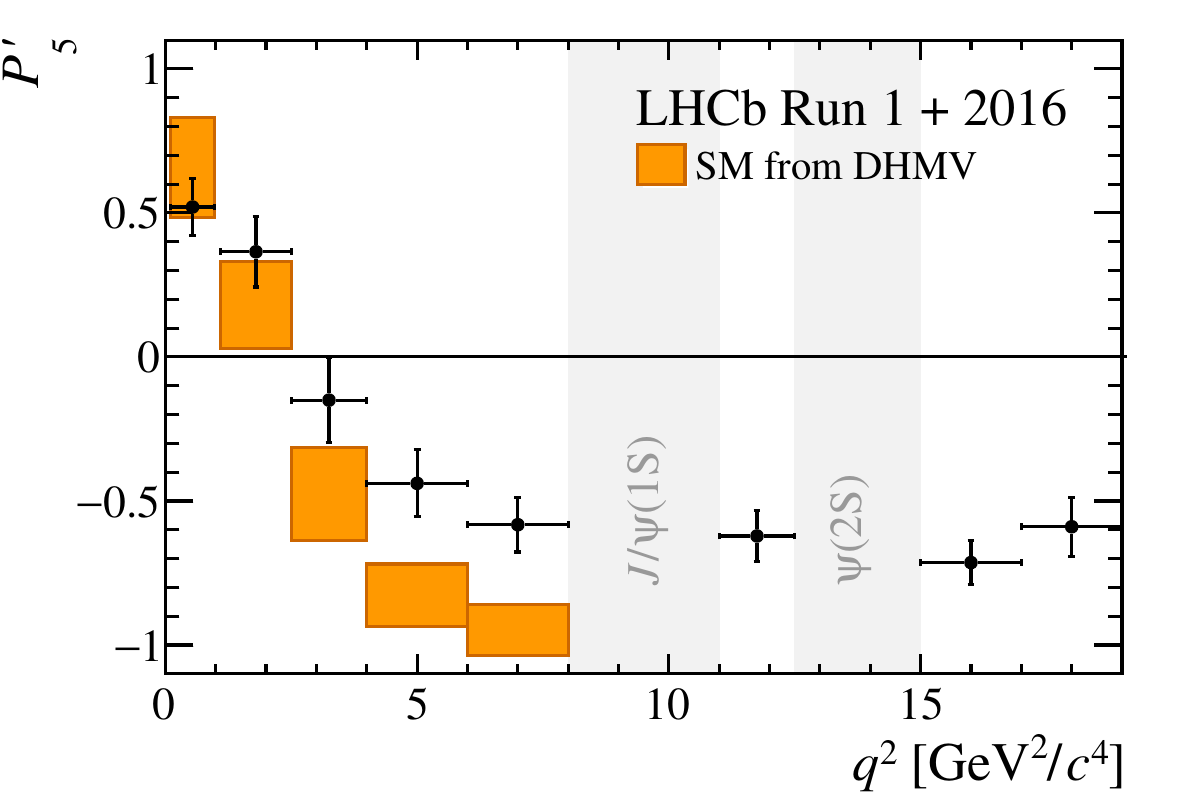}    
    \includegraphics[scale=0.19]{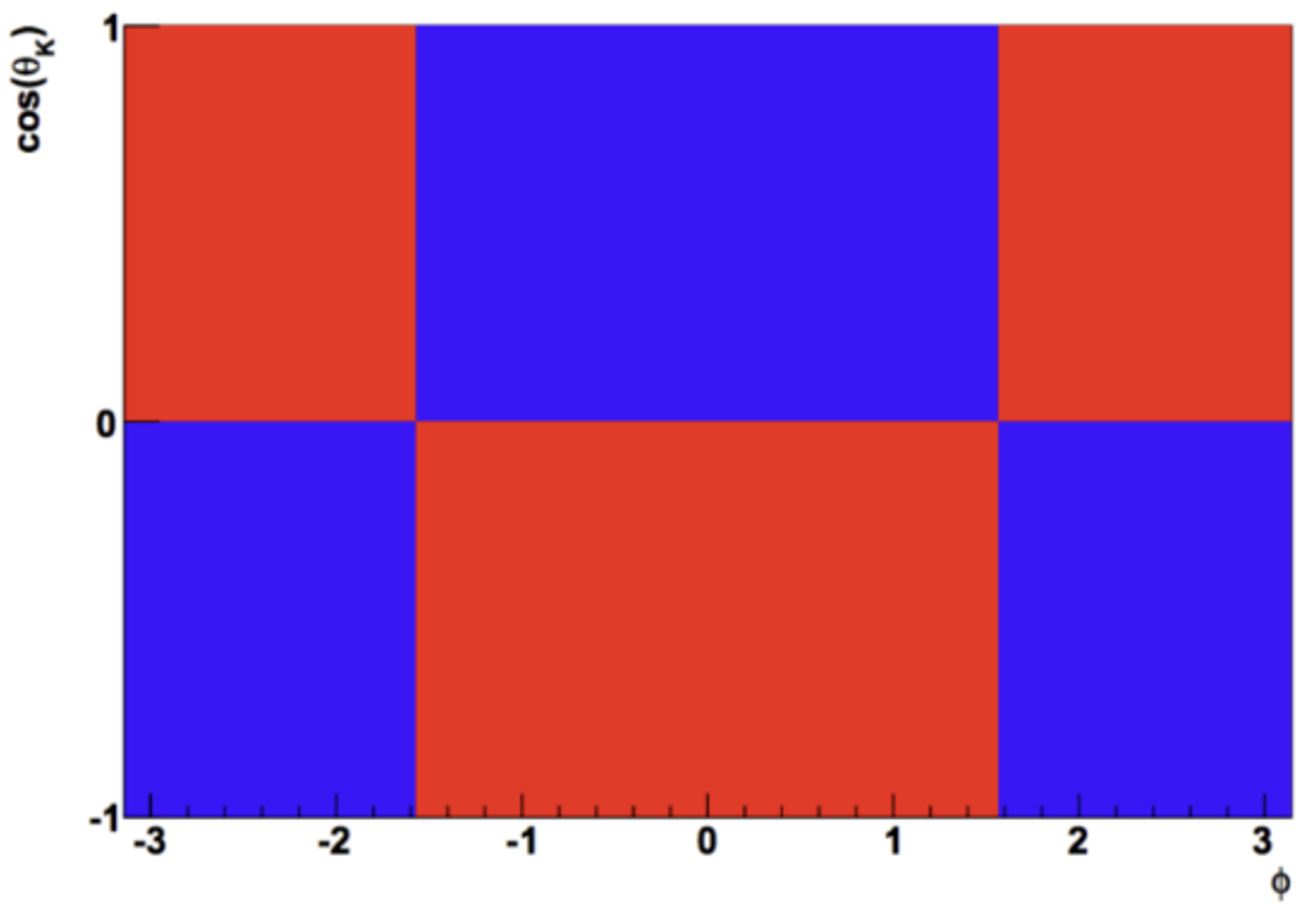}
	     \caption{
     Angular asymmetries $A_{FB}$ and $P_5'$, measured with $B^0 \to K^{*0}\mu^+\mu^-$ decays,
     and compared to the theoretical predictions~\cite{LHCb-PAPER-2020-002}. 
     {\em Right:} The $P_5^{'} =S_5/\sqrt{(1-F_L)F_L}$ observable is proportional to the asymmetry of events in the red area compared to the blue area. Similar to the case of Fig.~\ref{fig:JpsPhi2}, 
    the angle $\theta_K$ describes the angle of emission between the $K^+$ and $K^{*0}$ in the di-hadron rest frame, whereas $\phi$ corresponds to the angle between the di-lepton and di-meson planes.}
     \label{fig:Bkmm-ang}
\end{figure}

These observables in turn depend on the Wilson coefficients that quantify
the effective coupling strength for scalar, vector or axial couplings.
The most notable coefficients in the heavy-quark effective theory (HQET) expansion are ${\cal C}_7$ quantifying the $b\to s\gamma$ coupling,
and ${\cal C}_9$ and ${\cal C}_{10}$, quantifying the vector and axial 
$b\to s\ell^+\ell^-$ coupling, respectively.
A purely left-handed coupling to the leptons is thus quantified as
$({\cal C}_9-{\cal C}_{10})$ and right-handed couplings to leptons as $({\cal C}_9+{\cal C}_{10})$.
The right-handed couplings to the quark side are denoted as primed coefficients ${\cal C}^\prime_i$.
The observable $P_5^{'}= S_5/\sqrt{(1-F_L)F_L}$ 
has drawn most attention due the 
largest discrepancy with respect to Standard Model expectations.
The value of $P_5^{'}$ is determined as a function of the di-lepton invariant mass $q^2$, in the way that is illustrated in the right side panel of Fig.~\ref{fig:Bkmm-ang}.
LHCb's measurements of the observables $A_{FB}$, and $P_5^\prime$ as function of the di-muon $q^2$ for $B^0\rightarrow K^{*0}\mu^+\mu^-$ are shown in Fig.~\ref{fig:Bkmm-ang}~\cite{LHCb-PAPER-2020-002}.

\begin{figure}[!hb]
    \centering     
    \includegraphics[scale=0.275]{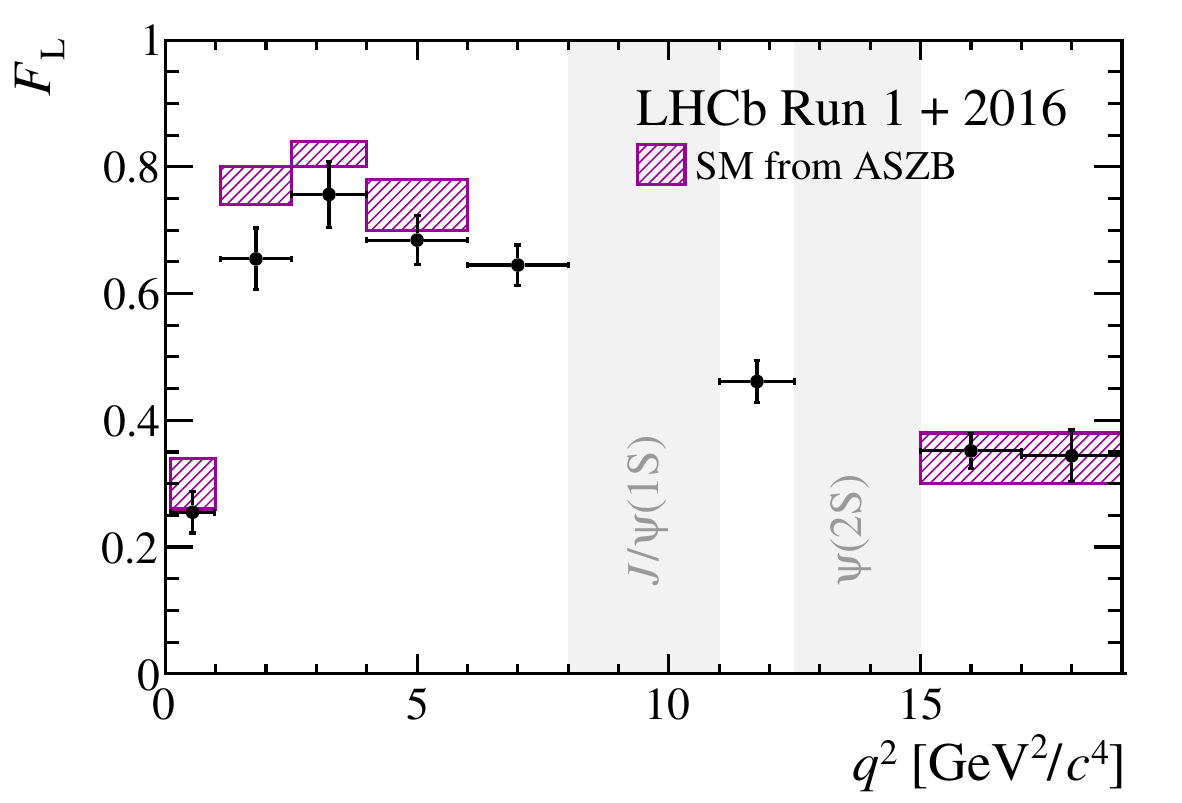}    
    \includegraphics[scale=0.25]{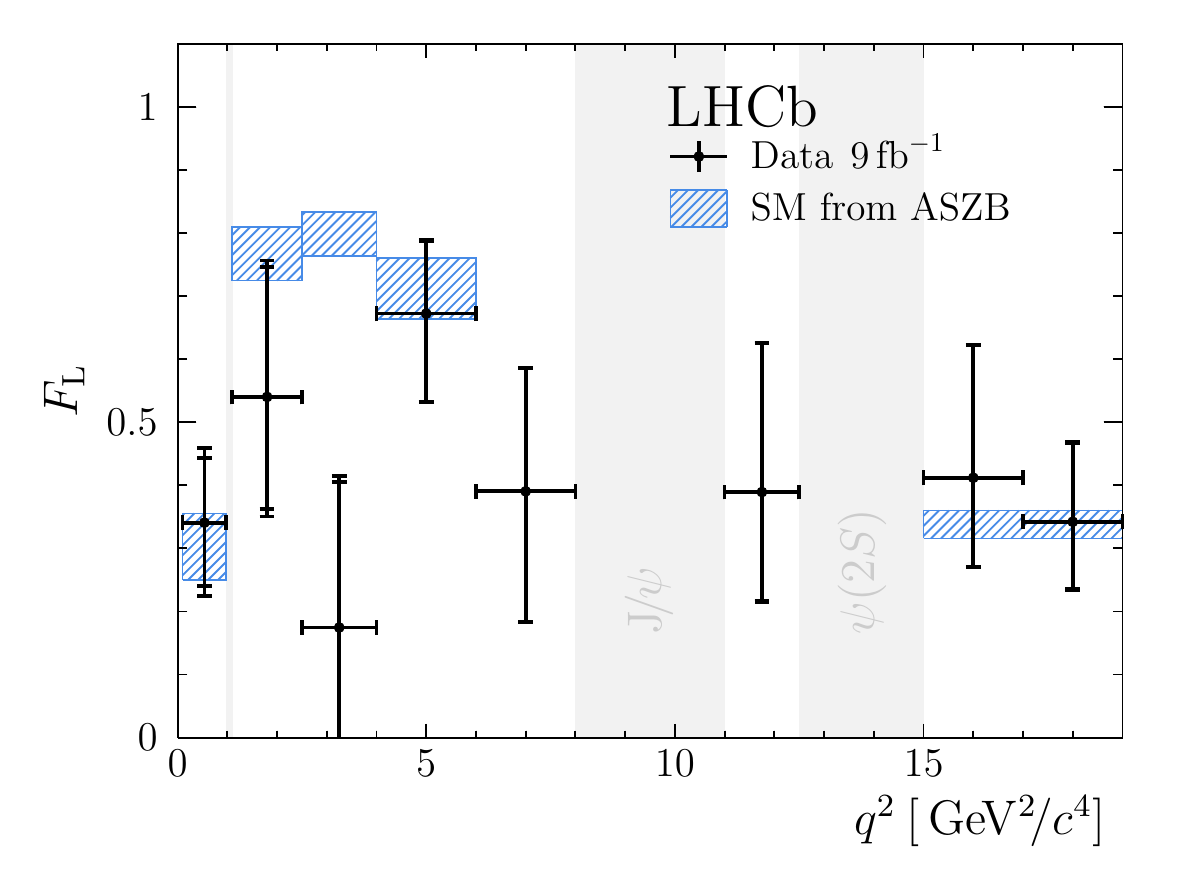}    
    \includegraphics[scale=0.275]{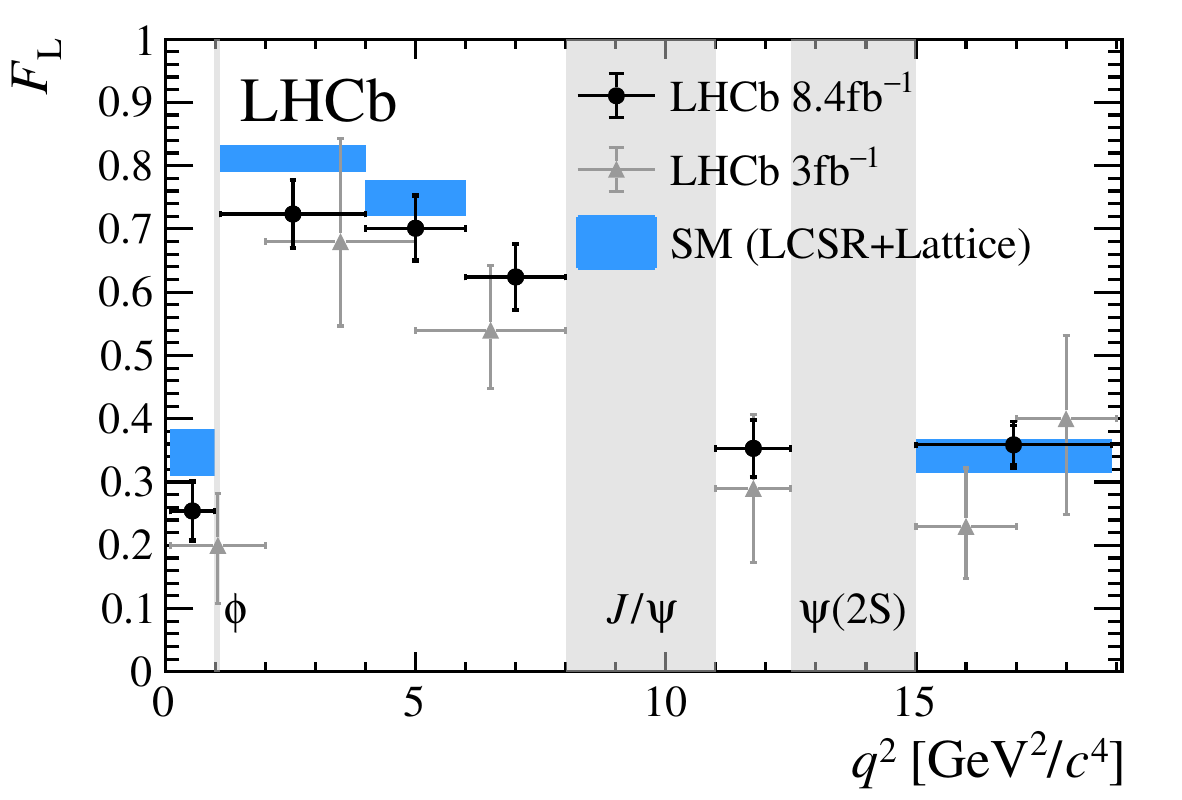}    
     \caption{Measurements of $F_L$ for the decays 
     $B^0 \to K^{*0}\mu^+\mu^-$~\cite{LHCb-PAPER-2020-002},
     $B^+\to K^{*+}\mu^+\mu^-$~\cite{LHCb-PAPER-2020-041} and 
     $B_s^0\to \phi \mu^+\mu^-$~\cite{LHCb-PAPER-2021-022}.}
     \label{fig:Bkmm-ang2}
\end{figure}

The angular analysis has also been performed
for the decays 
$B^+\to K^{*+}(\to K_S^0\pi^+) \mu^+\mu^-$~\cite{LHCb-PAPER-2020-041} and
$B_s^0\to \phi \mu^+\mu^-$~\cite{LHCb-PAPER-2021-022}.
However, in the latter case the decay flavour of the $B_s^0$ meson cannot be determined from the flavour-symmetric final state, and hence not all the angular observables are experimentally accessible, such as $P_5^{'}$. Fig.~\ref{fig:Bkmm-ang2} shows LHCb's measurements of $F_L$ for the decays $B^0 \to K^{*0}\mu^+\mu^-$, $B^+\to K^{*+}\mu^+\mu^-$ and $B_s^0\to \phi \mu^+\mu^-$, where $F_L$ is the fraction of the longitudinal polarisation of the $K^*$ or $\phi$ meson.

%\clearpage
\subsection{Very Rare Decays}
%------------------------------
\subsubsection{\texorpdfstring{$B^0_{(s)}\to \mu^+\mu^-$}{Bs->mumu}}

\begin{figure}[!hb]
    \centering
   \begin{picture}(400,210)(0,0)
        \put(0,120){\includegraphics[scale=0.95]{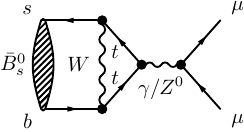}}
        \put(0,35){\includegraphics[scale=0.95]{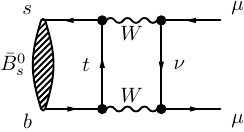}}
        \put(145,0){\includegraphics[scale=0.35]{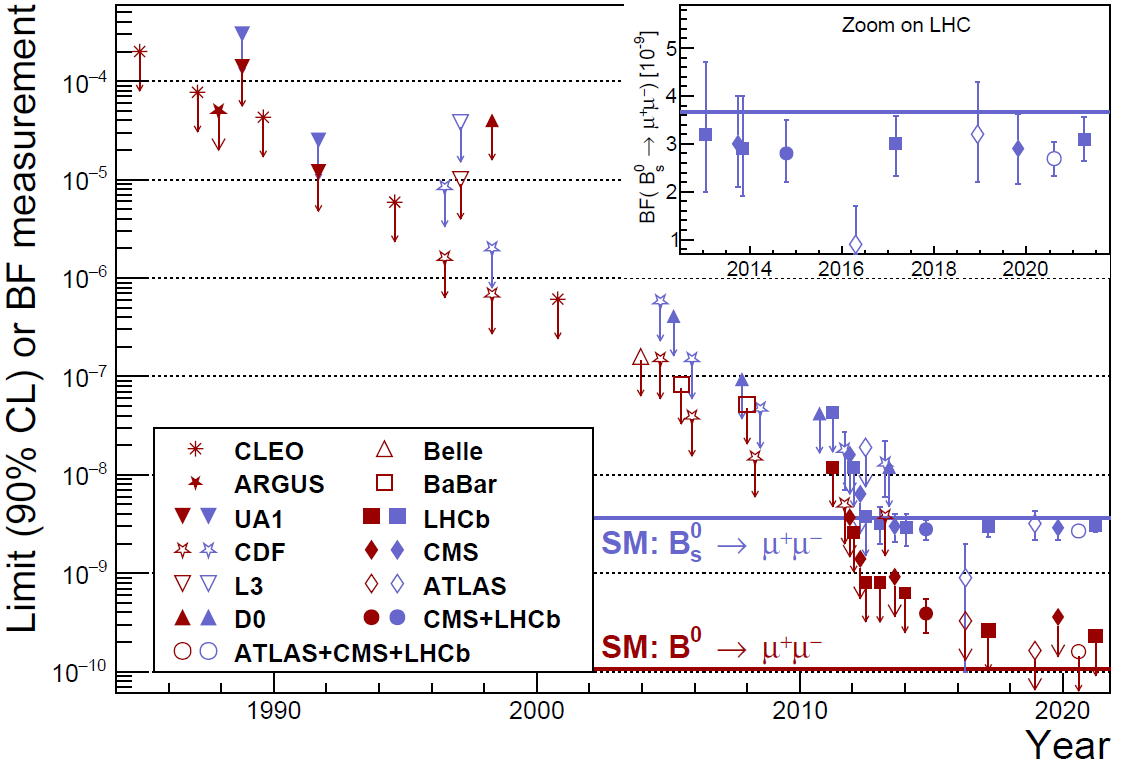}}
     \end{picture}
     \caption{{\em Left:} Diagrams of the decay $B_s^0\to \mu^+\mu^-$ shows the suppressed $b\to s\ell\ell$ FCNC Electroweak Penguin topology, which is suppressed in the SM. {\em Right:} The historical search for the decay $B_s^0\to \mu^+\mu^-$ led to its observation in 2012.}
     \label{fig:Bs2MuMu}
\end{figure}

\begin{figure}[!b]
    \centering  
    \begin{picture}(400,120)(0,0)
        \put(-39,12){\includegraphics[scale=0.145]{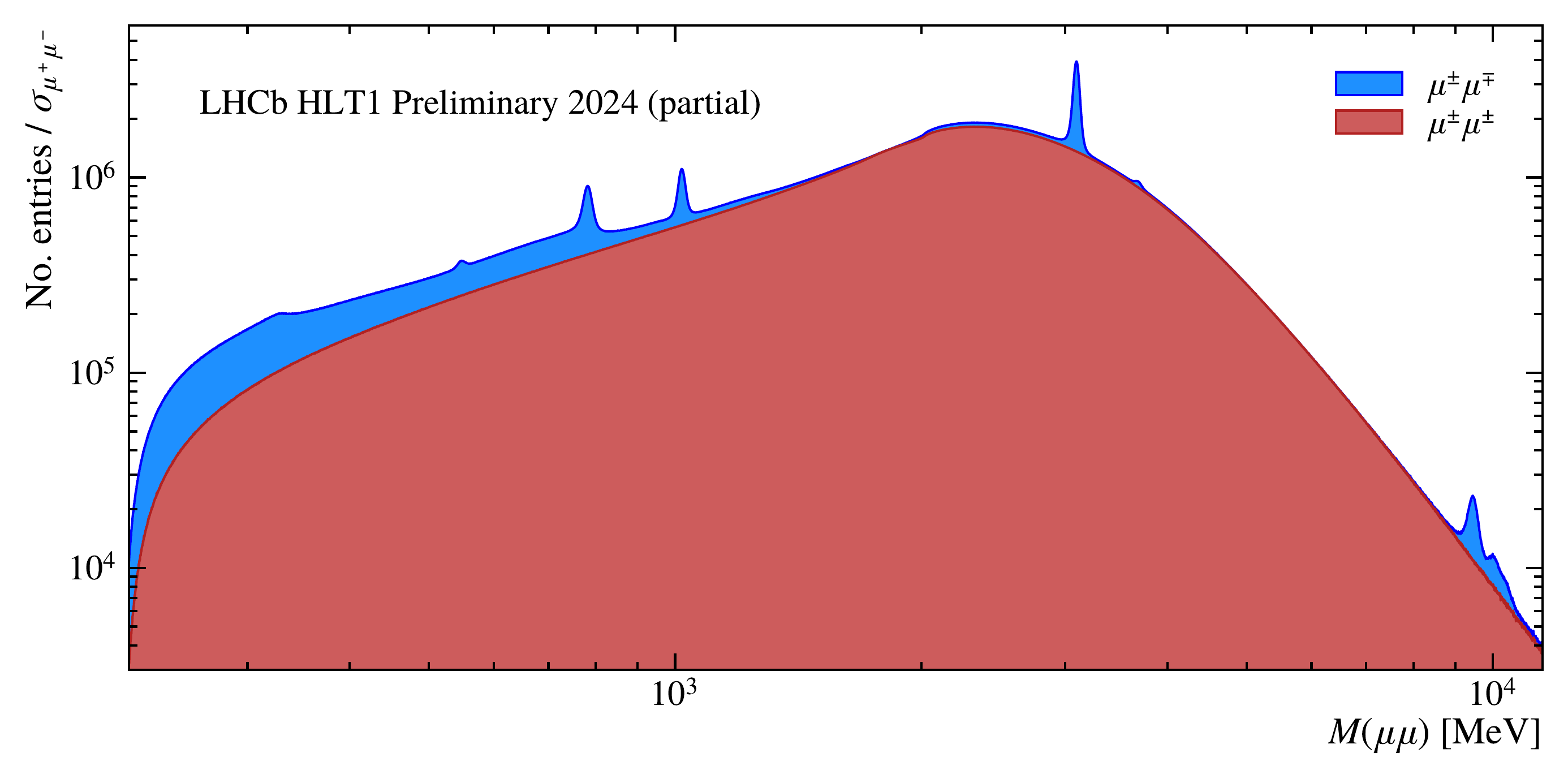}}
        \put(153,10){\includegraphics[scale=0.26]{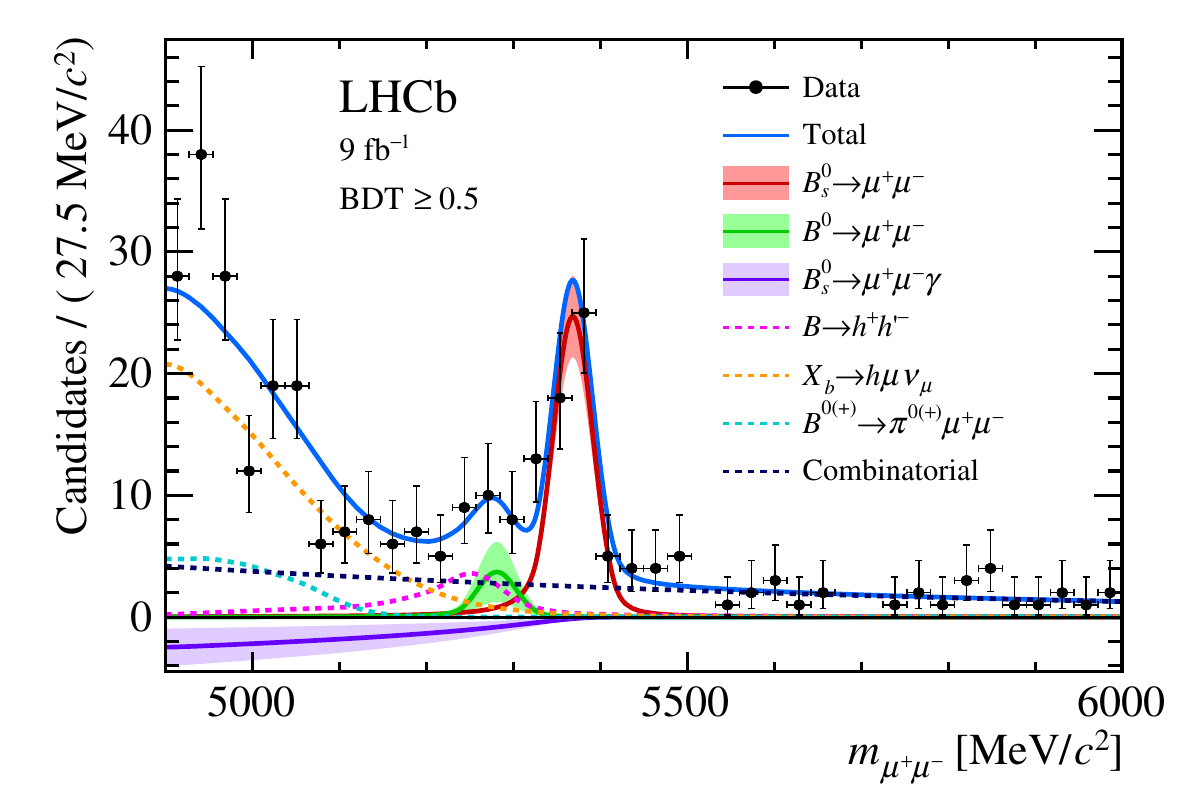}}
        \put(294,10){\includegraphics[scale=0.25]{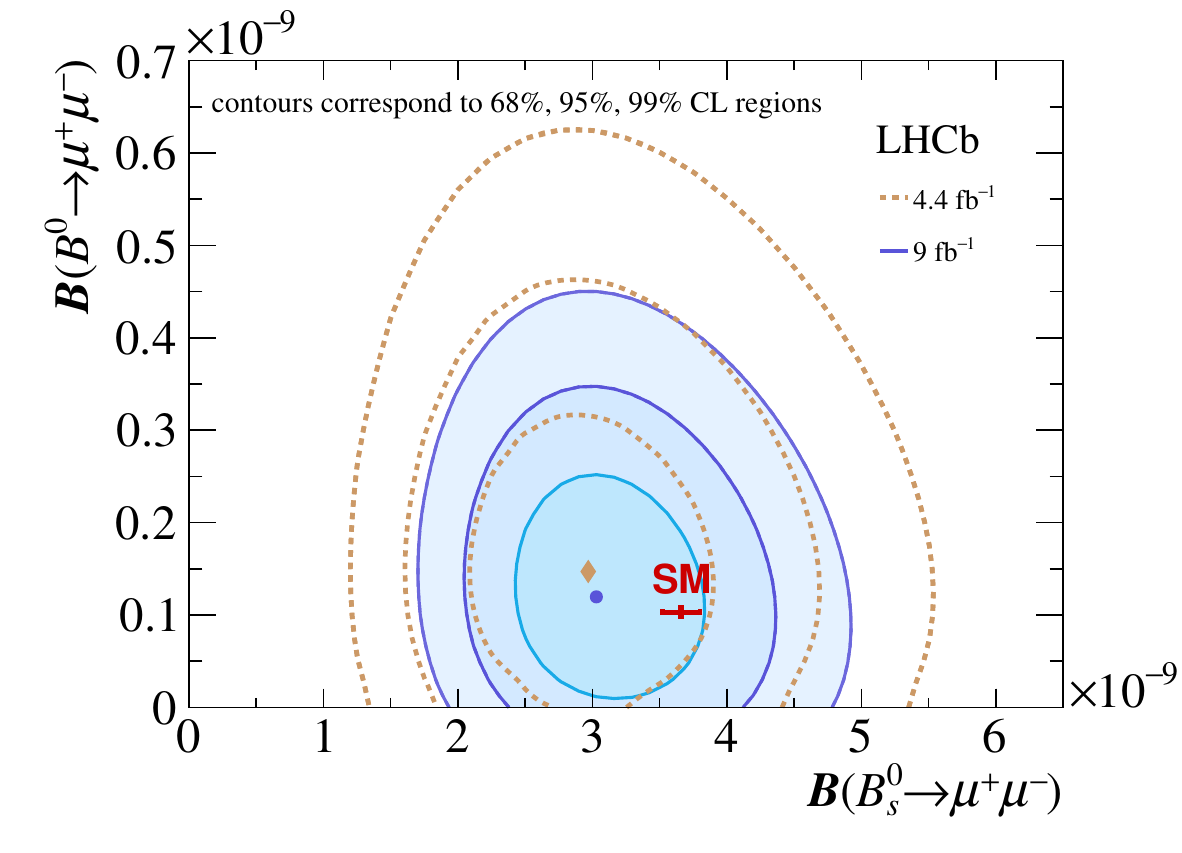}}
    \end{picture}
     \caption{{\em Left:} The di-muon spectrum as measured from an HLT1 sample shows the challenging task of isolating a handful di-muon candidates originating from a $B$ decay. {\em Middle:} The invariant mass distribution of $B^0\to \mu^+\mu^-$ and $B^0_s\to \mu^+\mu^-$ candidates. {\em Right:} The resulting value of the measured branching ratios~\cite{LHCb-PAPER-2021-007,LHCb-PAPER-2021-008}.
     }
     \label{fig:Bs2MuMu2}
\end{figure}

The very rare decay $B^0_{(s)}\to \mu^+\mu^-$ has been 
widely regarded as a sensitive probe to scrutinize the Standard Model,
and was searched for since several decades. 
%Its suppression is caused both by higher order diagrams as well as by helicity suppression $(m_\mu/m_{B_{(s)}})^2$ of the scalar $B_{(s)}^0$ decaying into a muon and anti-muon spinor-pair. 
Potential contributions from minimal supersymmetric extensions to the Standard Model were shown to largely modify the predicted value of the branching fraction~\cite{Isidori:2006pk}.
Fig.~\ref{fig:Bs2MuMu} displays the leading Feynman diagrams of the decay together with the search limits obtained by experiments over time, until it was observed by a combined result of CMS and LHCb in 2012 \cite{CMS:2014xfa}.
The data and result of LHCb are shown in Fig.~\ref{fig:Bs2MuMu2}, which respectively shows the rate of opposite sign and same sign di-muon pairs in the trigger, the offline selected $B_{(s)}\rightarrow \mu^+\mu^-$ candidates and the observed Branching Ratio contour plot for the $B^0$ and $B_s^0$ decays.

More recently, it was realized that also the measurement of the  effective lifetime can reveal new physics effects~\cite{DeBruyn:2012wk}.
The term {\em effective} lifetime is introduced to account for the fact that the experimental branching ratio is determined by integrating over all 
decay times, whereas the phenomenological decay amplitude is traditionally determined
independent of the decay time~\cite{DeBruyn:2012wj}.
In the Standard Model the decay $B^0_s\to \mu^+\mu^-$ proceeds to the 
CP-odd final state due to the helicities of the final state muons, and is thus dominated by the heavy lifetime eigenstate $B_{s,H}$ with a longer lifetime.
This leads to a larger measured branching fraction compared to the value from the decay amplitude at $t=0$, because the experimental measurement integrates over all decay times including $B$-$\bar{B}$ oscillations.
New scalar contributions in the decay process add a CP-even short living component and thus alter both the measured branching fraction and the effective lifetime. 
To determine the branching ratio, the production rate of $B_s^0$ mesons needs to be known.
This is done by normalizing to $B_s\rightarrow \mu^+\mu^-$ events to a known $B^0$ or $B^+$ decay mode, and by using a measurement of 
the relative production rate of $B_s^0$ over $B^0$ mesons, $f_s/f_d$~\cite{Fleischer:2010ay,LHCb-PAPER-2020-046}.

\subsubsection{\texorpdfstring{$B^0_{(s)}\to e^+e^-$}{Bs->ee} and 
\texorpdfstring{$B^0_{(s)}\to \tau^+\tau^-$}{Bs->tt}}

The decay $B^0_{(s)}\to \mu^+\mu^-$ is suppressed in the Standard Model due
to the higher order FCNC Electroweak Penguin loop process.
Additionally, it is subject to helicity suppression, similar to the decay $\pi^+\to e^+\nu$, which is also a spin-0 meson decaying to a fermion - antifermion pair.
This helicity suppression depends on the mass of the final-state leptons and scales as ($m_l^2/m_B^2$)).
As a result, the branching fractions for $B^0_{(s)}\to e^+e^-$ are expected to be even more suppressed, whereas those for $B^0_{(s)}\to \tau^+\tau^-$ are less suppressed due to the larger mass of the $\tau$-lepton \cite{Fleischer:2017ltw}.

Although searches for both $B^0_{(s)}\to e^+e^-$ and $B^0_{(s)}\to \tau^+\tau^-$  have been carried out, neither decay has been observed so far. The extreme helicity suppression significantly reduces the expected rate of $B^0_{(s)}\to e^+e^-$, while for $B^0_{(s)}\to \tau^+\tau^-$ the challenge lies in the low event yield due to the subsequent branching fractions of the $\tau $
leptons into favourable final states.

%\clearpage
\subsection{Combining the experimental results and effective field theory}
%------------------------------
%\textcolor{blue}{Niels, dit had ik naar boven gehaald - The true picture of a decaying hadron is more correctly described by the local operators, through heavy-quark effective theory (HQET)~\cite{Buras:1997fb}, where calculable short distance effects ($C_i$) are separated from non-perturbative long-distance operators $O_i$, $H_{eff}=G_F/\sqrt{2}V_{CKM}\Sigma_i  C_i O_i$.}

\begin{figure}[!hb]
	\centering
    \begin{picture}(450,180)(0,0)
        \put(10,110){\includegraphics[scale=0.7]{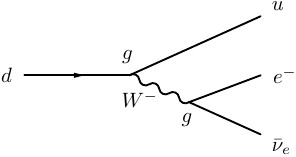}} 
        \put(120,110){\includegraphics[scale=0.7]{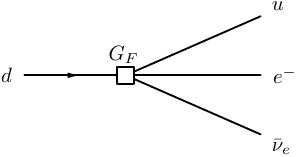}} 
        \put(10,30){\includegraphics[scale=0.7]{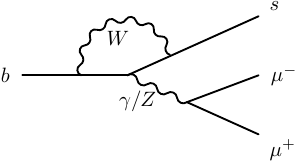}} 
        \put(120,30){\includegraphics[scale=0.7]{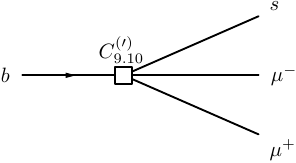}} 
        \put(245,0){\includegraphics[scale=0.31]{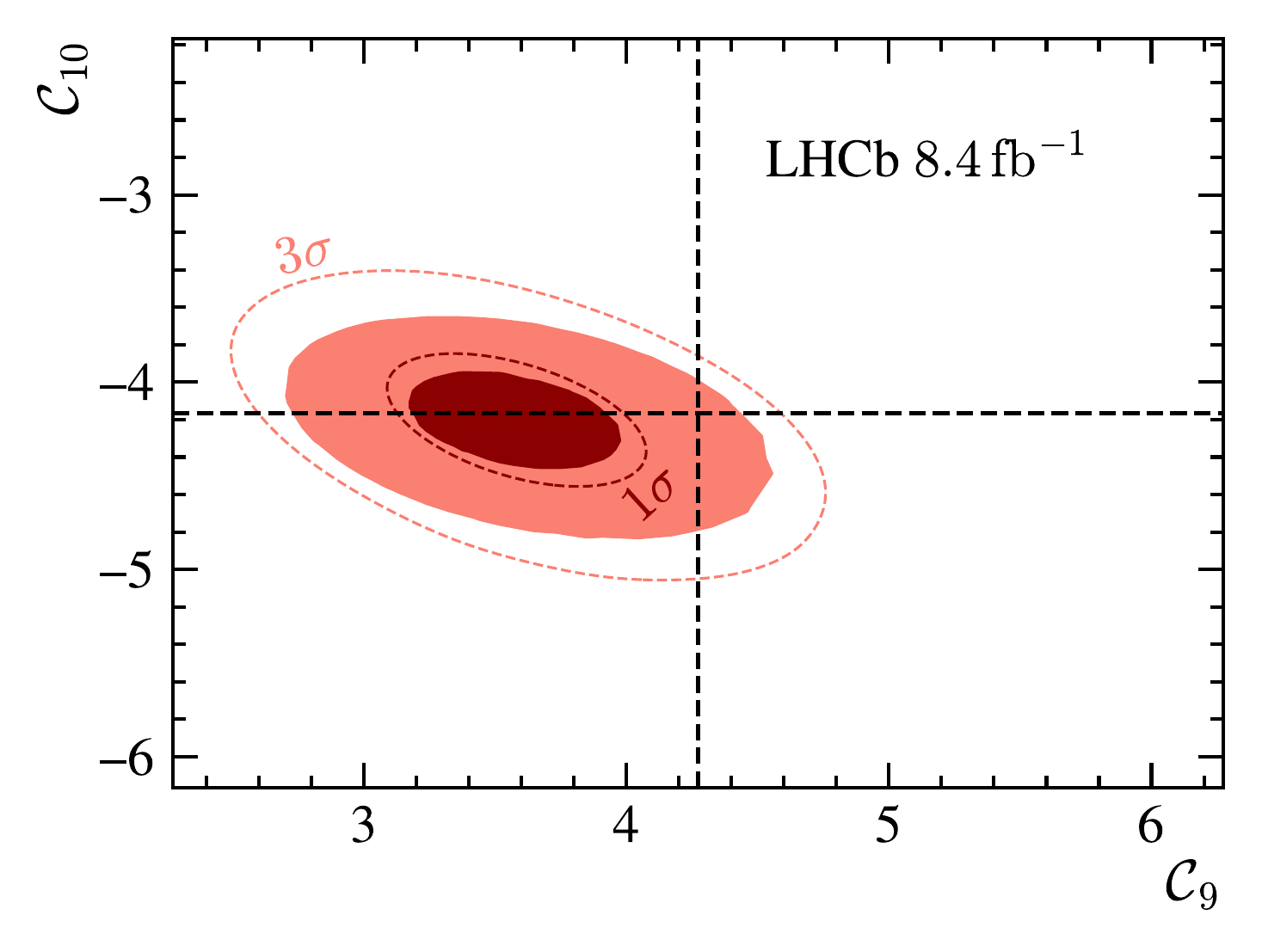}}
    \end{picture}
	\caption{{\em Left:} The diagrams depicting the weak interaction and the FCNC EWP $b\to s\ell\ell$ process. {\em Middle:} The corresponding diagrams in the HQET description where the Wilson coefficients quantify the various coupling types. {\em Right:} The fitted value for the axial and vector couplings $C_{9}$ and $C_{10}$ from $B^0\to K^{0*}\mu^+\mu^-$ decays, with the dashed lines indicating the SM value~\cite{LHCb-PAPER-2024-011}. }
	\label{fig:EFT}
\end{figure}

\begin{figure}[!b]
    \centering     
    \begin{picture}(450,120)(0,0)
        \put(0,5){\includegraphics[scale=0.345]{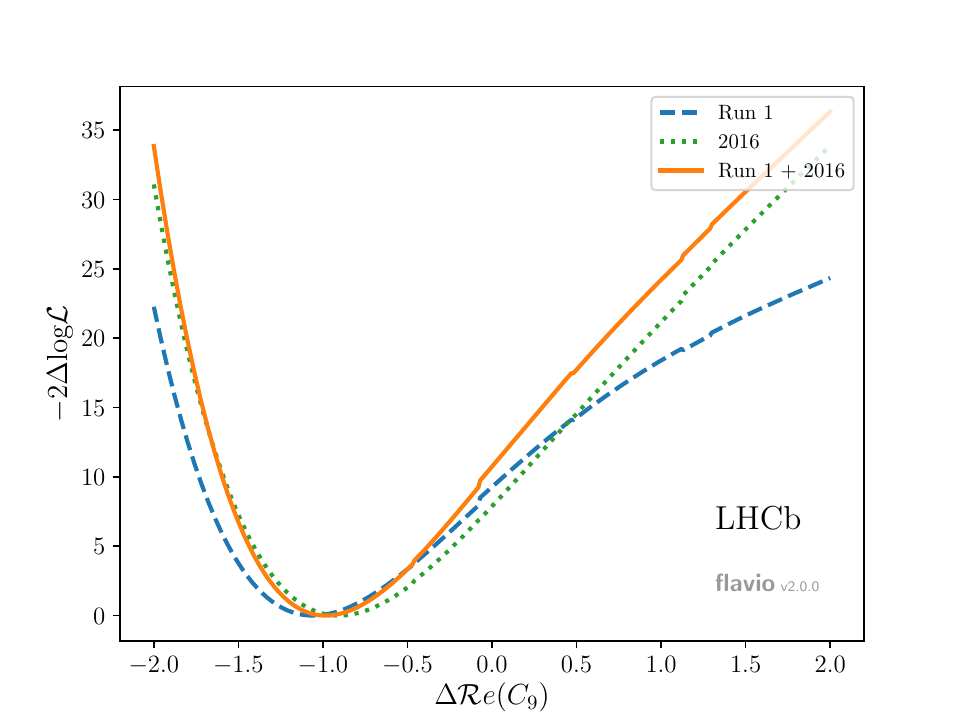}}
        \put(155,5){\includegraphics[scale=0.175]{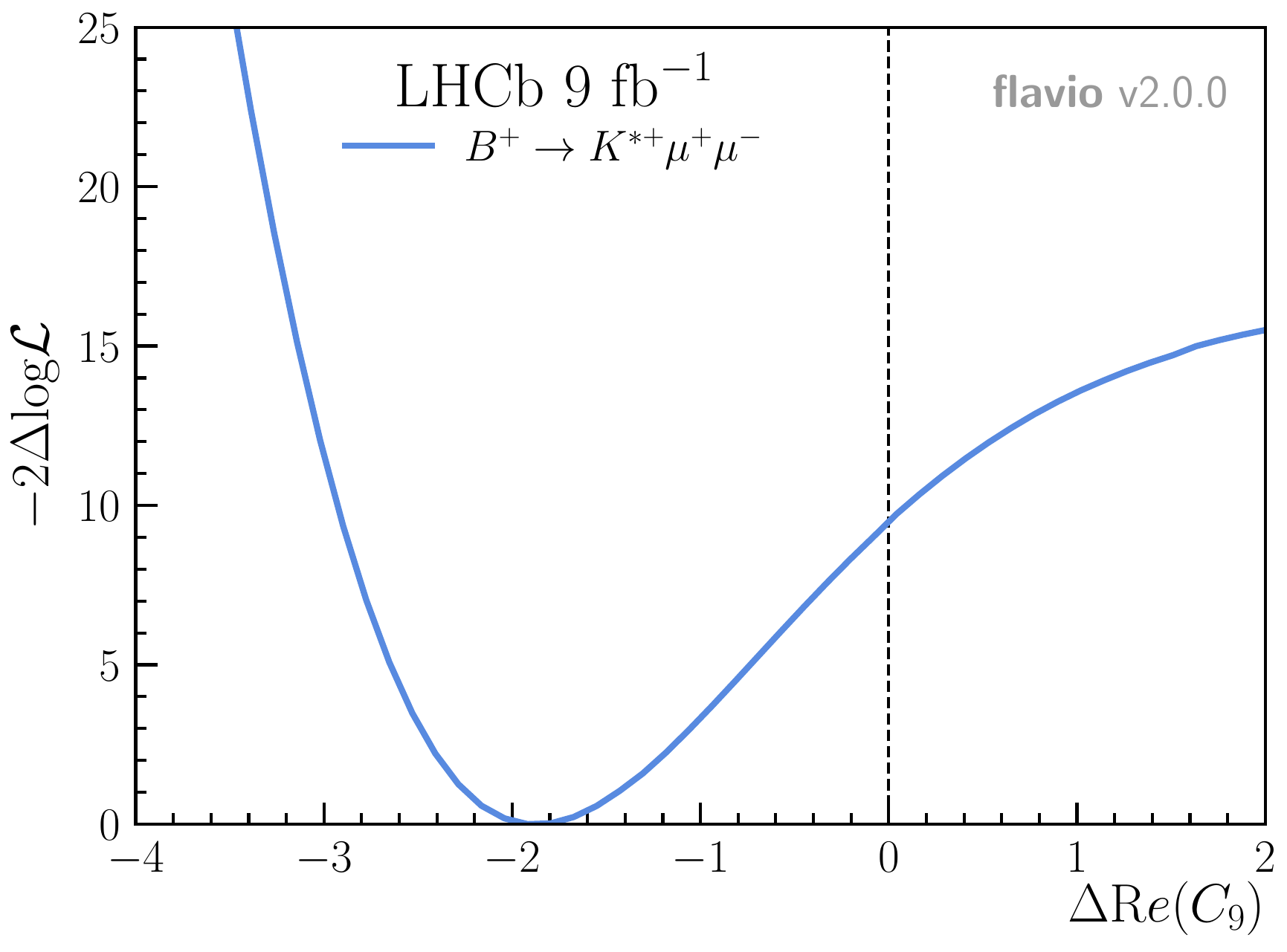}}    
        \put(300,5){\includegraphics[scale=0.34]{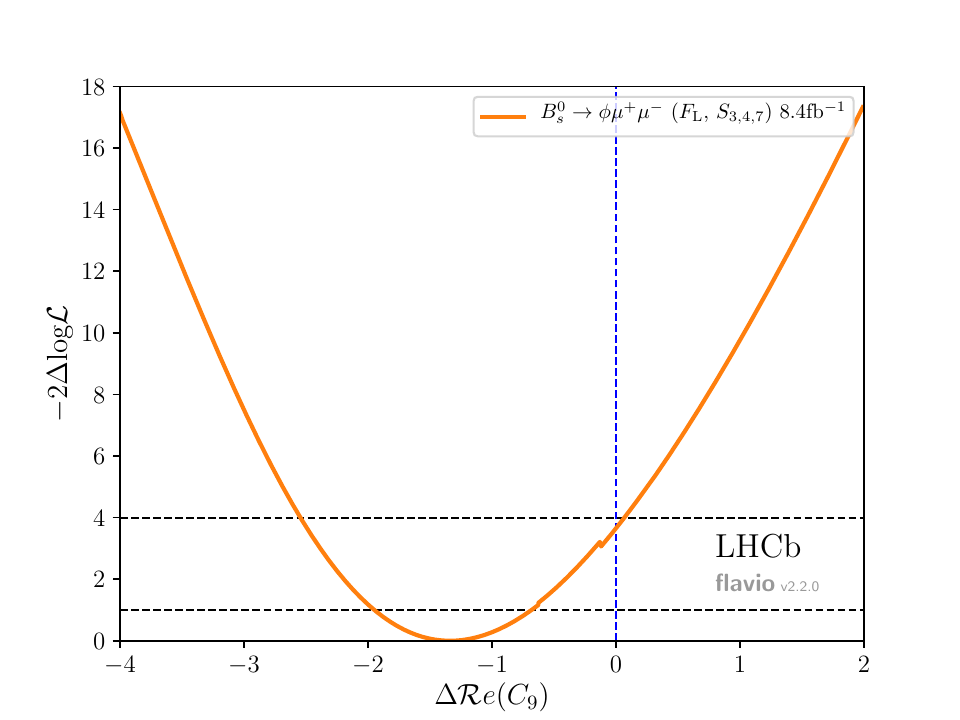}}    
        \put(30,113){$B^0 \to K^{*0}\mu^+\mu^-$}
        \put(180,113){$B^+\to K^+\mu^+\mu^-$}
        \put(330,113){$B^0_s\to \phi\mu^+\mu^-$}
     \end{picture}
     \caption{Comparison of the likelihood scan from the Wilson coefficient fit 
     to all angular observables for the decays 
     $B^0\to K^{*0}\mu^+\mu^-$~\cite{LHCb-PAPER-2020-002}, 
     $B^+\to K^{*+}\mu^+\mu^-$~\cite{LHCb-PAPER-2020-041} and 
     $B_s^0\to \phi \mu^+\mu^-$~\cite{LHCb-PAPER-2021-022}, leaving the vector coupling
     $C_9$ free in the fit, showing a remarkable agreement.}  
     \label{fig:BKmm-ang3}
\end{figure}
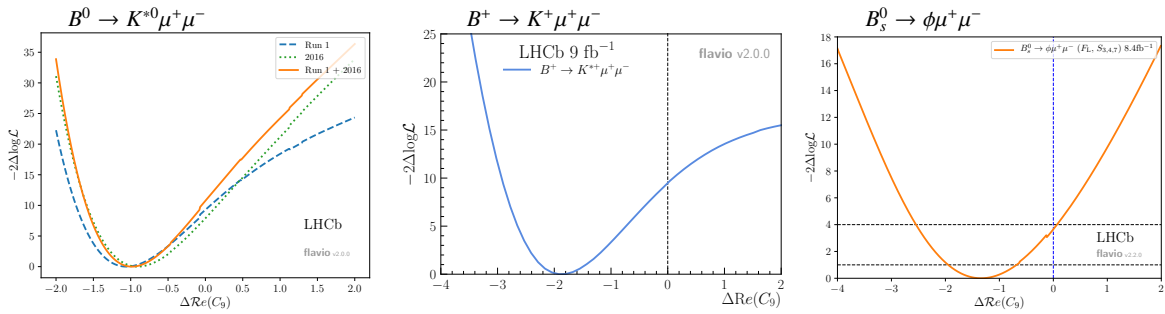

The separation of calculable short distance effects ${\cal C}_i$ from non-perturbative long distance operators ${\cal O}_i$ in heavy-quark effective theory (eq.~\ref{eq:Bdecay}) is analogous to the four-point model for neutron decay, as illustrated in Fig.~\ref{fig:EFT}.
Similarly to the effective description of the weak decay by Enrico Fermi 
with $G_F = \sqrt{2}g^2/8M^2_W$, here the $b\to s\ell\ell$ process can be described with the Wilson coefficients $C_{7,9,10}^{(')}$, which quantify the 
short distance part.
The left-hand side of Fig.~\ref{fig:EFT} illustrates how the effective theory integrates out the heavy propagators of the full SM theory.
%The contribution from the radiative decay $b\to s\gamma$ is quantified with $C_7^{(')}$, where the
%primed coefficient denotes potential right-handed coupling to the quarks.
%The coefficient $C_9$ quantifies the vector coupling to the leptons,
%whereas the $C_{10}$ quantifies their axial coupling. 
%(A left-handed $(V-A)$ coupling to the leptons is therefore equivalent to the coupling
%$(C_9 - C_{10})$). 
The right-hand side of the figure shows the contour plot for potential new physics contributions in a scan of the parameters $C_{10}$ vs $C_9$, where the SM expectation is indicated by the dashed lines. 
A deviation of $C_9$ from the predicted SM value, $\Delta Re(C_9)$, in decays $B^0 \rightarrow K^{*0}\mu^+\mu^-$, $B^+ \rightarrow K^{*+}\mu^+\mu^-$ and $B_s^0 \rightarrow \phi\mu^+\mu^-$, as shown in Fig.~\ref{fig:BKmm-ang3}, is referred to as the rare decay Flavour Anomaly.

%\clearpage

\subsection{Lepton Flavour Universality tests}
Measurements of ratios of processes with different lepton types in the final state 
are theoretically robust, as the hadronic uncertainties cancel, and lepton universality is manifest in the Standard Model.
The comparison of processes with final states electrons to those with muons is, however, experimentally challenging for LHCb due to material interactions of 
electrons leading to large energy loss from Bremsstrahlung.
The comparison of muons to tau-leptons, on the other hand, is hindered by the escaping neutrino in the tau decay.

\begin{figure}[!hb]
    \centering
%trim = <left> <bottom> <right> <top>
    \centering
    \includegraphics[scale=0.14]{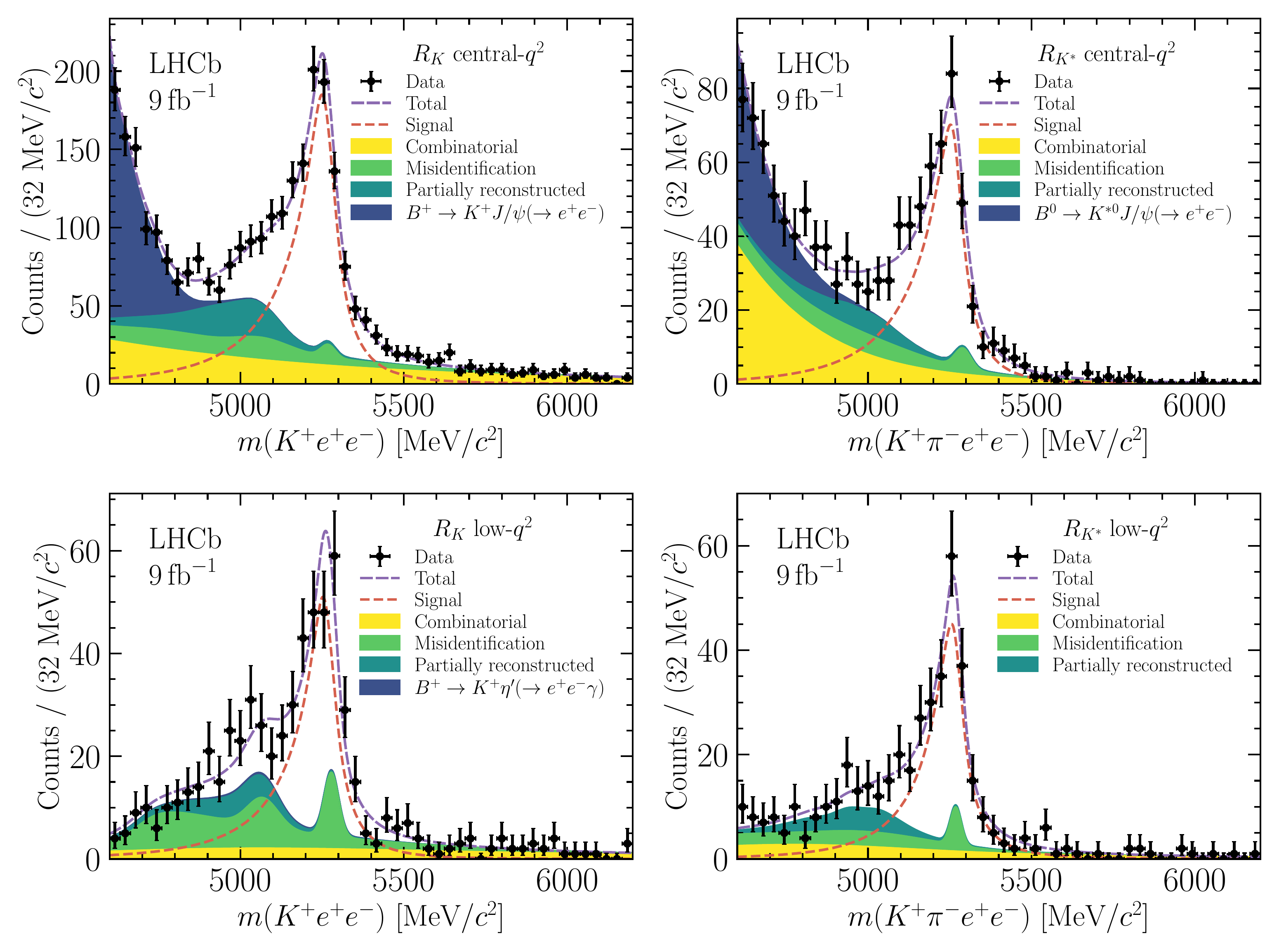} 
     \includegraphics[scale=0.24]{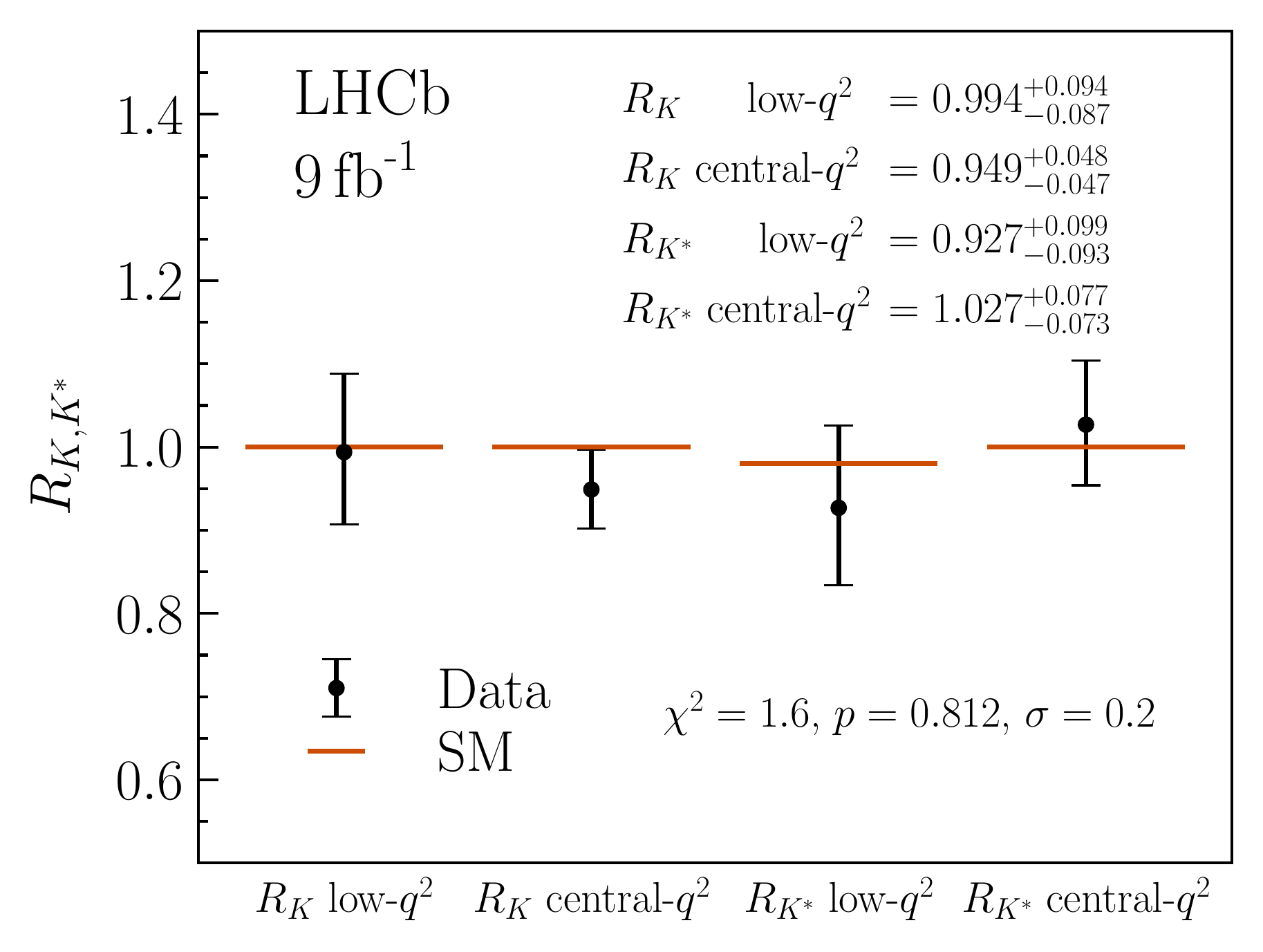}
     \caption{{\em Left:} The invariant mass distributions of $B^+\to K^+ e^+e^-$ decays in the two different $q^2$ regions. {\em Middle:} The invariant mass distributions of $B^0\to K^{*0} e^+e^-$ decays. {\em Right:} The ratio of decay rates with muons or electrons in the final state is consistent with unity, as predicted in the Standard Model~\cite{LHCb-PAPER-2022-046}.}
     \label{fig:b2sll}
\end{figure}

The ratio of decay rates of $b\rightarrow s$ with muons or electrons in the final state is known as
$R_K$,
\begin{equation}
    R_K = \frac{\Gamma(B\to K\mu^+\mu^-)}{\Gamma(B\to Ke^+e^-)}
\end{equation}
and similarly for $R_{K^*}$. The ratio as function of $q^2$ for $R_K$ and $R_{K^*}$ is determined separately in integrated bins of a low $q^2$ region of $0.1<q^2<1.1$~GeV$^2$, and in a central $q^2$ region of $1.1<q^2<6.0$~GeV$^2$.
The theoretical prediction for these four values is close to 1.0, with perhaps an exception of $R_{K^*}$ in the low $q^2$-region, which is slightly below unity due to a small increase of $B^0\to K^{0*}\gamma(\to e^+e^-)$, close to the photon pole.
Fig.~\ref{fig:b2sll} shows the mass distribution for $B^+\rightarrow K^+e^+e^-$ and $B^0\rightarrow K^{*0}e^+e^-$ decays in both the low $q^2$ and high $q^2$ region. In all the cases the background due to misidentified electrons shows a small peak inside the signal region. The right side of the figure graphically shows the results of the $R_{K^{(*)}}$ values. The corresponding mass plots for the muons (not shown) show very little background and are experimentally robust.

\begin{figure}[!ht]
    \begin{picture}(450,180)(0,0)
        \put(0,60){\includegraphics[scale=0.75]{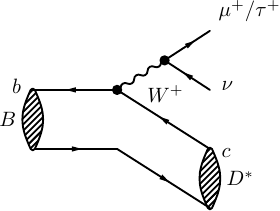}}
        \put(115,20){\includegraphics[scale=0.75]{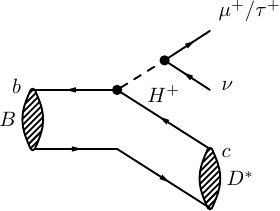}}
        \put(115,100){\includegraphics[scale=0.75]{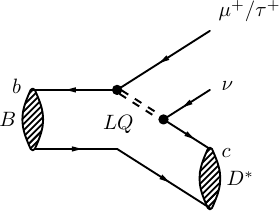}}
        \put(230,0){\includegraphics[scale=0.45]{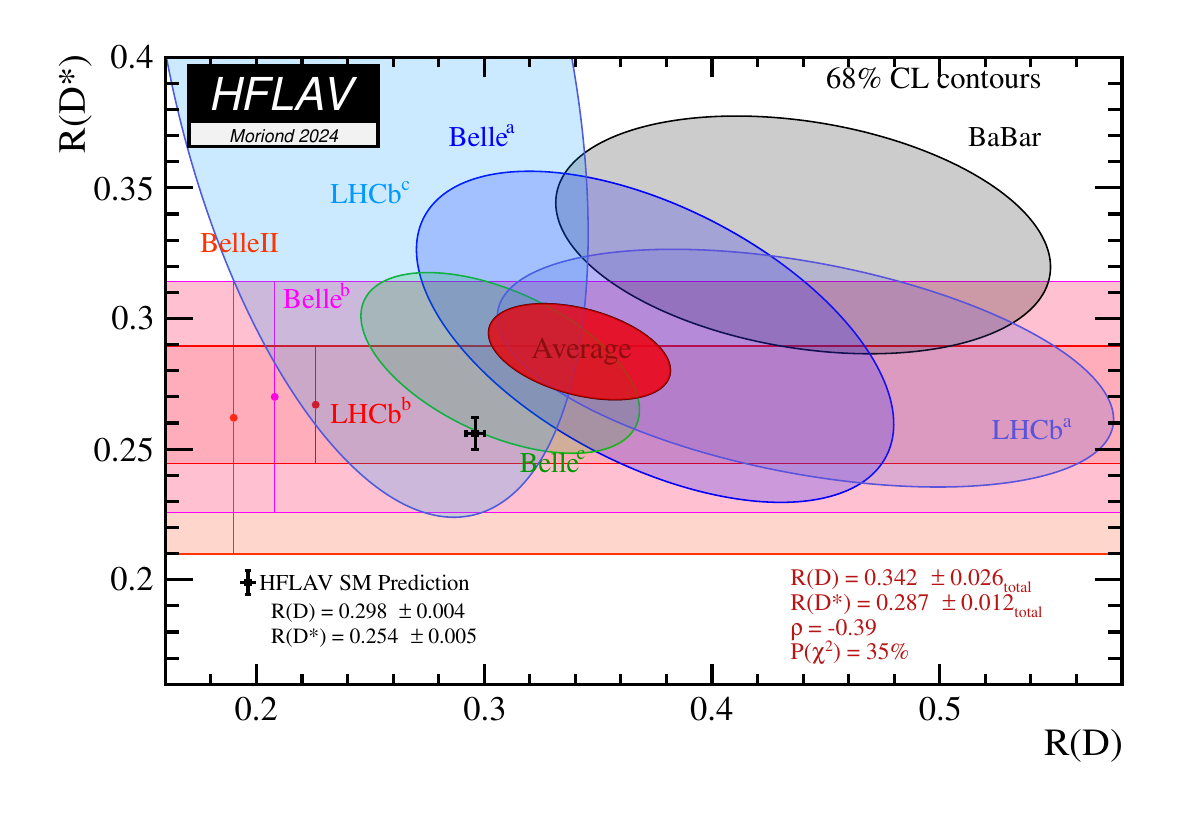}}
        \end{picture}
     \caption{{\em Left:} Feynman diagrams for $B\rightarrow D^*\bar{l}\nu$ with SM $W$-exchange on the left and BSM options of a leptoquark or a charge Higgs exchange on the right. {\em Right:} Lepton universality contour of $R(D^*)$ vs $R(D)$ in semileptonic decays to muons and $\tau$-leptons is measured by the Belle, BaBar and LHCb experiments~\cite{HeavyFlavorAveragingGroupHFLAV:2024ctg}.}
     \label{fig:rd}
\end{figure}

Lepton universality is also probed using $b\rightarrow c$ decays by comparing tree-level processes such as the decays $B\to D^{(*)}\mu^+\nu$ and $B\to D^{(*)}\tau^+\nu$, 
where the $\tau$-lepton can decay muonically ~\cite{LHCb-PAPER-2022-039,LHCb-PAPER-2024-007} or hadronically ~\cite{LHCb-PAPER-2022-052}.
Fig.~\ref{fig:rd} shows the SM Feynman diagram together with candidate NP physics diagrams with a charged Higgs or a leptoquark.
The ratio $R(D^{(*)}$
\begin{equation}
    R(D^{(*)}) = \frac{\Gamma(B\to D^{(*)}\tau^+\nu)}{\Gamma(B\to D^{(*)}\mu^+\nu)}
\end{equation}
is predicted to be smaller than unity due to large mass of the $\tau$-lepton compared to the muon, but is known to the percent level in the Standard Model.
%0.300 (0.252) to an accuracy of about 3(1)\%. 
The measurement combines decays of $\tau\rightarrow\mu\bar{\nu}_\mu \nu_\tau$ together with $\tau\rightarrow 3\pi\nu_\tau$ and is experimentally challenging to isolate the signal decays from backgrounds due to the missing neutrinos in the final state.
The resulting average value combining results from BaBar, Belle, BelleII and LHCb hint at a $\sim$3$\sigma$ separation from the Standard model~\cite{HeavyFlavorAveragingGroupHFLAV:2024ctg}.

%\clearpage
\subsection{Lepton Flavour Violation tests}
%--------------------------------------------

%\subsubsection{$B\to X e\mu$}
Lepton flavor is conserved in the Standard Model, which 
forbids decays such as $\mu\to e\gamma $. Any experimental observation of
lepton flavour violation would imply the presence of particles or interactions beyond the Standard Model and would open an intriguing view into the mechanisms of the three families of fermions.
An overview of the LFV searches are given in table~\ref{tab:LFV}.
The experimental searches with a $\tau$ lepton in the final state are significantly worse due to experimental difficulties from 
the missing neutrino(s) and from the branching fraction of the $\tau$ decay,
hence reducing the sensitivity to new physics.

% https://inspirehep.net/literature?sort=mostrecent&size=25&page=1&q=cn%20lhcb%20and%20t%20lepton%20violating
\begin{table}[!hb]
    \centering
    \begin{tabular}{llrr}
Category& Decay mode     & Limit at 90\% CL (LHCb) & World best limit \\ 
        \hline
 $\tau$ & $\tau^+\to \mu^+\mu^-\mu^+$   &  $<4.6 \times 10^{-8}$ \cite{LHCb-PAPER-2014-052}&  $<2.1 \times 10^{-8}$ \cite{Hayasaka:2010np} \\
        & $\tau^+\to p\mu^-\mu^+$       &  $<34 \times 10^{-8}$ \cite{LHCb-PAPER-2013-014} &  $<1.8 \times 10^{-8}$ \cite{Belle:2020lfn}   \\
        & $\tau^+\to p\mu^-\mu^-$       &  $<46 \times 10^{-8}$ \cite{LHCb-PAPER-2013-014} &  $<4.0 \times 10^{-8}$ \cite{Belle:2020lfn}   \\
        \hline
Same-sign&$B^+\to K^- \mu^+\mu^+$       & $<41  \times 10^{-9}$ \cite{LHCB-PAPER-2011-009} &  - \\
        & $B^+\to \pi^-\mu^+\mu^+$      & $<4.0 \times 10^{-9}$ \cite{LHCB-PAPER-2011-009} &  - \\
        \hline
$e\mu$  & $B^0\to K^{*0} e^+\mu^-$      & $<6.8 \times 10^{-9}$ \cite{LHCb-PAPER-2022-008} &  - \\
        & $B^0\to K^{*0} e^-\mu^+$      & $<5.7 \times 10^{-9}$ \cite{LHCb-PAPER-2022-008} &  - \\
        & $B^0\to K^{*0} e^\pm\mu^\mp$  & $<10.1 \times 10^{-9}$ \cite{LHCb-PAPER-2022-008}&  - \\
        & $B^0_s\to \phi e^\pm\mu^\mp$  & $<16.0 \times 10^{-9}$ \cite{LHCb-PAPER-2022-008}&  - \\
        & $B^0  \to e^\pm\mu^\mp$       & $<1.0 \times 10^{-9}$ \cite{LHCB-PAPER-2017-031} &  - \\
        & $B^0_s\to e^\pm\mu^\mp$       & $<6.0 \times 10^{-9}$ \cite{LHCB-PAPER-2017-031} &  - \\
        \hline
$\tau\mu$& $B^0\to K^{*0} \tau^+\mu^-$   & $<1.0 \times 10^{-5}$  \cite{LHCb-PAPER-2022-021}&  - \\
        & $B^0\to K^{*0} \tau^-\mu^+$   & $<0.82 \times 10^{-5}$ \cite{LHCb-PAPER-2022-021}&  - \\
        & $B^0_s\to \phi\tau^\pm\mu^\mp$& $<1.0 \times 10^{-5}$ \cite{LHCb-PAPER-2024-006} &  - \\
        & $B^0  \to \tau^\pm\mu^\mp$    & $<1.2 \times 10^{-5}$ \cite{LHCb-PAPER-2019-016} &  - \\
        & $B^0_s\to \tau^\pm\mu^\mp$    & $<3.4 \times 10^{-5}$ \cite{LHCb-PAPER-2019-016} &  - \\
        \hline
$p\mu$  & $B^0\to p\mu^-$               & $<2.6 \times 10^{-9}$ \cite{LHCb-PAPER-2022-022} &  - \\
        & $B^0_s\to p\mu^-$             & $<12.1 \times 10^{-9}$ \cite{LHCb-PAPER-2022-022}&  - \\
        \hline  
$D^0$   & $D^0  \to e^\pm\mu^\mp$       & $<1.3 \times 10^{-8}$ \cite{LHCB-PAPER-2015-048} &  - \\            
    \end{tabular}
    \caption{Overview of various lepton flavour violating decay modes searched for at LHCb. Apart from the limits on 
        the $\tau$ decay modes, these are all world best limits.}
    \label{tab:LFV}
\end{table}

%\clearpage
%\input{Semileptonics}

\clearpage

\section{Spectroscopy}\label{secSpectroscopy}
\begin{wrapfigure}{r}{6.0cm}
	\centering
    \vspace*{-0.5cm}
    \includegraphics[scale=0.24]{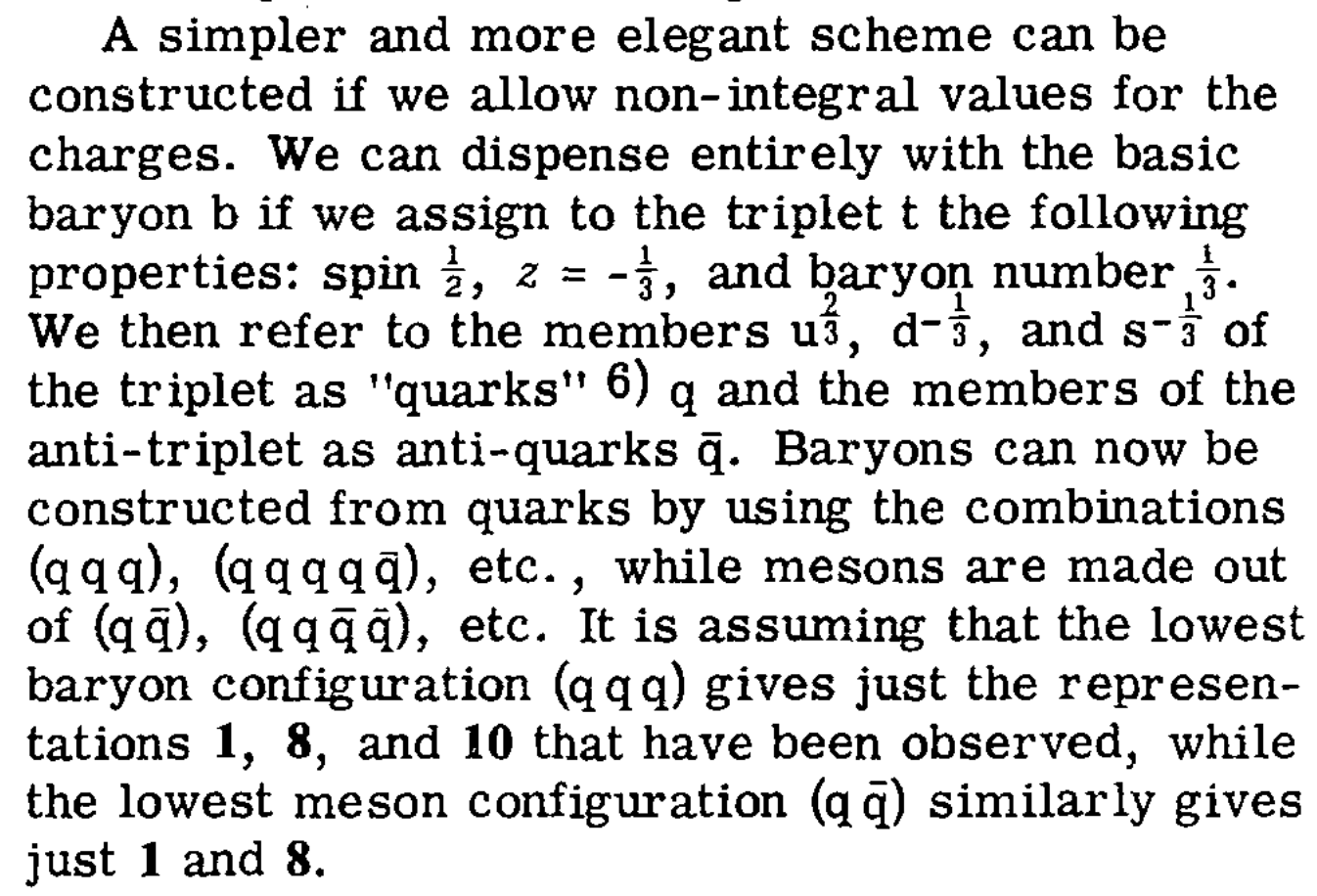}
%%    \begin{picture}(450,180)(0,0)
%%        \put(250,0){\includegraphics[scale=0.37]{Figures/RareDecays/C9C10Plot.pdf}}
%%    \end{picture}
	\caption{The quark model was proposed by Gell-Mann in 1964, hypothesizing the existence of tetra- and pentaquarks~\cite{Gell-Mann:1964ewy}.}
	\label{fig:GellMann}
\end{wrapfigure}

\begin{figure}[!b]
	\centering
    \includegraphics[scale=0.6]{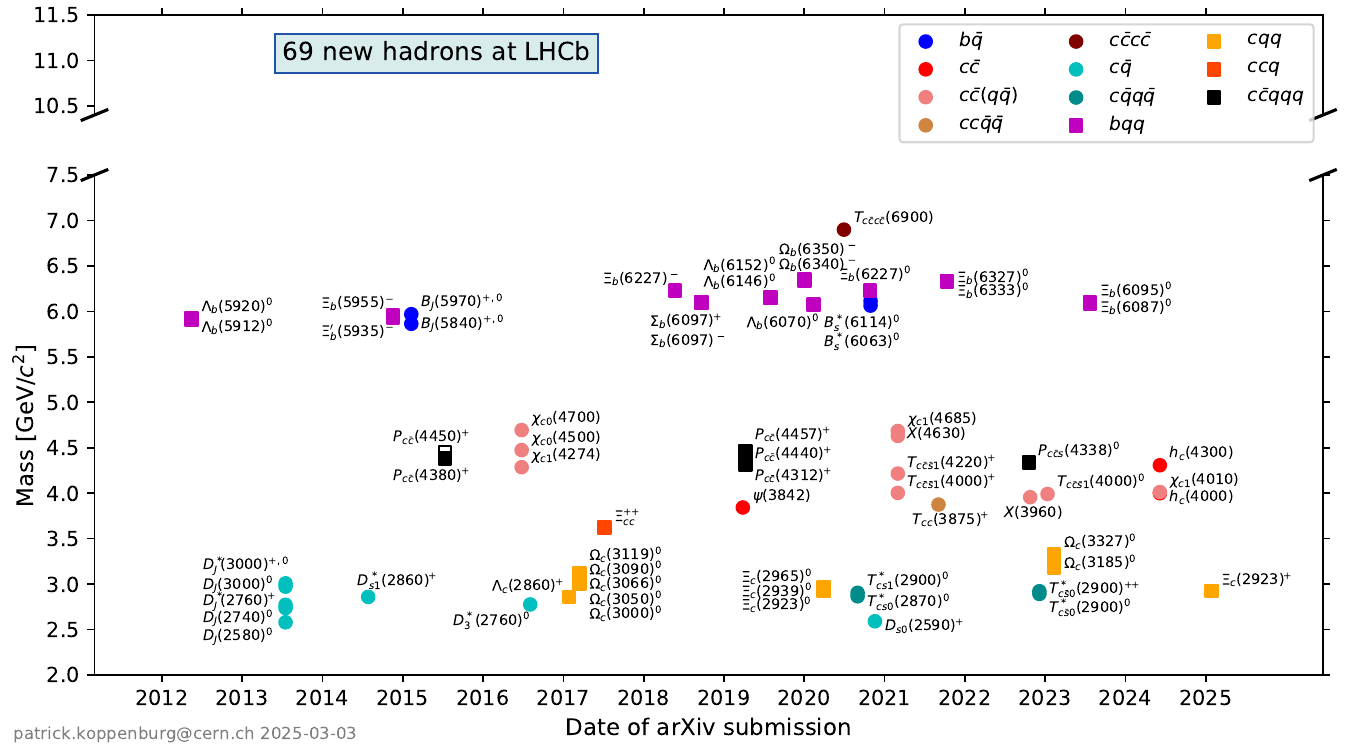}
	\caption{ 
    Overview of the sequence of newly observed hadrons by LHCb~\cite{LHCb-FIGURE-2021-001}.}
	\label{fig:newbhadrons}
\end{figure}

The perturbative dynamics of quarks and gluons is described by Quantum Chromodynamics (QCD), but breaks down for low energies where
the value of the coupling constant $\alpha_s$ becomes large.
The study of hadrons is largely driven by experimental observations, and started with the proposal of the quark model by Gell-Mann and Zweig~\cite{Gell-Mann:1964ewy,Zweig:1964jf}.
The spectrum of mesons and baryons show a regular pattern,
similar to atomic spectroscopy.
The simplest version of the quark model assumes 
bound states of strongly interacting quarks, representing the minimal quark content of the hadrons.
Until recently, the model of two- or three-quark bound states 
described well all the observations of mesons and baryons.
Lately, the picture is extended to include tetra- and pentaquarks, which might be tighly-bound states of quarks, or loosely-bound molecular states, or a combination.
The idea of such states goes back to 1964, in the quark model paper of Gell-Mann~\cite{Gell-Mann:1964ewy}, see Fig.~\ref{fig:GellMann}.
LHCb naturally studies such states including heavy charm or beauty quarks, which allow precise spectroscopy studies due to their relatively narrow widths.
Fig.~\ref{fig:newbhadrons} gives an overview of all the new states that have been discovered by LHCb. It includes classic mesonic and baryonic states, as well as tetraquark and pentaquark states.

\subsection{Classic spectroscopy}
The study of excited charm and beauty hadrons in LHCb delivered a wealth of newly discovered states, see Fig.~\ref{fig:newbhadrons}.
Numerous excited neutral and charged $D$ meson states, as well as excited $D_s^+$ and $\Lambda_c$ charm hadrons have been found.
Furthermore, three excited $\Xi_c^*$ baryons ~\cite{LHCb-PAPER-2020-004}
%$\Xi_c(2923)^0$ $\Xi_c(2939)^0$ $\Xi_c(2965)^0$
and  five excited $\Omega_c^*$ baryons~\cite{LHCb-PAPER-2017-002}
have been observed. These five states were predicted by the naive quark model, although the decay widths of two of these states are
surprisingly narrow.

\begin{figure}[!ht]
	\centering
    \hspace*{-0.3cm}
    \includegraphics[scale=0.19]{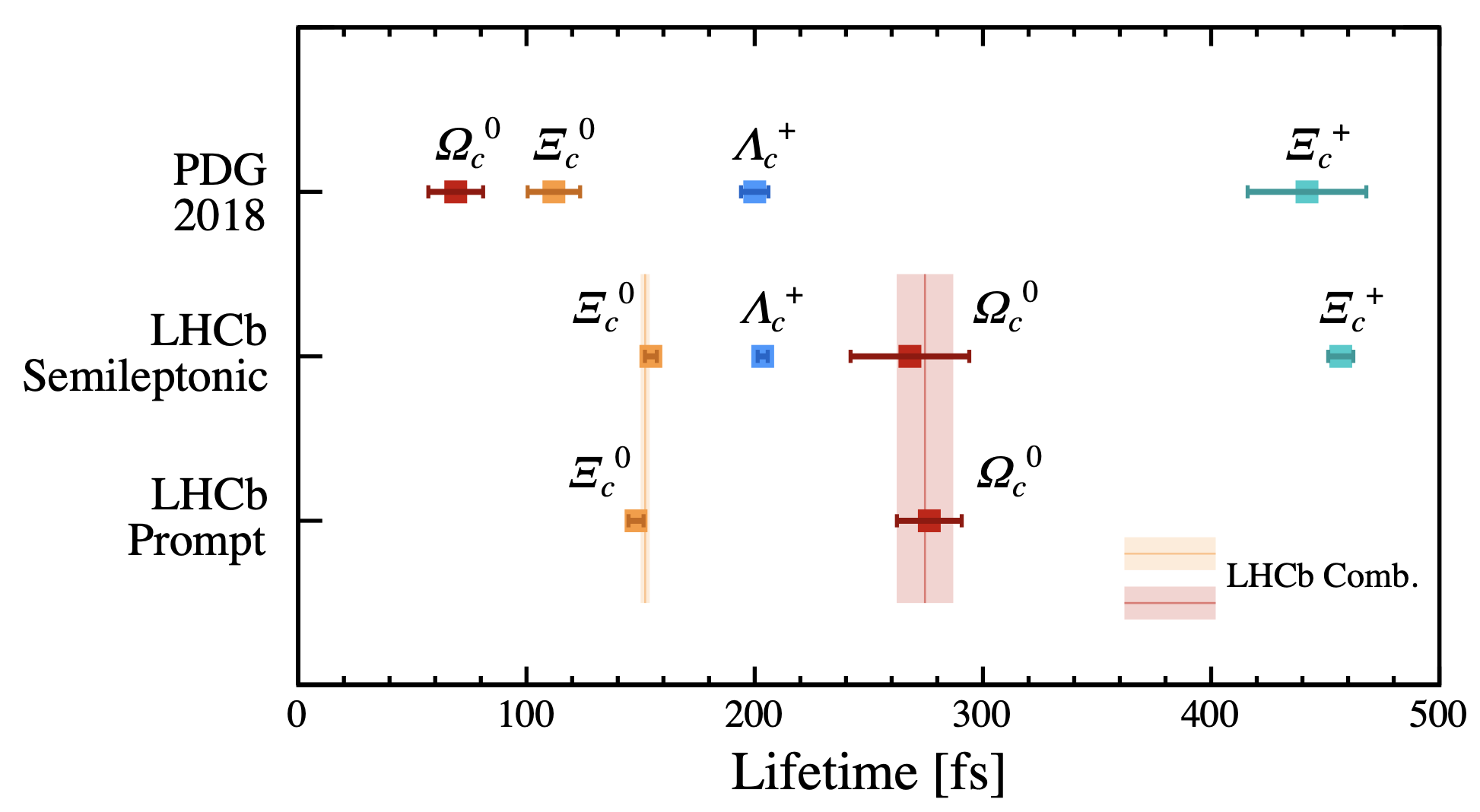}
    \hspace{0.1cm}
    \includegraphics[scale=0.24]{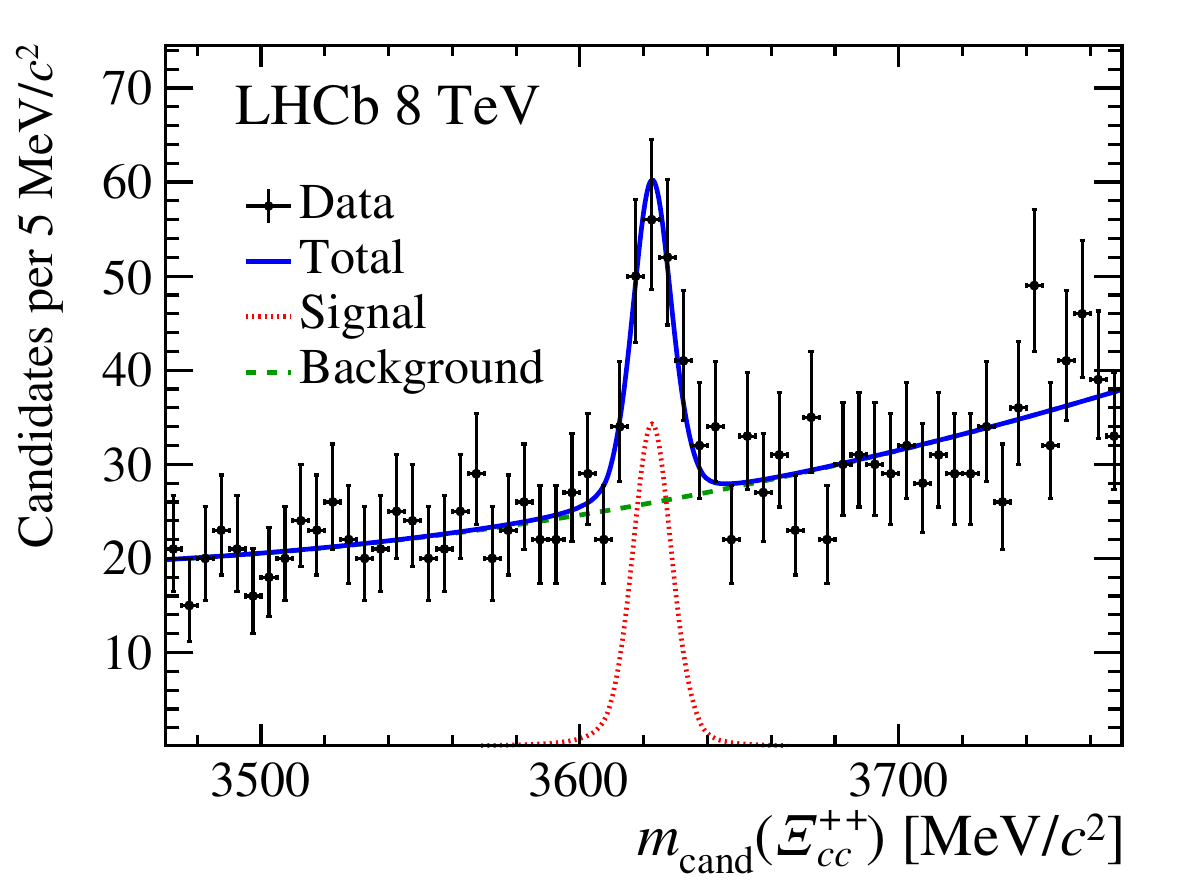}
    \hspace*{-0.2cm}
    \raisebox{0.2cm}
    {\includegraphics[scale=0.25]{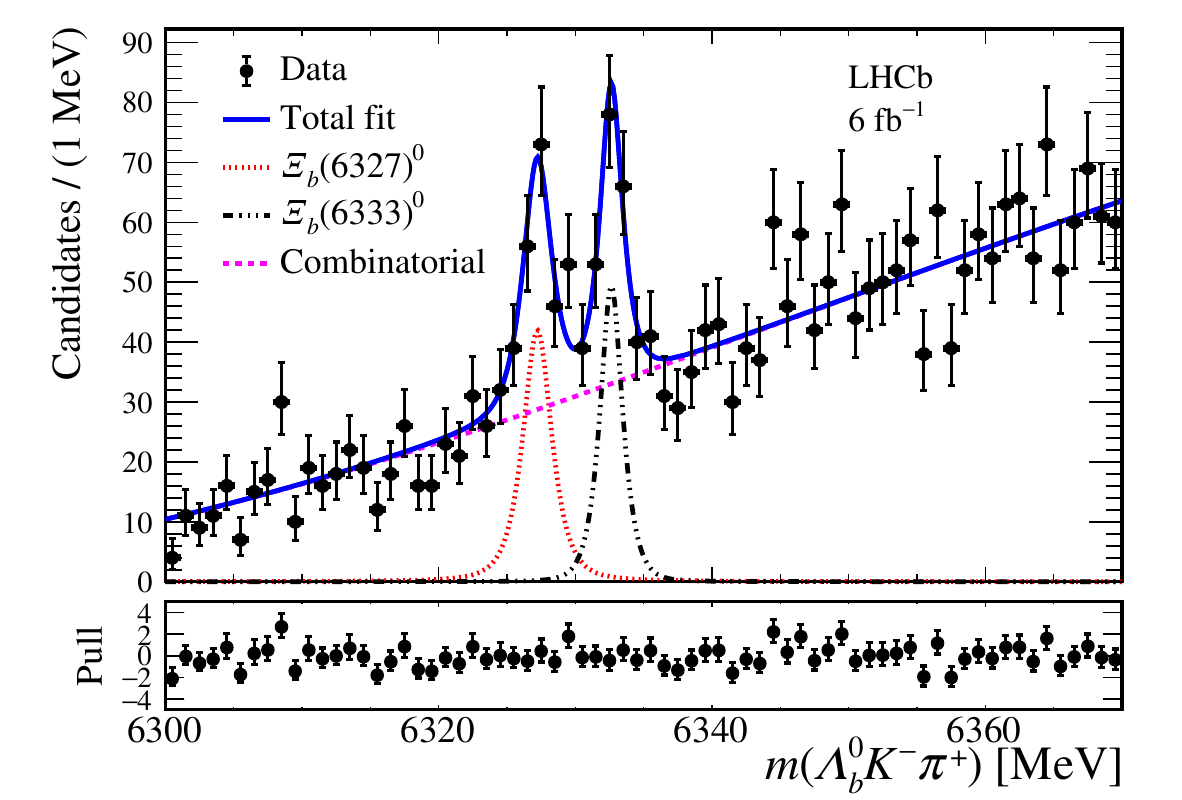}}
 \caption{
    %~\cite{LHCb-PAPER-2018-026}. 
    {\em Left:} Comparison of measured lifetimes from various charm baryons, showing the large shift of the  $\Omega_c^0$ lifetime observed by LHCb compared to the world average value~\cite{LHCb-PAPER-2021-021}. 
    {\em Center:} Invariant mass distribution of doubly charmed baryon $\Xi_{cc}^{++}$ candidates with minimal quark content $(ucc)$ observed in the decay $\Xi_{cc}^{++}\to \Lambda_c^+K^-\pi^+\pi^+$~\cite{LHCb-PAPER-2017-018}.
    {\em Right:} Invariant mass spectrum of $\Xi_b^0\to \Lambda_b^0K^-\pi^+$ decays showing the presence of $\Xi_b^0(6327)$ and  $\Xi_b^0(6333)$states.}
    %~\cite{LHCb-PAPER-2021-025}.}
	\label{fig:newcharmhadrons}
\end{figure}

In addition to the observation of new states, the long-lasting 
experimental puzzle of the $\Omega_c^0$ lifetime has been
settled with consistent measurements from
promptly produced $\Omega_c^0 (\to pK^-K^-\pi^+)$ decays, 
and from semileptonic $\Omega_b^+$ decays. As shown in the left panel
of Fig.~\ref{fig:newcharmhadrons}, a significant lifetime shift with respect to the earlier PDG value is observed.
%\textcolor{orange}{hier mis ik kennis om te kunnen begrijpen. Vragen we hier te veel voorkennis of weet ik te weinig?}
%
Furthermore, doubly charmed baryons such as the  $\Xi_{cc}^+$ and $\Xi_{cc}^{++}$ states attract particular interest. These states are theoretically well described, as heavy quark effective theory (HQET) approximations are expected to hold due to the presence of a heavy charm–charm diquark bound to a lighter quark - a configuration that resembles the structure of a hydrogen atom.
On the other hand, the $\Xi_{cc}^+\to \Lambda^+ K^-\pi^+$ decays are also experimentally interesting, given that the 
measurements of the $\Xi_{cc}^+$ mass
%\textcolor{orange}{kunnen we in twee woorden zeggen wat de measurement is? bv width, of lifetime of...?} 
by the SELEX experiment~\cite{SELEX:2004lln}
were not confirmed elsewhere~\cite{LHCb-PAPER-2019-029}.
The middle panel of
Fig.~\ref{fig:newcharmhadrons} shows the observed $\Xi_{cc}^{++}\to \Lambda_c^+K^-\pi^+\pi^+$ resonance by LHCb~\cite{LHCb-PAPER-2017-018}.
%\textcolor{purple}{The right side panel shows an overview of LHCb's observations from semileptonic and promptly produces charmed baryons}.
%

%\begin{figure}[!ht]
%	\centering
%    \includegraphics[scale=0.5]{Figures/Spectroscopy/Masses_conventional_hadrons.pdf}
%	\caption{Newly discovered conventional hadrons at the LHC, of which the majority by LHCb~\cite{pk}.}
%	\label{fig:newhadrons}
%\end{figure}

Similarly, various excited $b$-hadrons were discovered, such as
the $B_1(5721)$, $B_2(5747)$, $B_{s1}(5830)^0$, $B^*_{s2}(5840)^{0}$ and $B_c(2S)^{(*)+}$ mesons.
Furthermore, multiple excited $\Lambda_b^0$ ($udb$ content), $\Sigma_b$ ($uub$, $ddb$ and $udb$ content), $\Xi^0$ ($usb$ and $dsb$ content) and $\Omega_b^0$ ($ssb$ content) baryons have been seen, both with narrow and wide widths. 
The right side of Fig.~\ref{fig:newcharmhadrons} displays the mass spectrum of the $(\Lambda_b^0 K^-\pi^+)$ mass spectrum with two $\Xi_b$ resonances separated by 6 MeV that can be resolved due to the good mass resolution of LHCb. 
%The middle and right side panel of the same figure present an overview of theoretical predictions and experimental observations for $\Lambda_b$, $\Sigma_b$, $\Xi_b$ and $\Xi_b$ baryons.}

%\clearpage
\subsection{Tetraquarks}

The theoretical understanding of the conventional charmonium ($c\bar{c}$) spectrum is well-established, and the properties of many observed charmonium states show good agreement with theoretical predictions.
A notable exception is the $\chi_{c1}(3872)$ state, first discovered by the Belle collaboration. Given that its mass lies above the open charm threshold,
a significantly larger decay width would be expected than the measured value of
$\Gamma_{BW}=1.39 \pm 0.26$~MeV~\cite{LHCb-PAPER-2020-008}.
The proximity of its mass to the $D^0\bar{D}^{0*}$ threshold
has led to speculation that it may be a molecular state, loosely bound by pion exchange.
Investigated by many experiments, it is the most studied exotic hadron, with quantum numbers $J^{PC}=1^{++}$ and isospin zero. However, its mass deviates from expectations by 100 MeV.
The radiative decays $\chi_{c1}(3872)\to J/\psi\gamma$ $\chi_{c1}(3872)\to \psi(2S)\gamma$~\cite{LHCb-PAPER-2024-015}  shed further light on the structure of the state by examining the annihilation of the two light quarks. 
If the $\chi_{c1}(3872)$ were a compact tetraquark, one would also expect a charged partner, unless the isospin-one partners are suppressed.
Conversely, the prompt production in $pp$ collisions challenges the interpretation of the molecular state~\cite{LHCb-PAPER-2021-026}.
Additionaly, the increase of the ratio of cross-sections $\sigma^{\chi_{c1}(3872)}/\sigma^{\psi(2S)}$
from pp to pPb to PbPb collisions indicates that the exotic $\chi_{c1}(3872)$ state experiences different dynamics in the nuclear medium than the conventional charmonium state 
$\psi(2S)$~\cite{LHCb-PAPER-2023-026}.

% Saverio: is this X enhancement or ψ(2s) suppression?
% Nuclear modification factor shows X enhancement → coalescence dominating over breakup?
% Add column with threshold closeness!
\begin{figure}[!ht]
	\centering
    %trim = <left> <bottom> <right> <top>
   \begin{picture}(400,120)(0,0)
        \put(-20,7){\includegraphics[scale=0.31]{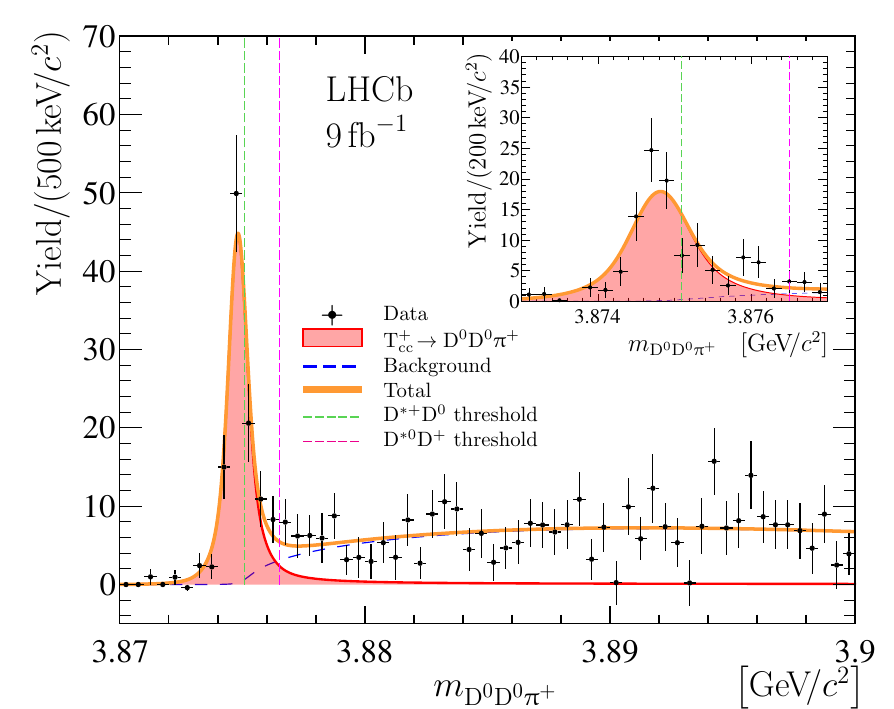}}
        \put(10,115){$T_{cc}^+(3875)\to D^0{D}^0\pi^+$}

%        \put(0,0){\includegraphics[scale=0.9]{Figures/Spectroscopy/fig4a-Tcs2900-paper-2020-024.pdf}}
 %       \put(20,115){$T_{cs0}(2900)^0\to D^+K^-$}
        
        \put(130,2){\includegraphics[scale=0.165]{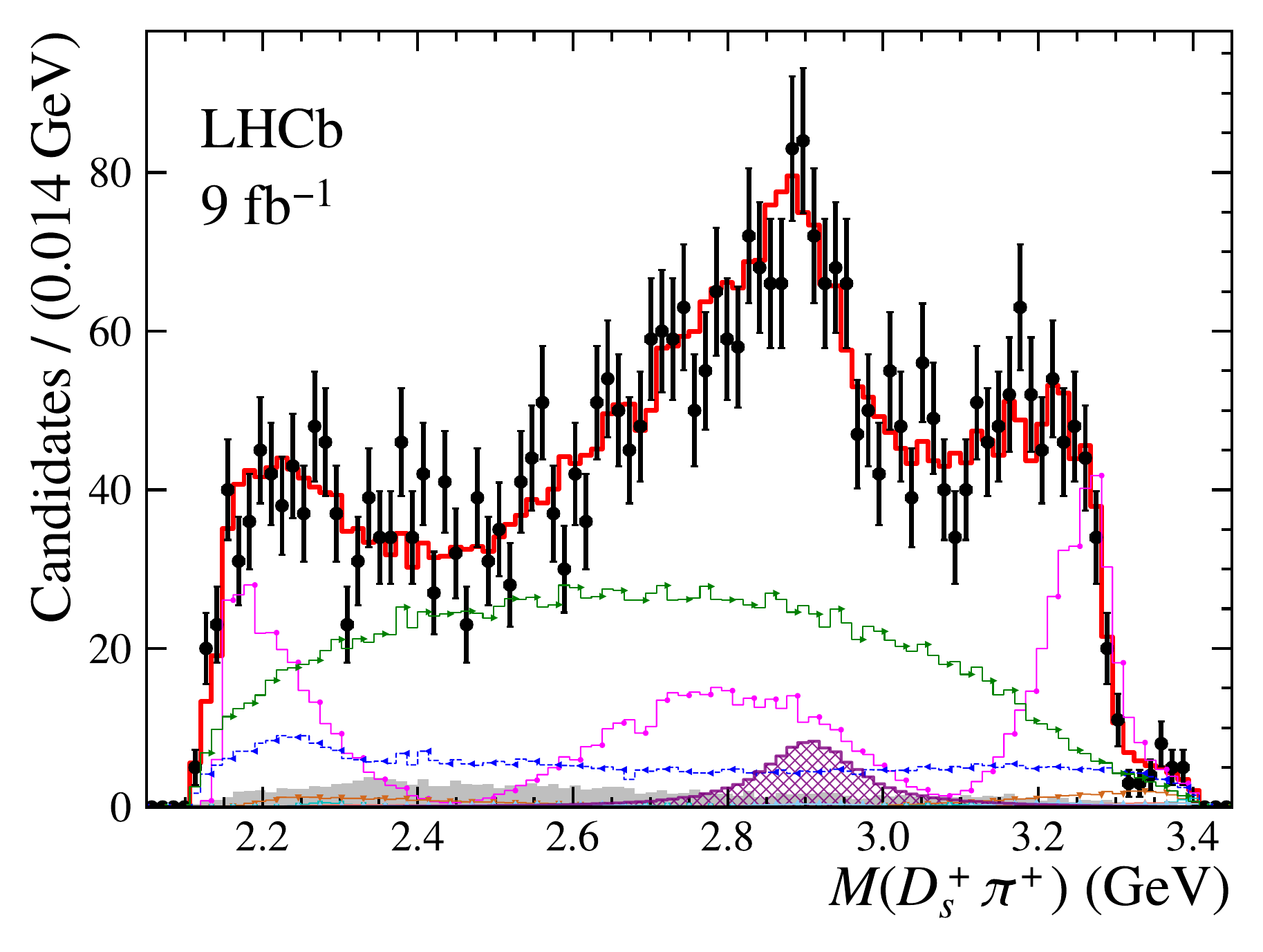}}
        \put(165,115){$T_{c\bar{s}0}(2900)^{++}\to D_s^+\pi^+$}

       \put(300,0){\includegraphics[trim=0.1cm 0.1cm 9.1cm 0.1cm,clip=,scale=0.42]{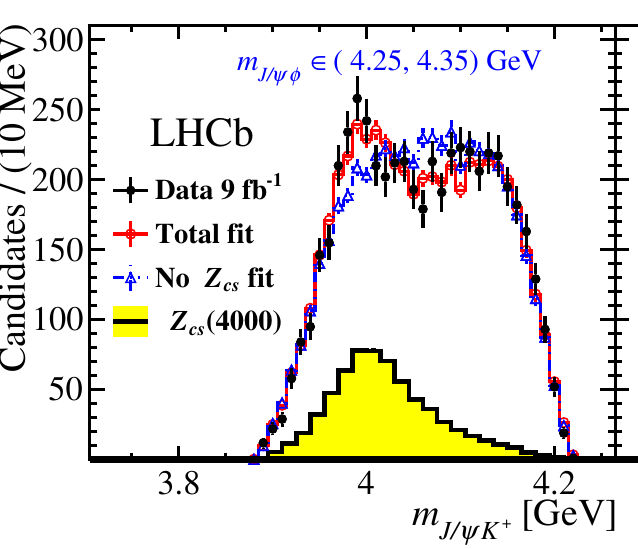}}
        \put(330,115){$T_{c\bar{c}s1}(4000)^+\to J/\psi K^+$}

     \end{picture}
	\caption{Invariant mass distributions showing four examples of manifestly exotic tetraquark states 
    ({\em left}) $T_{cc}^+(3875)$~\cite{LHCb-PAPER-2021-031},
    ({\em middle}) $T_{c\bar{s}0}(2900)^{++}$~\cite{LHCb-PAPER-2022-026}, and
    ({\em right}) $T_{c\bar{c}s1}(4000)^+$~\cite{LHCb-PAPER-2020-044}.
    %$T_{cs0}(2900)^0$~\cite{LHCb-PAPER-2020-024}.
%    \textcolor{orange}{Niels, wat denk je ervan om een van de twee Tcs0(2900) decays weg te laten om wat ruimte te winnen? Dan past het op een regel naast elkaar en de Tcccc (6900) laten we ook niet zien.}
    }
	\label{fig:tetra}
\end{figure}
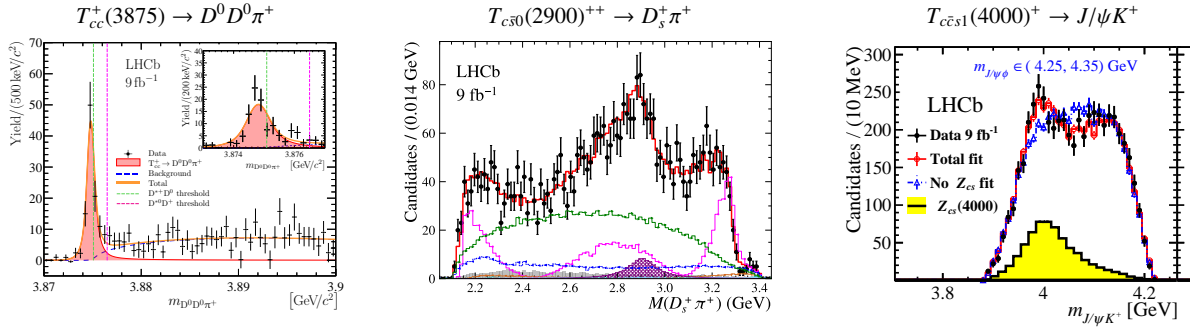

The discovery of the doubly charmed tetraquark for the decay $T_{cc}^+(3875)\to D^0 D^0\pi^+$ triggered large interest, 
because it is remarkably close, but below the $D^0\bar{D}^{+*}$ mass threshold~\cite{LHCb-PAPER-2021-031}.
It is hypothesized that the corresponding state with two beauty quarks could be even further from the 
$BB$ threshold, which would imply that it is stable for the strong interaction, and 
would only decay weakly.

\begin{table}[!bh]
    \centering
    \begin{tabular}{lcrlclc}
        Tetraquark      & Min. quark cont. & Production        & $I(J^{P})$    & Nearby threshold  & Comment        & Ref. \\ 
        \hline
$\chi_{c1}(3872)$            &  $c\bar{c}u\bar{u}$   & $b\to X(J/\psi\pi^+\pi^-)$ & $0(1^{+})$    & $D^{*0}\bar{D}^0$ & First; best studied; narrow width & \cite{LHCb-PAPER-2020-008}\\
$T_{cc}(3875)^+$          &  $cc\bar{u}\bar{d}$   & prompt $(D^0\bar{D}^0\pi^+)$& $0(1^{+})$    & $D^{*+}\bar{D}^{0}, D^{+}\bar{D}^{*0}$& Doubly charmed  & \cite{LHCb-PAPER-2021-031} \\
$T_{c\bar{c}s1}(4000)^+$  &  $c\bar{c}u\bar{s}$   & $B^+\to \phi(J/\psi K^+)$       & $1/2^+(1^{+})$& $D_s^{*+}\bar{D}^{0}, D_s^{+}\bar{D}^{*0}$& Open strangeness& \cite{LHCb-PAPER-2020-044} \\
$T_{cs0}(2900)^0$         &  $c{s}\bar{d}\bar{u}$ & $B^+\to D^+(D^-K^+)$           & $?(0^{+})$    & $D^*K^*$   & Open charm and strangeness & \cite{LHCb-PAPER-2020-024}\\
$T_{c\bar{s}0}(2900)^{0}$ & $ c\bar{s}d\bar{u}$   & $B^0\to \bar{D}^0(D_s^+\pi^-)$     & $1(0^{+})$    & $D^*K^*$   & Isospin multiplet  & \cite{LHCb-PAPER-2022-026} \\
$T_{c\bar{s}0}(2900)^{++}$& $ c\bar{s}u\bar{d}$   & $B^+\to D^-(D_s^+\pi^+)$       & $1(0^{+})$    & $D^*K^*$   & Isospin multiplet; doubly charged & \cite{LHCb-PAPER-2022-026} \\
$T_{c\bar{c}c\bar{c}}(6900)^0$&$c\bar{c}c\bar{c}$ & prompt $(J/\psi J/\psi)$    & ?           & $J/\psi J/\psi$ & Excited states  & \cite{LHCb-PAPER-2020-011}\\
    \end{tabular}
    \caption{A selection of tetraquark states studied at LHCb~\cite{Husken:2024rdk}.}
    \label{tab:tetra}
\end{table}

The first tetraquark with strangeness at LHCb was the observation of $T_{c\bar{c}s1}(4000)^+\to J/\psi K^+$~\cite{LHCb-PAPER-2020-044}, which is manifestly exotic, as it cannot be explained by conventional charmonium states, due to its charge and strangeness quantum numbers. 
Other states with open strangeness are 
$T_{cs0}(2900)^0$~\cite{LHCb-PAPER-2020-024} and 
$T_{c\bar{s}0}(2900)^{++}$~\cite{LHCb-PAPER-2022-026}, which both contain a single charm quark.
Note that the latter contains an anti-strange quark, and that also its isospin partner has been
observed.
The tetraquark with four charm quarks $T_{c\bar{c}c\bar{c}}(6900)^0$ is of particular interest,
because multiple excited states have been observed - including by the ATLAS and CMS experiments - supporting its
interpretation as a tightly bound 4-quark state.
Table~\ref{tab:tetra} lists specific states observed by LHCb and and their decay modes and Fig.~\ref{fig:tetra} shows the observed mass distributions of three Tetraquarks.

%\clearpage
\subsection{Pentaquarks}
% Ref to Ivan! arxiv:2410.06923

\begin{figure}[!ht]
	\centering
    %trim = <left> <bottom> <right> <top>
   \begin{picture}(400,180)(0,0)
        \put(0,10){\includegraphics[scale=0.27]{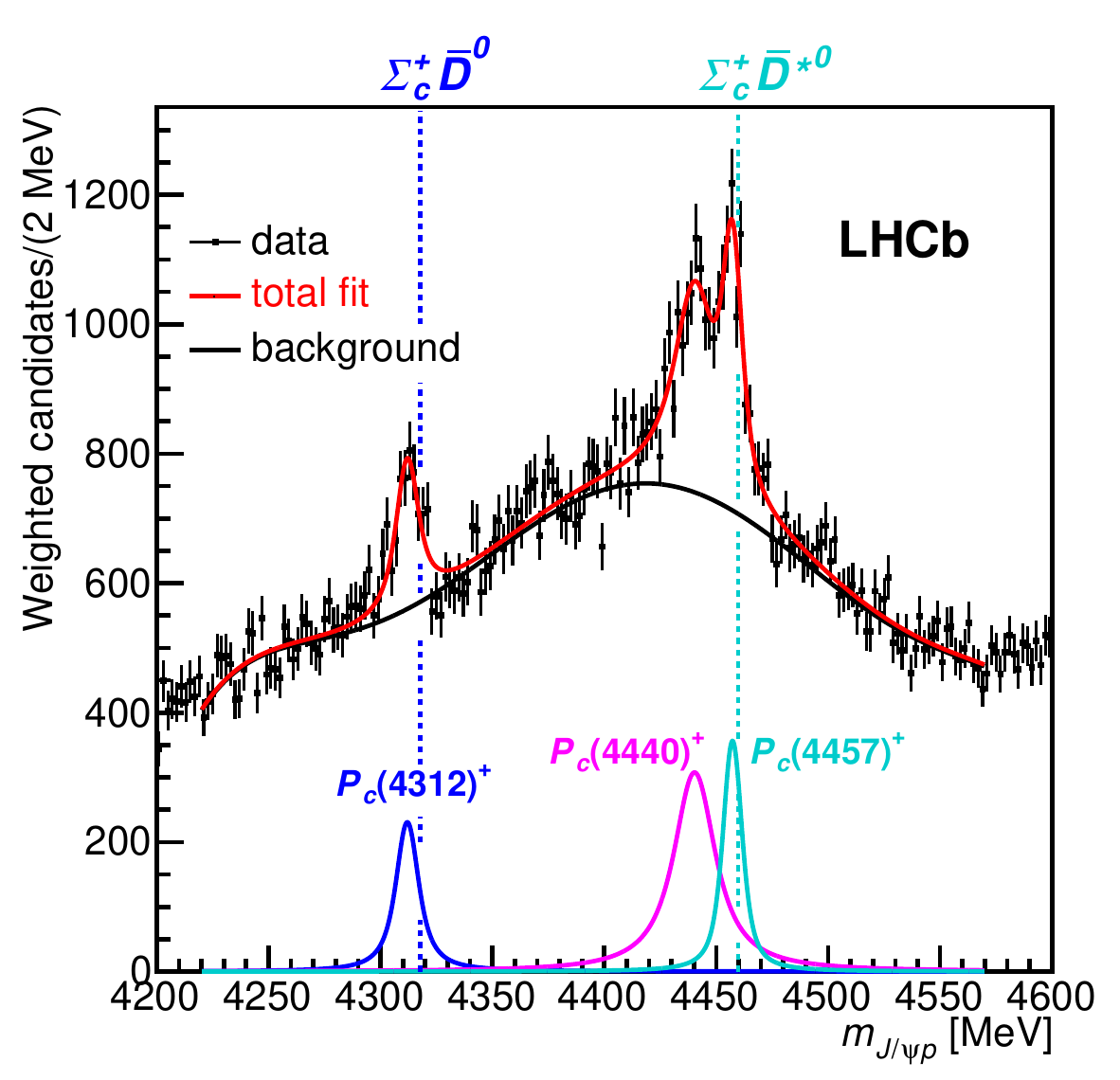}}
        \put(20,155){$P_{c\bar{c}}(4312)^+\to J/\psi p$}
        \put(200,0){\includegraphics[scale=0.13]{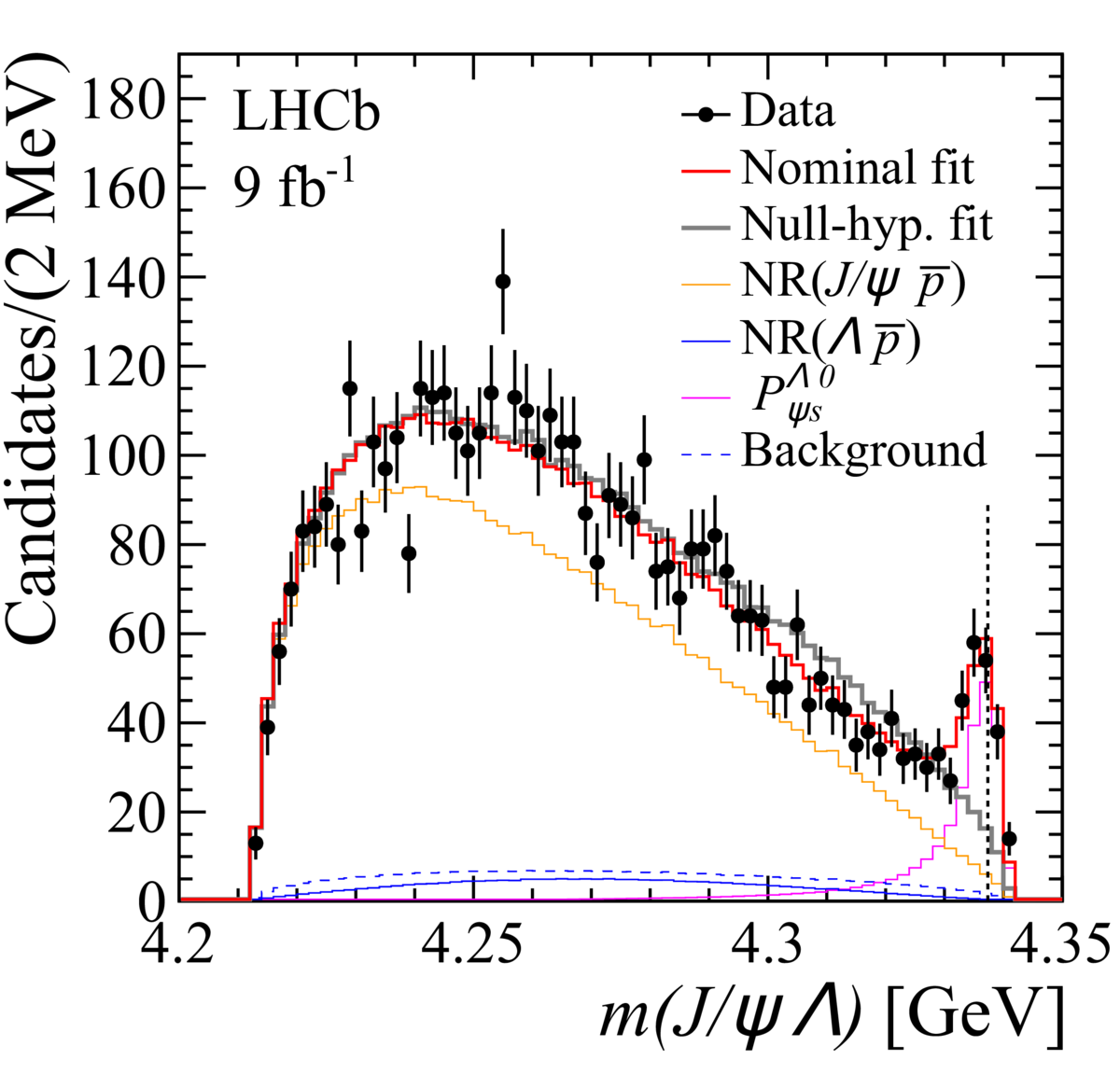}}
        \put(220,155){$P_{c\bar{c}s}(4338)^+\to J/\psi \Lambda$}
     \end{picture}
	\caption{Invariant mass distributions showing examples of pentaquark states 
   {\em (left)} $P_{c\bar{c}}(4312)^+$~\cite{LHCb-PAPER-2019-014}   and
   {\em (right)} $P_{c\bar{c}s}(4338)^+$~\cite{LHCb-PAPER-2022-031}.
    }
	\label{fig:penta}
\end{figure}

Similar resonances have been discovered in final states containing a baryon and a meson, rather than two mesons. These states thus carry baryon number and have a minimal quark content of five quarks, consistent with a pentaquark interpretation.
The first discovered pentaquarks were the $P_{c\bar{c}}(4312)^+$ state and a heavier state close to 
$4450$~MeV~\cite{LHCb-PAPER-2015-029}, both identified with a
$J/\psi p$ in the final state. The latter was later resolved into two distinct states 
$P_{c\bar{c}}(4440)^+$ and 
$P_{c\bar{c}}(4457)^+$~\cite{LHCb-PAPER-2019-014}.
These states lie close to the $\Sigma_c^+ D$ mass threshold, suggesting a molecular structure.
More recently, a pentaquark state with strangeness, $P_{c\bar{c}s}(4338)^+$, has been discovered
\cite{LHCb-PAPER-2022-031}, close to the $J/\psi \Lambda$ mass threshold.
Interestingly, a molecular interpretation of the $P_{cc}(4337)^+$, discovered in $B_s^0\to J/\psi \bar{p}p$ decays, is unlikely as it is
about 40 MeV above the $\Lambda_c^+ D^{*0}$ threshold.
The invariant mass spectrum of $J/\psi p$ and $J/\psi \Lambda$, showing the pentaquark resonance structures, are shown in Fig.~\ref{fig:penta}. Table~\ref{tab:penta} gives the properties for a selected number of pentaquark states observed in LHCb. 

\begin{table}[h!]
    \centering
    \begin{tabular}{lcrcclc}
        Pentaquark       & Min. quark content & Production             & $J^{P}$   & Nearby threshold        & Comment            & Ref. \\ 
        \hline
        $P_{c\bar{c}}(4312)^+$ & $c\bar{c}uud$ & $\Lambda_b^0\to K^-(J/\psi p)$  & $(3/2)^-$ & $\Sigma_c^+\bar{D}^0$   & First              & \cite{LHCb-PAPER-2015-029} \\
        $P_{c\bar{c}}(4440)^+$ & $c\bar{c}uud$ & $\Lambda_b^0\to K^-(J/\psi p)$  & $(5/2)^+$ & $\Sigma_c^+\bar{D}^{*0}$& Two-peak structure & \cite{LHCb-PAPER-2019-014} \\
        $P_{c\bar{c}}(4457)^+$ & $c\bar{c}uud$ & $\Lambda_b^0\to K^-(J/\psi p)$  & $(5/2)^-$ & $\Sigma_c^+\bar{D}^{*0}$& Two-peak structure & \cite{LHCb-PAPER-2019-014} \\
    \hline
        $P_{c\bar{c}}(4337)^+$ & $c\bar{c}uud$ & $B_s^0\to \bar{p}(J/\psi p)$ & $(1/2)^?$ &           ?                &Also $J/\psi \bar{p}$&\cite{LHCb-PAPER-2021-018} \\
    \hline
        $P_{c\bar{c}s}(4338)^0$ & $c\bar{c}uds$ & $B^-\to \bar{p}(J/\psi\Lambda)$& $(1/2)^-$ & $\Xi_c^+ D^-$           & Strangeness        & \cite{LHCb-PAPER-2022-031} \\
        $P_{c\bar{c}s}(4458)^0$ & $c\bar{c}uds$ & $\Xi_b^-\to K^-(J/\psi\Lambda)$& $(1/2)^-+ (3/2)^-?$ & $\Xi_c^+ \bar{D}^{*0}$& Strangeness, two-peak?  & \cite{LHCb-PAPER-2022-031} \\
    \end{tabular}
    \caption{A selection of pentaquark states studied at LHCb~\cite{Husken:2024rdk}.}
    \label{tab:penta}
\end{table}

%\subsection{Hexaquarks}

%Contrary to popular belief, Lorem Ipsum is not simply random text. It has roots in a piece of classical Latin literature from 45 BC, making it over 2000 years old. Richard McClintock, a Latin professor at Hampden-Sydney College in Virginia, looked up one of the more obscure Latin words, consectetur, from a Lorem Ipsum passage, and going through the cites of the word in classical literature, discovered the undoubtable source. Lorem Ipsum comes from sections 1.10.32 and 1.10.33 of "de Finibus Bonorum et Malorum" (The Extremes of Good and Evil) by Cicero, written in 45 BC. This book is a treatise on the theory of ethics, very popular during the Renaissance. The first line of Lorem Ipsum, "Lorem ipsum dolor sit amet..", comes from a line in section 1.10.32.

\clearpage
\section{Long-Lived Particle Searches}\label{secLongLived}

\begin{figure}[!ht]
    \begin{picture}(450,85)(0,0)
    \put(50,48){\includegraphics[scale=0.65]{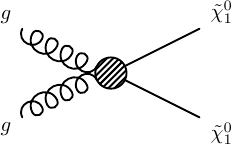}}
    \put(70,2){\includegraphics[scale=0.65]{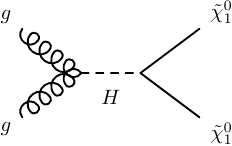}}
    \put(180,48){\includegraphics[scale=0.65]{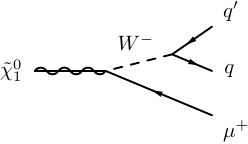}}
    \put(200,2){\includegraphics[scale=0.65]{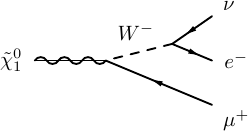}}
    \put(310,48){\includegraphics[scale=0.65]{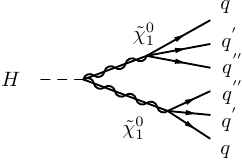}}
    \put(330,2){\includegraphics[scale=0.65]{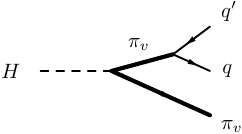}}
    \end{picture}
	\caption{
    {\em Left:} Production mechanism of long-lived particles (LLP) through direct pair production, or through the decay of a SM-like Higgs. 
    {\em Middle:} Decay of LLP to leptons~\cite{LHCb-PAPER-2020-027} or quarks~\cite{LHCb-PAPER-2021-028}. 
    {\em Right:} LLPs are also searched for originating from SM-like Higgs decay, and decaying to three quarks each~\cite{LHCb-PAPER-2016-014}, or to two neutral $\pi_v$ particle in hidden 
    valley (HV) models~\cite{LHCb-PAPER-2014-062}.
    }
	\label{fig:LLPdiagrams}
\end{figure}

Various new physics models predict the existence of massive particles with lifetimes long enough to produce macroscopic decay displacements within the LHCb detector, the so-called Long-Lived Particles, LLPs. Examples of these models include "hidden-valley" pions \cite{Strassler:2006im}, heavy neutral leptons \cite{Minkowski:1977sc}, dark photons \cite{Fayet:1980rr}, and supersymmetric neutralinos \cite{Graham:2012th}, see Fig.~\ref{fig:LLPdiagrams}.
LHCb has contributed~\cite{Alimena:2019zri} to the search of long lived particles with the following signatures:
\begin{itemize}
    \item {\em An LLP, here typically a neutralino, decaying semileptonically leading to two charged tracks accompanied by an isolated high $p_T$ muon.} 
    Two mechanisms for these are included. In the first, a Higgs-like boson with mass from 30 to 200 GeV/$c^2$ is produced by gluon fusion and decays into two LLPs. The analysis covers LLP mass values from 10 GeV/$c^2$ up to about one half the Higgs-like boson mass. In the second, the LLP production mode is directly from quark interactions, with LLP masses from 10 to 90 GeV/$c^2$. The LLP lifetimes considered range from 5 to 200 ps. The study uses LHCb data collected from proton-proton collisions at $\sqrt{s}=13$ TeV, corresponding to an integrated luminosity of 5.4 fb$^{-1}$ \cite{LHCb-PAPER-2021-028}. 
    \item {\em A displaced vertex decaying into $e^\pm\mu^\mp\nu$ final state.}
    The search is performed using an integrated luminosity of 5.4 fb$^{-1}$ at $\sqrt{s}=13$ TeV and covers LLP masses from 7 to 50 GeV/$c^2$, and lifetimes from 2 to 50 ps \cite{LHCb-PAPER-2020-027}.

    \item {\em A single long-lived particle decaying into two associated jets.}
    A search for long-lived hidden valley pion candidates with a mass between 25 and 50 GeV/$c^2$ and a lifetime between 1 and 200 ps, using an integrated luminosity of 0.62 fb$^{-1}$, at $\sqrt{s}=13$ at 7 TeV\cite{LHCb-PAPER-2014-062}. 
    \item {\em Two displaced high multiplicity vertices.}
    A search for massive long-lived particles, in the 20–60 GeV/c$^2$ mass range with lifetimes between 5 and 100 ps. The particles are assumed
    to be pair-produced by the decay of a Higgs-like boson with mass between 80 and 140 GeV/c$^2$
    The dataset used corresponds to 0.62 fb$^{-1}$ of data collected at $\sqrt{s}=7$ TeV \cite{LHCb-PAPER-2016-014}.

    \item {\em A low mass resonance in prompt and non-prompt di-muon pairs.}
    Searches are performed for both prompt and long-lived dark photon candidates decaying to two muon decays using 5.5 fb$^{-1}$ at 13 TeV. The long-lived search places world-leading constraints on low-mass dark photons with lifetimes of order 1 ps  \cite{LHCb-PAPER-2019-031}. 
\end{itemize}

%\vspace*{-0.3cm}

\begin{figure}[h!]
    \centering
    \includegraphics[width=7.3cm]{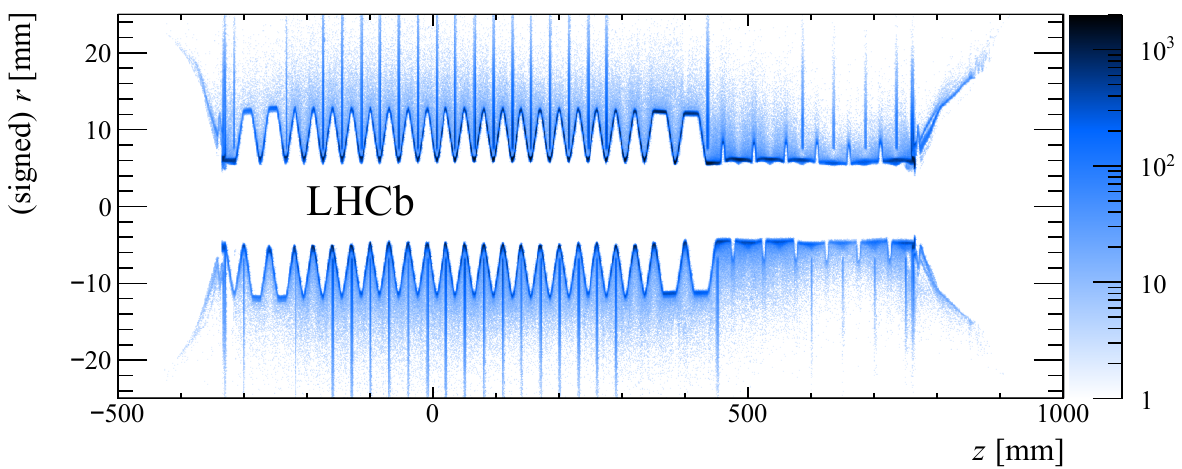}
    \includegraphics[width=4.3cm]{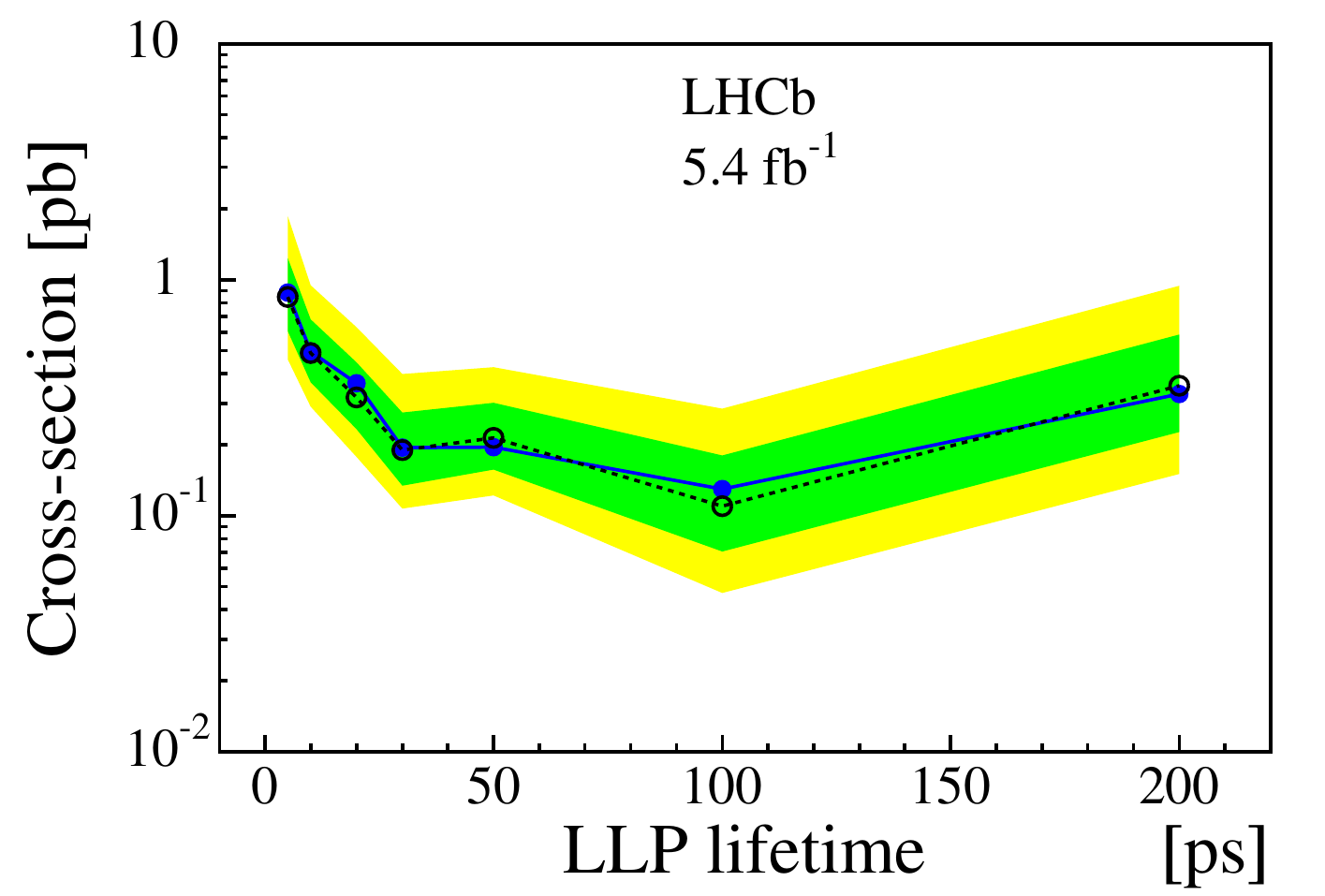}
    \includegraphics[width=4.3cm]{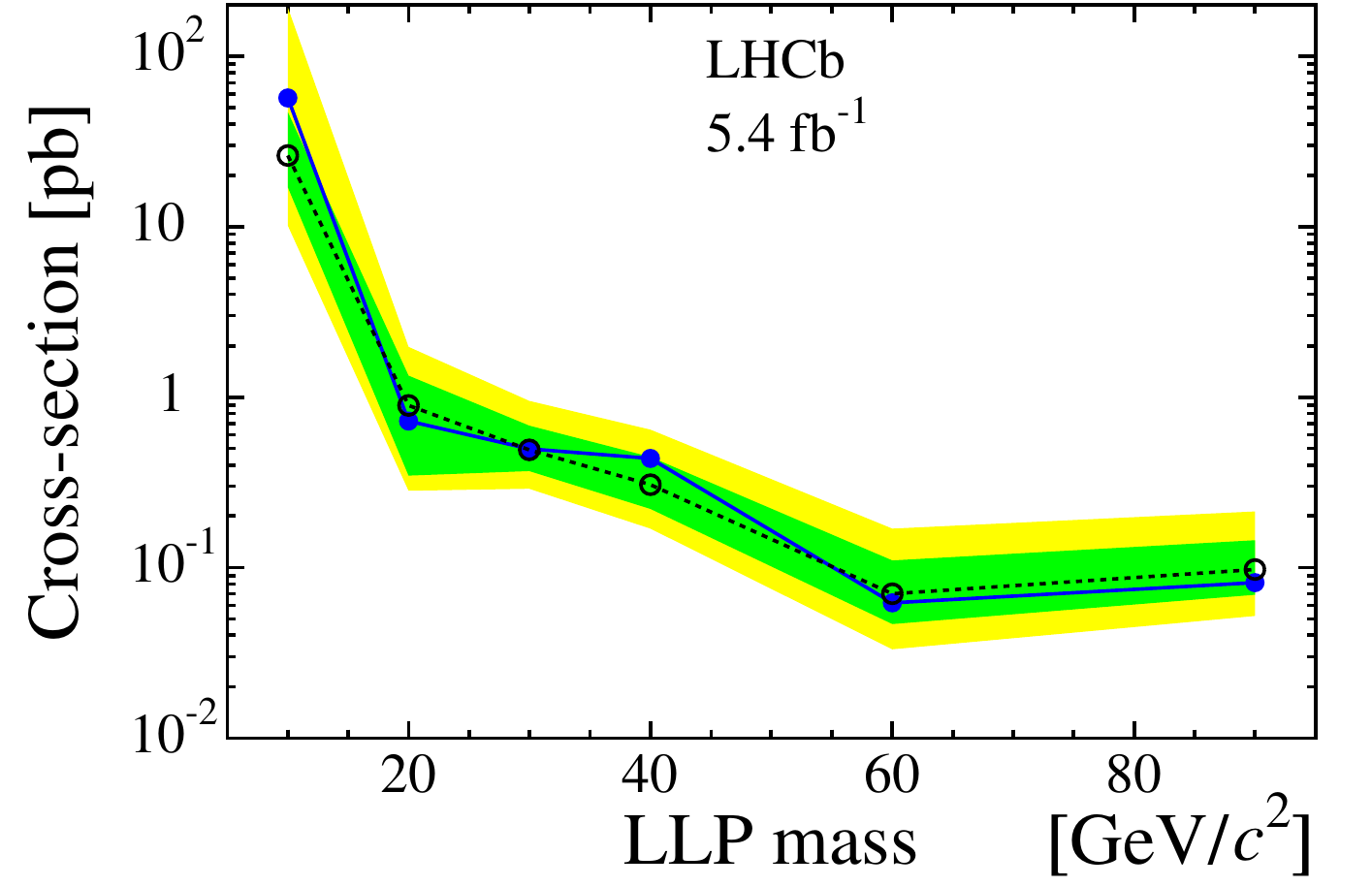}
    \caption{{\em Left:} An "x-ray" of the VELO detector. The reconstructed vertices show the outline of the RF-foil, seperating the detectors from the beam vacuum. The vertical lines indicate the material of the sensors. {\em Center and Right:} Cross section limits for semileptonic search plotted versus the Lifetime (for $m=30$ GeV/$c^2$) and mass (for $\tau=10$ ps) of the LLP candidate, respectively. Shown are the expected (open dots and 1$\sigma$ and 2$\sigma$bands) and observed (full dots) cross-section times branching fraction upper limits at 95\% CL~\cite{LHCb-PAPER-2021-028}.}
    \label{fig:LLP}
\end{figure}

To exclude backgrounds from collisions of particles with detector material it is essential that the regions where detector material is located are excluded from the displaced vertex search. 
The location of detector layers is known from the geometry database, but is also accurately calibrated with the data, as shown on the left of Fig.~\ref{fig:LLP}, in the so-called "x-ray" image of the VELO detector. The collection of observed vertices precisely maps the detector geometry. The middle and right side of the figure show the 95\% CL exclusion limit vs respectively the lifetime and the mass of the LLP, for the semileptonic search.

None of the searches for the signatures above observed a positive signal. 
For upcoming analyses in Run-3 and Run-4, searches for displayed vertices extending lifetimes to the region substantially outside the volume of the VELO detector will be performed using downstream tracks.

\vspace*{-0.3cm}

\clearpage

\section{Electroweak Physics}\label{secElectroweak}

LHCb's contribution to electroweak physics is particularly important because of its forward geometry.  This forward acceptance enables the LHCb detector to explore areas of electroweak physics that are complementary to the central detectors ATLAS and CMS. LHCb thus plays a unique role in probing processes such as the production and properties of the $Z$ and $W$ bosons.

\begin{figure}[!hb]
    \centering
     \includegraphics[width=7cm]{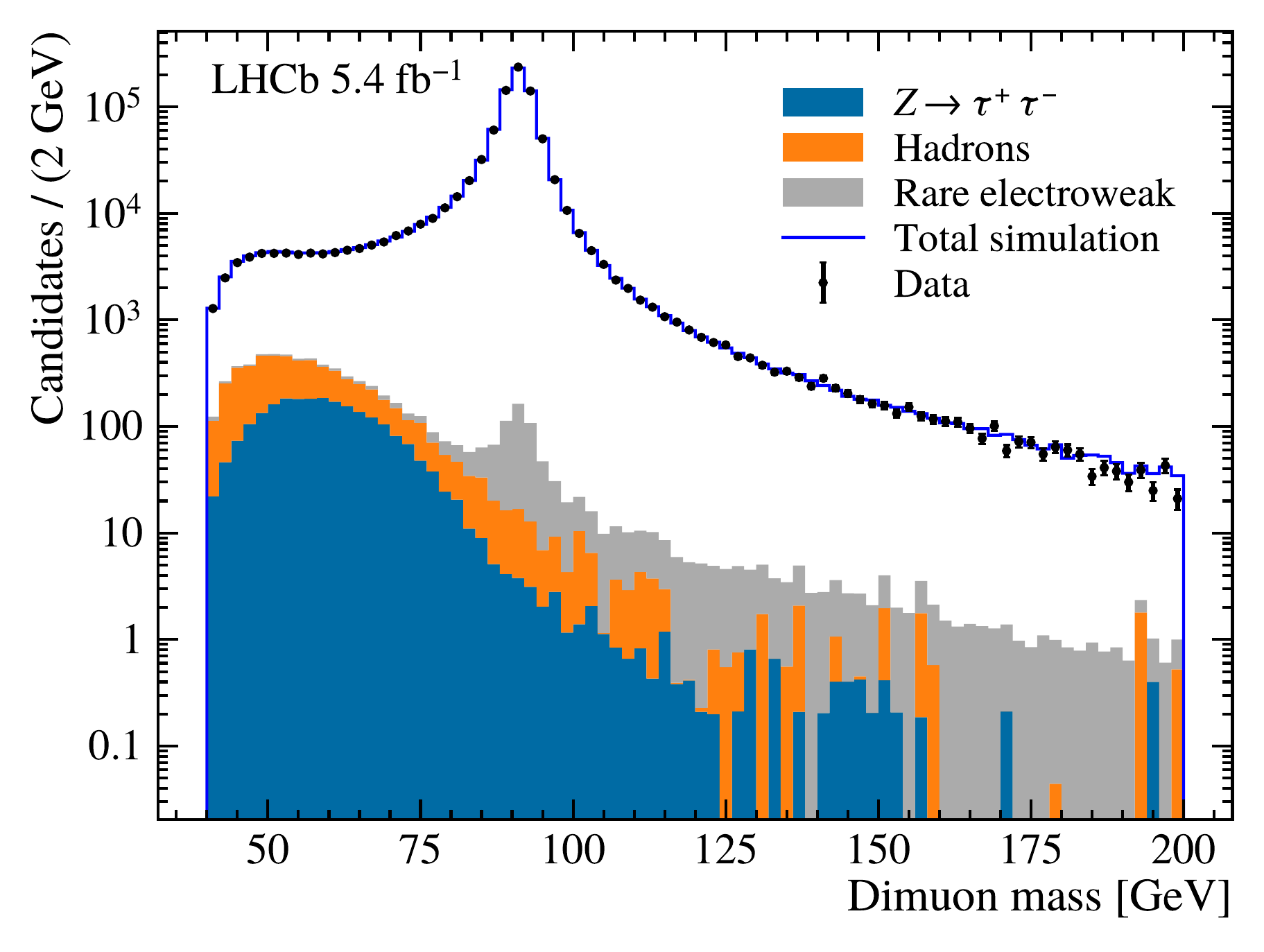} 
     \includegraphics[width=7cm]{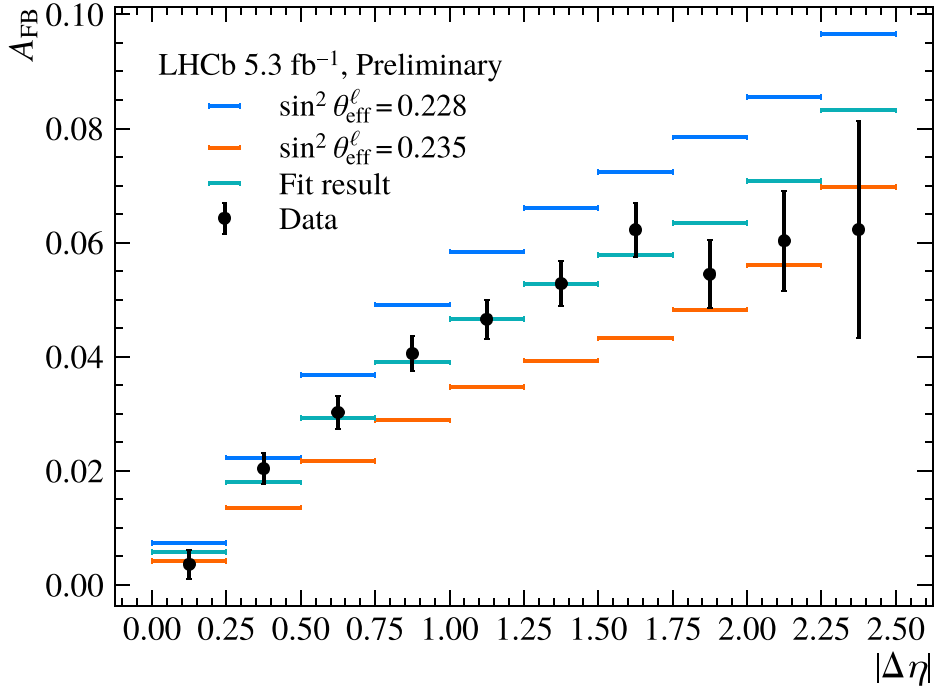} 
     \caption{{\em Left:} Invariant mass distribution of $Z\to \mu^+\mu^-$ candidates, showing the purity of the sample with a small peaking background from rare production processes of the $Z$ boson. {\em Right:} The forward-backward asymmetry increases with the rapidity difference $|\Delta\eta|$. The lines indicate different values of the weak mixing angle $\theta_W$ showing the sensitivity of the measurement~\cite{LHCb-PAPER-2024-028}.}
     \label{fig:SignTheta}
\end{figure}

A pure sample of about 860,000 $pp \to Z^*/\gamma \to \mu^+\mu^-$ events was used to determine the forward-backward asymmetry,
$$
A_{FB}=
\frac{\sigma_F-\sigma_B}{\sigma_F+\sigma_B}=
\frac{N_{\eta^->\eta^+}-N_{\eta^-<\eta^+}}{N_{\eta^->\eta^+}-N_{\eta^-<\eta^+}} 
$$
where $\eta^+$ and $\eta^-$ are the pseudorapidities of the muons. The asymmetry is determined with per-event weights depending on the difference $|\Delta\eta|$ between the rapidities of the two muons.
The asymmetry allows to fit the coefficient $\alpha$ in the differential cross section
$d\sigma/d\cos\theta^* = 1 +\cos^2\theta^* + \alpha\cos\theta^*$~\footnote{$\theta^*$ is the polar angle in a suitable reference frame \cite{Collins:1977iv}}, which in turn depends on the weak mixing angle $\theta_W$. 
Fig.~\ref{fig:SignTheta} shows the invariant mass of the $Z\rightarrow\mu^+\mu^-$ decays together with the obtained fit result $\sin^2\theta^l_{\text{eff}} = 0.23147 \pm 0.00044 \pm 0.00005 \pm 0.00023$, where the first uncertainty is due to statistics, the second due to systematics in measurement of $A_{FB}$ and the third the systematic error in the fit model. The measurement is in good agreement with the global electroweak fit.

\begin{figure}[!ht]
    \centering
     \includegraphics[scale=0.4]{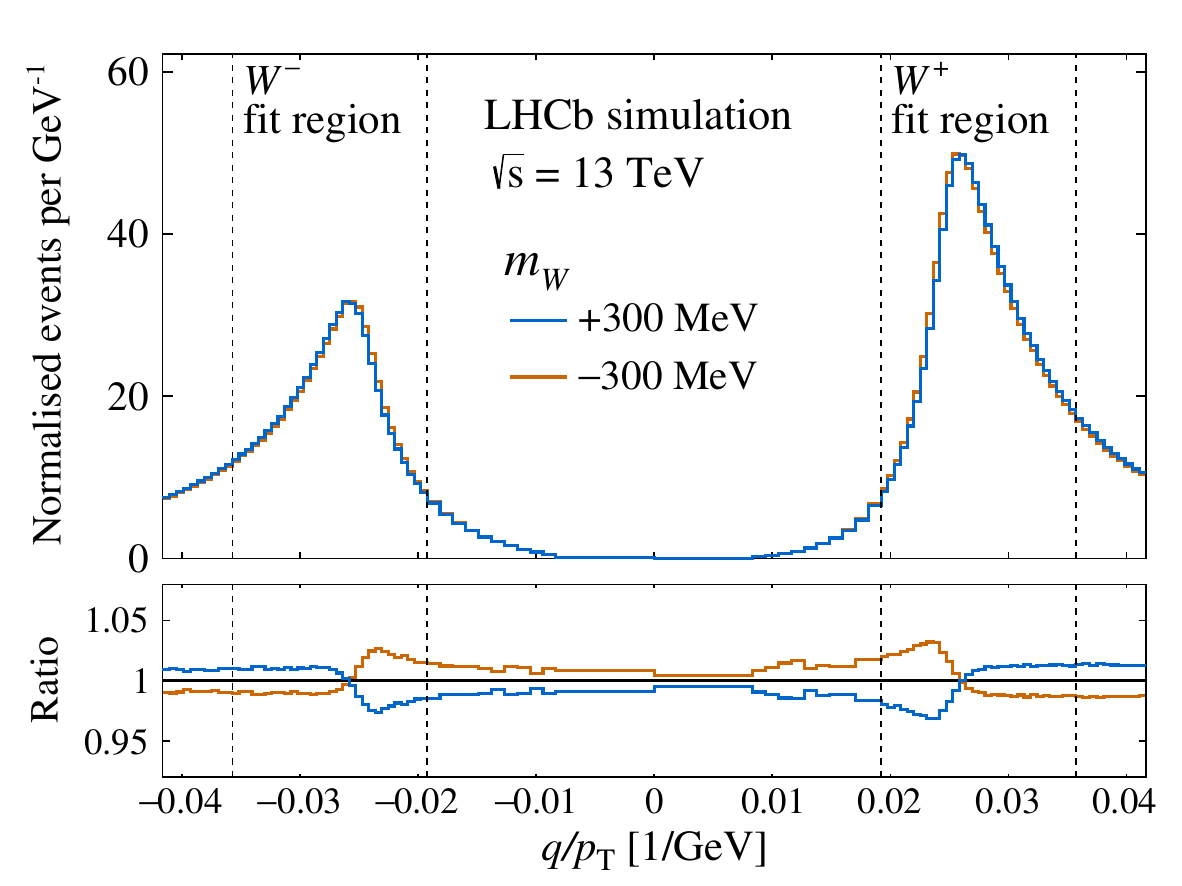} 
     \includegraphics[scale=0.4]{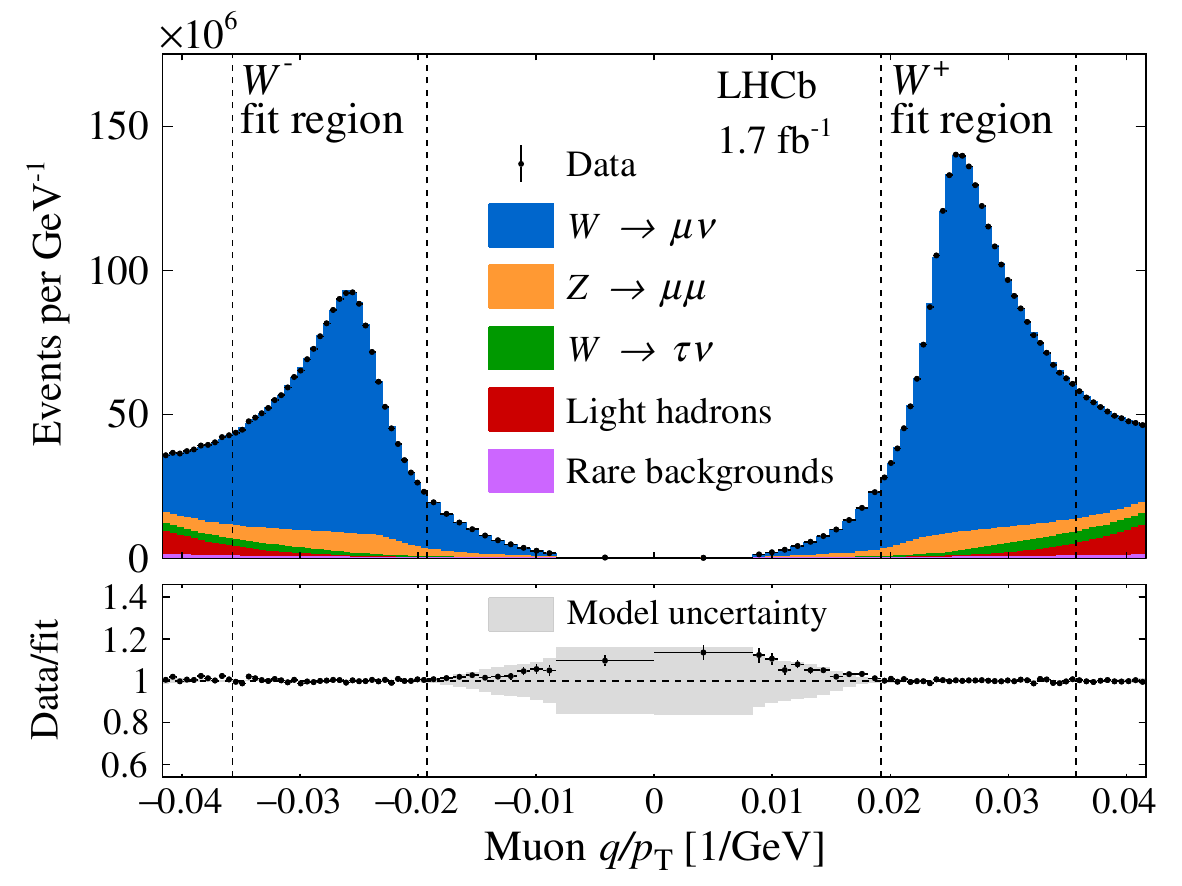} 
    \caption{
    {\em Left:} The distribution of $1/p_T$ of the muons from simulation, and the effect on the spectrum for different values of the $W$ mass.
    {\em Right:} The measured distribution of $1/p_T$ of the muons that is used to determine the mass of the $W$ boson~\cite{LHCb-PAPER-2021-024}. 
    }
     \label{fig:mW}
\end{figure}

The mass of the $W$-boson is determined from $W\rightarrow\mu\nu$ decays using the shape of the $q/p_T$-distribution of the muons, see Fig.~\ref{fig:mW}.
Despite the lower integrated luminosity collected by the LHCb experiment
as compared to the ATLAS and CMS experiments, important sensitivity to the $W$ boson mass is obtained, with a statistical uncertainty of only 23~MeV on the 2016 data alone. An extra 9~MeV uncertainty arises from uncertainties from the parton density functions (pdf's) of the proton,
$m_W=80354\pm 23 (stat) \pm 10 (exp) \pm 17 (theory) \pm 9 (PDF)$~MeV~\cite{LHCb-PAPER-2021-024}.
Interestingly, the uncertainty 
from the pdf's is anti-correlated with the
pdf uncertainty on the $W$ mass measurement at ATLAS and CMS, which will lead to an interesting improvement of the overall result when the measurements are combined.
Recently, the first measurement of the $Z$-mass at the LHC was performed
by LHCb, $m_Z=91184.2\pm 8.5 (stat) \pm 4.3 (syst)$~MeV~\cite{LHCb:2025nob}
%\cite{LHCb-PAPER-2025-008-sem}.

\begin{figure}[!ht]
    \centering
    \includegraphics[scale=0.45]{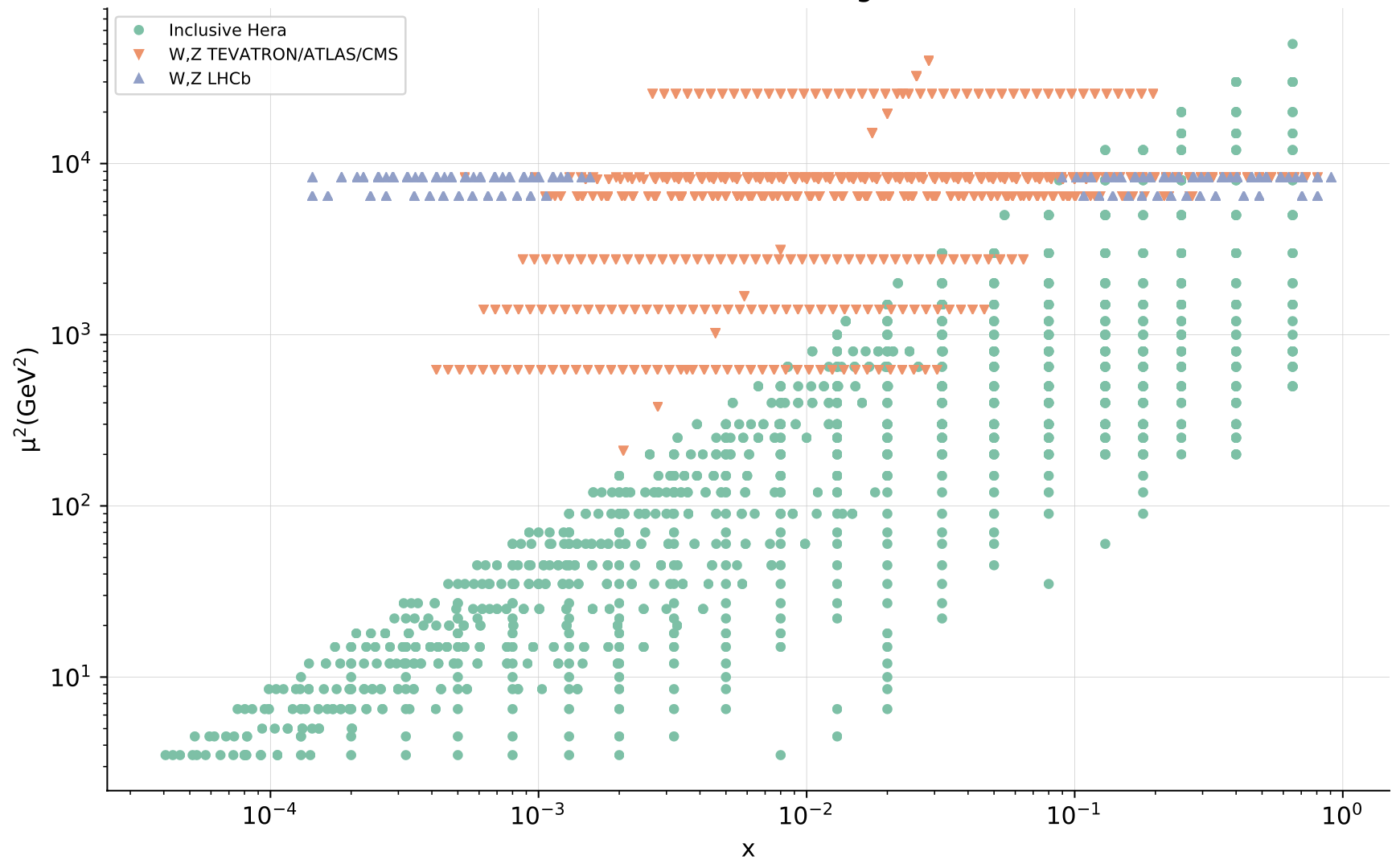} 
     \caption{The kinematic region $(x,Q^2)$ (here displayed with $\mu^2\equiv Q^2$) shows the added value of $W$ and $Z$ production measurements at LHCb that extend the coverage to very low Bjorken $x$ at high $Q^2$~\cite{Rojo:2017xpe}.}
     \label{fig:xQ2}
\end{figure}

The $W$ and $Z$ bosons are produced from 
quark interactions (unlike the Higgs boson, which is
predominantly produced through gluon fusion), and therefore the differential cross section carries information on the proton pdf's~\cite{NNPDF:2017mvq}.
The forward acceptance of the LHCb experiment implies
a large asymmetry of the value of Bjorken-$x$ 
of the interacting quarks, and therefore the
$W$ and $Z$ production measurements form LHCb 
provides valuable information in the extremes of the kinematic plane $(x,Q^2)$. Fig.~\ref{fig:xQ2} shows the complementarity of the LHC measurements to the HERA results, where LHCb extends the low x coverage for high $Q^2$.

%\begin{figure}[h!]
%    \centering
%     \includegraphics[height=8cm]{Figures/Electroweak/Fig5-theta-2024-028.pdf} 
%    \includegraphics[height=8cm]{Figures/Electroweak/mW.pdf} 
%     \caption{Comparison of the electroweak measurements of the weak mixing angle $\theta_W$~\cite{LHCb-PAPER-2024-028} and the $W$-boson mass (from Keira's talk at Blois).}
%     \label{fig:SignTheta}
%\end{figure}

\clearpage

\section{Heavy Ion Physics}\label{secHeavyIon}

\begin{figure}[!htb]  
	\centering
	\includegraphics[width=6.4cm]{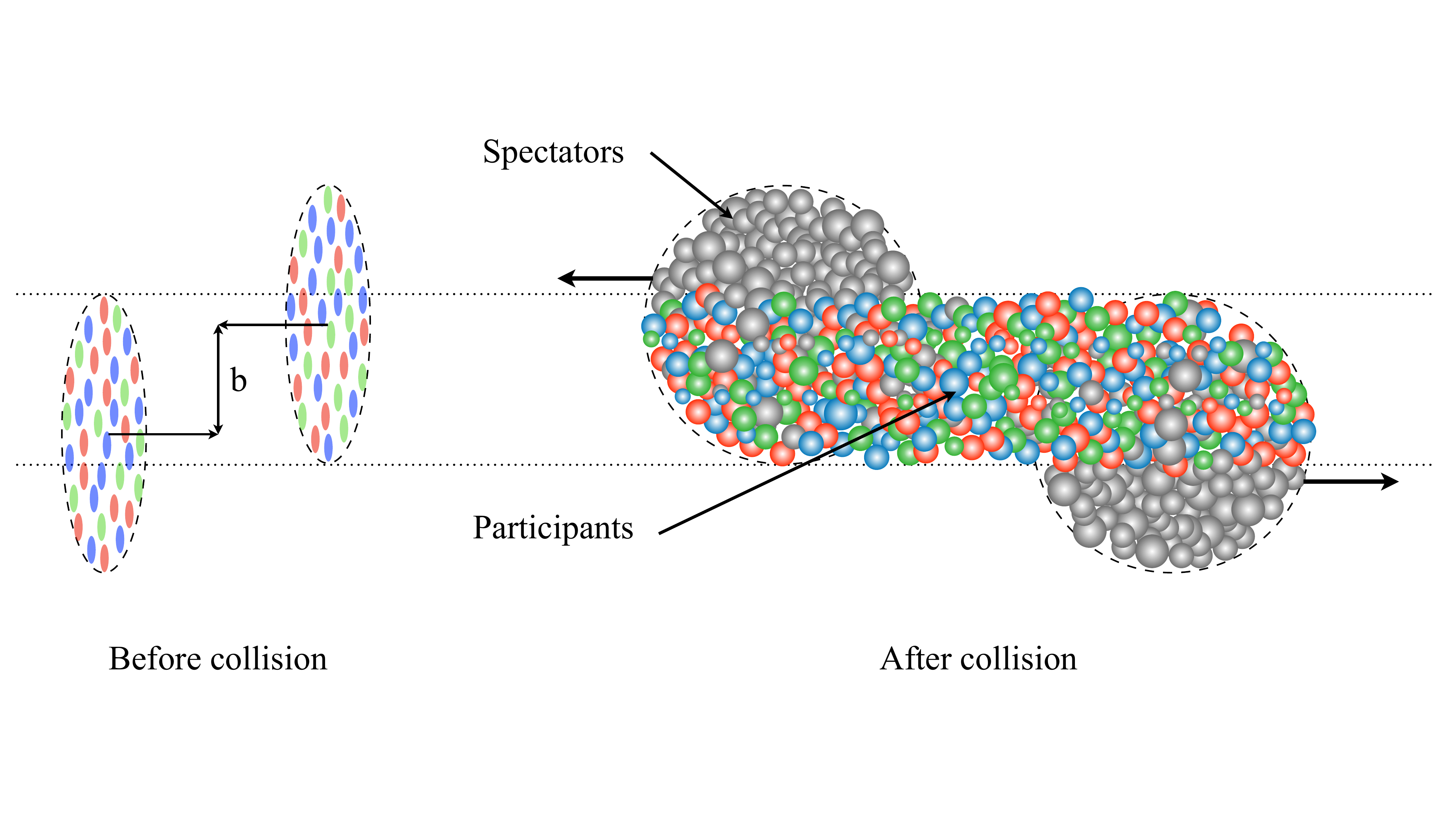}
 	\includegraphics[width=4.7cm]{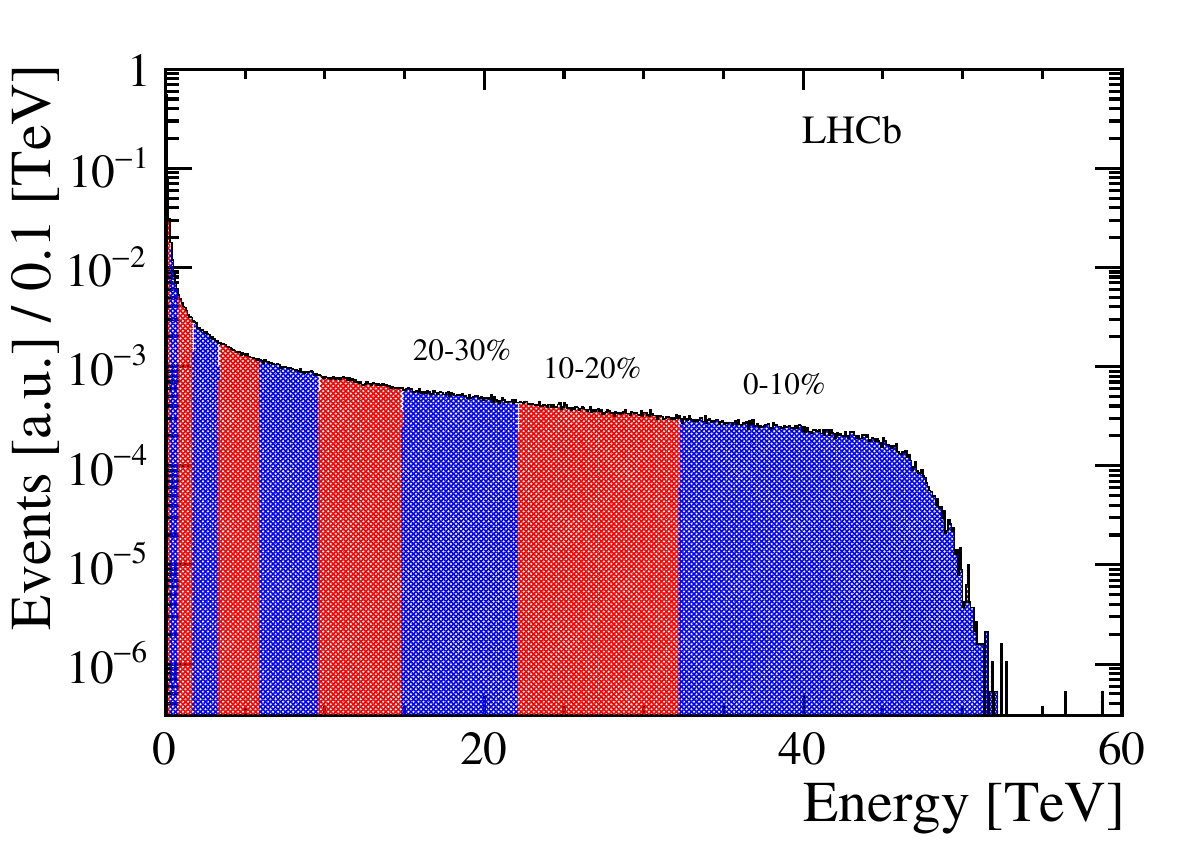}
	\includegraphics[width=5.4cm]{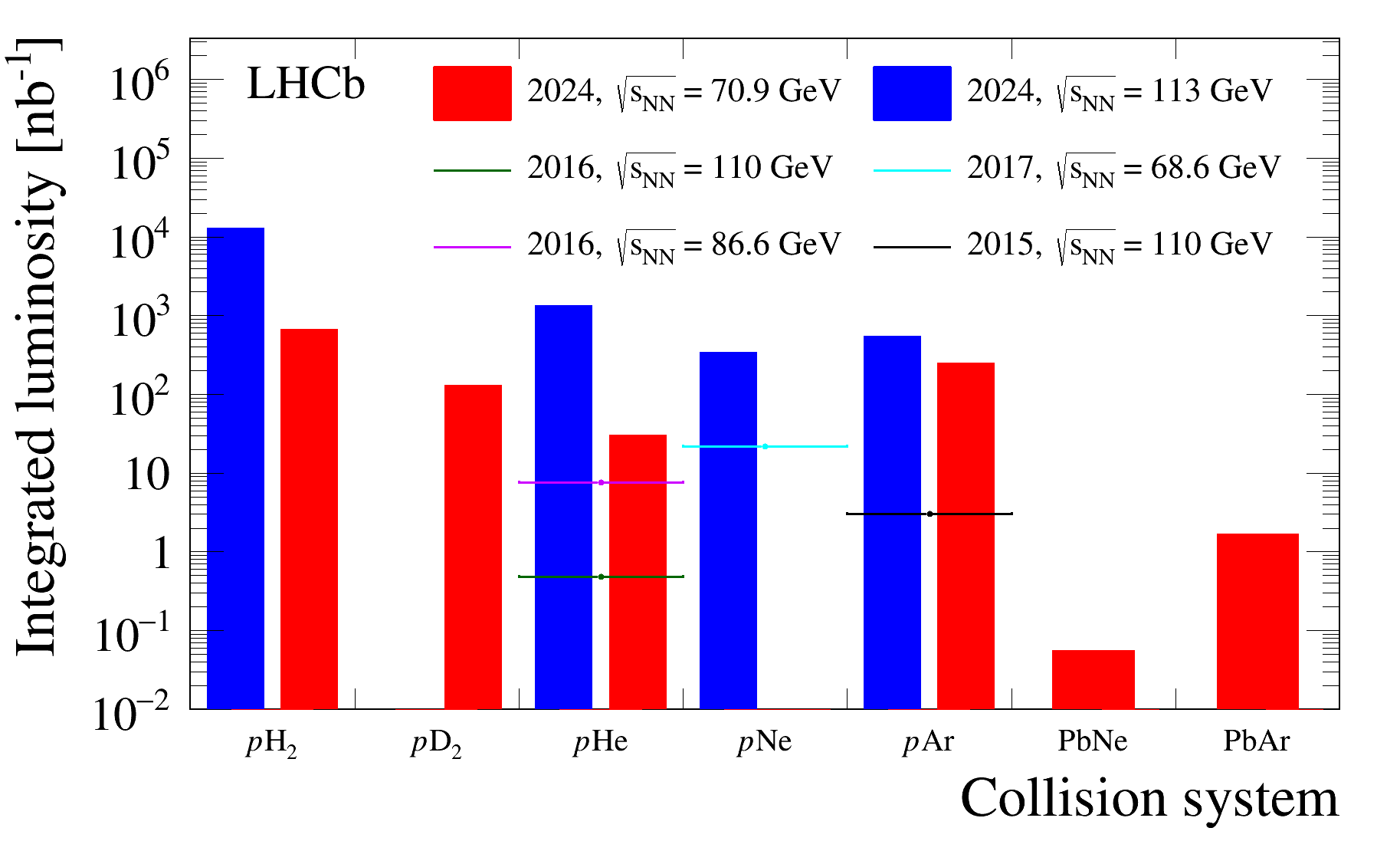}
    \caption{{\em Left:} Sketch of a heavy-ion collision illustrating a collision at half centrality. 
    {\em Middle:} The centrality classes can be defined in terms of the energy deposited in the ECAL~\cite{LHCb-DP-2021-002}. 
    The number of nucleons participating in the collision depoends on the impact parameter of the colliding nuclei. The Glauber model is used to relate the energy deposited in the ECAL to the centrality and the number of participating nucleons in the collision~\cite{Miller:2007ri}.
    {\em Right:} Comparison between Run 2 and Run 3 of the accumulated data sets of fixed-target samples for the different gas targets. }
	\label{fig:pp-vs-hi-event}
\end{figure}

Heavy-ion collisions produce extreme temperatures and energy densities by colliding
multiple nucleons from within the nuclei, up to about 100 participating nucleons at
50\% centrality~\cite{LHCb-DP-2021-002}, as illustrated in the left panel of Fig.~\ref{fig:pp-vs-hi-event}.
These conditions enable the formation of a quark-gluon plasma (QGP), a state of matter in which quarks and gluons are deconfined and can interact freely over extended ranges. This environment allows the study of the phase transition from deconfined to confined matter as the system expands and thermalizes.
At the LHC the species $^{208}$Pb is 
chosen for its large nucleus size, with 82 protons and 126 neutrons. In addition,
$^{208}$Pb is a doubly-magic nucleus with fully filled proton and neutron shells,
leading to a stable nucleus resulting in reduced theoretical uncertainties.
($^{197}$Au has a slightly higher charge over mass ratio $Z/A$, which allows for higher 
acceleration, which justified the species choice at the RHIC facility.)
The beam energies at the LHC for the different running periods are listed in Tab.~\ref{tab:beam}.

%\begin{figure}[!h]  
%	\centering
%	\includegraphics[width=7.5cm]{Figures/HeavyIon/Proton-Proton-Collision.jpg}
% 	\includegraphics[width=7.5cm]{Figures/HeavyIon/Lead-Argon-Collision.jpg}
%    \caption{Comparison of a proton-proton (L) and a Lead-Argon (R) collision in the LHCb spectrometer. The heavy ion collision results in a higher hit occupancy in the detector. \textcolor{orange}{Do we actually need this figure? It is just to say that the occupancy is higher, but perhaps it is a waste of space?} }
%	\label{fig:pp-vs-hi-event}
%\end{figure}

\begin{table}[h!]
    \centering
    \begin{tabular}{ll|cl|cl|cl}
     $E_b$ (Z TeV) & Comment & \multicolumn{2}{c|}{pp} & \multicolumn{2}{c|}{p-Pb}    &  \multicolumn{2}{c}{Pb-Pb}  \\ 
                   &         & $\sqrt{s}=2E$ & Year    & $\sqrt{s}=2E\sqrt{r}$ & Year & $\sqrt{s}=2Er$ & Year       \\
        \hline
         1.38   & pp-ref Run 1    & \bf{2.76}  & 2013       &    -       &           &    -       &       \\
         2.51   & pp-ref Run 2    & \bf{5.02}  & 2015, 2017 &    -       &           &    -       &       \\
         2.68   & pp-ref Run 3    & \bf{5.36}  & 2023, 2024 &    -       &           &    -       &       \\
        \hline
          3.5   & pp nominal Run 1& 7          & 2011       &    -       &           &   \bf{2.76}& 2010  \\
           4    & pp nominal Run 1& \bf{8}     & 2012       &  \bf{5.02} & 2013, 2016&   -        &       \\
          6.37  &                 &  -         &            &   -        &           &   \bf{5.02}& 2015, 2018  \\
          6.5   & pp nominal Run 2& 13         & 2015-2018  &  \bf{8.16} &  2016     &   -        &        \\
          6.8   & pp nominal Run 3& 13.6       & 2021-      &   -        &           &   \bf{5.36}& 2023, 2024  \\
      \end{tabular}
    \caption{Overview of the beam energy and corresponding nucleon-nucleon center-of-mass energy at the LHC, and the year of data taking for the various running periods. Dedicated proton-proton {\em reference} runs were performed corresponding to the same center-of-mass energy as the nucleon-nucleon center-of-mass energy (numbers in bold face), which allows for a direct comparison between small and large systems.
    The relevant center-of-mass for p-Pb and Pb-Pb is calculated using the charge to mass ratio $r=82/208$. 
    }
    \label{tab:beam}
\end{table}

The LHCb experiment's forward kinematic reach enables probing of QGP properties in the forward rapidity region, complementing studies by the ALICE and, to some extend, ATLAS and CMS detectors, which are optimized for more central rapidity regions.
The LHCb heavy-ion program focuses on charmonium and open-charm production, on multiplicity measurements and on ultra-peripheral collisions.
The forward geometry is well-suited for studying ultra-peripheral collisions (UPC), where heavy ions interact electromagnetically rather than strongly. These interactions, dominated by photon exchanges, allow for studies of vector meson production such as $J/\psi$ mesons~\cite{LHCb-PAPER-2020-043}, providing insight into photon-induced reactions and nuclear gluon distributions.

\begin{figure}[!t]  
    \begin{picture}(400,150)(0,0)
     \put(0,5){\includegraphics[scale=0.34]{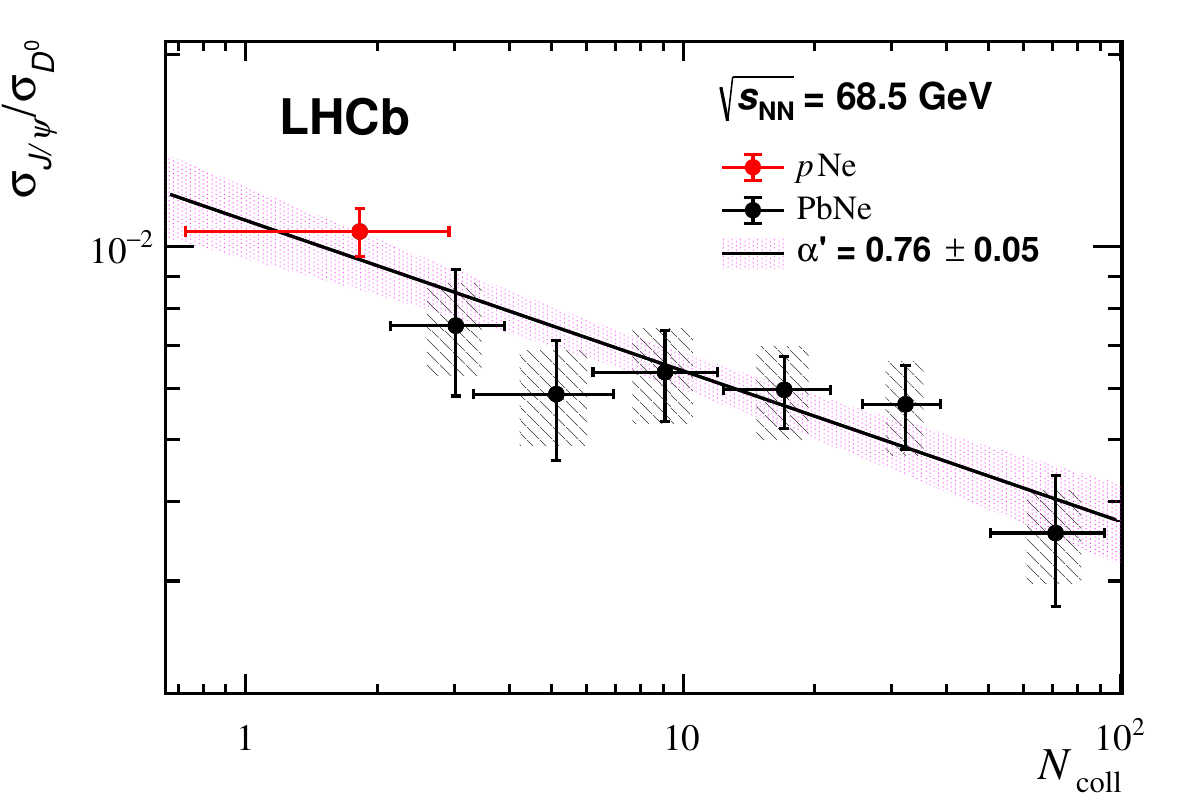}}
    \put(195,0){\includegraphics[scale=0.52]{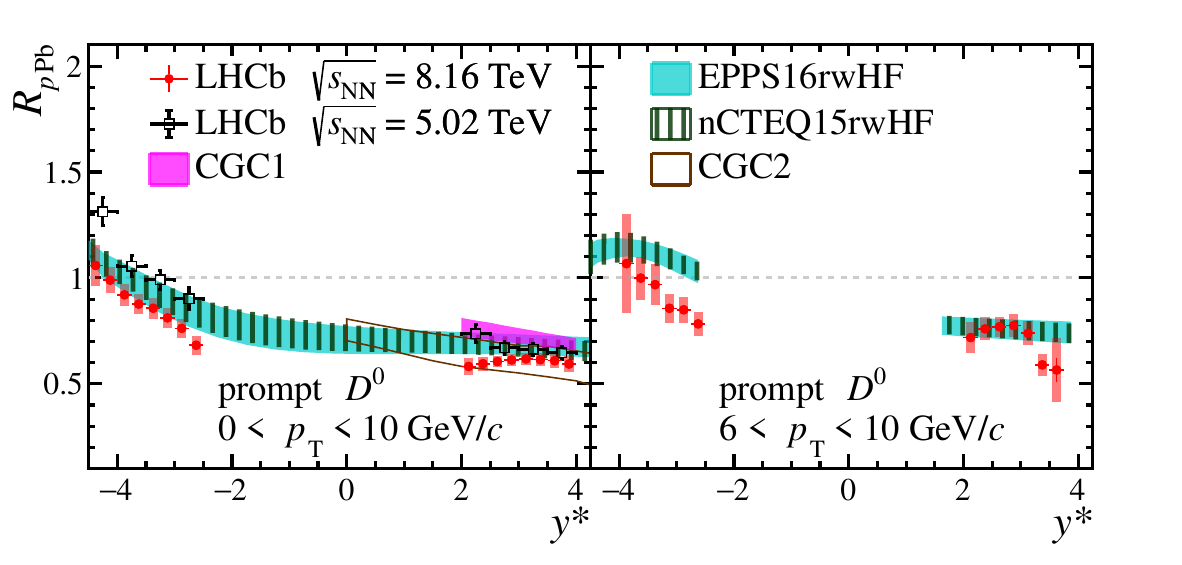}}
    %\put(300,0){\includegraphics[scale=0.5]{Figures/HeavyIon/PAPER-2023-021-Ds.pdf}}
    %\put(300,0){\includegraphics[scale=0.5]{Figures/HeavyIon/PAPER-2023-031-Flow.pdf}}
    \end{picture}
   \caption{
    {\em Left:} The ratio of $J/\psi$ over $D^0$ production as a function of number of colliding nucleons in fixed target p-Ne and Pb-Ne collisions. A reduction of $J/\psi$ production is seen due to cold nuclear matter effects~\cite{LHCb-PAPER-2022-011}.
    {\em Right:} The $D^0$ production in p-Pb collisions relative to pp collisions as a function of rapidity is compared to predictions using nuclear pdfs EPPS and nCTEQ, and the color-glass-condensate (CGC) models.
    Here $y^*$ is related to the rapidity in the laboratory frame by $y^* = y_{lab}-0.465$ due to the boosted frame of the asymmetric p-Pb collisions~\cite{LHCb-PAPER-2022-007}.
    %\textcolor{orange}{Wat zijn CGC1 en 2? andere modellen? end EPPS.. en CTEQ...} 
%Strangeness enhancement with $D_s$ in p-Pb \cite{LHCb-PAPER-2023-021} \\
   }
	\label{fig:HI-results}
\end{figure}

%\begin{itemize}
%\item {\em Quarkonium production and suppression.}
Heavy quarkonia, such as $J/\psi$ mesons, are key probes of the QGP. In the QGP, these bound states of heavy quark–antiquark pairs can dissociate due to color screening  (analogous to 'melting' in the plasma), a phenomenon predicted by lattice QCD~\cite{Matsui:1986dk}. By measuring the suppression patterns of quarkonia across different states and rapidities, insights into QGP formation is obtained. The forward coverage offers complementary data  of quarkonium production and suppression across a broader kinematic range.
The relative production of $J/\psi$ compared to $D^0$ mesons has been studied in small systems from p-Ne collisions, 
and in bigger systems from Pb-Ne collisions, and a suppression has been observed due to cold nuclear matter effects~\cite{LHCb-PAPER-2022-011,LHCb-PAPER-2022-015},
as shown in the left panel of Fig.~\ref{fig:HI-results}.
No further suppression from QGP effects has been observed yet, but with the 2024 Pb-Pb data results to lower centrality should become available, with more sensitivity to the deconfined regime.
%\item {\em Open heavy-flavor production.}

The production of open heavy-flavor hadrons such as $D$ mesons, also serves as a direct probe of QGP dynamics
~\cite{LHCb-PAPER-2021-046,LHCb-PAPER-2023-006}. Heavy quarks are produced early in the collision and propagate through the QGP, experiencing energy loss through interactions with the medium. 
The unique gas injection system at LHCb (known as the SMOG system~\cite{LHCb-DP-2022-002}), allows to perform these studies at different values of the center-of-mass, and for different system-sizes by varying the gas target species (H$_2$, Ne and Ar). The fixed-target data samples are shown in
the right panel of Fig.~\ref{fig:pp-vs-hi-event}.
Through the measurement of nuclear modification factors $R$ of open heavy-flavor hadrons, the energy loss mechanisms in dense hadronic environments is studied, such as in 
p-Pb collisions~\cite{LHCb-PAPER-2022-007} (see right panel of
Fig.~\ref{fig:HI-results}), leading to improved knowledge of the
nuclear parton distribution functions~\cite{AbdulKhalek:2022fyi}.
This measurement shows that the nuclear parton distribution functions (PDFs) are suppressed relative to the proton PDFs ("shadowing") due to cold nuclear matter effects at small Bjorken-$x$ at high $y^*$, which can be described as a color glass condensate (CGC) with the high gluon densities at low-$x$. 

\begin{wrapfigure}{R}{6cm}
%\vspace*{-0.2cm}
\includegraphics[width=6cm, trim=0cm 7.6cm 9.5cm 0cm, clip]{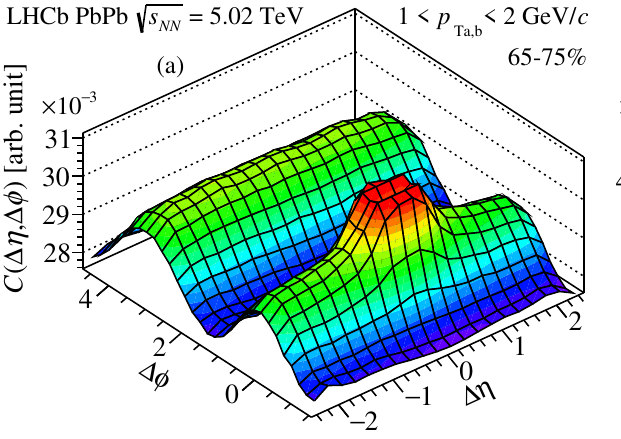}
\caption{The ridge effect in Pb-Pb collisions: the two-particle correlation function shown for an example of a $p_T$ and centrality bin~\cite{LHCB-PAPER-2023-031}.}
\label{fig:ridge}
\vspace*{-0.2cm}
\end{wrapfigure}
Studies of high-multiplicity \mbox{Pb-Pb}, \mbox{p-Pb} and \mbox{p-p} collisions extend the understanding of QGP-like effects from bigger to smaller systems, such as potential signatures of collective flow, ridge correlations, and particle suppression patterns~\cite{LHCB-PAPER-2015-040,LHCB-PAPER-2023-031}. 
LHCb studied flow by comparing the two-dimensional correlation functions from two particles in the same event ($S(\Delta\eta,\Delta\phi)$), where $\Delta\eta$ is the particle rapidity difference and $\Delta\phi$ the azimuthal difference. A two-dimensional angular correlation function
\begin{equation}
C\left(\Delta\eta,\Delta\phi\right) = \frac
{S\left(\Delta\eta,\Delta\phi\right)}
{B\left(\Delta\eta,\Delta\phi\right)}
\end{equation}
where $S$ combines tracks in the same event and $B$ uses uncorrelated track combinations from different events.
The two-dimensional correlations in Pb-Pb in Fig.~\ref{fig:ridge} show pronounced near- and away-side ridges compared to the published LHCb p-Pb and Pb-p results~\cite{LHCB-PAPER-2015-040}, indicating
stronger forward particle flow in Pb-Pb events than in p-Pb and Pb-p events.
These flow measurements in the forward region are important to understand the “cooler” region where freeze-out is dominant.
The coefficients measured using the two-particle angular correlation analysis method are smaller than the central-pseudorapidity measurements at ALICE and ATLAS from the same collision system but share similar features.
Observing these effects in the forward region provides tests for models that attempt to describe collective phenomena in non-central collisions and contributes to debates on the limits of QGP formation in small systems~\cite{Giacalone:2018fbc}.
%Recently,  flow harmonics have been studied to understand the temperature dependence of the shear viscosity to entropy density ratio of the QGP~\cite{LHCb-PAPER-2023-031}.
%\item {\em Ultra-Peripheral collisions.}

%PbPb:
%J/psi in UPC \cite{LHCb-PAPER-2020-043}
%J/psi vs centrality \cite{LHCb-PAPER-2024-041}; poor precision....
%\end{itemize}

The LHCb heavy-ion program exploits its unique detector characteristics to contribute to a wide range of high-precision measurements that are vital for understanding both hot and cold nuclear matter effects in a previously under-explored kinematic regime. By offering complementary measurements to those performed by the central detectors, LHCb strengthens the LHC’s collective efforts in exploring QGP properties, heavy quark dynamics, and nuclear structure.
Finally, the fixed-target program offers valuable measurements
to the understanding of cosmic ray spectra by measuring the production
rate of anti-protons in p-He collisions~\cite{LHCb-PAPER-2018-031,LHCb-PAPER-2022-006}.

%\begin{figure}[b]  
%	\centering
%	\includegraphics[width=12cm]{Figures/HeavyIon/HI-Kinematics.png}
%    \caption{Kinematics of Heavy Ion.LHCb public page Oct 30 2023}
%	\label{fig:HI-Kinematics}
%\end{figure}

\clearpage
\section{Complementarity with other experiments in the field}\label{secComparisons}
\begin{figure}[!hb]
    \centering
     \includegraphics[width=7.5cm]{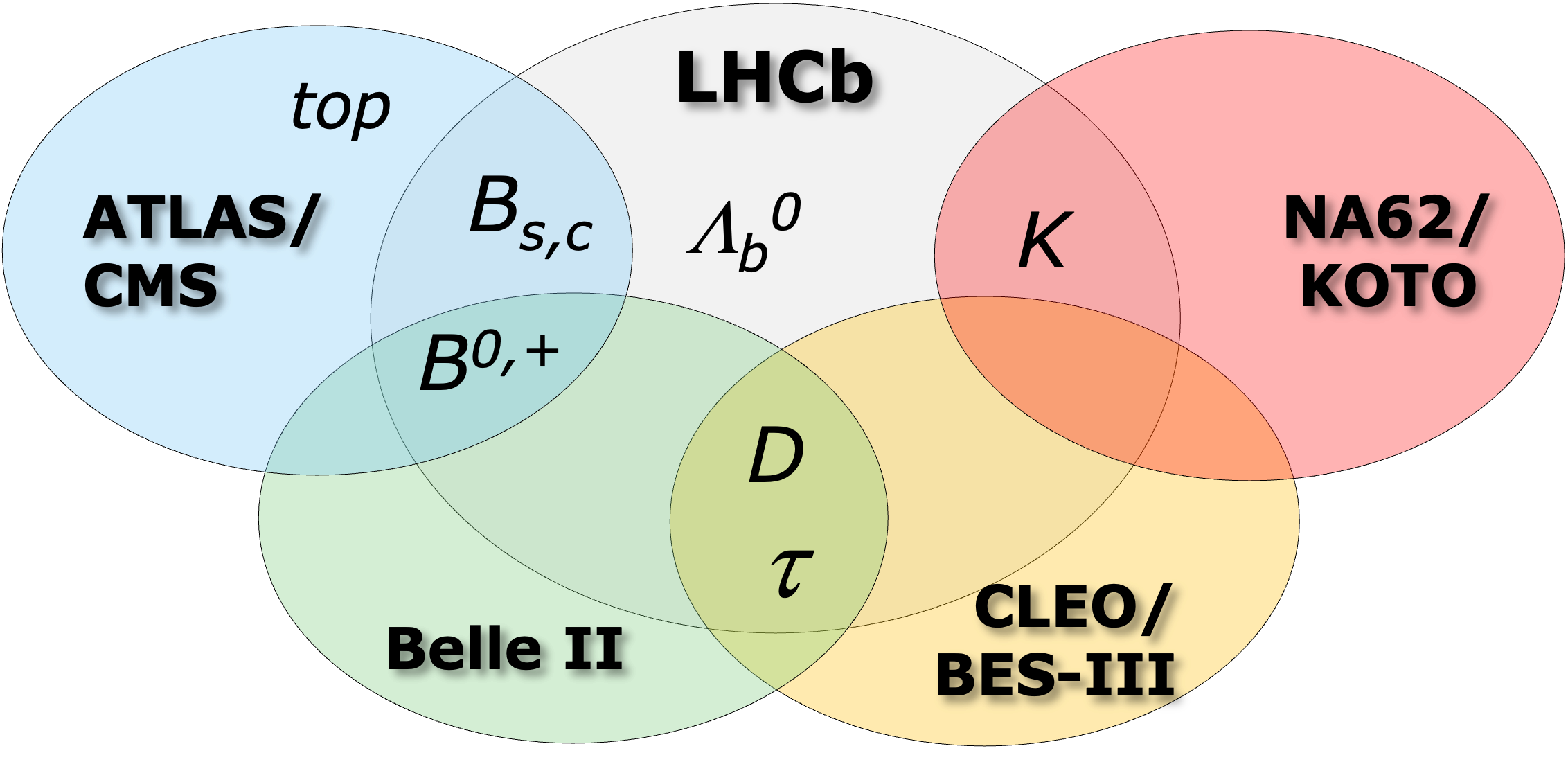} 
     \hspace{1.0cm}
     \includegraphics[width=5.5cm]{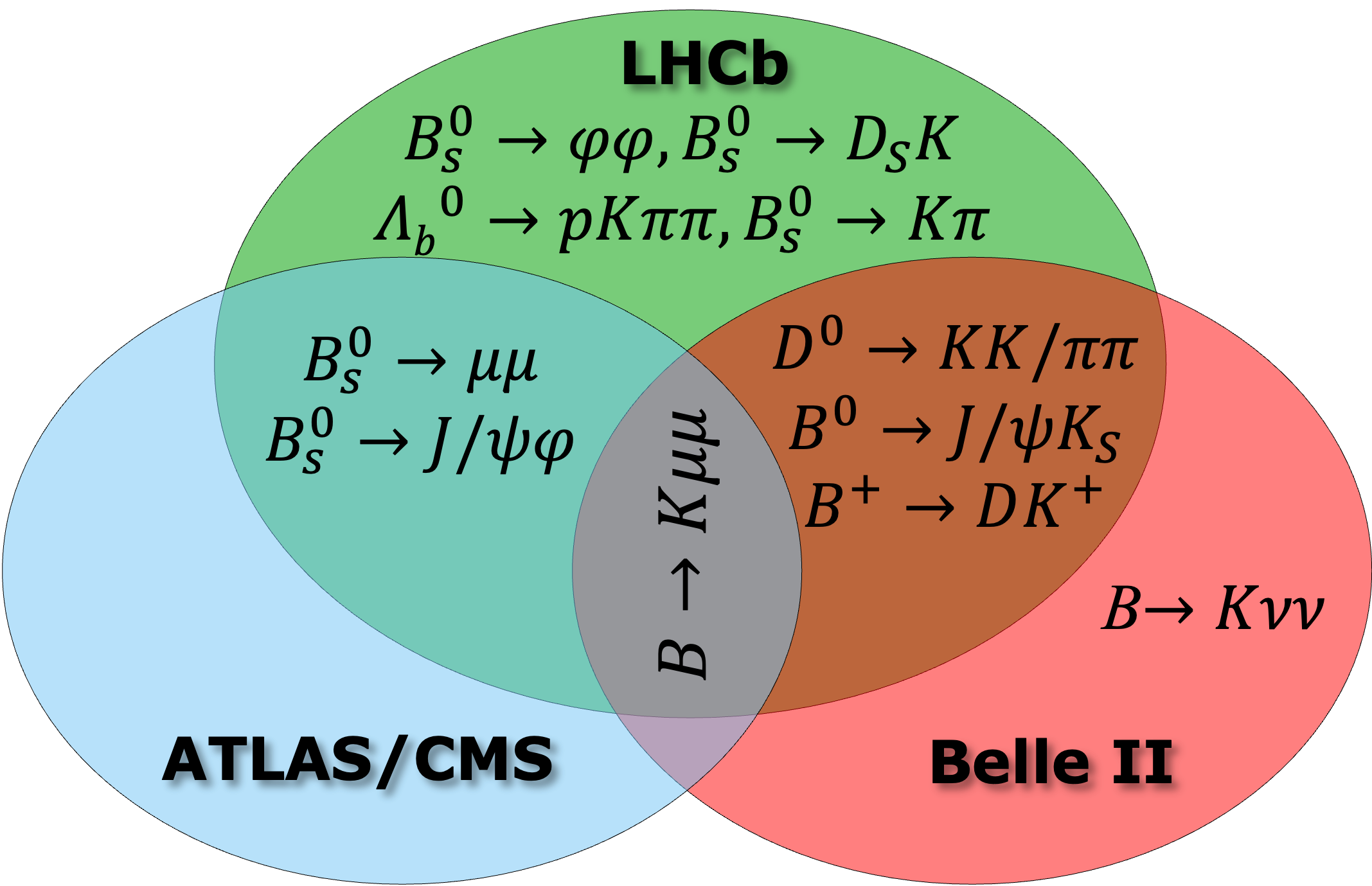} 
     \caption{{\em Left:} Landscape of flavour experiments, studying quark interactions, ranging from strange to top. {\em Right:} The bottom quark has the richest experimental and phenomenological reach and is currently studied at the LHC and at Belle-II. }
     \label{fig:flavour}
\end{figure}

%Please explain relations to other experiments in the field.
Over the years, multiple experimental facilities have been designed and built to exploit the opportunity of scrutinizing the Standard Model through precision measurements of flavour processes.
For example, the NA62 and KOTO experiments
aim to measure rare kaon decays from fixed target proton interactions, 
the BES-III experiment, and formerly CLEO, collide electrons and positrons to produce large quantities
of charm pairs, whereas the Belle-II experiment, and formerly BaBar, were optimized to measure b-quark pairs. ATLAS and CMS add a unique top flavour programme. A sketch of the flavour landscape for these experiments is given in Fig.~\ref{fig:flavour}.
In the following the focus will be on the comparison of measurements of 
$B$-meson decays.

\subsection{Comparison with asymmetric \texorpdfstring{$e^+e^-$}{e+e-} machines}

In 1999 the two asymmetric $e^+e^-$ colliders, dedicated to $B$-physics, started operation at the SLAC laboratory in the USA and at the KEK laboratory in Japan, with the BaBar detector at the PEP-II
collider, and the Belle detector at the KEK collider, respectively.
The center-of-mass energy at these $B$-factories is 10.82~GeV, which corresponds to the mass of the
$\Upsilon(4S)$ resonance, also known as the $\Upsilon(10580)$. The $\Upsilon(4S)$ decays
predominantly to $B$-meson pairs, almost equally to charged and neutral $B$-mesons.
The branching fractions are measured to be $B(\Upsilon(4S)\to B^0\bar{B}^0)=(48.6\pm0.6)\%$ 
and $B(\Upsilon(4S)\to B^+B^-)=(51.4\pm 0.6)\%$, but care needs to be taken whether these measured $\Upsilon(4S)$ branching ratios are assumed in the determination
of $B$-meson branching fractions, or whether 
isospin symmetry is assumed with equal branching ratios of the $\Upsilon(4S)$ to charged or neutral $B$-mesons~\cite{Jung:2015yma}.
%\textcolor{orange}{Ik begrijp de boodschap niet: is dit for use door bv LHCb metingen? Of intern in BelleII. Ik mis iets denk ik}
 
The asymmetric beam energies cause a boost of the produced $\Upsilon(4S)$, which in turn
leads to a boost for the $B$-mesons in the detector. The measurement of the flight distance between the two $B$ meson decay points is required to determine $B$-lifetime and oscillation measurements.

The cross section for $b\bar{b}$ production at the $\Upsilon(4S)$ resonance is about 1.05~nb.
The high instantaneous luminosity and the almost perfect trigger and selection efficiencies
compensate for the relatively low production cross section in comparison to the strong production at a hadron machine, and make the $B$-factories highly competitive. Furthermore, at the $B$-factories the events include only particles from $B$ decays, whereas in a hadron machine the final state includes many particles from the underlying event.
The $B$-factories are particularly performant for decays with electrons (less Bremsstrahlung),
semi-leptonic decays (full acceptance) and for events with $\tau$-decays (less background due to a missing neutrino).
In decay-time dependent CP violation analyses they profit from the coherent oscillations of the $B$-$\bar{B}$ pairs, which results in about a factor 10 better flavour tagging performance and subsequent gain in statistical power per observed event.
Hadron machines on the other hand benefit from the large production cross section and hence unique sensitivity for observing very rare decays, and have sufficient energy for
production of heavier $b$-species such as $B_s^0$, $B_c^+$ and $\Lambda_b^0$ hadrons. Note that, although $B_s$ mesons can be produced at the $\Upsilon(5S)$ resonance in Belle-II, their oscillations are too fast to be resolved.
Table~\ref{tab:Bfact} summarizes main parameters of the $B$-factory experiments.

\begin{table}[!hb]
    \centering
    \begin{tabular}{lllccccc}
        Experiment & Accelerator & Laboratory & $E_{beam 1,2}$ (GeV) & Boost ($\beta\gamma$)
          & $c\tau_B$ (mm) &  $L_{inst}^{max}($cm$^{-2}$s$^{-1})$ &  $L_{int}$(fb$^{-1}$)\\ 
        \hline
        BaBar   & PEP-II   & SLAC & 3.1, 9 & 0.558 & 0.26 & $1.2\times 10^{34}$ & 424 \\
        Belle   & KEKB     & KEK  & 3.5, 8 & 0.425 & 0.19 & $2.1\times 10^{34}$ &1041 \\
        Belle-II& SuperKEKB& KEK  & 4, 7   & 0.284 & 0.13 & $4.7\times 10^{34}$ & 424 \\        
    \end{tabular}
    \caption{Comparison of the beam energies at the B-facories, and the 
    resulting boost from the asymmetric beams, together with the instantaneous and integrated luminosities.
    The values for Belle-II are as of the end of 2024~\cite{Han:2024duu}.}
    \label{tab:Bfact}
\end{table}

\subsection{Comparison with hadron machines}
Flavour physics at hadron machines profits from the large $b$ production cross section.
The total $b\bar{b}$ production cross section at the LHC at 
13~TeV is estimated to be 500~$\mu$b~\cite{LHCB-PAPER-2015-037}.
The cross section scales approximately with the center-of-mass energy~\cite{Catani:2000xk}, and thus the production cross section of $b$ quark pairs at the  
Tevatron was about 70~$\mu$b.

Due to the large increase of the gluon density at low values of 
Bjorken-$x$, $b$-quarks are more likely to be asymmetrically produced by 
a medium-$x$ gluon from one proton, and a low-$x$ gluon from the other proton.
As a result, the $b$-quark pair is produced in a forward direction.
The design of general purpose detectors such as CDF, D0, ATLAS and CMS maximizes the 
sensitivity for high-$p_T$ processes such as top, electro-weak and Higgs physics.
The production of the lighter $b$-quarks follows a broad rapidity distribution and is predominantly in the forward region of LHCb: $2 < \eta < 5$.
Due to a broad physics programme of the general purpose experiments, limitations on the data bandwidth requires a selection on the minimal transverse momenta of the trigger particles. Taking it together, the resulting available data samples of $b$-hadrons follows the 
 $b\bar{b}$ production cross section as listed in Table~\ref{tab:Bhadron}. 

\begin{table}[!hb]
    \centering
    \begin{tabular}{llllc|cr}
        Experiment & $\sqrt{s}$       & Process & Phase space  & Meas. $\sigma$  & Scaled $\sigma(pp\to b)$ & Ref.        \\ 
                   &            (TeV) &         &               &  ($\mu$b)      &      ($\mu$b)            &         \\ 
        \hline
        CDF/D0     & 1.96 & $\sigma(pp\to bX\to J/\psi X$ & $|y|<0.6$               & 0.33 & 17.6 & \cite{CDF:2004jtw}  \\
        \hline
   %     CMS       &  7    & $\sigma(pp\to b\bar{b}X \to \mu\mu X)$  & $|\eta|<2.1$, $p_T>4$~GeV& 0.026& - & \cite{CMS:2012xsp} \\
        CMS       &  7    & $\sigma(pp\to bX \to \mu X)$  & $|\eta|<2.1$, $p_T>6$~GeV& 1.32& 12.1 & \cite{CMS:2011xhf}   \\
        CMS       &  7    & $\sigma(pp\to B^+X)$   & $|\eta|<2.5$, $p_T>5$~GeV      & 28.1 & 69.2 & \cite{CMS:2011oft}   \\
        ATLAS     &  7    & $\sigma(pp\to B^+X)$   & $|\eta|<2.25$, $9<p_T<120$~GeV & 10.6 & 26.1 & \cite{ATLAS:2013cia} \\
        ATLAS     &  7    & $\sigma(pp\to b)$   & $|\eta|<2.5$, $p_T>9$~GeV & 32.7 & 32.7 & \cite{ATLAS:2012sfc} \\
        LHCb      &  7    & $\sigma(pp\to b)$ & $2<\eta<5$                 & 72.0 & 72.0 & \cite{LHCb-PAPER-2016-031} \\
        LHCb      &  7    & $\sigma(pp\to B^+X)$    & $2.0<y<4.5$, $0<p_T<40$~GeV   & 43.0 & 105.9 & \cite{LHCb-PAPER-2017-037} \\
        \hline
        LHCb       & 13    & $\sigma(pp\to b)$       & $2.0<\eta<5$                  & 144.0 & 144.0 & \cite{LHCb-PAPER-2016-031}   \\
         LHCb      & 13    & $\sigma(pp\to B^+X)$           & $2.0<y<4.5$, $0<p_T<40$~GeV   &  86.6 & 213.3 & \cite{LHCb-PAPER-2017-037} \\
        LHCb       & 13   & $\sigma(pp\to bX\to J/\psi X)$ & $2.0<y<4.5$,  $0<p_T<14$~GeV& 2.25  & 194.0 & \cite{LHCB-PAPER-2015-037}\\
        \hline
        LHCb       & 13    & $\sigma(pp\to b)$ & (extrapolated w. Pythia)                         & 495  & 495 & \cite{LHCB-PAPER-2015-037}   \\
    \end{tabular}
    \caption{Comparison of the measured $b\bar{b}$ production cross section at hadron machines. 
    To compare the $b\bar{b}$ production cross section directly, the values are corrected for the 
    branching fractions $f(b\to B^+)=40.6\%$ , $BR(b\to \mu X)=10.95\%$ and $BR(b\to J/\psi X)=1.16\%$.
    The forward rapidity, and the low $p_T$ range at LHCb leads to a large $b\bar{b}$ production cross section.}
    \label{tab:Bhadron}
\end{table}

% sigma(b->B+) LHCb-PAPER-2017-037

\subsection{Comparison of sensitivities}

The experimental sensitivity to the flavour physics observables such as
the CKM-angles $\sin 2\beta$, $\gamma$ and $\phi_s$ or the 
lengths of the unitarity triangle via $\Delta m_s$ and $V_{ub}$,
or observables from rare decays such as $BR(B_s^0\to \mu^+\mu^-)$, $P_5^\prime$
and $R_K^{(*)}$, are all statistically limited and hence depend on the size of the signal sample.
This in turn depends on the $b\bar{b}$ production cross section as 
listed in Tab.~\ref{tab:Bhadron}, combined with the integrated luminosity, trigger efficiency and the offline event selection.
A comparison of the resulting event yields is listed in Tab.~\ref{tab:yields}.
Note, however, that the sensitivity for specific physics observable not only depends on the signal sample size, 
but also also on experimental aspects such as momentum and decay time resolution, flavour tagging and background levels, which are not included in this comparison.

\begin{table}[!hb]
    \centering
    \begin{tabular}{ll|rrrr}
    Decay mode & Observable & \multicolumn{4}{c}{Event yields} \\              
               &         & $B$-factories   & Tevatron      & ATLAS/CMS         & LHCb          \\
%               &         & per ab$^{-1}$ & per 10 fb$^{-1}$ & per ~100 fb$^{-1}$ & per ~10 fb$^{-1}$ \\
    \hline
    $B^0 \to J/\psi K_S^0$ & $\sin 2\beta$ & 12 649 (per 772M) & -    & -                     &  306 090 (/6 fb$^{-1}$)\\
    $B^+ \to D^0(KK)K^+$   & $\gamma$      &  1 131 (per 275M) & -    & -                     &  16 107 (/9 fb$^{-1}$)  \\
    $B^0_s\to J/\psi\phi$  & $\phi_s$      & - & 11 000 (/10 fb$^{-1}$)& 491 270 (/96 fb$^{-1}$)& 349 000 (/6 fb$^{-1}$)  \\
    $B^0_s\to D_s^-\pi^+$  & $\Delta m_s$  & - & 11 000 (/10 fb$^{-1}$)&   -                  & 378 700 (/6 fb$^{-1}$)  \\
    $B_s^0\to \mu^+\mu^-$  & BR            & - & 1.5   (/10 fb$^{-1}$)& 295 (/140 fb$^{-1}$)  & 104 (/9 fb$^{-1}$)     \\    
    $B^0\to K^{*0}\mu^+\mu^-$& $P_5^{'}$   & 140 (per 772M)  & - &  1430 (/20 fb$^{-1}$)      & 2741 (/9 fb$^{-1}$)\\    
    $B^0\to K^{*0}e^+e^-$    & LFNU        & 103 (per 772M)  & - &   -                       &  690 (/9 fb$^{-1}$)\\    
    $B^+\to K^{+}\mu^+\mu^-$& $R_K$        & 137 (per 657M)  & - & 1270 (/40 fb$^{-1}$)        & 5940 (/9 fb$^{-1}$)\\    
    $B^+\to K^{+}e^+e^-$    & LFNU         & 138 (per 657M)  & - &   21 (/40 fb$^{-1}$)        & 1495 (/9 fb$^{-1}$)\\    
    $B^0\to \pi^- \ell^+\nu$& $V_{ub}$     &  463  (per 722M)& - &  -                        & - \\
    \end{tabular}
    \caption{Comparison of published event yields for the different $B$ decay modes for the different experiments. The size of the data set used is given in brackets,
    listing the total number of produced $B$ pairs at the $B$-factories, and the 
    integrated luminosity for the LHC experiments. 
}
    \label{tab:yields}
\end{table}

\begin{figure}[h!]
    \centering
    \includegraphics[scale=0.25]{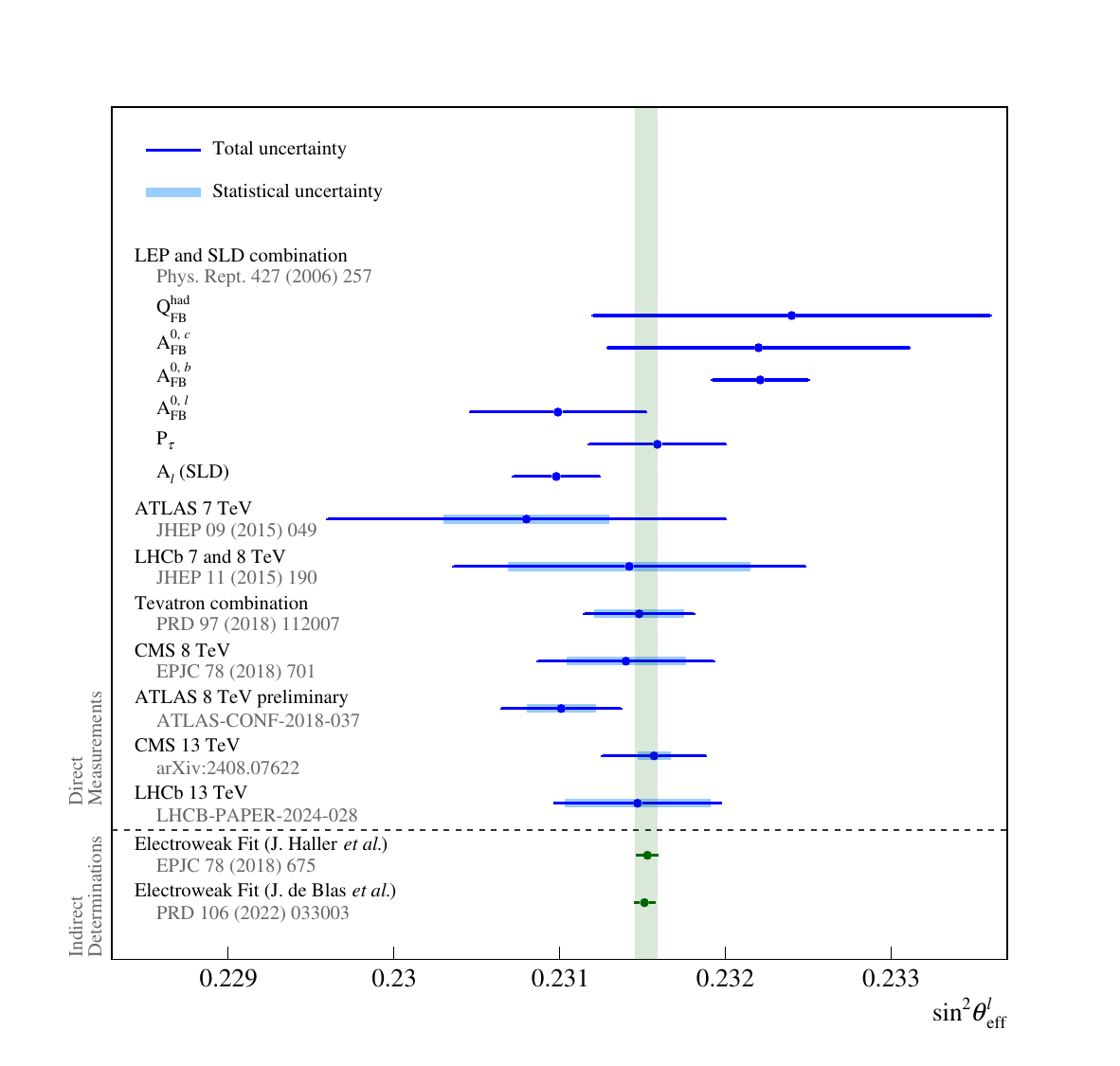} 
    \includegraphics[scale=0.27]{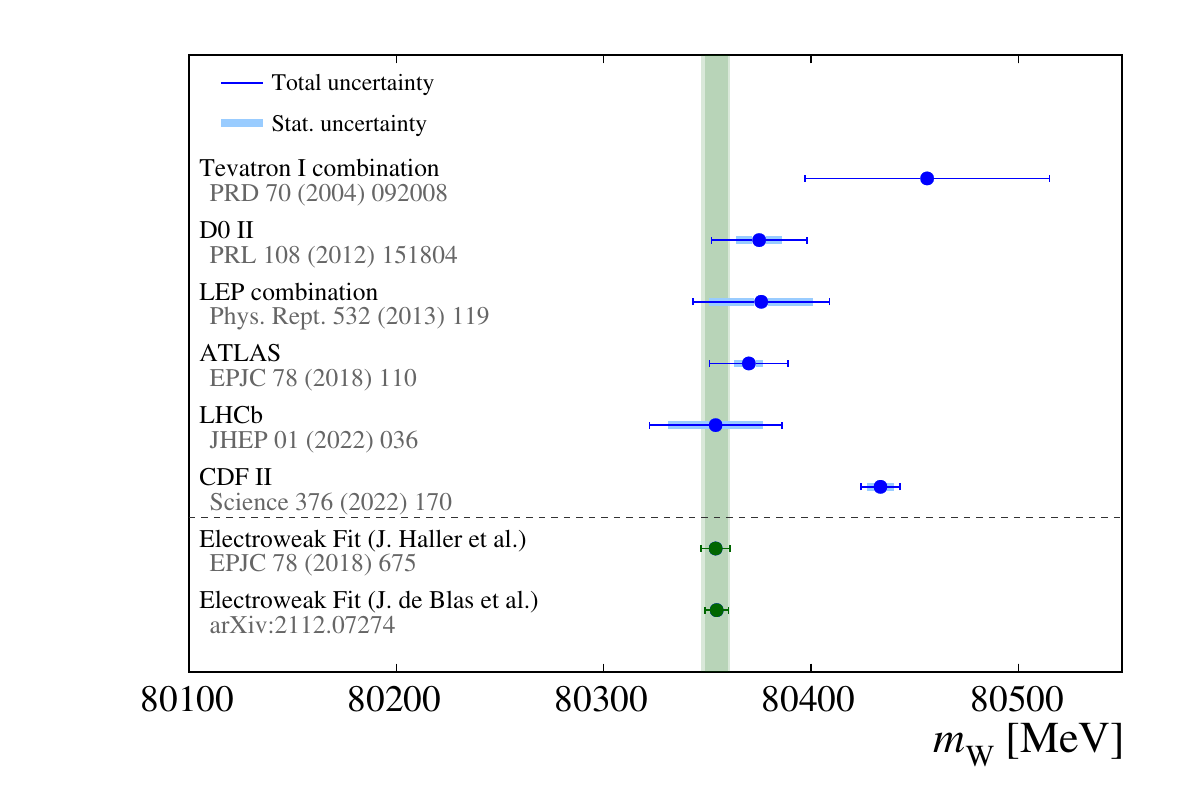} 
    \hspace{0.5cm}
    \includegraphics[scale=0.25]{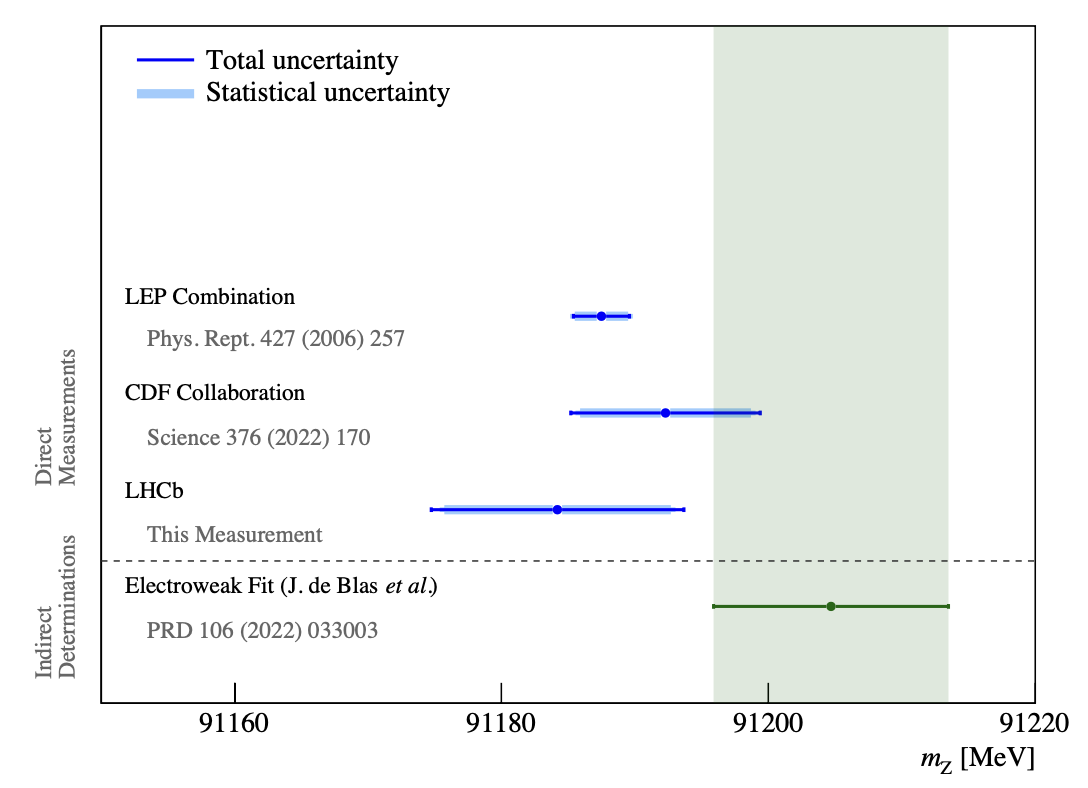} 
    \caption{{\em Left:} Comparison of the measurement of $\sin\theta_W$.
    {\em Middle:} Comparison of the electroweak measurements of the $W$-boson mass 
    ~\cite{LHCb-PAPER-2021-024} and ({\em right}) $Z$-boson mass~\cite{LHCb-PAPER-2025-008-sem}.
    %From LHC seminar 18 Mar 2025: https://indico.cern.ch/event/1525997/
    }
     \label{fig:EW}
\end{figure}

Although the LHCb experiment is specialized in beauty and charm physics,
recently precise electroweak measurements have been performed on 
the $W$ and $Z$ mass and the weak mixing angle $\theta_W$, traditionally the 
realm of the general purpose detectors ATLAS and CMS. A comparison of the 
achieved precision is shown in Fig.~\ref{fig:EW}.

\clearpage
\section{Future development}\label{secFuture}
%Please explain how the experiment will develop in the future

\subsection{LHCb Upgrade-2 at the HL-LHC}
The European Strategy for Particle Physics envisions a continued and expanded  Flavour Physics program at the High-Luminosity LHC (HL-LHC) to fully exploit its potential. This effort includes efforts from the general purpose experiments ATLAS and CMS, but also from a major, second, upgrade of the LHCb experiment, referred to here as "LHCb-U2". 

As the ultimate flavour physics precision experiment at the LHC, LHCb-U2 will enable searches for subtle virtual effects that new particles may have on Standard Model processes. Such indirect searches provide sensitivities to energy scales much higher than what is directly accessible at the energy frontier.  
The flavour program is complementary to the high-$p_T$ physics programmes of ATLAS and CMS, which add top decays and diagonal Higgs Yukawa couplings - areas not covered by the forward spectrometer of LHCb. 
A comprehensive discussion of the opportunities for flavour physics at the HL-LHC, including weak decays of beauty, charm, and strange hadrons, as well as studies of top quarks, the $\tau$ lepton and the Higgs boson, is presented in \cite{FlavourOpportunities-HL-LHC}.

ATLAS and CMS foresee to operate at an instantaneous luminosity about 5 to 10 times higher than LHCb, and will be competitive in flavour physics measurements involving muons and electrons in the final state. Meanwhile, LHCb-U2 will continue to benefit from superior decay-time resolution and dedicated particle identification capabilities, making it the optimal experiment for flavour physics at the LHC.

%\subsection{LHCb Upgrade 2}
The LHCb-U2 experiment plans to collect at least 300 fb$^{-1}$ of data, operating at an instantaneous luminosity of $1$ - $2$ $\times10^{34}$ cm$^{-2}$ s$^{-1}$. Under these conditions, about 50 pp collisions will occur simultaneously per bunch crossing, implying that even pile-up of heavy-flavour interactions will occur. This environment presents unique challenges for efficient data collection and event reconstruction.
A detailed overview of the planned physics potential can be found in the report "Physics case for an LHCb Upgrade II" \cite{LHCbU2-Physics-case}, while the Technical Design Report  \cite{LHCbU2-FrameworkTDR} provides an extensive description of the proposed detector upgrades.

\begin{figure}[b]
	\centering
    \begin{picture}(450,160)(0,0)
        \put(-30,15){\includegraphics[scale=0.43]{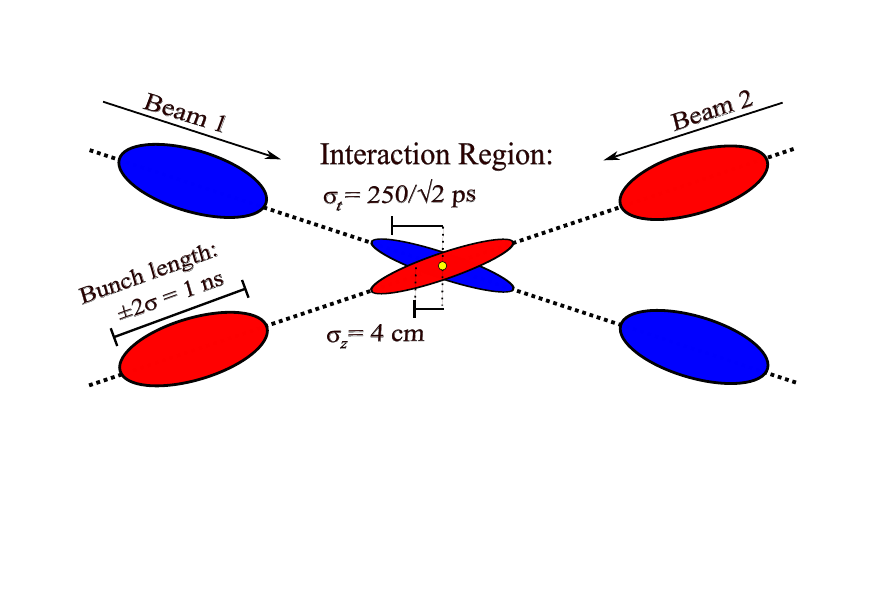}}
        \put(138,10){\includegraphics[scale=0.15, trim= 35cm 0cm 0cm 0cm, clip]{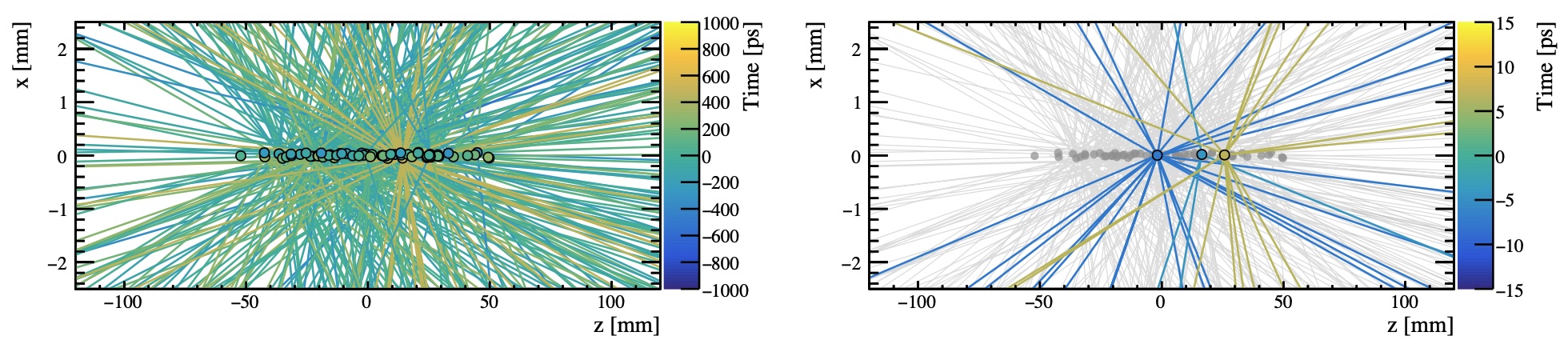}}
        \put(138,80){\includegraphics[scale=0.15, trim= 0cm 0cm 35cm 0cm, clip]{Figures/Detector/FastTimingVelo.jpg}}
         \put(282,0){\includegraphics[scale=0.18]{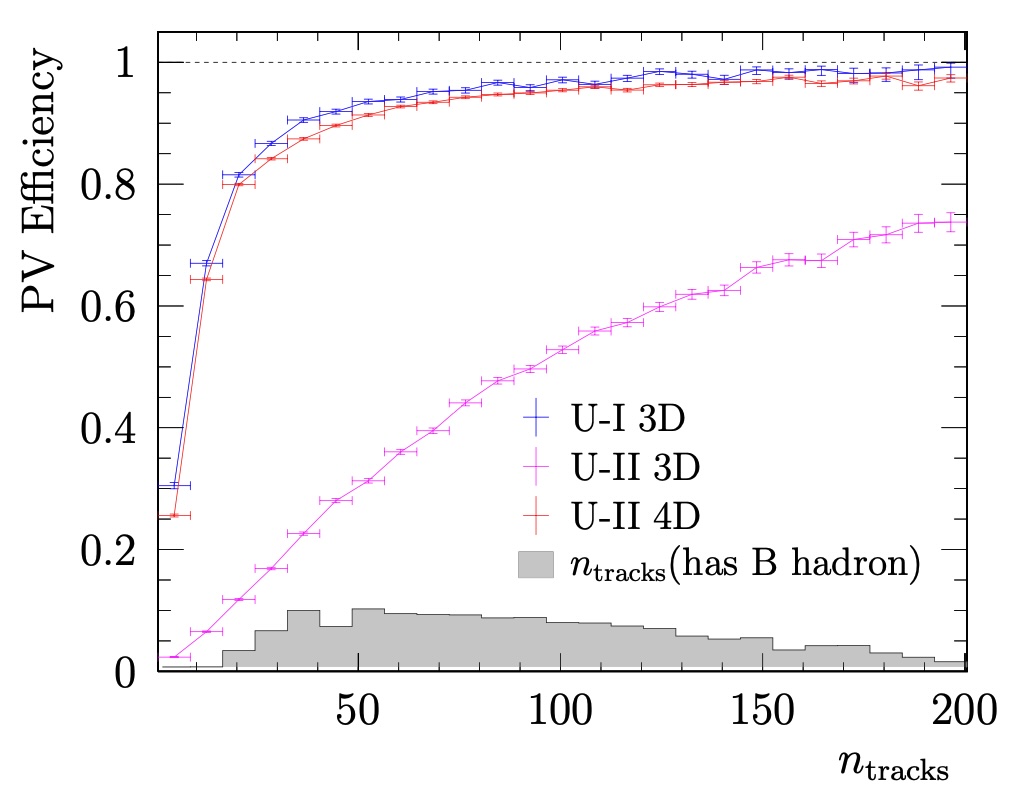}}
    \end{picture}
    \caption{The application of fast-timing measurements in the VELO. 
    {\em Left}: Sketch of the beam dynamics: the spread in the time of collisions, $\sigma_t$, depends on the bunch length (about 1 ns), whereas the spread in the PV position, $\sigma_t$, also depends on the crossing angle. 
    {\em Middle-top}: Illustration of VELO tracks in the $x$-$z$ plane over a full (2ns) time range, originating from $\sim$ 50 pile-up interactions within a bunch crossing. {\em Middle-bottom}: Tracks that fall within is 30 ps time-window, where then only few pile-up interactions contribute. {\em Right}: Reconstruction eﬃciency against the number of tracks per primary vertex, comparing the nominal 3D reconstruction in LHCb-U (U-I 3D) with the same reconstruction method for LHCb-U2 (U-II 3D) and with the 4D fast-timing reconstruction in LHCb-U2 (U-II 4D), showing that the use of fast-timing recovers the efficiency loss caused by pile-up.  The grey solid area shows the distribution of the number of tracks per primary vertex~\cite{LHCbU2-FrameworkTDR}.}
    \label{fig:fasttiming}
\end{figure}

\subsection{Detector Options and Innovations}
The overall layout of the LHCb-U2 spectrometer will remain similar to LHCb-U, but to cope with the challenging experimental conditions, novel detector technologies will be introduced. The main innovation to mitigate the effects of high collision pile-up is the implementation of {\em fast timing} technology. Adding a precise time measurement to the spatial hit information provides 4-dimensional (spatial and temporal) hit information, which allows the separation of tracks belonging to different primary vertices, both in time and in space. Fast timing capabilities will be integrated into the VELO vertex detector, the RICH-1 and RICH-2 detectors, the ECAL and in a novel subsystem called TORCH (Time Of Flight Resolve CHerenkov photons). The TORCH will be installed directly downstream of RICH-2 adding low PID for low momenta particles though a high-precision time-of-flight measurement. 
Fig.~\ref{fig:fasttiming} shows the idea behind the application of fast timing and the resulting improvement of vertex reconstruction.

The fast-timing information aims at a resolution of a few tens of picoseconds per particle, allowing individual collisions and their produced particles to be distinguished within the 250 ps bunch crossing time. This is expected to provide a factor $\sim$ 10 gain in separation power in comparison to spatial information alone, enabling a similar increase in luminosity while maintaining the same physics analysis power per $pp$ collision.

In addition to fast-timing sensors LHCb-U2 also introduces the following additional improvements in the spectrometer:
\begin{itemize}
    \item {\em VELO}: The material of the RF-foil, currently present between the collision point and the sensors, will be reduced or even removed, leading to less scattering and better vertexing performance. 
    \item {\em Upstream Tracker and Inner Part Downstream Tracker}: New detectors will implement Monolythic Active Pixel Sensors (MAPS) enhancing further the capabilities of downstream tracking.
    \item {\em Magnet Stations}: Inside the magnet, a scintillator based tracking stations will be installed at the side walls of the magnet, improving the momentum measurement of upstream tracks from a current resolution $\delta(p)/p\approx12\%$ to about 1\%.
    \item {\em ECAL}: To maintain the good energy resolution while operating under increased occupancy requires a finer cell granularity and higher radiation tolerance for the inner region. The baseline solution is to install a Spaghetti Calorimeter (SPACAL) in the inner region combined with the current Shashlik technology in the outer region.
    \item {\em HCAL}: As the HCAL is no longer used in the trigger, it is being considered to be removed for LHCb-U2, in order to provide better shielding for the muon detectors in the high radiation environment of the HL-LHC.  
    \item {\em Muon Stations}: In the inner region the Multi Wire Proportional Chambers must be replaced, for which a new type of Micro-Pattern Gaseous Detectors is being considered \cite{muRwell}. The outer part of the muon  stations sees less radiation and the current Multi-Wire Proportional Chambers (MWPC) can be re-used, but with new electronics to reduce readout dead-time.
\end{itemize}

\subsubsection{Physics Reach and Expected Sensitivities}

\begin{figure}[hb]
	\centering
    \raisebox{0.05cm}{\includegraphics[width=7.5cm, trim=0cm 23cm 0cm 0cm, clip=true ]{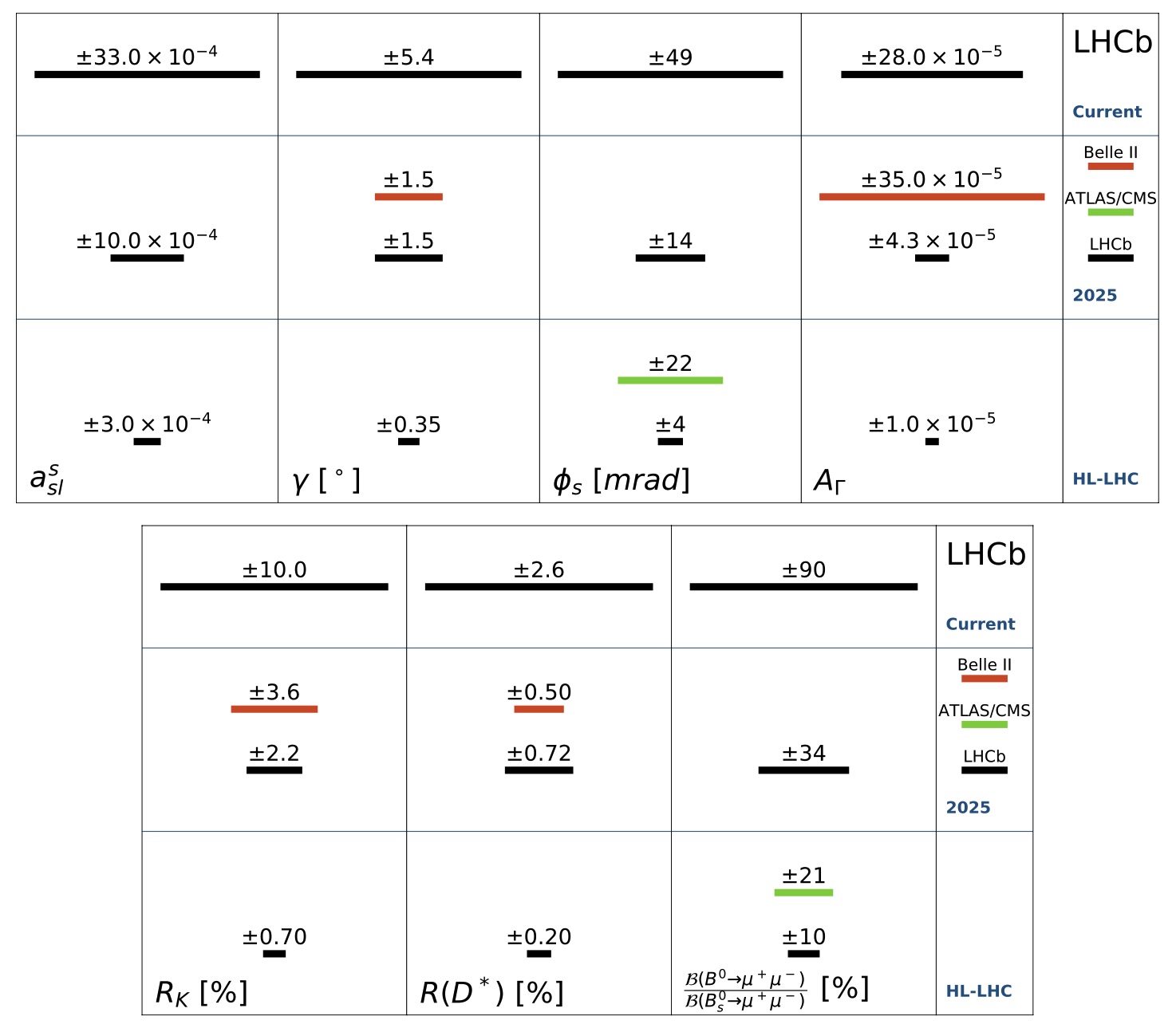}} 
    %\hspace*{0.5cm}
    \includegraphics[width=7.5cm, trim=2cm 0cm 2cm 23cm, clip=true ]{Figures/Detector/LHCbU2-Sensitivities.jpg} 
	\caption{{\em Left}: Projected sensitivities for LHCb-U2 for CP violating observables. {\em Right}: Projected sensitivities LHCb-U2 sensitivities for rare decays and lepton universality tests. Taken from \cite{LHCbU2-Physics-case}.}
	\label{fig:LHCbU2-Sensitivities}
\end{figure}

\begin{figure}[h!]
	\centering
    \includegraphics[width=8.0cm]{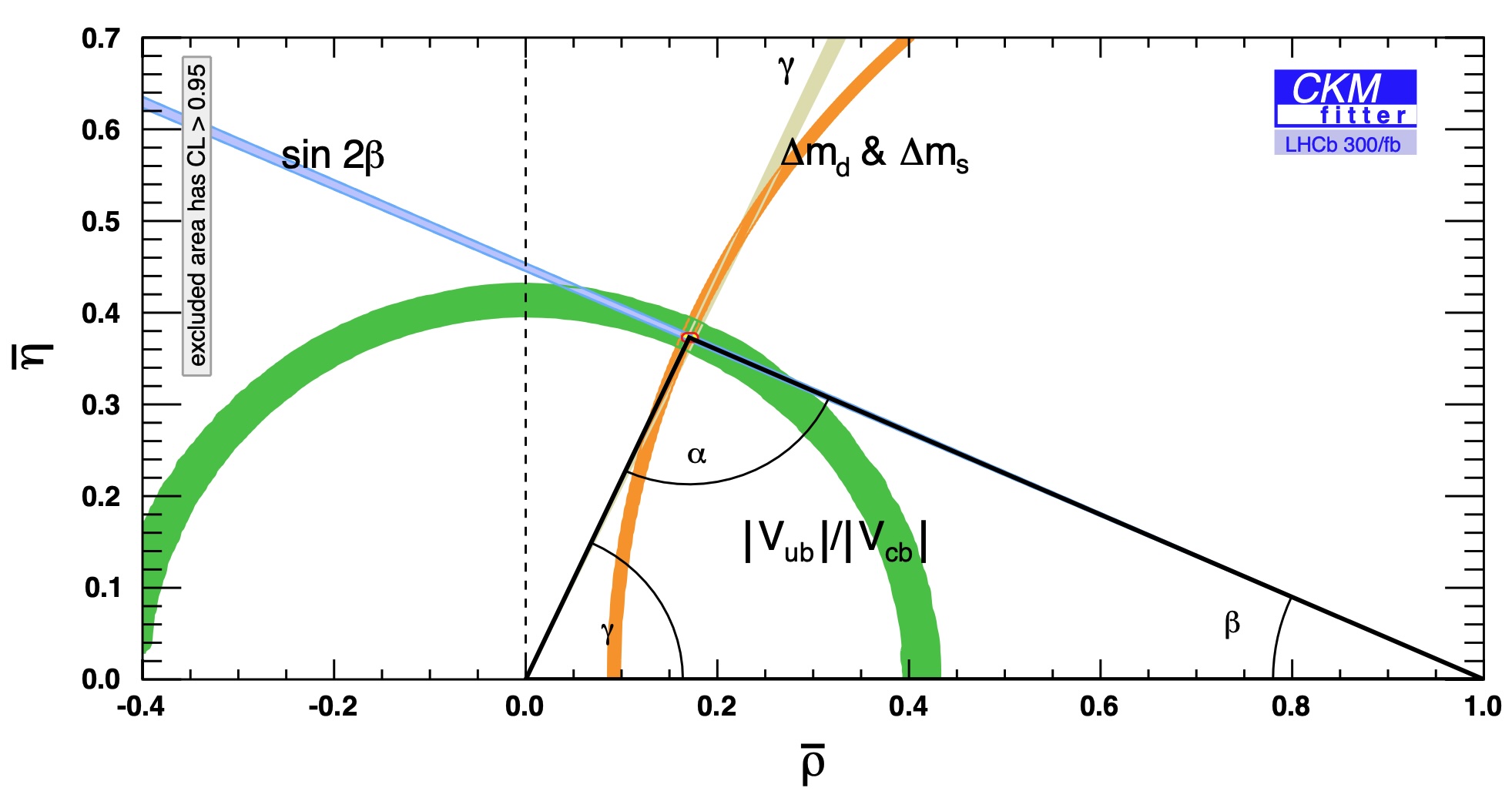}
	\caption{LHCb constraints from the dominant CKM observables to the apex of the unitarity triangle with anticipated improvements from 300 fb$^{-1}$, assuming consistency with the SM. Taken from \cite{LHCbU2-Physics-case}.}
	\label{fig:LHCbU2-UnitarityTriangle}
\end{figure}

The new detector attributes will further enhance LHCb's capabilities to a wide range of physics topics, while maintaining primary focus on heavy-flavour physics.  The expected performance of LHCb-U2 is described in \cite{LHCbU2-Physics-case} and compared to ATLAS, Belle-II and CMS in \cite{ATLAS:2025lrr}. 
Here, the projected sensitivities of key measurements are summarized in Fig.~\ref{fig:LHCbU2-Sensitivities} and Fig.~\ref{fig:LHCbU2-UnitarityTriangle}, which shows the anticipated precision of CKM triangle measurements. 
%and Fig.~\ref{fig:LHCbU2-Sensitivities}, which compares the expected LHCb-U2 sensitivities for key CP-violating observables and rare-decays with those of LHCb-I and other experiments. 
These projections are based on extrapolations from current measurements, assuming that the upgraded detector maintains similar efficiencies per unit of luminosity as LHCb-I. However, an important improvement comes from the real-time reconstruction and selection, which is expected to double the efficiency for hadronic modes, as already observed at the beginning of Run-3.

The key sensitivities include \cite{LHCbU2-Physics-case}:
\begin{itemize}
    \item CP Violation in beauty decays: determinations of the phase $\gamma$ with a precisions of $0.35^\circ$, the $B^0_s$ mixing phase $\phi_s$ with 4 mrad, and the semileptonic mixing parameter $a_{sl}^s$ with an uncertainty of $3 \times 10^{-4}$.
    \item CP violation in charm decays: measurement of both parameters $\Delta A_{CP}(KK-\pi\pi)$ and $A_\Gamma$ with a precision of $10^{-5}$. 
    \item The CKM triangle: measurement of the ratio $\left|V_{ub}\right|/\left|V_{cb}\right|$ with a precision of 1\% to precisely determine the apex of the unitarity triangle.
    \item Very rare decays: determining the relative branching ratios of  $B^0\rightarrow \mu^+\mu^-$ and $B_s^0\rightarrow \mu^+\mu^-$ with a precision of 10\%.
    \item Rare decays: Measurements of the rare decay transitions $b\rightarrow s l^+ l^-$ and $b\rightarrow d l^+l^-$ for electrons and muons leading to a precision of $R_K$ to 0.7\%. 
    \item Universality tests with $b\rightarrow c l^- \overline{\nu}_l$: determining the ration $R(D^*)$ to a precision of 0.2\%. 
\end{itemize}

\vspace*{-0.5cm}

\vspace*{-0.5cm}

\clearpage
%\input{Template}

%%%%%%%%%%%%%%%%%%%%%%%%%%%%%%%%%%%%%%%%%
%% Mandatory: A concluding paragraph summing up your main points in the chapter
%% Optional: Also include big questions in the field that are still to be answered. What topics/methods/questions are researchers like to focus on next?
\section{Conclusions}
\label{sec:conclusions}

The LHCb experiment was conceived as an experiment to study CP violation in $B$-decays and successfully developed a comprehensive program to do so in both direct CP violation as well as decay-time dependent CP violation measurements. In addition to charged and neutral $B$-mesons, CP violation was observed in charm particles and in baryons.
An initially limited palette of rare decays developed into a very rich research area searching for indirect signals of physics from beyond the SM, where apart from tantalizing hints in the $b\rightarrow s$ transition, no single evidence has been observed to date. As the initial LHCb-I detector developed into the LHCb-U upgrade, its physics programme grew into a general purpose detector by adding more functionality due to a full software trigger scheme. The programme includes QCD, spectroscopy, electroweak physics, and searches for exotic particles, as well as a growing heavy ion programme both via collider mode and fixed target physics. At the time of writing a second upgrade of LHCb is being proposed, intended to be the ultimate flavour physics experiment at the HL-LHC collider. The aim of the flavour physics experiment is to probe the energy frontier of virtual particles by means of the precision measurements, exploring the quantum aspect of the forces of nature.

\begin{ack}[Acknowledgements]%

We consider ourselves privileged to have been part of the exciting scientific journey of the LHCb experiment over many years. We extend our heartfelt thanks to all our colleagues in the collaboration with whom we shared this remarkable experience. Contributing to the detector’s design, construction, commissioning, and physics exploitation has been both an honour and a pleasure.

In particular, we warmly thank our colleagues at the Nikhef "bfys" and the Maastricht "GWFP" teams for providing a pleasant, supportive, and intellectually stimulating research environment.

%We thank our employers, the NWO-i research institute Nikhef and the University of Maastricht in the Netherlands. In addition we express gratitude towards CERN for hosting one of us, to the LHC team for providing a wealth of collisions, an to the LHCb online and data-taking teams for collecting the detector data. Finally, also a warm thanks to our colleagues in LHCb collaboration, at the Nikhef bfys and the Maastricht GWFP teams for providing a pleasant and stimulating research environment.
\end{ack}

%%%%%%%%%%%%%%%%%%%%%%%%%%%%%%%%%%%%%%%%%%%%
%% Optional: A list of references to other relevant works/articles/websites which are not cited in the text but that would further enhance a readers understanding of this topic
%\seealso{article title article title}

%%%%%%%%%%%%%%%%%%%%%%%%%%%%%%%%%%%%%%%%%
%% Mandatory: Bibliography using bibtex 
%\bibliographystyle{Numbered-Style} %% for Numbered Reference Style
%\bibliographystyle{unsrt} %NT: sort in order of appearance!
\bibliographystyle{unsrtnat} %NT: 04/05/2026 needed to resolve error for arxiv submission
\bibliography{reference,LHCb-PAPER,LHCb-DP,LHCb-CONF}

\end{document}